\documentclass[11pt]{article}
\pdfoutput=1
\usepackage{graphicx}
\usepackage{subfigure,amssymb,amsmath,slashed}
\usepackage{cite,color,xcolor,url,cancel}
\usepackage{multirow}
\usepackage{bm}
\usepackage{float}
\usepackage{jheppub}
\usepackage[normalem]{ulem}
\newcommand{\be}{\begin{equation}}
\newcommand{\ee}{\end{equation}}
\newcommand{\bea}{\begin{eqnarray}}
\newcommand{\eea}{\end{eqnarray}}
\newcommand{\noi}{\noindent}
\newcommand{\bse}{\begin{subequations}}
\newcommand{\ese}{\end{subequations}}
\newcommand{\tp}{T^\prime}
\newcommand{\xp}{x^\prime}
\newcommand{\nn}{\nonumber}
\newcommand{\ra}{\rightarrow}
\newcommand{\zp}{Z^\prime}
\newcommand{\hd}{h_d}
\newcommand{\x}{\chi}
\newcommand{\xlo}{{\xi_{{}_1}}_L}
\newcommand{\xlt}{{\xi_{{}_2}}_L}
\newcommand{\chio}{{\chi_{{}_1}}}
\newcommand{\chit}{{\chi_{{}_2}}}
\newcommand{\chii}{{\chi_{{}_i}}}
\newcommand{\nzp}{n_{{{}_{\zp}}}}
\newcommand{\nhd}{n_{\hd}}
\newcommand{\hub}{\mathcal{H}}
\newcommand{\ecm}{\mathfrak{s}}
\newcommand{\smgauge}{${\rm SU(3)}_c \otimes {\rm SU(2)}_L \otimes {\rm U(1)}_Y$}

\title{When Freeze-out occurs due to a non-Boltzmann
suppression: A study of degenerate dark sector}
\author[a,b]{Anirban Biswas,}
\affiliation[a]{School of Physical Sciences,
Indian Association for the Cultivation of Science,\\  
2A \& 2B Raja S.C. Mullick Road, Jadavpur, 
Kolkata 700 032, India}
\affiliation[b]{Centre of Excellence in Theoretical
and Mathematical Sciences,
Siksha `O'Anusandhan (Deemed to be University),
Khandagiri Square, Bhubaneswar 751030,
Odisha, India}
\author[a]{Sougata Ganguly,}
\author[a]{Sourov Roy}
\emailAdd{anirban.biswas.sinp@gamil.com}
\emailAdd{tpsg4@iacs.res.in}
\emailAdd{tpsr@iacs.res.in}
\abstract{Exponential suppression or commonly
known as the Boltzmann suppression in the number density
of dark matter is the key ingredient for creating
chemical imbalance prior to the usual thermal freeze-out.
A degenerate/quasi-degenerate dark sector can experience
a different exponential suppression in the number density
analogous to the radioactive decay law leading to a delayed
freeze-out mechanism of dark matter known as the co-decaying
dark matter. In this work, we study the dynamics of a multicomponent
dark matter from thermally decoupled degenerate dark sector in a 
{\it hidden} U$(1)_{X}$ extension of the Standard Model.
We compute the relic density of dark matter frozen-out through
the co-decaying mechanism by solving four coupled
Boltzmann equations. We demonstrate how temperature $T^\prime $ of
the dark sector changes due to all types of $3\ra2$
and $2\ra2$ interactions along with the eternal expansion
of the Universe. We find that $3\ra 2$ interactions enhance
$T^\prime$ by producing energetic particles in the dark sector
while the excess heat is transferred by $2\ra 2$ interactions
to the entire dark sector. As the direct detection is
possible only through the feeble portal couplings, 
we investigate the neutrino and $\gamma$-ray
signals from dark matter annihilation
via one step cascade processes and compare
our results with the measured fluxes of atmospheric
neutrinos by Super-Kamiokande and diffuse $\gamma$-rays
by Fermi-LAT, EGRET, INTEGRAL collaborations. 
We find that the present scenario easily evades all the existing
bounds from atmospheric neutrino and diffuse $\gamma$-ray
observations for degenerate dark sector. However, the constraints
are significant for quasi degenerate scenario.
}
     
\vspace{1cm}
\begin{document}
\maketitle
\section{Introduction}
\label{Intro}
The gravitational effects established the fact that
almost $25\%$ energy budget of our Universe is made
of an unseen matter commonly known as the dark matter (DM).
In particular, the satellite borne experiments
like WMAP \cite{Hinshaw:2012aka} and Planck \cite{Aghanim:2018eyx}
while measuring the temperature anisotropy in the Cosmic Microwave
Background radiation (CMB) have established the present
value of dark matter relic density $\Omega_{\rm DM} h^2
= 0.120 \pm 0.001$ \cite{Aghanim:2018eyx}. Other indirect evidences like 
the rotation curves of spiral galaxies \cite{Sofue:2000jx}, the gravitational 
lensing of distant objects \cite{Bartelmann:1999yn}, the bullet cluster
observation \cite{Clowe:2006eq} etc. also strongly indicate
the presence of more matter than only the visible matter.
In spite of having clear evidences, the nature and properties
of dark matter is an open problem to date. There exist many
particle physics models in the literature, which have one or more candidates
of dark matter in their particle spectrum. Most of these
beyond Standard Model (BSM) theories have focused on
weakly interacting massive particle (WIMP) dark matter
\cite{Srednicki:1988ce,Gondolo:1990dk,Jungman:1995df,Bertone:2004pz},
a popular and well studied class of dark matter candidates. In the WIMP paradigm, 
it is assumed that dark matter was in thermal equilibrium (both chemical and
kinetic equilibrium) with the Standard Model (SM) particles
and the chemical equilibrium was lost as dark matter became
non-relativistic. When expansion rate of the Universe becomes
dominant over the $DM ~DM \rightarrow SM ~SM$ interaction rate, the dark matter number density
gets frozen-out and changes only by expansion
of the Universe afterwards. The most important
property of a WIMP dark matter is that it has
sizeable interaction strength (in the weak scale) 
with the SM particles that predicts observable number of 
events in the experiments over the possible backgrounds.
However, the parameter space of this theoretically well
motivated scenario are now acutely constrained due to
non-observations of any signature at various direct detection
experiments over the last two decades
\cite{Sanglard:2005we, Lebedenko:2009xe, Ahmed:2009zw, Angle:2011th,
Aprile:2012nq, Amole:2015lsj, Akerib:2016vxi, Cui:2017nnn, Aprile:2018dbl}.
Several near future experiments like DARWIN \cite{Aalbers:2016jon},
XENONnT \cite{Aprile:2015uzo}, LUX-ZEPLIN \cite{Akerib:2018lyp}
will have the required sensitivities to probe the remaining WIMP
parameter space above the \textit{neutrino floor} \cite{Boehm:2018sux},
a region dominated by coherent neutrino-nucleon scatterings.  

In the light of these observations, many well motivated proposals
have been suggested which can explain the null results of
direct detection experiments and also remain compatible with the
current Planck result.  The ``secluded sector dark matter'' 
\cite{Pospelov:2007mp, Ko:2014gha, Berlin:2014pya, Escudero:2017yia, Banik:2015aya, 
Feng:2008mu, Chu:2011be, Hambye:2019dwd, Evans:2019vxr,Fitzpatrick:2020vba} 
is one of such alternatives. In the secluded sector scenario, dark matter
is connected with SM through a metastable mediator and the coupling between the
mediator and SM is small to evade the constraints from the direct searches
and also from collider experiments. However, the coupling is sufficient enough to establish
the kinetic equilibrium with the SM sector. Since the metastable mediator
can decay into the SM species, it should decay well before the onset of
big bang nucleosynthesis (BBN) to ensure that the observables
(e.g. effective number of neutrino species, abundance of deuterium and helium)
during the BBN era remains unaltered. In this scenario, if the dark
matter mass is greater than the mediator mass, annihilation of dark
matter into the metastable mediators is possible and that determines the
dark matter relic abundance at the present era while the dark matter is
still a thermal relic. This type of dark matter though \textit{secluded}
from the SM sector due to the tiny portal coupling with the SM, 
can be detected via indirect searches.  Most of the work on
secluded sector dark matter have been done on the basis of
the assumption that the dark sector is in kinetic equilibrium
with the SM bath. In general, this condition can be relaxed
by assuming that the dark sector is decoupled from the SM
sector in the early Universe while it was relativistic.
Since, the dark sector is not in kinetic equilibrium with the SM, 
the entropy is conserved separately in each sector and
the temperature evolution in the dark sector is in general
different from the SM bath. Now, if dark sector has 
sufficiently strong interactions among its species
then beyond $2\ra 2$ scatterings e.g. inelastic scatterings
like $3\ra2$, $4\ra 2$ processes etc. can be active. 
As a result, the dark sector enters into the \textit{cannibal} phase 
\cite{Carlson:1992fn, Pappadopulo:2016pkp, Berlin:2016gtr,
Yang:2019bvg} and during this phase, the dynamics of temperature
evolution of the dark sector is drastically different from
the evolution of the SM temperature.
  
In this work, our principal objective is to carry out a detailed phenomenological
analysis of the secluded sector dark matter involving both
$2\ra 2$ and $3\ra 2$ scatterings. In order to implement this, we have considered
a dark sector which has a U(1)$_{X}$ gauge invariance.    
In this dark sector, two left chiral Weyl fermions
$\xlo$ and $\xlt$ singlet under \smgauge ~gauge group
and having U(1)$_{X}$ charges $+1$ and $-1$
respectively are added. Moreover, we have included a
complex scalar $\eta$ which is also singlet under the
SM gauge group but has a nonzero U(1)$_{X}$ charge $+2$
to break this additional U(1) symmetry spontaneously. After
symmetry breaking in the dark sector, we have two Majorana
fermions $\chio$ and $\chit$, a dark scalar $\hd$ and a
neutral gauge boson $\zp$. Among them, the Majorana fermions
are absolutely stable and will play the role of dark matter.
However, other two species $\hd$ and $\zp$ are feebly
coupled to the SM sector through the kinetic mixing $\epsilon$
between two U(1) groups (one of them is U(1)$_{\rm Y}$ of the SM)
and the scalar mixing $\alpha$ between the SM Higgs boson and $\eta$.  
However, these portal couplings are such that two sectors
are not in kinetic equilibrium and $\zp$, $\hd$ can decay
out of equilibrium to the SM particles. Since we do not
have any prior knowledge about the species
of dark sector (except dark matter) and the
scale of interactions among the members 
of dark sector, it is natural to assume
all the dark sector couplings are of same order. This
simplifying assumption results in a degenerate/quasi-degenerate
dark sector which can drastically modify the entire
freeze-out mechanism of the dark matter. It has been observed
that due to degeneracy among the species there is
no Boltzmann suppression in number densities of
non-relativistic dark species which is a key
ingredient for a dark matter proceeding towards
freeze-out. Instead, here each species develops a
nonzero chemical potential which discards the Boltzmann
suppression. However, the necessary suppression in
number density arises when either $\zp$ or $\hd$
(or both) starts to decay into the SM particles and
thereby introduces a new kind of exponential suppression
$e^{-\frac{\Gamma}{4\hub}}$ to the number
densities of dark species, where $\Gamma$ is the
decay width and $\hub$ is the Hubble parameter. This
results in the violation of chemical equilibrium in
the dark sector and consequently triggers the freeze-out
process of dark matter. As the freeze-out is initiated
only after a sufficient time interval from the beginning
of decay of species other than dark matter, which are
feebly coupled to the SM fields, the process naturally
leads to a delayed freeze-out of dark matter. 

This novel mechanism was proposed
in \cite{Dror:2016rxc} under the name of ``co-decaying dark matter''. Thereafter
a few studies have been performed on this topic concentrating
on astrophysical implications like small scale
structure formation \cite{Dror:2017gjq} and formation of
primordial black hole \cite{Georg:2019jld}. In this co-decaying
framework, we have considered a temperature $\tp$, different from the
SM temperature $T$, in the dark sector, which is a natural
choice as the two sectors are not in thermal contact. We have
done a detailed analysis of temperature evolution of the dark
sector taking into account all types of $2\ra 2$ and $3\ra 2$
scattering processes among the dark species. In this process,
we have numerically solved four coupled Boltzmann equations,
among which three equations are for evolution of number densities of
$\chio+\chit$, $\zp$ and $\hd$ while the last one
is for $\tp$ and have computed the relic density of
dark matter frozen-out via the co-decaying mechanism.
We have found that $3\ra 2$ processes have important
implications in the evolution of $\tp$ as they increase
the temperature by introducing highly energetic
particles in the dark sector. The excess heat thus
generated via $3\ra 2$ scatterings spreads across all
the species through the $2\ra 2$ scatterings so that
the entire dark sector has a common temperature.
Here, we have chosen the portal couplings $\epsilon$ 
and $\alpha$ which connect the dark sector with
the visible sector in such a way that the observations
during the BBN era remain unaltered. Moreover, in this
framework, we have also studied extensively the prospects
of indirect signatures of our dark matter candidates ($\chio$
and $\chit$) through 
neutrinos and $\gamma$-rays. 
For that we have considered the most dominant processes which
are one step cascade processes and have the following generic
structure like $\chii \chii \ra A (A\ra X\, Y$). The intermediate
states are $\zp$ for neutrinos and $\hd$ for $\gamma$-rays. This
type of cascade process generally produces box-shaped spectrum.
However, as we are considering degenerate/quasi degenerate dark
species, the resulting spectrum takes a line like shape with
energy $m_{\x}/2$ of the outgoings particles. 
Moreover, $\gamma$ ray flux from
the Final State Radiation (FSR) and the Inverse Compton
Scattering (ICS) has also been discussed.
We have compared our results with the observed neutrino
flux by the Super-Kamiokande \cite{Richard:2015aua}
detector while the $gamma$-ray flux has been compared with
diffuse $\gamma$-ray background measured by the Fermi-LAT \cite{Abdo:2010nz}, 
the EGRET \cite{Strong:2004de}, and INTEGRAL \cite{Bouchet:2011fn} 
collaborations. We have found that in both cases the present scenario easily
satisfies all the existing bounds arising from diffuse background
$\gamma$-rays and atmospheric neutrinos for degenerate dark sector.
However for quasi degenerate dark sector, the parameter space is constrained
from CMB observations and measurement of positron flux at AMS-02 experiment.

The rest of the paper has been organised in the following manner.
In Section\,\,\ref{sec:model} we have discussed the present
model describing our dark sector. A detailed discussion on
the dynamics of the dark sector following co-decaying mechanism
is given in the first part of Section\,\,\ref{sec:dynamics}.
The necessary Boltzmann equations and related discussions
are given in Subsection\,\,\ref{sec:Boltz}.
Various constraints on the portal couplings have been
shown in Subsection\,\,\ref{sec:constraints}. Numerical results
that we have obtained by solving four coupled Boltzmann equations
are presented in great detail in Subsection\,\,\ref{sec:Nres}.
The indirect detection constraints on this framework from astrophysical
neutrinos and diffuse background $\gamma$-rays are discussed
in Section\,\,\ref{sec:indirect}. Finally, we summarise
in Section\,\,\ref{sec:conclu}. A comprehensive discussions
on the Boltzmann equations, all {2$\ra$2} 
and $3\ra 2$ inelastic scatterings, necessary vertex
factors, an approximate analytical form of the dark
matter relic density and a short note on the thermal averaged
cross section in the degenerate limit  
have been given in Appendices\,\,\ref{App:A}-\ref{App:F}. 
\section{Model}
\label{sec:model}
In this section, we discuss an anomaly free U(1) extension of the
SM which has two Majorana dark matter candidates. In order
to proceed further, let us discuss the model in detail. We extend
the fermionic sector of the SM by two left-chiral Weyl fermions
($\xlo,\,\xlt$) which are singlet under the SM gauge group but
are charged under the new U(1)$_X$ gauge group. Besides the
two fermions, we need at least one complex scalar ($\eta$) with
nonzero U(1)$_X$ charge for breaking of new gauge symmetry spontaneously
and this results in a massive neutral gauge boson in the particle
spectrum. The complete field contents and their individual charges\footnote{Here
we have used the definition of electric charge as $Q_{EM} = T_3 + \frac{Y}{2}$.}
under  \smgauge$\otimes {\rm U(1)}_X$ symmetry are listed
in Table \ref{tab1}. The U(1)$_X$ charges of $\xlo$ and $\xlt$ are equal to
$+1$ nd $-1$ respectively and this is an economic choice leading to the 
cancellations of both [U(1)$_X]^3$ and
[Gravity$]^2$U(1)$_X$ anomalies. This can easily be understood as follows. 
Suppose, $q_1$ and $q_2$ are the charges of $\xlo$ and $\xlt$ respectively
under U(1)$_X$. Then the anomaly cancellation conditions for axial vector
anomaly and mixed gauge-gravitational anomaly are given by
\bea
\left[{\rm U(1)}_X\right]^3
&:& 
q_1^3 + q_2^3 = 0\,\,,\nn \\
\left[\rm Gravity\right]^2 {\rm U(1)}_X
&:&
q_1 + q_2 = 0\,\,\nn.
\label{ano}
\eea
The general solution (real) for the above set of equations is $q_1 = -q_2$ and we choose
$q_2 = -1$  and $q_1 = -q_2 = 1$. On the other hand, the gauge invariance of the
Majorana mass terms for both $\xlo$ and $\xlt$ demands the charge of $\eta$ to be -2.
Now, we shall write the Lagrangian of our model which is invariant under the
entire symmetry groups i.e. \smgauge$\otimes {\rm U(1)}_X$. The total
Lagrangian $\mathcal{L}$, composed of the SM Lagrangian ($\mathcal{L}_{\rm SM}$)
and the additional part involving newly added gauge, Yukawa and scalar
sectors respectively, is given by  
\begin{center}
\begin{table}[hbt!]
\begin{tabular}{||c|c||}
\hline
Field content & Charge under \smgauge$\otimes {\rm U(1)}_X$ symmetry \\
\hline
\hline
$\ell_{L} = \begin{pmatrix}
\nu_e \\e
\end{pmatrix}_{L},\,
\begin{pmatrix}
\nu_\mu \\ \mu
\end{pmatrix}_{L},\,
\begin{pmatrix}
\nu_\tau \\ \tau
\end{pmatrix}_{L}$ & $(1,\,2,\,-1,\,0)$ \\
\hline
$\ell_{R}=e_R,\,\mu_R,\,\tau_R$ & $(1,\,1,\,-2,\,0)$\\
\hline
$Q_L$=$\begin{pmatrix}
u \\d
\end{pmatrix}_L,\,
\begin{pmatrix}
c \\ s
\end{pmatrix}_L,\,
\begin{pmatrix}
t \\ b
\end{pmatrix}_L$ & $(3,\,2,\,\frac{1}{3},\,0)$\\
\hline
$U_R=u_R,\,c_R,\,t_R$ & $(1,\,1,\,\frac{4}{3},\,0)$\\
\hline
$D_R=d_R,\,s_R,\,b_R$ & $(1,\,1,\,-\frac{2}{3},\,0)$\\
\hline
$\Phi = \begin{pmatrix}
\phi^+ \\ \phi^0
\end{pmatrix}$ & $(1,\,2,\,1,\,0)$\\
\hline
$\xlo$ & $(1,\,1,\,0,\,1)$\\
\hline
\vspace{0.2 pt}
$\xlt$ & $(1,\,1,\,0,\,-1)$\\
\hline
$\eta$ & $(1,\,1,\,0,\,-2)$\\
\hline
\end{tabular}\\
\textit{}
\caption{Field contents of our model and their charges
under \smgauge$\otimes {\rm U(1)}_X$.}
\label{tab1}
\end{table}
\end{center}

\bea
\mathcal{L}
&=&
\mathcal{L}_{\rm SM} + \mathcal{L}_{\rm gauge} + \mathcal{L}_{\rm DM-gauge} + 
\mathcal{L}_{\rm DM-Yukawa} + \mathcal{L}_{\rm scalar}\,\,,\nn
\label{lag}
\eea
where,
\bea
\mathcal{L}_{\rm gauge}
&=&
-\dfrac{1}{4} X_{\mu \nu} X^{\mu \nu} - 
\dfrac{\epsilon}{2} B_{\mu \nu} X^{\mu \nu}\,\,,
\label{gauge}
\eea
\bea
\mathcal{L}_{\rm DM-gauge}
&=&
 i\,\overline{\xlo} \slashed{D} \xlo+ 
i\,\overline{\xlt} \slashed{D} \xlt\,\,,
\label{DM-gauge}
\eea
\bea
\mathcal{L}_{\rm DM-Yukawa}
&=&
-\left(\dfrac{y_1}{2}\,\overline{{\xlo}^c}\xlo\,\eta + 
\dfrac{y_2}{2}\,\overline{{\xlt}^c} \xlt\,\eta^\dagger 
+h.c.\right) \,\,,
\label{DM-Yukawa}
\eea
\bea
\mathcal{L}_{\rm scalar}
&=&
(D_\mu \eta^\dagger)(D^\mu \eta) 
 + \mu^2 (\Phi^\dagger\Phi) +  \mu_X^2 (\eta^\dagger\eta)
 - \lambda (\Phi^\dagger\Phi)^2 
 -\lambda_X (\eta^\dagger\eta)^2 
 -\lambda^\prime (\eta^\dagger\eta)(\Phi^\dagger\Phi)\,\,.
\label{scalar}
\eea
Where, $\Phi$ is the SM Higgs doublet, ${\xi_{{}_i}}_L = 
\mathcal{C} \,\overline{{\xi _{{}_i}}_L}^T$ , $(i =1,\,2 )$
and $\mathcal{C}$ is the charge
conjugation operator. The field strength tensor of the extra U(1)$_X$ gauge symmetry
is $X_{\mu\nu} = \partial_\mu X_\nu - \partial_\nu X_\mu$ while the
corresponding tensor for the U(1) part of the SM is $B_{\mu\nu}$. The
Covariant derivatives in the above Lagrangian, needed to restore
the gauge invariance under the local U(1)$_X$ symmetry, have the usual definition as  
$D_\mu = \partial_\mu + i g_X q_X  X_\mu$ where, $q_X$ is the U(1)$_X$ charge
of that particular field on which $D_{\mu}$ is acting and $g_X$ is the
new gauge coupling. In Eq.\,\ref{DM-Yukawa}, the first two terms
are responsible for the Majorana masses for both $\xlo$ and $\xlt$ respectively when
$\eta$ gets a nonzero VEV. There could also be a Dirac mass term
like $m_{12}\,\overline{{\xlo}^c} \xlt$
involving both $\xlo$ and $\xlt$, 
which conserves U(1)$_{X}$ charge. However,
since the Dirac mass term triggers mixing between $\xlo$ and $\xlt$, for simplicity,
one can easily avoid such term by assuming one of
the two chiral fermions as $\mathbb{Z}_2$ odd.

The electroweak symmetry and U(1)$_X$ symmetry are broken
when both $\Phi$ and $\eta$ acquire VEVs $v$ and $v_X$ respectively
and in the unitary gauge they can be represented as
\bea
\Phi = 
\begin{pmatrix}
0\\
\dfrac{H + v}{\sqrt{2}}\\
\label{scalar-component}
\end{pmatrix}
&,&
\eta = \dfrac{\zeta + v_X}{\sqrt{2}} \,\,\nn.
\eea 

The second term in Eq.\,\ref{gauge} with coefficient
$\epsilon/2$, represents the kinetic
mixing between U(1)$_Y$ and U(1)$_X$ and is not forbidden by any symmetry
of the Lagrangian. Therefore, before proceeding further, we shall write
Eq.\,\ref{gauge} in the canonical form. To do this, we perform a basis
transformation from the ``un-hatted" basis to the ``hatted" basis by a
non-orthogonal transformation, which is given by
\bea
\begin{pmatrix}
B_\mu\\
X_\mu
\end{pmatrix}
&=&
\begin{pmatrix}
1 & - \dfrac{\epsilon}{\sqrt{1-\epsilon^2}}\\
0 & \dfrac{1}{\sqrt{1-\epsilon^2}}
\end{pmatrix}
\begin{pmatrix}
\hat{B}_\mu\\
\hat{X}_\mu\\
\end{pmatrix}.
\label{kin_mixing}
\eea 
Assuming $\epsilon << 1$ (supported by various experimental observations
\cite{Rizzo:2006nw, Langacker:2008yv, Erler:2009jh, Cline:2014dwa}),
one can write $B_\mu \approx \hat{B}_\mu -  \epsilon \hat{X}_\mu$ and
$X_\mu \approx \hat{X}_{\mu}$. In spite of restoring the canonical form
in the Lagrangian for $\hat{B}_{\mu}$ and $\hat{X}_{\mu}$, the mixing among
the neutral gauge bosons $W_{\mu}^3$, $\hat{B}_{\mu}$ and $\hat{X}_{\mu}$
will again reappear when we substitute the transformation relation
for $B_{\mu}$ in the covariant derivative of $\Phi$. After spontaneous symmetry breaking
these mixing terms are solely responsible for all four off-diagonal elements (except
the mixing between $W^3_{\mu}$ and $\hat{B}_{\mu}$) of the $3\times 3$ neutral gauge
boson mass matrix which has the following form in the basis
$\left(\hat{B}_\mu \,  W_\mu^3  \, \hat{X}_\mu\right)$:
\bea
\mathcal{M}^2_{\rm GB} 
&=&
\begin{pmatrix}
\dfrac{g_1^2v^2}{4} & -\dfrac{g_1 g_2 v^2}{4} & -\dfrac{g_1^2 \epsilon v^2 }{4}\\
\vspace{0.1cm}\\
-\dfrac{g_1 g_2 v^2}{4} & \dfrac{g_2^2 v^2}{4} & \dfrac{g_1 g_2 \epsilon v^2}{4}\\
\vspace{0.1cm}\\
-\dfrac{g_1^2 \epsilon v^2}{4} & \dfrac{g_1 g_2 \epsilon v^2}{4} & 4 g_X^2 v_X^2
\end{pmatrix}\,\,.
\label{GB_mass_matrix}
\eea
Here, $g_1$ and $g_2$ are the gauge couplings of U(1)$_Y$ and SU(2)$_L$
respectively while as mentioned earlier $\epsilon$ is the co-efficient of
the kinetic mixing term in Eq.\,\ref{gauge}. To diagonalize $\mathcal{M}^2_{GB}$,
we first rotate the basis vector
$\left(\hat{B}_\mu \,  W_\mu^3  \, \hat{X}_\mu\right)^T$ at 
the Weinberg angle ($\theta_W$) in the
$\hat{B}_\mu \, -\, W_\mu^3$ plane. After this rotation,
both $W^3_\mu$ and $\hat{B}_{\mu}$ have changed to
$\mathcal{Z}_{\mu} = {\rm cos}\theta_W W_\mu^3 - {\rm sin}\theta_W \hat{B}_\mu$
and the orthogonal state $A_{\mu} = 
{\rm sin}\theta_W W_\mu^3 + {\rm cos}\theta_W \hat{B}_\mu$
with $m^2_{A} = 0$ while the third state $\hat{X}_{\mu}$ remains unaffected. Finally,
we apply another rotation in the $\mathcal{Z}_\mu - \hat{X}_\mu$ plane 
at an angle $\theta_{1}$ and get the diagonal matrix in the basis
$\left(A_\mu \,  Z_\mu  \, {Z^{\prime}}_\mu \right)$:
\bea
\mathcal{M}_{GB}^2{^{(\rm dia.)}}
&=&
\begin{pmatrix}
0 & 0 & 0\\
0 & m_Z^2 & 0\\
0 & 0 & m_{\zp}^2
\end{pmatrix}\,\,,\nn
\eea
where, masses of the new physical states are  
\bea
m_Z^2 &=& \dfrac{\left( g_1^2 + g_2^2\right) v^2 {\rm cos}^2\theta_1}{4}
+ 4 g_X^2 v_X^2 {\rm sin}^2\theta_1
+\dfrac{\epsilon \,v^2 g_1 \sqrt{g_1^2 + g_2^2}\,{\rm sin}2\theta_1}{4}\,\,,\nn\\
m_{\zp}^2 &=& \dfrac{\left( g_1^2 + g_2^2\right) v^2 {\rm sin}^2\theta_1}{4}
+ 4 g_X^2 v_X^2 {\rm cos}^2\theta_1
-\dfrac{\epsilon \,v^2 g_1 \sqrt{g_1^2 + g_2^2}\,{\rm sin}2\theta_1}{4}\,\,\,,
\eea
where, $Z_{\mu}$ is the usual SM $Z$-boson having mass
$M_Z = 91.1876\pm0.0021$ GeV \cite{Tanabashi:2018oca}
while the new gauge boson corresponding to $U(1)_X$ is represented
by $Z^\prime_{\mu}$. Unlike the SM, here we have one more mixing angle $\theta_1$
along with the Weinberg angle $\theta_W$. The two mixing angles $\theta_W$ and
$\theta_1$ are given by
\bea
\theta_W = {\rm tan}^{-1}\left(\dfrac{g_1}{g_2}\right)
&,&
\theta_1 = \dfrac{1}{2} {\rm tan}^{-1}
\left( \dfrac{\dfrac{2 \epsilon g_1}{\sqrt{g_1^2 + g_2^2}}}
{1-\dfrac{16\,g_X^2}{(g_1^2 + g_2^2)}\dfrac{v^2_X}{v^2}}\right)\,\,.
\eea
For completeness, we have written below the
transformation relation between the physical
basis and the gauge basis in a matrix form as
\bea
\begin{pmatrix}
A_\mu\\
Z_\mu\\
\zp_{\mu}
\end{pmatrix}
&=&
\begin{pmatrix}
1 & 0 & 0\\
0 & {\rm cos}\,\theta_1 & {\rm sin}\,\theta_1\\
0 & -{\rm sin}\,\theta_1 & {\rm cos}\,\theta_1
\end{pmatrix}
\begin{pmatrix}
{\rm cos}\,\theta_W & {\rm sin}\,\theta_W & 0 \\
-{\rm sin}\,\theta_W & {\rm cos}\,\theta_W & 0 \\
0 & 0 & 1
\end{pmatrix}
\begin{pmatrix}
\hat{B}_\mu\\
W_\mu^3\\
\hat{X}_\mu
\end{pmatrix}\,\,.
\label{GB-trans}
\eea

Let us now look at the fermionic sector. After the breaking
of U(1)$_{X}$ one can easily rewrite
Eq.\,\ref{DM-Yukawa} in terms of two Majorana fermions
$\chio$ and $\chit$ as
 \bea
 \mathcal{L}_{\rm DM-Yukawa}
 &=&
 -\dfrac{1}{2}\,m_1\,\overline{\x_{1}}\x_{1}
 -\dfrac{1}{2}\,m_2\,\overline{\x_{2}}\x_{2} 
 - \dfrac{m_1}{2\,v_X} \overline{\x_{1}}\x_{1}\,\zeta
 -\dfrac{m_2}{2\,v_X} \overline{\x_{2}}\x_{2}\,\zeta 
 \,\,,
 \label{DM-mass-matrix-II}
\eea
where, each Majorana field is defined as $\chii = \xi_{iL} + \xi_{iL}^c \,\,,$
and the corresponding Majorana mass $m_i = \dfrac{y_i\,v_X}{\sqrt{2}}$.

In the scalar sector, after the EWSB and U(1)$_X$ breaking, 
the masses of two physical CP-even scalars, which are the admixtures of $H$ and $\zeta$,
are generated. On the other hand, both the CP-odd scalars become
would-be Goldstone bosons corresponding to
$Z$ and $Z^\prime$ bosons respectively. The CP-even scalar mass matrix in
the basis $\left(H\,\,\zeta \right)$, using Eq.\,\ref{scalar} and
Eq.\,\ref{scalar-component}, is given by
\bea
\mathcal{L}_{\rm scalar}^{mass} 
&=& 
-\dfrac{1}{2}
\begin{pmatrix}
H \,&\, \zeta
\end{pmatrix}
\begin{pmatrix}
2\lambda v^2 & \lambda^\prime v v_X\\
\lambda^\prime v v_X & 2\lambda_X v_X^2
\end{pmatrix}
\begin{pmatrix}
H\\
\zeta
\end{pmatrix}\,\,\nn\\
&=&
-\dfrac{1}{2}
\begin{pmatrix}
H \,&\, \zeta
\end{pmatrix}
\mathcal{M}^2_{\rm scalar}
\begin{pmatrix}
H\\
\zeta
\end{pmatrix}\,\,.
\label{scalar-mass-I}
\eea
The scalar mass matrix $\mathcal{M}^2_{\rm scalar}$
can be diagonalised using an orthogonal transformation 
\bea
\begin{pmatrix}
h\\
h_d
\end{pmatrix}
&=&
\begin{pmatrix}
{\rm cos}\alpha & {\rm sin}\alpha\\
-{\rm sin}\alpha & {\rm cos}\alpha
\end{pmatrix}
\begin{pmatrix}
H\\
\zeta
\end{pmatrix}\,\,.
\label{scalar-trans}
\eea 
The two physical CP-even scalars $h$ and $\hd$ have masses
$m_h$ and $m_{\hd}$ respectively and are given by
\bea
m_h^2
&=&
\lambda v^2 + \lambda_X v_X^2 + \sqrt{\lambda^\prime{^2} v^2 v_X^2 
+ \left(\lambda v^2 -\lambda_X v_X^2\right)^2}\,\,,\nn\\
m_{h_d}^2
&=&
\lambda v^2 + \lambda_X v_X^2 - \sqrt{\lambda^\prime{^2} v^2 v_X^2 
+ \left(\lambda v^2 - \lambda_X v_X^2\right)^2}\,\,.
\eea
The new scalar sector mixing angle $\alpha$ is given by
\bea
\alpha
&=&
\dfrac{1}{2}{\rm tan}^{-1}\left(\dfrac{\lambda^\prime v v_X}
{ \lambda\,v^2 - \lambda_X\,v_X^2}\right)\,\,.
\eea
Here, $h$ is our SM-like Higgs boson having mass $m_h = 125.10 \pm 0.14$ GeV
\cite{Tanabashi:2018oca, Aad:2012tfa, Chatrchyan:2012ufa}. 
Using the transformation relations in Eq.\,\ref{GB-trans} and
Eq.\,\ref{scalar-trans} along with the definition of two Majorana fermions
$\chii = \xi_{iL} + \xi_{iL}^c \,$, we can write 
the Lagrangian (given in Eq.\,\ref{gauge}-Eq.\,\ref{scalar}) in terms of physical
fields (e.g. $A_{\mu}$, $Z_{\mu}$, $Z^\prime_{\mu}$, $\chii$, $h$
and $\hd$) and all the relevant vertex factors are listed in Appendix \ref{App:c}.
\section{Dynamics of the dark sector}
\label{sec:dynamics}
As discussed in the previous section, the present model has a very
rich dark sector having a scalar $\hd$, two Majorana fermions
$\chio$, $\chit$ and a dark gauge boson $Z^\prime$.
Among these particles, both $\chio$ and $\chit$ are absolutely stable and thus they are
automatic choice for our dark matter candidates. The remaining two species $\hd$ and
$Z^\prime$ have tiny couplings with the SM particles through mixings (parametrised
by the mixing angles $\theta_1$ and $\alpha$ respectively) which are assumed to be
extremely small $\theta_1,\,\alpha << 1$. In this situation,
if we consider a ``democratic choice''
that all the dark sector couplings are of the same order\footnote{In an
unknown dark sector, the assumption of equal order for all couplings 
is a realistic one. Typically here we need
$y_i/\sqrt{2} \simeq 2\,g_X \simeq \sqrt{2}\lambda_X$.},
a degenerate dark sector can easily be achieved. However, such a simplifying
assumption has deep impact on the cosmological evolution of dark matter and
associated particles. In this case, instead of the usual freeze-out process
with the Boltzmann suppressed dark matter number density in the non-relativistic regime, 
we are encountered with a different type of exponential
suppression $\propto e^{-\Gamma t}$ in the number density followed
by late freeze-out of dark matter. Here, $\Gamma$ is the decay width of associated dark
sector particles into the SM particles. This is known as co-decaying dark matter
scenario as first proposed in \cite{Dror:2016rxc}. Before going to the detailed
phenomenological analysis of the dark sector, we would first like to discuss
on the co-decaying dark matter scenario briefly. 

Here we need at least two species in the dark sector which are
either degenerate or quasi-degenerate in mass and the dark sector has decoupled
from the SM while it was relativistic. Now, let us consider
two species {\bf A} and {\bf B} in the dark sector and
furthermore let us assume {\bf B} is feebly connected to the SM, so that later on
it can {\it slowly} decay ($\Gamma_B \ll {\mathcal H}$)
into the SM particles. In our model, both $\chio$ and
$\chit$ are playing the role of ${\bf A}$ while $\hd$ and $\zp$
are mimicking the species ${\bf B}$. Initially, after decoupling
from the SM sector at a temperature $T_d$, the dark sector particles {\bf A} and {\bf B}
remain in thermal equilibrium among themselves due to the scattering
${\bf A}{\bf A} \rightleftarrows {\bf B}{\bf B}$ and strong
individual self-interactions respectively. Moreover, they maintain
the thermal equilibrium even after the dark sector temperature
$T^\prime$ drops below $m$, the common mass scale of ${\bf A}$ and ${\bf B}$.
This is particularly due to the reason that as ${\bf A}$ and ${\bf B}$
are degenerate in mass, the scattering ${\bf A}{\bf A} \rightleftarrows {\bf B}{\bf B}$
remains kinematically viable even after $T^\prime$ becomes less than $m$.
Besides, there can be $3\rightarrow 2$ interactions as well
involving both ${\bf A}$ and ${\bf B}$ species, which are
also in chemical equilibrium\footnote{Although,
$2\rightarrow 3$ scatterings are kinematically
suppressed for $T^\prime<m$, the interaction rate for forward and backward
processes still could-be of same order.}.
During this phase, due to these {\it cannibalism}, the
chemical potential of each species in the dark sector is zero.
However, $3 \rightarrow 2$ processes decouple before $2\rightarrow 2$
scatterings as interaction rates of the former are quadratically
suppressed by both number density and velocity of the initial state
particles. Let us assume that the temperature at which all
$3\rightarrow 2$ interactions are frozen-out is $T^\prime_c$ and
the corresponding SM temperature is $T_c$. Therefore
for $T^\prime<T^\prime_c$, both ${\bf A}$ and ${\bf B}$
develop a nonzero chemical potential which helps them to get
rid of the Boltzmann suppression in the non-relativistic
regime. Now, the chemical equilibrium of
${\bf A}{\bf A} {\rightleftarrows} {\bf B}{\bf B}$ interactions
will be lost if the species ${\bf B}$ starts decaying into the SM particles
and thus reducing the number density of ${\bf B}$ by an exponential
factor $e^{-\Gamma_{\bf B}\,t}$, where $\Gamma_{\bf B}$ is the
decay width of ${\bf B}$. Once the chemical equilibrium is lost
(${\bf A}{\bf A} \slashed{\rightleftarrows} {\bf B}{\bf B}$)
at $T^\prime_{\Gamma}$, the freeze-out
of dark matter (${\bf A}$) occurs when the interaction rate
of ${\bf A}{\bf A} \rightarrow {\bf B}{\bf B}$ at a
particular temperature $T_f^\prime$ goes below
the corresponding expansion rate of the Universe controlled by the Hubble parameter
$\hub$. Therefore, the final abundance of ${\bf A}$ in this co-decaying
scenario depends on both the annihilation cross section of
${\bf A}{\bf A} \rightarrow {\bf B}{\bf B}$ as well as the
decay width of ${\bf B}$ into the SM particles.
Moreover, as we will see later the freeze-out of ${\bf A}$,
for certain values of model parameters, occurs well after it
enters into the non-relativistic regime (i.e.\,$x^\prime_f=m/T^{\prime}>>20$)
if ${\bf B}$ has longer lifetime. This requires
large annihilation cross section of ${\bf A}$
compared to that of the usual freeze-out cases ($\sim 3\times 10^{-26}$ cm$^3$/s)
and this opens the possibility of indirect detection prospects
of ${\bf A}$ in the present and upcoming neutrino and $\gamma$-ray detectors, which
has been discussed elaborately in Section \ref{sec:indirect}.

Now, we will try to understand the dynamics of the dark sector in more details
by computing certain thermodynamic quantities analytically. We will
compare our analytical predictions with those obtained from numerical solutions
of the Boltzmann equations in the next section. As we mentioned earlier,
the dark sector has decoupled from the visible sector when they
are relativistic. As a result, the entropies of the dark sector and 
the visible sector are separately conserved until either
$Z^\prime$ or $\hd$ or both start to decay out-of equilibrium
into the SM species. Suppose, $t_d$, $t_c$ and $t_\Gamma$ are the time scales
for decoupling of the dark sector from the SM, freeze-out of $3\rightarrow 2$
processes and beginning of decay of metastable mediators ($\hd$, $\zp$).
The corresponding hidden sector temperatures\footnote{Here we would like note that,
throughout this work we have used the convention to denote the dark sector
temperature with a prime of the corresponding visible sector temperature.}
are already defined above as $\tp_d$, $\tp_c$ and $\tp_\Gamma$
respectively with $\tp_d = T_d$. Before the freeze-out of the
$3\rightarrow 2$ processes\,i.e.\,during the {\it cannibal} phase with
temperature $T^\prime > T_c^\prime$, chemical potential of each dark sector
species is zero. Now, using the entropy conservation and the conservation
of total number of dark sector species in a co-moving volume for a
temperature $\tp$ lying in the range $\tp_c > \tp > \tp_{\Gamma}$,
one can have, 
\bea
s^\prime(\tp) \,a(T)^3 = s^\prime(\tp_c) a(T_c)^3
\,\,\,\,\,\,\,\,&\text{and}&\,\,\,\,\,\,\,\,
n^\prime(\tp) \,a(T)^3 = n^\prime(\tp_c) a(T_c)^3\,\,,
\label{entro_number_cons}
\eea
where $s^\prime$ is the total entropy density of the dark sector
and $n^\prime = n_{\chio} + n_{\chit} + \nzp + \nhd $ is the total
number density of all dark species. Therefore, from Eq.\,\ref{entro_number_cons},
we can write
\begin{eqnarray}
\dfrac{s^\prime(\tp)}{n^\prime(\tp)} = \dfrac{s^\prime(\tp_c)}{n^\prime(\tp_c)}\,.
\label{entro_number_rat}
\end{eqnarray}
Here $a(T)$ is the cosmic scale factor at a temperature $T$.  
Now, using the second law of thermodynamics for a non-relativistic species, we
can easily find the ratio of entropy density to number density at an arbitrary
temperature $\tp$ as
\begin{eqnarray}
{s^\prime(\tp)} = \dfrac{\rho^\prime(\tp) + 
P^\prime(\tp) - \mu^\prime(\tp) n^\prime(\tp)}{\tp}
\label{2ndlaw}
\end{eqnarray}
where, $\rho^\prime(\tp)$ and $P^\prime(\tp)$ are total energy density and
total pressure of all the dark sector particles at $\tp$. Moreover, in the above
we have considered that chemical potential for each species is same\footnote{This
assumption is indeed a realistic one since the chemical equilibrium of the process
${\bf A}{\bf A} \rightleftarrows {\bf B} {\bf B}$ demands $\mu_{\bf A} = \mu_{\bf B}$.}
and equal to $\mu^\prime (\tp)$. Substituting the expressions of $\rho^\prime(\tp)$,
$P^\prime(\tp)$ for a non-relativistic dark sector in Eq.\,\ref{2ndlaw}, we get
\begin{eqnarray}
\dfrac{s^\prime(\tp)}{n^\prime(\tp)} = 
\dfrac{m + \frac{5}{2} \tp - \mu^\prime(\tp)}{\tp}\,.
\label{sbyn_rat}
\end{eqnarray}
Here $m$ is the common mass scale between all degenerate dark sector
species. Finally, substituting Eq.\,\ref{sbyn_rat} in Eq.\,\ref{entro_number_rat} and
considering $\mu^\prime (\tp_c) = 0$, we get the following expression of
chemical potential for each species at any arbitrary temperature
$\tp$ ($\tp_c > \tp > \tp_\Gamma$),
\begin{eqnarray}
\mu^\prime (\tp) = m\left(1-\dfrac{\tp}{\tp_c}\right)\,.
\label{mutp}
\end{eqnarray}
Eq.\,\ref{mutp} clearly shows that each species in the dark sector
develops a temperature dependent chemical potential which pauses
the Boltzmann suppression in number density for a non-relativistic species
between $\tp_c$ and $\tp_{\Gamma}$.

Now we want to calculate how $T^\prime$ changes with respect to the
SM temperature $T$ between $T_d$ and $T_\Gamma$. The nature of $T^\prime$
between $\tp_d$ and $\tp_\Gamma$ is vastly different in two regimes
separated by $\tp_c$ (this is the temperature where all $3\rightarrow2$
interactions are frozen-out). In these two regimes, the chemical potential
$\mu^\prime(\tp)$ for each species behaves differently with $\tp$.
In the first domain bounded between $\tp_d$ and $\tp_c$, $\mu^\prime =0$ while
the temperature dependence in the second domain is given in Eq.\,\ref{mutp}.  
Let us first consider $\tp$ lying between $\tp_d$ and $\tp_c$. At the time
of decoupling of the dark sector from the SM, both the sectors
are relativistic and they have a common temperature $T_d$ . Now, using the entropy
conservation separately for both the sectors at a temperature $T<T_d$, we have
\bea
s^\prime(\tp(T)) = \dfrac{g^\prime_s(T_d)}{g_s(T_d)} \dfrac{2\pi^2}{45}\,g_s(T)\,T^3 \,,
\label{entrpy_cons_td_t}
\eea
where $g_s(T)$ is the number of relativistic degrees of freedom in the
visible sector contributing to the entropy density at $T$ while
the corresponding quantity for the dark sector has been denoted by $g^\prime_s(\tp)$.
The SM at $T$ is radiation dominated however, the dark sector is
non-relativistic at $\tp$ ($\tp<m$). Using $s^\prime(\tp)$ given in
Eq.\,\ref{2ndlaw} for $\mu^\prime(\tp) = 0$, we get 
\bea
8\left(\xp + \frac{5}{2}\right)\left(\frac{m^2}{2\pi\xp}\right)^{3/2} e^{-\xp}
= \dfrac{g^\prime_s(\frac{m}{x_d})}{g_s(\frac{m}{x_d})} \dfrac{2\pi^2}{45}
\,g_s\left(\frac{m}{x}\right)\,\frac{m^3}{x^3} \,,
\eea
where, $\xp = m/\tp$ and $x = m/T$. After a few mathematical simplifications
we obtain the following expression of $\xp$ valid for $\xp\gtrsim 1$,
\begin{eqnarray}
\xp \simeq -\log\left(\frac{2\pi^2}{45}\left(\frac{\pi}{2}\right)^{3/2}\,
\dfrac{g^\prime_s(\frac{m}{x_d})}{g_s(\frac{m}{x_d})}
\,g_s\left(\frac{m}{x}\right)\,{x^{-3}}\,\right)\,.
\label{tp_cannibal}
\end{eqnarray}
Therefore, due to the logarithmic dependence, the hidden sector
temperature before $\tp_c$ varies slowly with respect to the SM temperature.
This can be understood in the following way. In this regime, due to the
expansion of the Universe there is a rapid decrease of $\tp$ with
respect to $T$ as the former is proportional to $T^2$. However,
the decrement in $\tp$ due to expansion has been compensated partially
by heat generation in the dark sector via all the  $3\rightarrow 2$
scatterings active at temperature $\tp>\tp_c$. The resultant effect
is a logarithmic dependence on $T$. 

Finally, we will show the behaviour of $\tp$ with respect to $T$
in the other domain where $\tp$ lies between $\tp_c$ and $\tp_\Gamma$.
Following the similar procedure as we have done in the previous
case, i.e. using the entropy conservation at $\tp_c$ and $\tp$ respectively
for both the sectors and substituting the expression of $\mu^\prime(\tp)$
as given in Eq.\,\ref{mutp}, we find the following expression
of $\xp$ as given bellow
\bea
\xp = \dfrac{\xp_c}{x^2_c} \left(\dfrac{g_s(\frac{m}{x_c})}
{g_s(\frac{m}{x})}\right)^{2/3} x^2\,.
\eea
As there is no other source active for heating the
dark sector, the variation of $\xp$ follows the usual
redshift of temperature of a non-relativistic species due
to expansion of the Universe, which scales as $a^{-2}(T)$.
\subsection{The Boltzmann equations}
\label{sec:Boltz}
\begin{figure}[h!]
\centering
\includegraphics[height=3cm,width=17cm,angle=0]{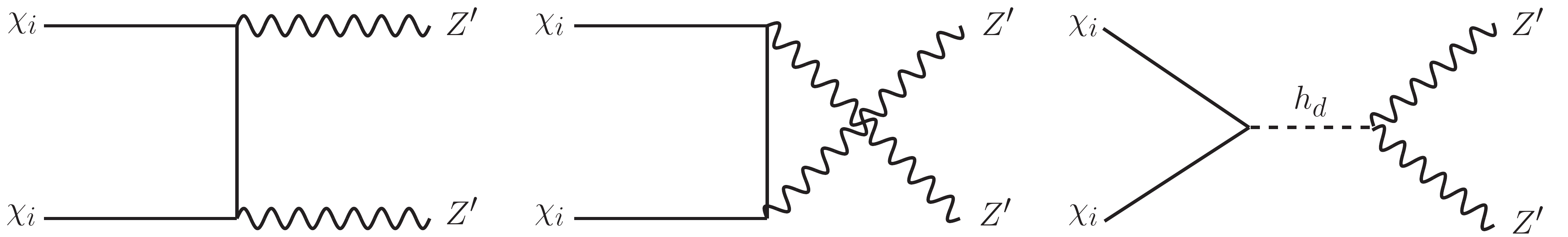}
\includegraphics[height=3cm,width=17cm,angle=0]{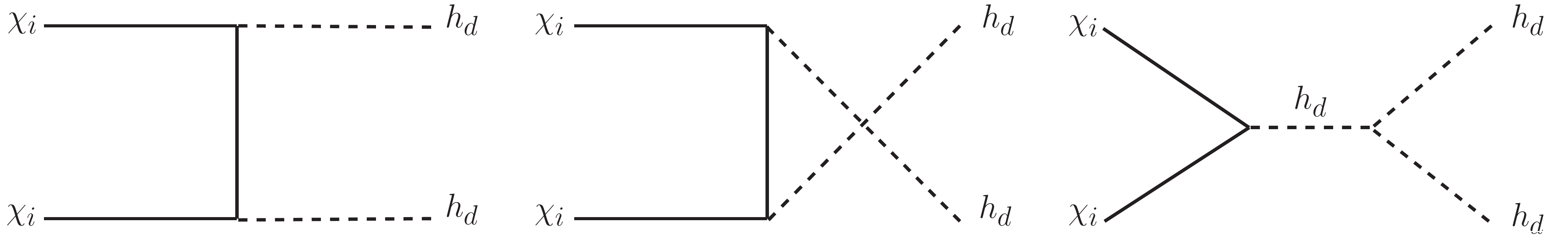}
\includegraphics[height=6cm,width=17cm,angle=0]{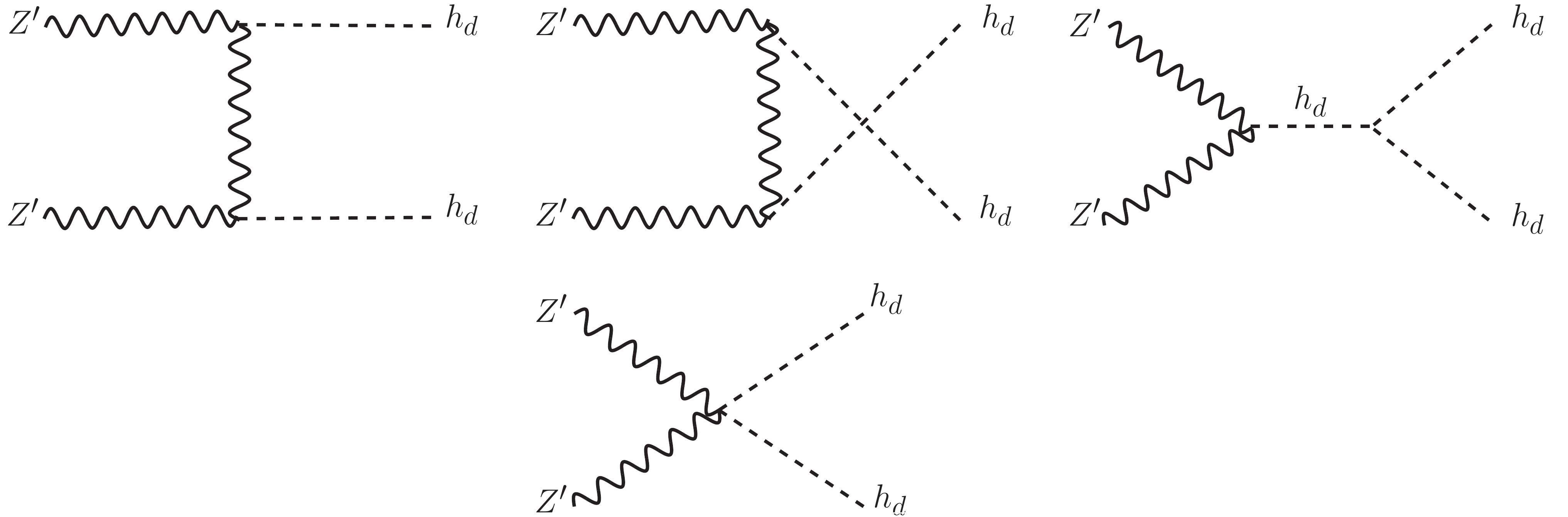}
\caption{$2\ra2$ annihilation processes significant for the evolutions
of temperature and number densities of different species within
the dark sector.}
\label{Fig:2to2_Feyn_dia}
\end{figure}

\begin{figure}[h!]
\centering
\includegraphics[height=6cm,width=9cm,angle=0]{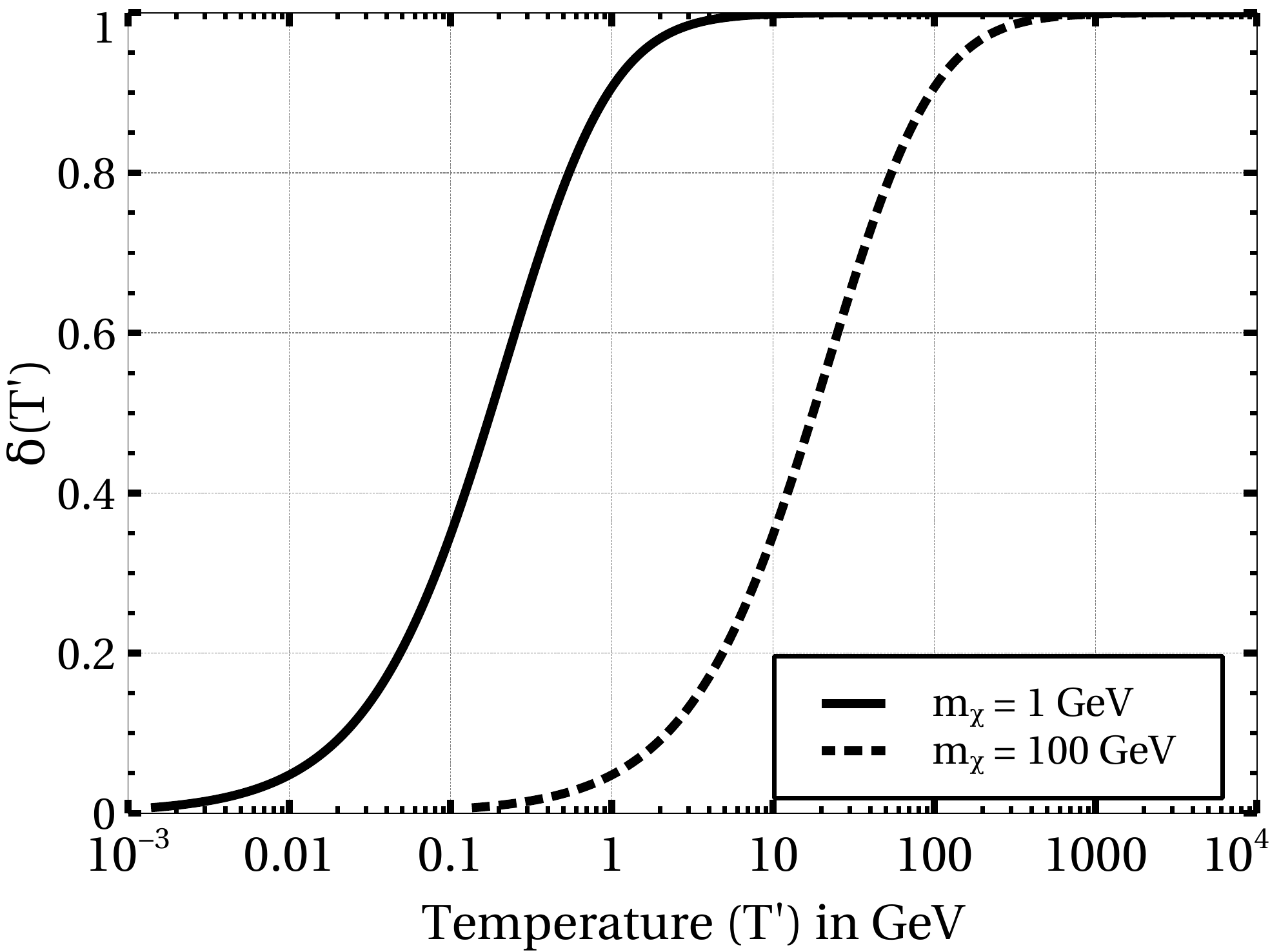}
\caption{Variation of $\delta(\tp)$ with temperature $\tp$ for
two different values of $m_{\x}$.}
\label{Fig:deltatp}
\end{figure}

Now, we will formulate the system of Boltzmann equations
for our dark sector described in Section \ref{sec:model}.
The dark sector particles $\zp$ and $\hd$ have important
roles on the thermal evolution of the dark matter candidates $\chio$
and $\chit$ as the freeze-out of the latter is initiated only after the beginning
of decay of the former into the SM particles. Moreover, since our
dark sector has decoupled from the SM at a temperature $T_d$, it
is not in kinetic equilibrium with the SM thereafter and its
temperature $\tp$, beyond $T_d$, is different from that
of the SM denoted by $T$. Thus, we have four coupled Boltzmann equations,
where first three equations are corresponding to the evolution of number
densities of $\chio + \chit$, $\zp$ and $\hd$ respectively while the last
equation describes the evolution of $\tp$ for $T<T_d$. The four coupled
Boltzmann equations which we need to solve to extract
the physics of the dark sector are given below,

\bea
&&\dfrac{dn_{\x}}{dt} + 3\,\hub\,n_{\x} 
=
-\dfrac{1}{4}\sum_{j=\zp,\,\hd}
\langle {\sigma {\rm v}}_{\x \x \ra j j }\rangle^{\tp}
\left[n_{\x}^2 - \left(\dfrac{n_{\x}^{\rm eq}(\tp)}
{n_{j}^{\rm eq}(\tp)}\right)^2 n_{j}^2 \right]  
 \,\,,
\label{ndm-BE} \\
\nn \\
&&\dfrac{d \nzp}{dt} + 3\,\hub\,\nzp
=
\dfrac{1}{4}\langle {\sigma {\rm v}}_{\x \x \ra \zp \zp }\rangle^{\tp}
 \left[n_{\x}^2 - \left(\dfrac{n_{\x}^{\rm eq}(\tp)}
 {\nzp^{\rm eq}(\tp)}\right)^2 \nzp^2 \right]\,- \nn \\ &&~~~~~~~~~~~~~~
 \langle {\sigma {\rm v}}_{\zp \zp \ra \hd \hd}\rangle^{\tp}
 \left[\nzp^2 - \left(\dfrac{\nzp^{\rm eq}(\tp)}
 {\nhd^{\rm eq}(\tp)}\right)^2 \nhd^2 \right] 
 \,+\,
 \langle \Gamma_{\zp}\rangle^{T}\,\nzp^{\rm eq}(T)\,-\, 
 \langle \Gamma_{\zp}\rangle^{\tp} \nzp,
 \label{z-BE} \\
 \nn \\
&& \dfrac{d \nhd}{dt} + 3\,\hub\,\nhd 
 =
 \dfrac{1}{4}\langle {\sigma {\rm v}}_{\x \x \ra \hd \hd }\rangle^{\tp}
 \left[n_{\x}^2 - \left(\dfrac{n_{\x}^{\rm eq}(\tp)}
 {\nhd^{\rm eq}(\tp)}\right)^2 \nhd^2 \right]\,+ \nn \\ &&~~~~~~~~~~~~~~
 \langle {\sigma {\rm v}}_{ \zp \zp \ra \hd \hd}\rangle^{\tp}
 \left[\nzp^2 - \left(\dfrac{\nzp^{\rm eq}(\tp)}
 {\nhd^{\rm eq}(\tp)}\right)^2 \nhd^2 \right] 
 +
 \langle \Gamma_{\hd}\rangle^{T}\,\nhd^{\rm eq}(T)\, -\, 
 \langle \Gamma_{\hd}\rangle^{\tp}\,\nhd,
 \label{hd-BE} \\
 \nn \\
 &&\dfrac{d \tp}{dt} + \left(2 - \delta(\tp)\right)\hub\,\tp
 =
 - 
\dfrac{\tp}{n_\chi}\left(\dfrac{dn_\chi}{dt} + 3\,\hub\,n_\chi \right)
+ \dfrac{1}{n_\chi} \left[\mathcal{F}(\tp)_{2\ra2} + \mathcal{F}(\tp)_{3\ra2}\right]
\nn \\
&&~~~~~~~~~~~~~~~~~~~~~~~~~~~~~~= 
\dfrac{\tp}{4\,n_\chi}\sum_{j=\zp,\,\hd}
\langle {\sigma {\rm v}}_{\chi \chi \ra j j }\rangle^{\tp}
\left[n_{\x}^2 - \left(\dfrac{n_{\x}^{\rm eq}(\tp)}
{n_{j}^{\rm eq}(\tp)}\right)^2 n_{j}^2 \right]\,+\, \nn \\
&&~~~~~~~~~~~~~~~~~~~~~~~~~~~~~~~~~~
 \dfrac{1}{n_\chi} \left[\mathcal{F}(\tp)_{2\ra2} 
+ \mathcal{F}(\tp)_{3\ra2}\right] 
\label{Tp-BE}
\eea
where, in the last equation\footnote{In non-relativistic limit, 
this equation can also be derived from the more commonly used equation of energy density
(see Eq.\,\,12 of \cite{Dror:2016rxc}) by putting
$\rho_\x \simeq m_\x n_\x + \dfrac{3 n_\x \tp}{2}$.}
we have substituted the right hand side 
of Eq.\,\ref{ndm-BE}. Here $n_\x = n_{\chio} + n_{\chit}$ is the total dark
matter number density while that of $\zp$ and $\hd$ are denoted by
$\nzp$, $\nhd$ respectively. The corresponding equilibrium number density
of a species $i$ at temperature $\tp$ is shown by $n^{\rm eq}_{i}(\tp)$
and $\hub$ is the Hubble parameter. 
The quantity $\langle \sigma {\rm v} \rangle_{a\,b\ra c\,d}^{\tp}$
is the thermal averaged annihilation cross section for the process
$a\,b\ra c\,d$ and it depends on the temperature of the species
appearing in the initial state which in our case is $\tp$.
Here ${\rm v}$ is the magnitude of the
relative velocity between $a$ and $b$.
The $2\ra 2$ annihilation processes in the dark sector are
$\chii\chii\ra\zp\zp$, $\chii\chii\ra\hd\hd$ and $\zp\zp\ra\hd\hd$.
The Feynman diagrams and the corresponding cross
sections for these processes are given in
Fig.\,\,\ref{Fig:2to2_Feyn_dia} and
Appendix \ref{App:E} respectively. 
Here We have taken contributions
from both the dark matter candidates for
each annihilation channels and these have
been incorporated by the quantity
$\langle \sigma{\rm v}_{\x\x\ra jj}\rangle^{\tp} = \sum_{i=1}^2
\langle \sigma{\rm v}_{\x_i\x_i\ra jj}\rangle^{\tp}$.
On the other hand, $\Gamma_i$ ($i=\zp,\,\hd$) is the total decay width of the
species $i$ and the thermal average of $\Gamma_i$ has been indicated
by $\langle \Gamma_i \rangle^{\tp}$ which like the thermal average of
annihilation cross section depends on the temperature of the parent
particle. However, in Eq.\,\ref{z-BE} and Eq.\,\ref{hd-BE}, we also have
the quantities like $\langle \Gamma_i \rangle^{T}$ for $\zp$ and $\hd$ respectively.
These terms actually represent the contributions coming from
inverse decay. In that case the initial state particles
are the SM particles whose temperature is different from $\tp$, as both
$\zp$ and $\hd$ can decay only into the SM particles. Moreover,
$\delta(\tp)$ in Eq.\,\ref{Tp-BE} has the following expression
\begin{equation}
\delta \left(\tp \right) = 1-\dfrac{g_\x}
{n_\chi(\tp) \tp}\int  \dfrac{d^3 \vec{p}}{(2\pi)^3} 
\dfrac{p^2 m_\chi^2}{3 E_p^3} f_\chi(p,\tp)\,\,,
\label{delta}
\end{equation}
where $m_{\x}$ is the mass of any of the
dark matter species (i.e.\,\,$m_{\x} = m_1 = m_2$)
and its distribution function is denoted by $f_{\x}(p,\,\tp)$ with $p$ and $E_p$
being the magnitude of three momentum and energy respectively. $g_\x$ is
the internal degrees of freedom which in our case is equal to 2.\,\,In
Fig.\,\ref{Fig:deltatp}, we have shown how $\delta(\tp)$ varies
with the temperature $\tp$ for two different values of $m_{\x}$ such as
$m_{\x}=1$ GeV and 100 GeV respectively. From this figure, it is clearly
evident that in both cases $\delta(\tp)$ tends to a very small
value when $\tp<<m_{\x}$ (i.e. in the non-relativistic regime). However,
$\delta(\tp)$ increases as $\tp$ increases and finally saturates to a value
of unity in the ultra-relativistic limit (i.e.\,\,$\tp>>m_{\chi}$). In the
present work, as we have studied the dynamics of a dark sector in the
non-relativistic regime, we have set $\delta(\tp)$ to be equal to zero while
solving the above set of Boltzmann equations. A detailed derivation
of all the four Boltzmann equations are given in Appendix \ref{App:A}.
Finally, the last term in the right hand side of Eq.\,\ref{Tp-BE}
contains all the $2\ra2$ and $3\ra2$ processes having significant impact on the
evolution of dark sector temperature. The contribution from
relevant $2\ra 2$ scatterings is denoted by the
function $\mathcal{F}_{2\ra2}(\tp)$. In Appendix\,\,\ref{App:B1},
we have derived the analytical expression of
$\mathcal{F}_{2\ra2}(\tp)$ for a generic process like
$\chii \chii \ra jj$. More importantly, the dark sector temperature
$\tp$ has significant effect from inelastic $3\ra 2$ scatterings
which are shown in Figs.\,\,\ref{Fig:3to2_2}-\ref{Fig:3to2_11} (Appendix\,\,\ref{App:B})
and a detailed discussion on this topic is given in Appendix\,\,\ref{App:B2}.

Let us define three dimensionless variables namely, co-moving number density
$Y_i = n_i(\tp)/{s}(T)$, $x = m_0/T$ and $\xi = \tp/T$, where $m_0$
is a reference mass scale which can also be the mass of any dark sector
particle. As expected, the final solutions are independent of the choice
of $m_0$. In terms of these dimensionless variables, the above
set of Boltzmann equations take the following form.
\bea
&&\dfrac{d Y_\x}{dx} 
=
-\dfrac{s(x)}{4\,\hub(x)\,x}g^{1/2}_{*}\dfrac{\sqrt{g_{\rho}}}{g_s}
\sum_{j=\zp,\,\hd}
\langle {\sigma {\rm v}}_{\x \x \ra j j }\rangle (\xi,x)
\left[Y_{\x}^2 - \left(\dfrac{Y_{\x}^{\rm eq}(\xi,x)}
{Y_{j}^{\rm eq}(\xi,x)}\right)^2 Y_{j}^2 \right],
\label{Ydm-BE}\\
\nn \\
&&\dfrac{d Y_{\zp}}{dx} 
=
\dfrac{s(x)}{\hub(x)\,x}g^{1/2}_{*}\dfrac{\sqrt{g_{\rho}}}{g_s}
\Bigg[\dfrac{1}{4}\langle {\sigma {\rm v}}_{\x \x \ra \zp \zp }\rangle(\xi,x)
\left\{Y_{\x}^2 - \left(\dfrac{Y_{\x}^{\rm eq}(\tp)}
{Y_{\zp}^{\rm eq}(\tp)}\right)^2 Y_{\zp}^2 \right\} \,-\, 
 \langle {\sigma {\rm v}}_{\zp \zp \ra \hd \hd}\rangle (\xi,x) \times
\nn \\ &&~~~~~~~~~
 \left\{Y_{\zp}^2 - \left(\dfrac{Y_{\zp}^{\rm eq}(\xi,x)}
 {Y_{\hd}^{\rm eq}(\xi,x)}\right)^2 Y_{\hd}^2 \right\} \,+\,
\dfrac{1}{s(x)}\left( \langle \Gamma_{\zp}\rangle(x)\,Y_{\zp}^{\rm eq}(x)\,-\, 
 \langle \Gamma_{\zp}\rangle(\xi,x)\,Y_{\zp}\right) \Bigg], 
 \label{Yzp-BE} \\
\nn \\
 &&\dfrac{d Y_{\hd}}{dx} 
=
\dfrac{s(x)}{\hub(x)\,x}g^{1/2}_{*}\dfrac{\sqrt{g_{\rho}}}{g_s}
\Bigg[\dfrac{1}{4}\langle {\sigma {\rm v}}_{\x \x \ra \hd \hd }\rangle (\xi,x)
\left\{Y_{\x}^2 - \left(\dfrac{Y_{\x}^{\rm eq}(\xi,x)}
{Y_{\hd}^{\rm eq}(\xi,x)}\right)^2 Y_{\hd}^2 \right\} \,+\, 
 \langle {\sigma {\rm v}}_{\zp \zp \ra \hd \hd}\rangle(\xi,x) \times 
\nn \\ &&~~~~~~~~~
 \left\{Y_{\zp}^2 - \left(\dfrac{Y_{\zp}^{\rm eq}(\xi,x)}
 {Y_{\hd}^{\rm eq}(\xi,x)}\right)^2 Y_{\hd}^2 \right\} \,+\,
\dfrac{1}{s(x)}\left( \langle \Gamma_{\hd}\rangle(x)\,Y_{\hd}^{\rm eq}(x)\,-\, 
 \langle \Gamma_{\hd}\rangle(\xi,x)\,Y_{\hd}\right) \Bigg],
 \label{Yhd-BE}\\
 \nn \\
&& x\,\dfrac{d\xi}{dx} + \left\{\left(2-\delta\left(\xi,x\right)\right)
 g^{1/2}_{*}\dfrac{\sqrt{g_{\rho}}}{g_s}
-1\right\} \xi
 = 
 g^{1/2}_{*}\dfrac{\sqrt{g_{\rho}}}{g_s}\,\,\dfrac{s(x)}{Y_{\x}\,\hub(x)}
 \Bigg[-\dfrac{1}{4}
 \sum_{j=\zp\,\hd} \langle {\sigma {\rm v}}_{\x \x \ra j j }
 \rangle^{\prime} (\xi,x) \times
 \nn  \\ && ~~~~~~~~
\left\{Y_{\x}^2 - \left(\dfrac{Y_{\x}^{\rm eq}(\xi,x)}
{Y_{j}^{\rm eq}(\xi,x)}\right)^2 Y_{j}^2 \right\}
+ \dfrac{x\,s(x)}{m_0}\,\mathcal{F}^{Y}_{3\ra2}(\tp)
\Bigg]. 
\label{xi-BE}
\eea
In the above, $g_{\rho}$ and $g_s$ are effective number of degrees
of freedoms associated with the energy and entropy densities
of the Universe while $g^{1/2}_{*}$ is defined as
\bea
g^{1/2}_{*}(x) = \dfrac{g_s(x)}{\sqrt{g_{\rho}(x)}}
\left(1 - \dfrac{1}{3}\dfrac{d \ln{g_{s}(x)}}{d \ln x}\right)\,.
\eea
\begin{figure}[h!]
\centering
\includegraphics[height=6.5cm,width=8.0cm]{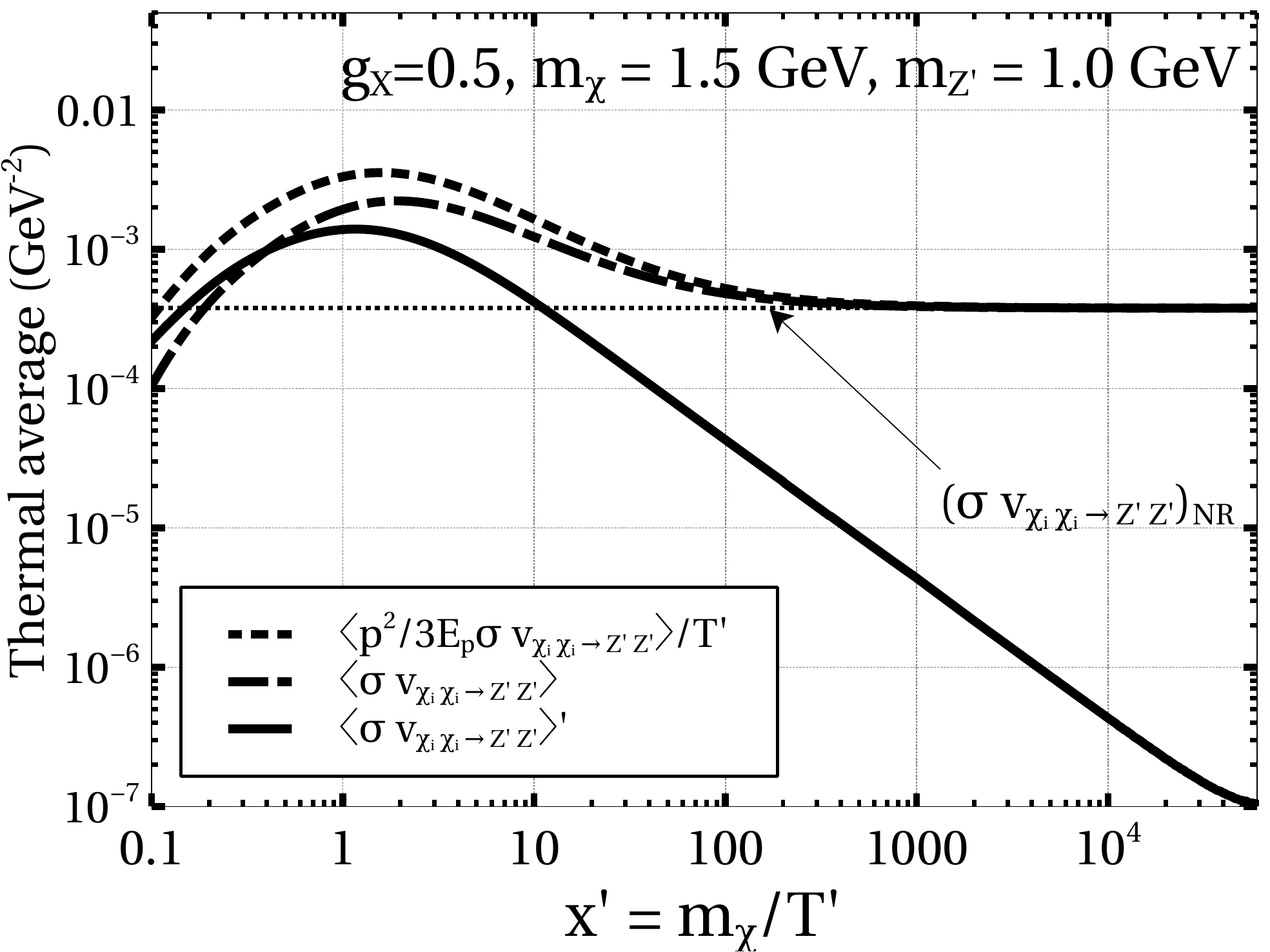}
\includegraphics[height=6.5cm,width=8.0cm]{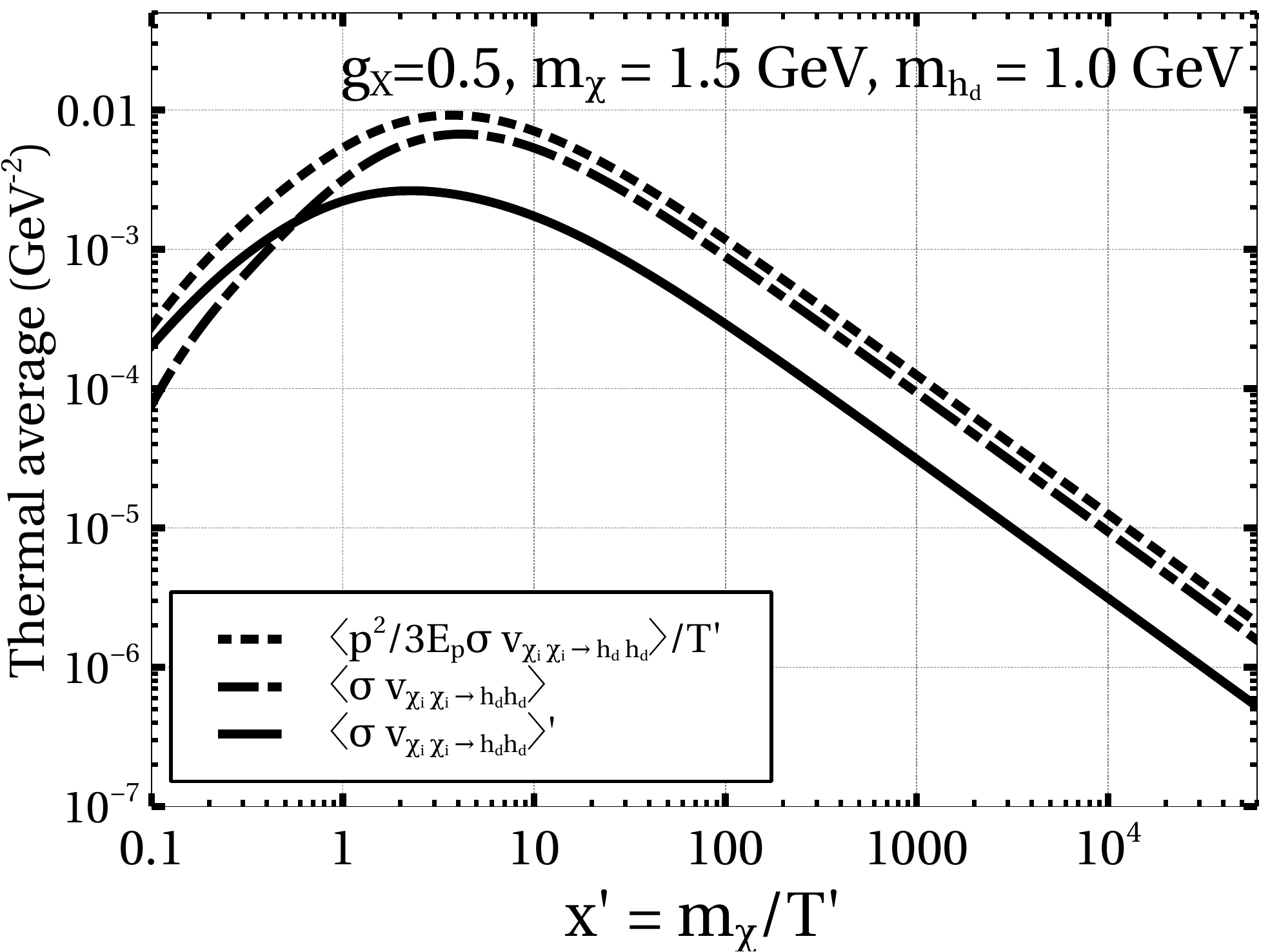}
\caption{Behaviour of $\dfrac{1}{\tp}\langle\dfrac{p^2}{3\,E_p}\,
\sigma {\rm v} \rangle$ (dotted line),
$\langle \sigma {\rm v} \rangle$ (dash-dotted line),
and $\langle \sigma {\rm v} \rangle^\prime$ (solid line) with
$x^\prime$ for a s-wave (left panel) and a p-wave annihilation
process.}
\label{Fig:sigmaVp}
\end{figure}
In Eq.\,\ref{Ydm-BE}-Eq.\ref{xi-BE},
we have replaced the dark sector temperature $\tp$ by the newly
defined variables $\xi$ and $x$. The right hand side of Eq.\,\ref{xi-BE}
has been obtained after substituting the expression
of $\mathcal{F}(\tp)_{2\ra2}$ given in Appendix \ref{App:B1} into Eq.\,\ref{Tp-BE}
and defining a new quantity
$\langle {\sigma {\rm v}}_{\x \x \ra j j } \rangle^\prime$ for
an interaction process $\chii \chii \ra j j$ ($i=1,\,2$)
as
\bea
\langle {\sigma {\rm v}}_{\x \x \ra j j } \rangle^\prime(\xi, x) = \sum_{i=1}^{2}
\dfrac{x}{m_0} \langle\dfrac{p^2}{3E_p}\,\sigma {\rm v}_{\chii \chii \ra j j }
\rangle (\xi, x) -\xi\,\langle \sigma {\rm v}_{\chii \chii \ra j j }\rangle (\xi, x)\,,
\label{sigmavp}
\eea
where, $\langle\dfrac{p^2}{3\,E_p}\,\sigma {\rm v}_{\chii \chii \ra j j} \rangle$
is the thermal average of $\sigma {\rm v}_{\chii \chii \ra j j} 
\times \dfrac{p^2}{3\,E_p}$
with $p$ and $E_p$ are magnitude of 3-momentum and energy of any of
the initial state particles respectively. The integral form
of $\langle\dfrac{p^2}{3\,E_p}\,\sigma {\rm v}_{\chii \chii \ra j j } \rangle$
in terms of annihilation cross section of $\chii \chii \ra j j $ has
been given in Eq.\,\ref{sigmaVp2by3E} of Appendix \ref{App:B}. In order to understand
the quantity $\langle\sigma {\rm v}_{\chii \chii \ra j j } \rangle^\prime$
in more details, let us look at the plots in both panels of Fig.\,\ref{Fig:sigmaVp}. 
In the left panel of Fig.\,\ref{Fig:sigmaVp}, we have shown the
variations of $\dfrac{1}{\tp}\langle\dfrac{p^2}{3\,E_p}\,
\sigma {\rm v} \rangle$ (dotted line),
$\langle \sigma {\rm v} \rangle$ (dash-dotted line),
and $\langle \sigma {\rm v} \rangle^\prime$ (solid line) for the
annihilation process $\x_i\x_i \rightarrow \zp\zp$ 
with $x^\prime$ concurrently. Here, $\x_i$ can be
any of the dark matter candidates $\chio$ or
$\chit$. From this plot, we can see that for large
$x^\prime$ (i.e\,\,for smaller temperature $T^\prime$),
the thermal averaged annihilation cross section becomes
independent of $x^\prime$ and it coincides with the value of
$\langle \sigma {\rm v}_{\chii\chii\rightarrow\zp\zp}\rangle$ in
the non-relativistic limit denoted by
$(\langle \sigma {\rm v}\rangle_{\chii\chii\rightarrow\zp\zp})_{\rm NR}$.
It is mainly due to the fact that for larger values of $x^\prime$, the velocity
independent term (i.e\,\,the s-wave term) of
$\langle \sigma {\rm v}_{\chii\chii \rightarrow \zp\zp} \rangle$
dominates over the other terms which, at that regime, are heavily
velocity suppressed (e.g.\,\,p-wave, d-wave terms). However,
unlike $\langle \sigma {\rm v}_{\chii\chii \rightarrow \zp\zp} \rangle$,
the quantity $\langle \sigma {\rm v}_{\chii\chii\rightarrow \zp\zp}\rangle^\prime$
keeps on decreasing with increasing $x^\prime$ after attaining
a peak at $T^\prime \sim m_{\x}$. The continuous decreasing nature of 
$\langle \sigma {\rm v}_{\chii\chii\rightarrow \zp\zp}\rangle^\prime$
can be easily understood if we notice the dotted line
indicating the behaviour of $\dfrac{1}{\tp}\langle\dfrac{p^2}{3\,E_p}\,
\sigma {\rm v}_{\chii \chii \ra \zp\zp} \rangle$
with $x^\prime$. Here, in the large $x^\prime$ limit
both dotted and dash-dotted curves are overlapping with each other
which implies that in the non-relativistic regime (i.e.\,\,for
$x^\prime>>1$), the quantity $\langle\dfrac{p^2}{3\,E_p}\,
\sigma {\rm v}_{\chii \chii \ra \zp \zp } \rangle \sim \tp \times
\langle \sigma {\rm v}_{\chii \chii \ra \zp \zp } \rangle$.
Therefore, from Eq.\,\ref{sigmavp} it becomes very
straightforward to figure out the decreasing nature of
$\langle \sigma {\rm v}_{\chii\chii\rightarrow \zp\zp}\rangle^\prime$.
The similar thing for a p-wave annihilation cross section
$\chii\chii\rightarrow\hd\hd$ has been depicted
in the right panel of Fig.\,\ref{Fig:sigmaVp}. Here, unlike
the previous case of a s-wave process, the thermal averaged
annihilation cross section $\langle {\sigma{\rm v}}_{\chii\chii\ra\hd\hd}\rangle$
decreases in the large $x^\prime$ limit as the former is proportional
to ${\rm v}^2$ (neglecting $\mathcal{O}({\rm v}^4)$ term and beyond)
which is getting lower with $x^\prime$. Nevertheless, here also
$\dfrac{1}{\tp}\langle\dfrac{p^2}{3\,E_p}\,
\sigma {\rm v}_{\chii \chii \ra \hd \hd} \rangle$ follows 
$\langle {\sigma{\rm v}}_{\chii\chii\ra\hd\hd}\rangle$ closely
in the non-relativistic regime and the decreasing nature of
$\langle {\sigma{\rm v}}\rangle^\prime$ remains.

Now, if we look at Eq.\,\,\ref{xi-BE} once again, we can understand
the effects of various terms on $\xi$ (and hence on $\tp$).
Let us start from the right hand side. The second term
in the right hand side is the only source
term which raises the temperature and this is due to $3\ra2$
inelastic scatterings. These scatterings convert non-relativistic
species into a fewer number of relativistic species
having more kinetic energy compared to the mean
kinetic energy which is $\sim \tp$. Thus, the term
$\mathcal{F}^Y_{3\ra 2}(\tp)$ heats up the dark sector
and shuts off when all inelastic scatterings are frozen-out.
Here, the function $\mathcal{F}^Y_{3\ra 2}(\tp)$
represents the contribution from all relevant $3\ra 2$
scatterings as encountered previously in Eq.\,\,\ref{Tp-BE},
except all the number densities in $\mathcal{F}_{3\ra 2}(\tp)$ should
now be replaced by the corresponding comoving number densities.
The excess heat generated due to these inelastic scatterings
is smeared over the entire dark sector by the $2\ra 2$ scatterings
so that all the species have a common temperature. This has been
incorporated in the first term on the right hand side of
Eq.\,\,\ref{xi-BE}. Therefore, as the $2\ra2$ scatterings
reduce the rate of increase of temperature,
the first term appears with a -ve sign.
Moreover, it is already evident from the discussions given above along
with Fig.\,\,\ref{Fig:sigmaVp} that the effect of
$2\ra2$ scatterings become sub-leading at large $x$. Finally,
there is another source of reduction of dark sector temperature
which has always been present in the background. It is the expansion
of the Universe and the second term in the left hand side
confirms this effect. In the large $x$ limit when all the
terms in the right hand side become insignificant, $\xi$
varies as $1/x$ (or $\tp \propto a^{-2}(T)$) due to the
effect of expansion of the Universe. 
\subsection{Constraints on the portal couplings}
\label{sec:constraints}
The portal couplings ($\epsilon$, $\alpha$) and 
the mass of the mediators ($m_{\zp}$, $m_{\hd}$) 
can be constrained from various experimental observations
(See \cite{Anchordoqui:2015fra} for example). In this
section we discuss the relevant constraints in
$m_{\zp}$ - $\epsilon$ and $m_{\hd}$ - $\sin \alpha$ planes.
\subsubsection{BBN constraint}
\label{subsubsec:BBN}
In our model, $\zp$ and $\hd$ can decay into the
SM fields through the kinetic mixing parameter $\epsilon$
and $h-\hd$ mixing angle $\alpha$ respectively.
Therefore, the lifetime of $\zp$ and $\hd$ are 
controlled by these portal couplings. To ensure
the observations during the BBN epoch remain
unaltered, the lifetime of these fields must be
less than one second i.e. they must decay
before the BBN substantially. This sets a lower
limit on each portal coupling. The constraints from
the BBN on $\epsilon$ and $\alpha$ are shown in Fig.\,\,\ref{Fig:cosmo}
by the light blue coloured regions.
\subsubsection{Thermalisation condition}
\label{subsubsec:thermalisation}
On the other hand, we cannot choose very large value for the portal couplings
since the upper limits of these couplings are set from
the requirement of early kinetic decoupling of the
dark sector from the visible sector. Now, at the
time of decay of $\zp$ and $\hd$, 
$\Gamma_{\zp (h_d)} \simeq \hub(x_{\Gamma})$, where 
$x_\Gamma = m_0/T_{\Gamma}$ and $T_{\Gamma}$ is the temperature
of the SM bath at the time of decay of $\zp$($\hd$).
From this relation, we can estimate $x_{\Gamma} \simeq 
\left(\Gamma_{\zp(\hd)}/\hub(m_0)\right)^{-1/2}$. As
we require out-of-equilibrium decay of both $\zp$ and $\hd$
into the visible sector, hence $\Gamma_{\zp,\hd}/H(m_{j})< 1$
($j=m_{\zp}, m_{\hd}$) and this implies $x_{\Gamma} > m_0/m_{j}$.
However, in order to ensure that the dark sector
will not re-equilibrate with the SM bath, we have adopted
a more conservative lower limit $x_{\Gamma} > 5 m_0/m_{j}$,\footnote{As
in this work we are considering a degenerate dark sector, the
condition in the present case is effectively $x_{\Gamma}>5$.}
as mentioned in \cite{Dror:2016rxc}.
Therefore, the requirement $x_{\Gamma}>5$ sets an upper limit
on the portal couplings $\epsilon$ and $\alpha$. The disallowed region
of the parameter spaces are indicated by light green color
in Fig.\,\,\ref{Fig:cosmo}.
\subsubsection{Direct Detection}
\label{DD}
Due to the presence of kinetic mixing portal, 
our DM candidates $\x_i$ can interact with 
the nucleon exchanging $\zp$ and $Z$ bosons.
Since $\x_i$ is a Majorana fermion, therefore
only the axial vector couplings are present in our model.
Consequently, both $Z$ and $\zp$ mediated DM-nucleon scattering
contributes to the spin dependent (SD) scattering cross
section. Moreover, the DM-nucleon scattering can 
be mediated through $h$ and $\hd$ also as there is
$h$-$\hd$ mixing in our model and 
it contributes to the spin independent (SI) 
elastic scattering cross section.
\\
The spin dependent DM-nucleon scattering cross section
mediated by $Z$ and $\zp$ is given by the following
expression \cite{Gehrlein:2019iwl, Arcadi:2013qia}.
\bea
\label{SD cross}
\sigma_{\rm SD} &=& \dfrac{3 \mu^2_{\x n}}{\pi\left(S_p + S_n\right)^2} 
\left[
\dfrac{C_{A}^{\bar{\x_i}\x_i \zp} \sum_{\rm q = u, d, s} 
C_A^{\, \bar{q}q \zp}\left(S_p \Delta_q^p + S_n 
\Delta_q^n\right)}{m_{\zp}^2} \right. \nn\\
&&~~~~~~~~~~~~~~
\left. + \dfrac{C_{A}^{\bar{\x_i}\x_i Z} \sum_{\rm q = u,d,s}
C_A^{\bar{q}q Z}\left(S_p \Delta_q^p + S_n
\Delta_q^n\right)}{m_{Z}^2}
\right]^2\,\,.
\eea
In the above, $C_{A}^{\bar{\x_i}\x_i \zp}\, 
(C_{A}^{\bar{q} q \zp})$ and 
$C_{A}^{\bar{\x_i}\x_i Z}\,(C_{A}^{\bar{q}q Z})$ are the
axial vector couplings of $\x_i\,(q)$ with $\zp$ and $Z$ bosons
respectively. The quantity $\mu_{\x n}$ is the reduced mass
of the DM-nucleon system. The contribution of proton (neutron)
to the nuclear spin is denoted by $S_{p\,(n)}$.	
The values of $\Delta_q^n$ and $\Delta_q^p$ are:
$\Delta_u^p = \Delta_d^n = 0.84, \,
\Delta_u^n = \Delta_d^p = -0.43, \,
\Delta_s^p = \Delta_s^n = -0.09 \,$ \cite{Berlin:2014tja}.
\\
\\
The spin independent DM-nucleon scattering cross section
mediated by $h$ and $\hd$ is given by \cite{Gehrlein:2019iwl}
\bea
\label{SI cross}
\sigma_{\rm SI} = \dfrac{\mu_{\x n}^2}{\pi} 
\left(\sin \alpha \cos \alpha\right)^2 \left(\dfrac{m_\x}{v_X}\right)^2
\left(\dfrac{1}{m_h^2} -\dfrac{1}{m_{\hd}^2} \right)^2
\left[\dfrac{Z}{A} f_p + \dfrac{A-Z}{A} f_n
\right]^2\,\,\,.
\eea
In the above expression $Z(A)$ is the atomic\,(mass) number of a nucleus
while $m_p$ and $m_n$ are the mass of proton and neutron respectively.
Moreover, $f_{p(n)} = \frac{m_{p(n)}}{v} \left(1 - \frac{7}{9} f_{TG}\right)$
with $f_{TG} = 0.91$ \cite{Berlin:2014tja}.
\\
\\
For both the cases discussed above an effective DM-nucleon
scattering cross section $ f_{\x_i} \,\sigma_{\rm SD (\rm SI)}$ 
($f_{\x_i} = \Omega_{\x_i}/\sum_i \Omega_{\x_i}$)
is defined to compare with the current bounds
from XENON1T \cite{Aprile:2018dbl,Aprile:2019dbj}, 
CDMSlite \cite{Agnese:2017jvy,Agnese:2015nto},
CRESST-III \cite{Abdelhameed:2019hmk}. In Fig. \ref{Fig:cosmo},
we show the spin dependent (left panel) and spin
independent (right panel) direct detection
constraint for $g_X = 1$ by two purple coloured regions.
\subsubsection{Supernovae cooling}
\label{SN cooling}
Dark sector particles with mass $\lesssim 200 \, {\rm MeV}$
can be constrained from the observation of supernova
SN 1987A \cite{Chang:2016ntp, Dreiner:2013mua}.
The constraints are derived based on the following
two conditions.
\vspace{10pt}

\noi
i) The emissivity of the hidden sector
particles must be smaller than that of the observed neutrino signal.
This is known as ``\textit{Raffelt Criterion}" \cite{Raffelt:1996wa} 
and the constraint derived from this condition is known as 
``\textit{cooling constraint}". This condition 
puts an upper bound on the portal couplings.

\vspace{10pt}

\noi
ii) The mean free path of the hidden sector particle must be
smaller than the radius of the supernova (SN). The constraints derived
from this condition put a lower bound on the portal coupling
and this is known as \textit{``Trapping condition"}.

\vspace{8pt}
In this work, we have considered the bound 
from SN 1987A cooling \cite{Dent:2012mx,Rrapaj:2015wgs,Chang:2016ntp}
relevant for the dark gauge boson $\zp$ in our model. 
Similarly $\hd$ can also contribute to the SN 1987A
cooling because of the presences of $h-\hd$ mixing angle 
$\alpha$ \cite{Krnjaic:2015mbs}. In both panels of Fig.\,\,\ref{Fig:cosmo},
the red coloured regions indicate the excluded parameter
space from the SN 1987A cooling.
\subsubsection{Beam dump experiments}
\label{BD expt}
The dark gauge boson $\zp$ with mass $\lesssim 1\rm GeV$ and $\epsilon$
$\simeq$ $10^{-8} - 10^{-2}$ can be constrained
from the result of beam dump experiments.
The excluded parameter space \cite{Bauer:2018onh}
is shown in the left panel of Fig.\,\,\ref{Fig:cosmo}
using magenta color.
The dark scalar $\hd$ of mass $\lesssim 1 \, \rm GeV$ and 
portal coupling $\sin \alpha \simeq 10^{-4}-10^{-2}$
can be constrained from proton beam dump experiments
by CHARM collaboration\cite{Bergsma:1985qz}. 
Following \cite{Clarke:2013aya,Bezrukov:2009yw}, we have derived the
bound in $m_{\hd}-\sin \alpha$ plane and the excluded
parameter spaces is shown in the right panel of Fig. \ref{Fig:cosmo} 
with magenta color.
\subsubsection{Electroweak Precision Observables}
\label{EWPO}
\begin{figure}
	\centering
	\includegraphics[height=7cm,width=8.0cm]{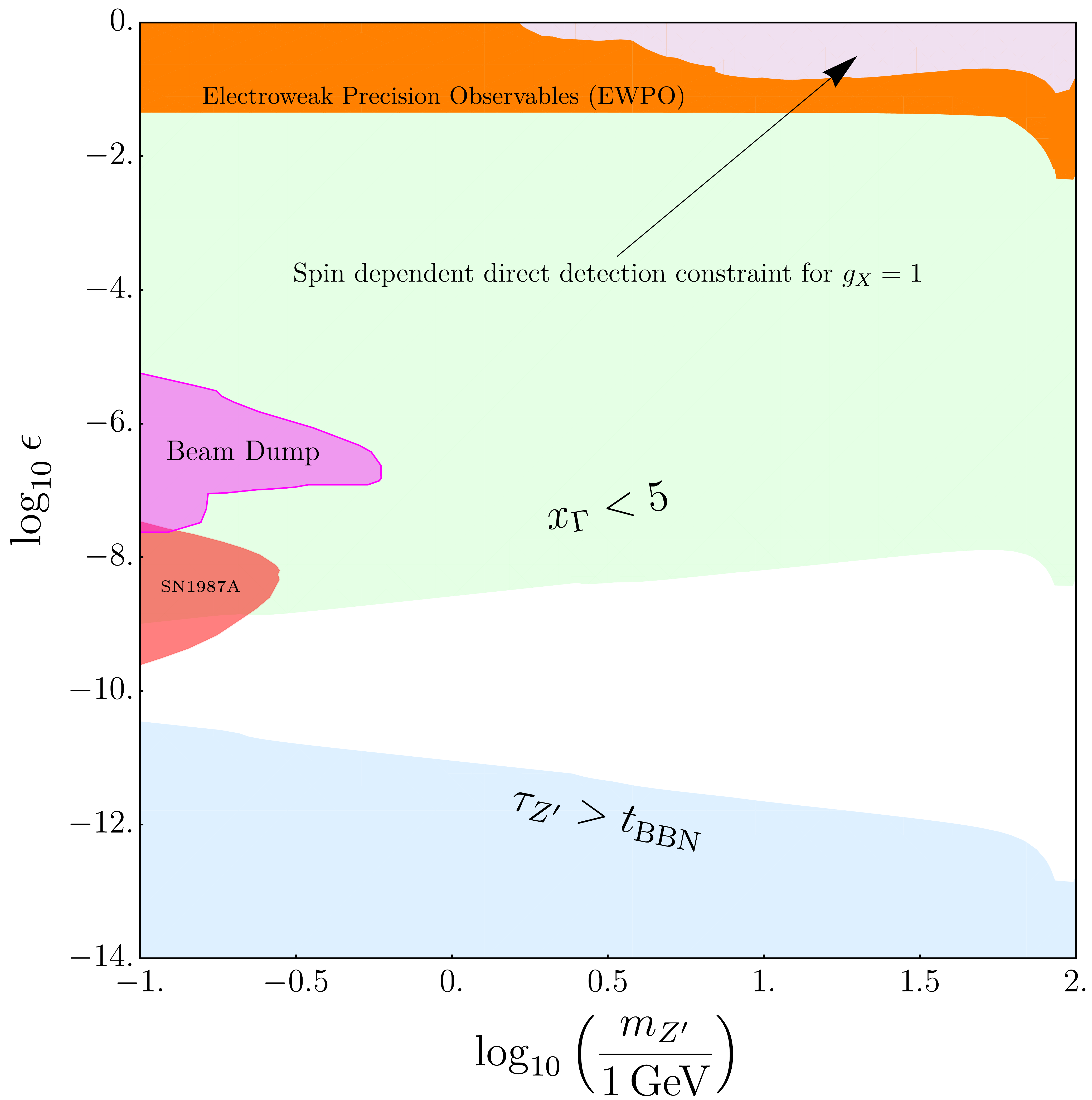}
	\includegraphics[height=7cm,width=8.0cm]{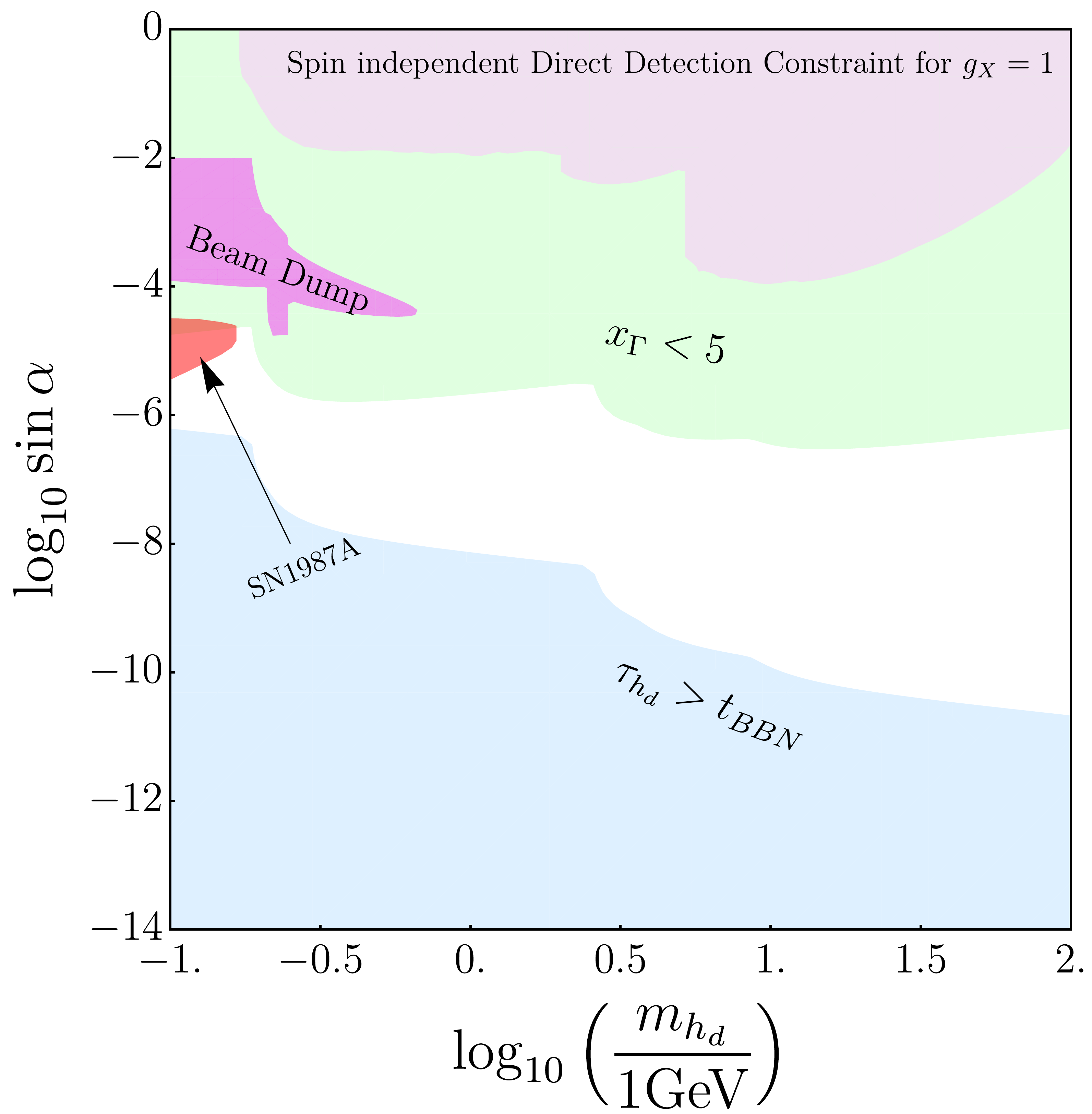}
	\caption{\textbf{\textit{Left panel:}} Allowed parameter space in the 
	$m_{\zp}-\epsilon$ plane. The light blue shaded region is disallowed from the BBN and
	the light green region is disallowed from the requirement of 
	a kinetically decoupled dark sector. The Magenta and red 
	shaded regions are disallowed from Beam dump experiments
        and SN 1987A cooling respectively. 
	The Constraints from EWPO and direct detection cross section are
	shown by the orange and the purple coloured regions respectively.
	\textbf{\textit{Right panel:}} Allowed parameter space 
	in $m_{\hd}-\sin \alpha$ plane. Colour code
	is same as the left panel.}
	\label{Fig:cosmo}
\end{figure}

The presence of kinetic mixing parameter $\epsilon$ 
can alter the decay width of $Z$ boson and also
all the couplings involving $Z$ boson. To study the
constraints from Electroweak Precision Observables (EWPO),
we calculate the $S$ and $T$ parameters for our model in terms of
the measured $Z$ boson mass and the Weinberg angle by comparing
our Lagrangian with the effective Lagrangian
formulation of $Z\bar{f}f$ current \cite{Babu:1997st,Lao:2020inc}. 
The best fit values of $S =  0.06 \pm 0.09$ 
and $T = 0.10 \pm +0.07$ \cite{Baak:2014ora} are used
in deriving the constraint and it has been shown 
in the left panel of Fig.\,\,\ref{Fig:cosmo} by the
orange coloured contour.
\subsubsection{Dark matter Self Interaction}
\label{SIDM}
We have calculated the momentum transfer cross section $\sigma_T$
for DM self interaction processes mediated by $\zp$ and $\hd$.
For multicomponent DM, we define an effective momentum transfer
cross section $\sigma_T^{\rm eff} = f_{\x_i}^2 \sigma_T$ 
($f_{\x_i}$ is defined in Section \ref{DD})
and from the Bullet cluster observation it is bounded as 
$\sigma_T^{\rm eff} < 1.25 \, \rm cm^2 (m_\x/\rm g)$ \cite{Randall:2007ph}.
Using this relation, we have found 
$g_X < 1.87 \times 10^4 \,(m_\x/1 \, \rm GeV)^{3/4}$ and
our parameter space of interest is well below the exclusion
limit.
\subsection{Numerical results}
\label{sec:Nres}
\begin{figure}[h!]
\centering
\includegraphics[height=4.5cm,width=5.5cm]{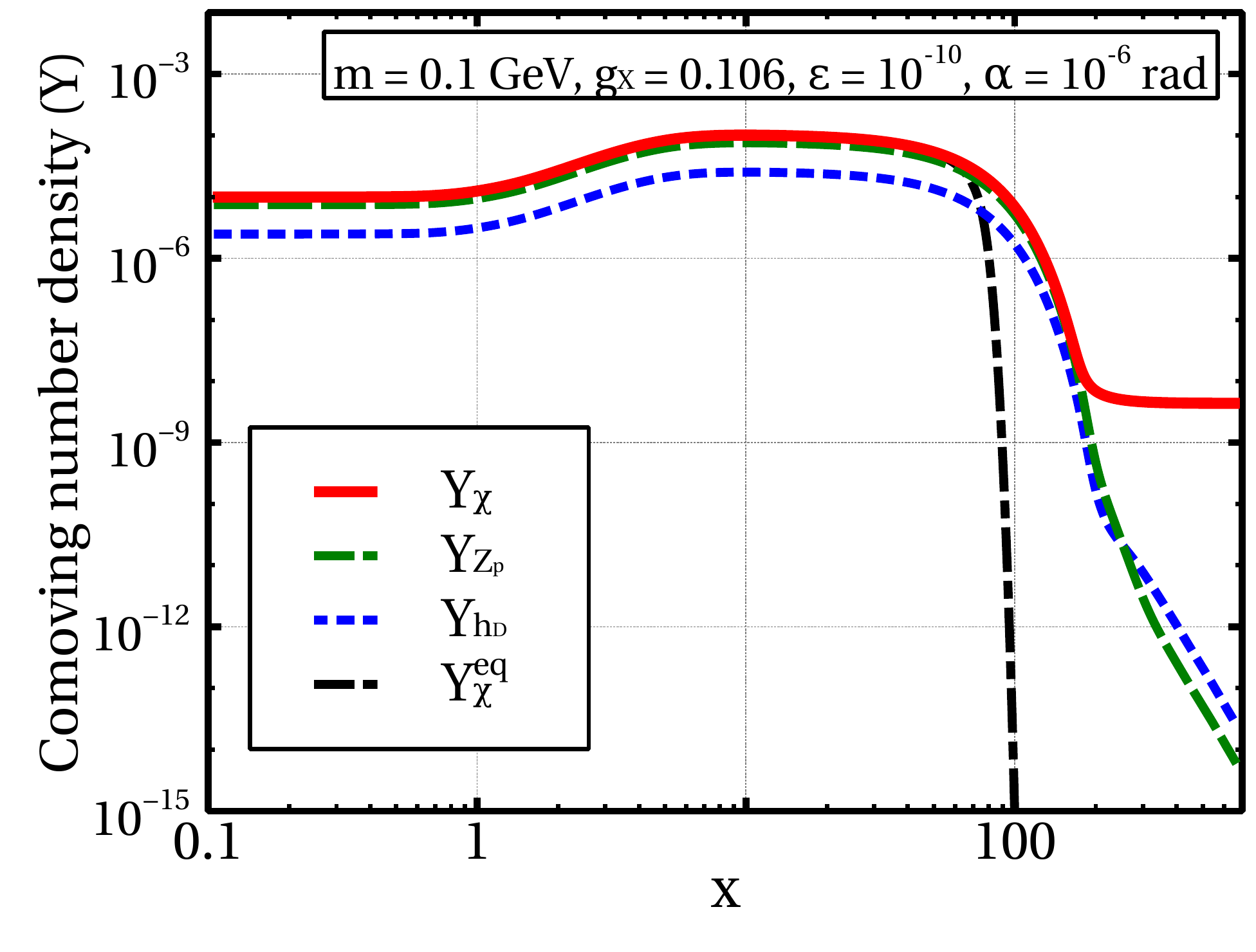}
\includegraphics[height=4.5cm,width=5.25cm]{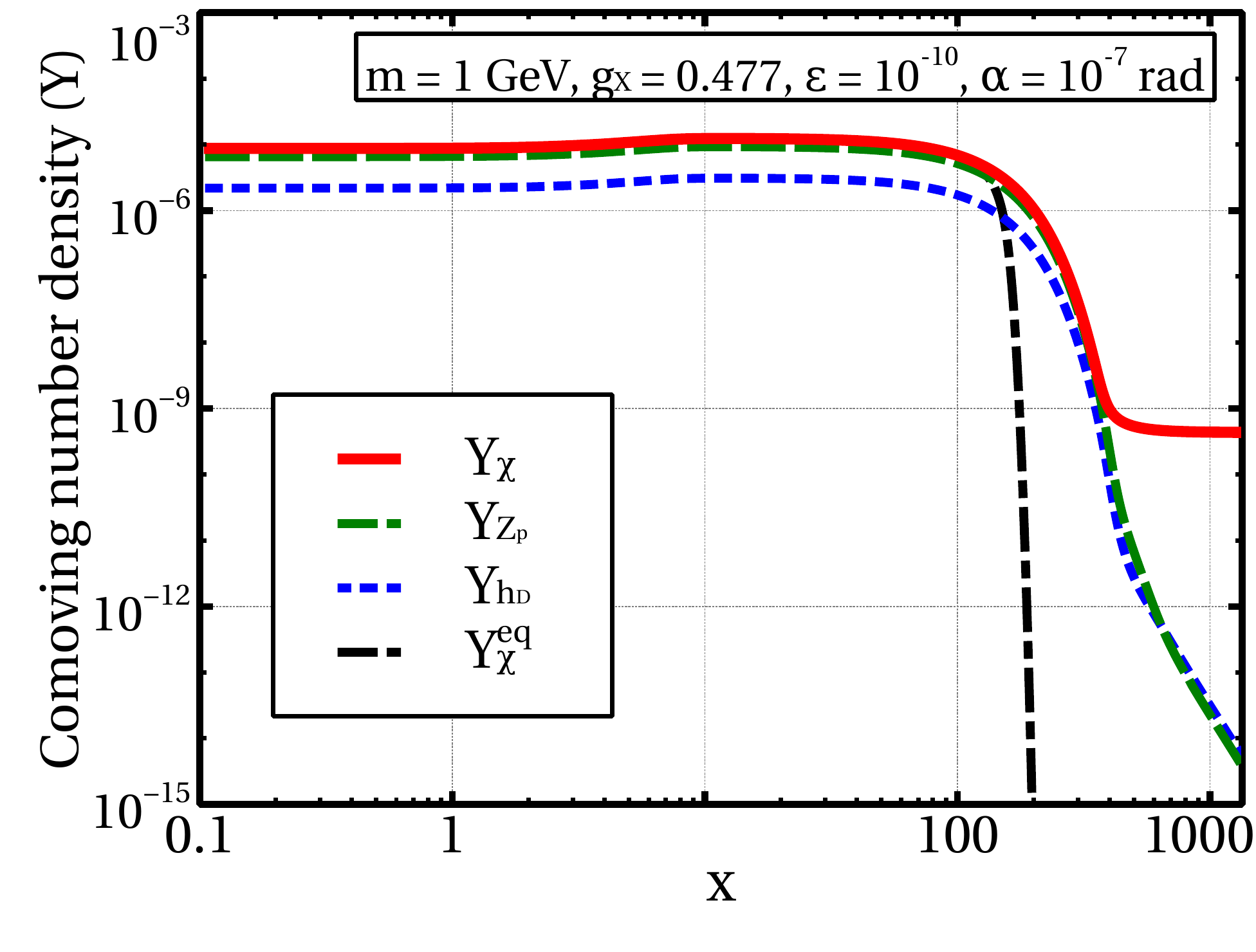}
\includegraphics[height=4.5cm,width=5.5cm]{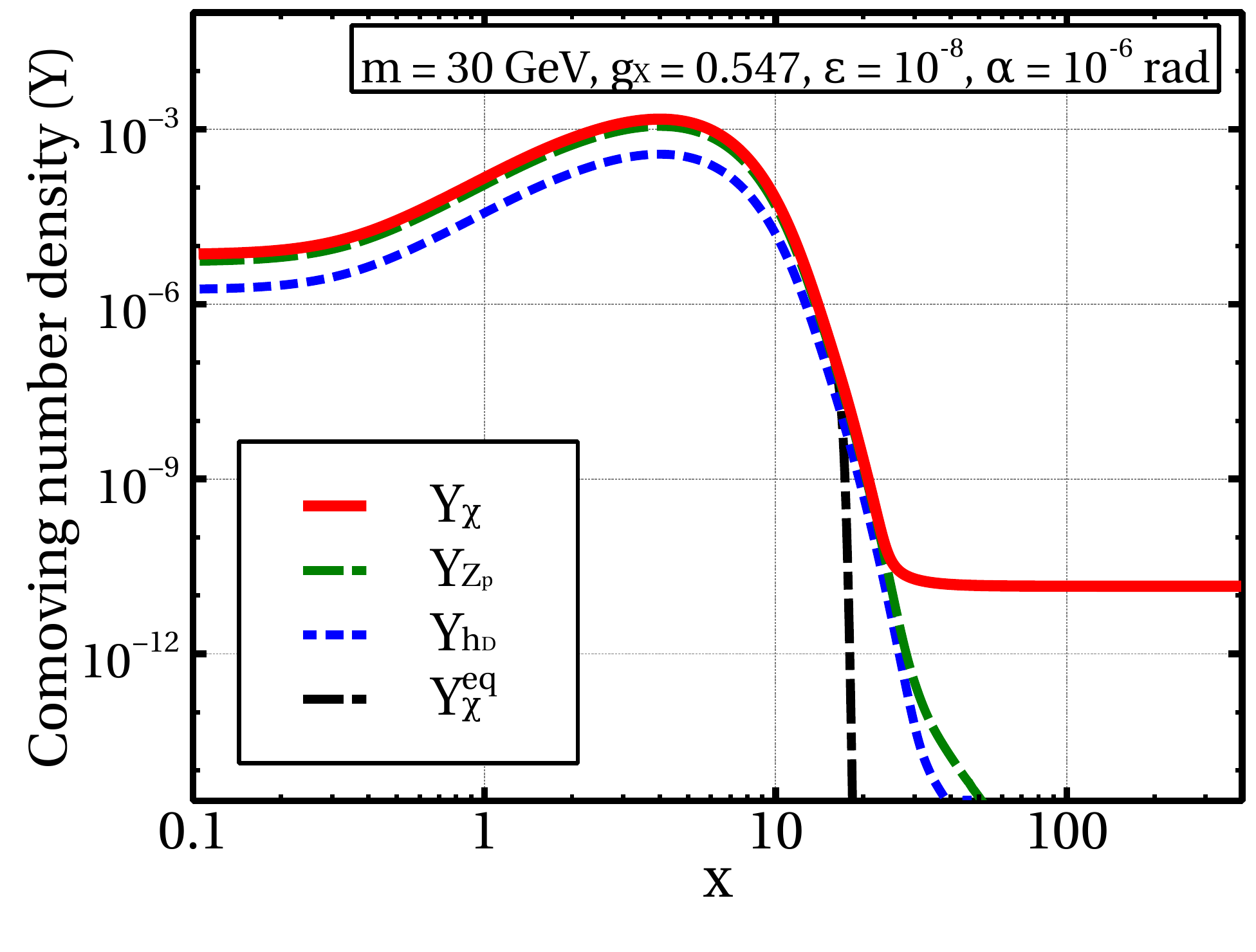}\\
\vskip 0.3in
\includegraphics[height=4.5cm,width=5.5cm]{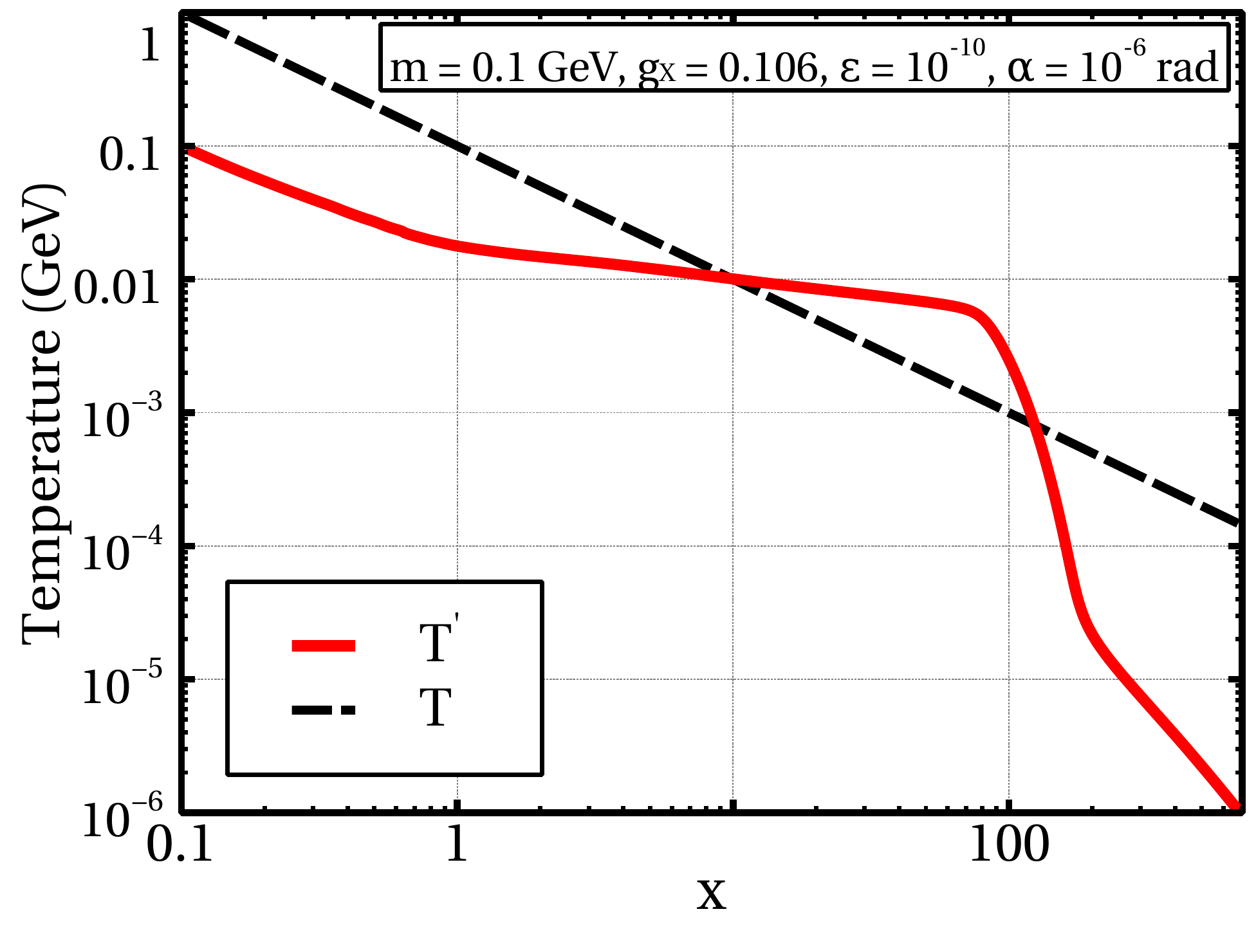}
\includegraphics[height=4.5cm,width=5.25cm]{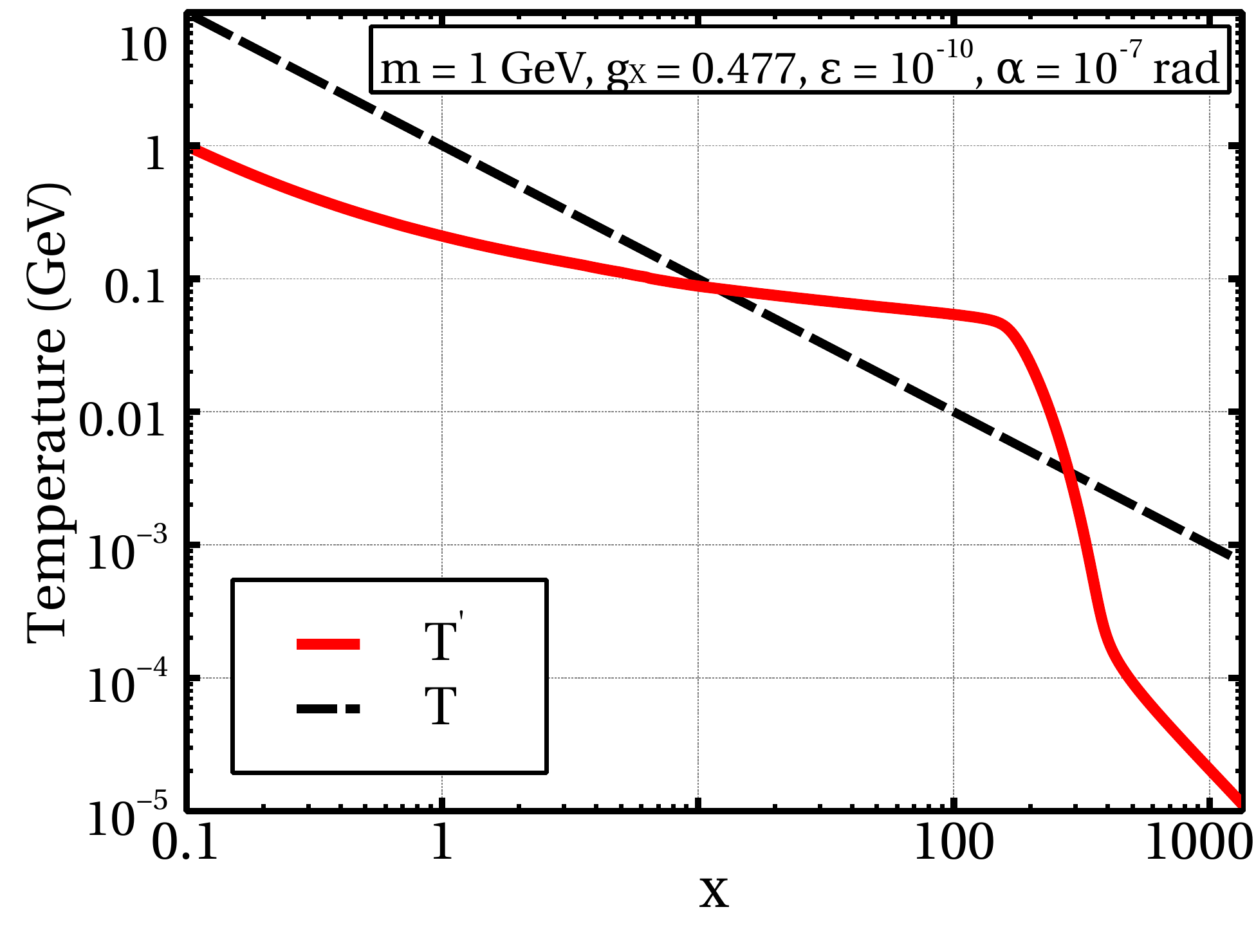}
\includegraphics[height=4.5cm,width=5.5cm]{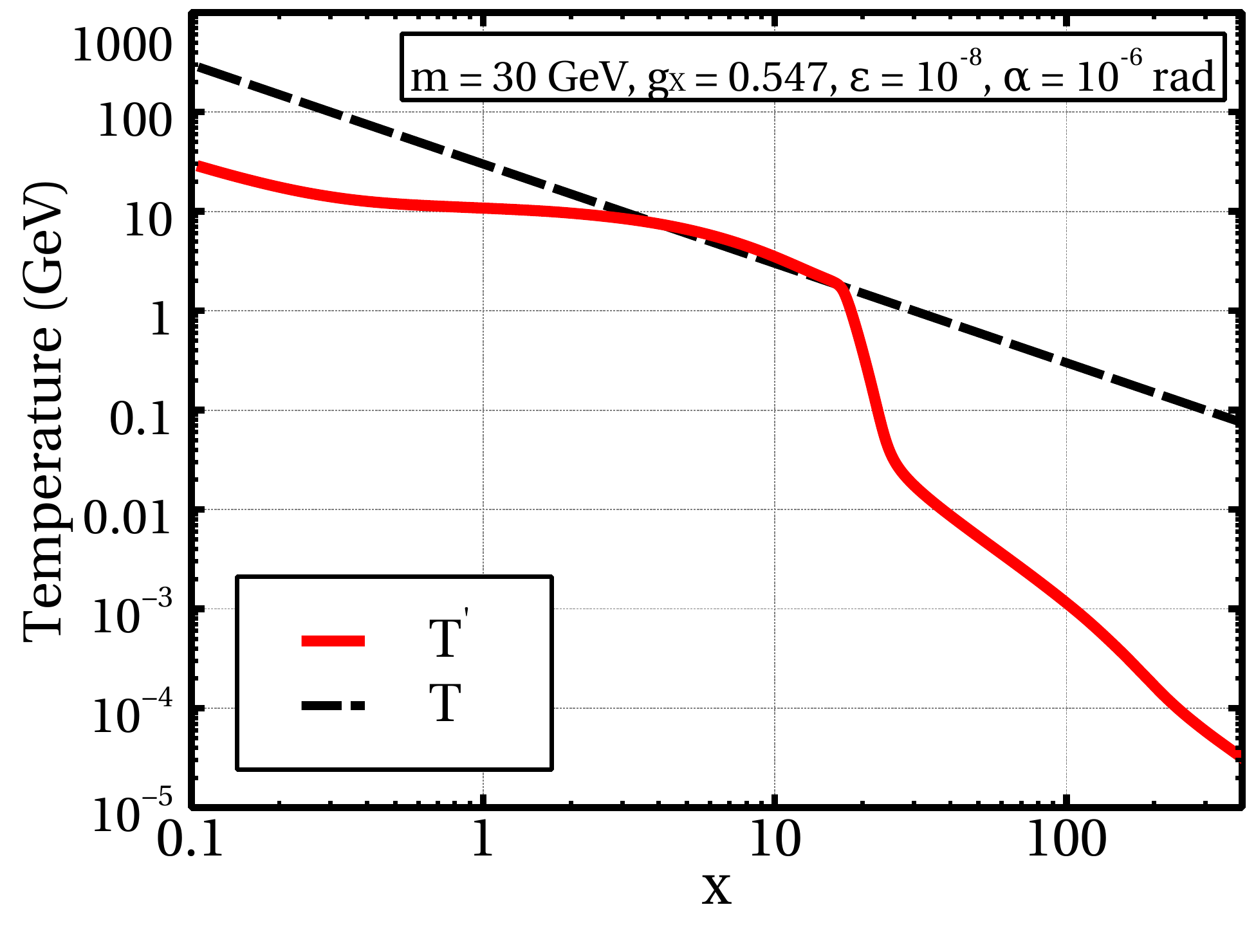}
\caption{\textbf{\textit{Upper panel:}} Variation of comoving number densities of $\chio + \chit$,
$\zp$ and $\hd$ with $x=m/T$ for three different values of masses $m = 0.1$ GeV, 1 GeV
and 30 GeV. \textbf{\textit{Lower panel:}} Evolution of dark sector
temperature $\tp$ with $x$ for same choices 
of masses. For reference, we have shown the SM temperature
in each plot by the black dashed line.}
\label{Fig:line_plot_1}
\end{figure}

We have numerically solved four coupled Boltzmann equations given in
Eq.\,\,\ref{Ydm-BE}-\ref{xi-BE} to find the evolution
of comoving number density of each dark sector species
with temperature $\tp$. 
The initial condition\footnote{The choice of
initial condition is by no means unique. However we have checked 
that the final relic abundance is independent
of the initial condition.}
for solving these equations is at $x=0.1$, $\xi=0.1$ (i.e. $x^\prime =1$)
and $Y_{i}=Y^{\rm eq}_{i}$. We have checked that with
$g_X \sim 1$ both sectors are kinetically decoupled
from each other at $x=0.1$ for the considered ranges of portal
couplings. Using the solutions of
coupled Boltzmann equations, one can easily estimate
the observable parameter namely the dark matter relic
density from the following well known relationship $\Omega_{\x}h^2=
2.755 \times 10^{8}\,\left(m_{\x}/{\rm GeV}\right)\,Y^0_{\x}$, where
$Y^0_{\x}$ is the asymptotic value of $Y_{\x}$ at the present era. 
The numerical results are presented in Fig.\,\ref{Fig:line_plot_1}.
As we have discussed in the Section\,\,\ref{sec:dynamics} that the degeneracy among
species of the dark sector leads to a completely different
freeze-out dynamics compared to the usual thermal freeze-out
of dark matter and this has been known as the co-decaying
framework. Therefore, in this numerical analysis we have
considered mass degeneracy in the dark sector by appropriately
tuning (see footnote 3) the relevant couplings. In the upper panel of 
Fig.\,\,\ref{Fig:line_plot_1}, we have shown that how
the comoving number densities for $\chio+\chit$, $\zp$ and $\hd$
are changing as a function of\,\,$x$ (inverse of $T$)
for three different values of $m=0.1$ GeV,
1 GeV and 30 GeV respectively. Here, in each
plot, red solid line, green dashed line and blue dotted line
represent $Y_{\x}$, $Y_{\zp}$ and $Y_{\hd}$ respectively.
The equilibrium value of dark matter comoving number
density $Y^{\rm eq}_{\chi}$ is also shown by the black
dash-dotted line.  Initially, the number density of dark matter
follows the equilibrium number density. 
Thereafter it deviates from equilibrium, starts depleting and finally 
freezes out to a particular density. The time of departure from the equilibrium
depends on when the decay modes of $\zp$ and/or $\hd$
into the SM particles open up. This creates chemical
imbalance in the $2\ra2$ interactions between $\chii$
and other species ($\zp$, $\hd$) leading to the freeze-out
of $\chio$ and $\chit$. For example, in the left and
middle plot for $m=0.1$ GeV and 1 GeV respectively,
the freeze-out occurs at larger values of $x$ ($x > 100$)
and much larger values of $x^\prime$ (depending on the
parameter $\xi$ which is at least 0.1 or even less
for $x>100$ as seen from the plots in the lower panel)
while in the right plot dark matter
freezes-out relatively early. This is because,
in the plot for $m=30$ GeV we have chosen higher
values of $g_X$ and $\epsilon$ and hence decay
of $\zp$ starts much earlier compared to the other
two cases. The relevant model parameters
such as mass ($m$), gauge coupling ($g_{X}$), kinetic
mixing parameter ($\epsilon$) and scalar sector
mixing angle ($\alpha$) in all three plots are adopted
in such a way so that we achieve the right dark matter
relic abundance.   

\begin{figure}[h!]
\centering
\includegraphics[height=6cm, width=9cm,angle=0]{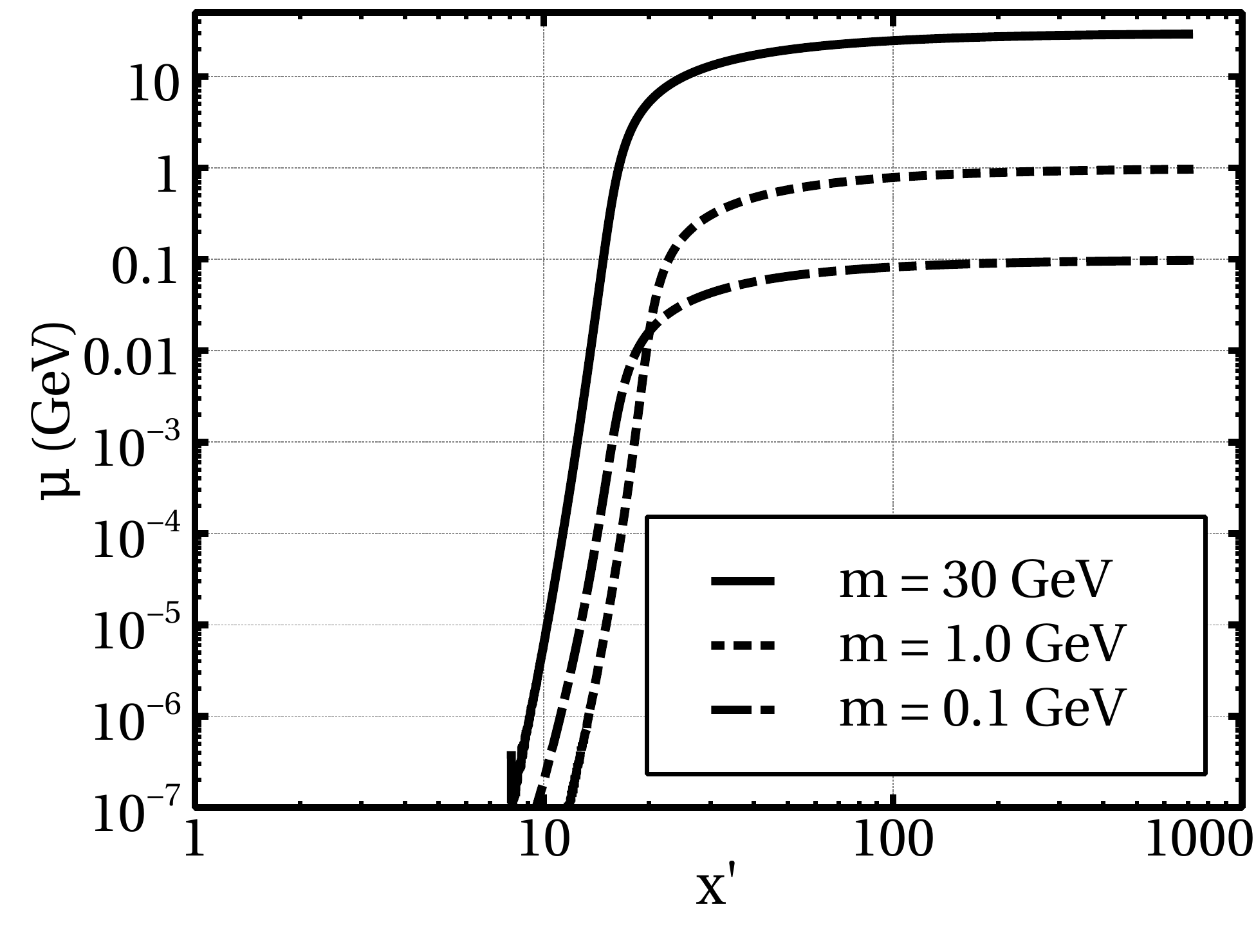}
\caption{Variation of chemical potential with $x^\prime$ for three different values of
$m$.}
\label{Fig:chem_pot}
\end{figure}

The evolution of dark sector temperature
with $x$ for the same choices of masses
have been depicted in the lower panel of Fig.\,\,\ref{Fig:line_plot_1}.
The trivial variation of the SM temperature $T$ with $x$ (inverse of $T$)
is also shown by a black dashed line in each plot. From
these plots it is seen that at first $\tp$ decreases with
$x$ and thereafter the variation of $\tp$ with $x$ becomes
almost constant between $1 \lesssim x\lesssim 100$ (in the
left and middle plot) and $0.3 \lesssim x\lesssim 10$ (in
the right plot) respectively. This is the ``phase of cannibalism''
where all the inelastic $3\ra2$ scatterings among the dark sector
species are active and generating excess heat in the dark sector.
A detailed discussion about these inelastic scatterings are given in
Appendix\,\,\ref{App:B2} along with all possible Feynman diagrams
in Figs.\,\,\ref{Fig:3to2_2}-\ref{Fig:3to2_11}. As mentioned earlier, the heat generated
due to $3\ra 2$ scatterings spreads equally over the
entire dark sector by the $2\ra 2$ scatterings so
that all species share a common temperature. 
However, expansion of the Universe red-shifts the
temperature and it is always present in the background. Due to these
two opposite effects, the dark sector temperature becomes a
slowly varying function of $x$. Actually, in this regime $\tp$ varies
logarithmically with $T$ as shown in Eq.\,\ref{tp_cannibal}. 
Moreover, it is also seen from these plots in the lower panel
of Fig.\,\,\ref{Fig:line_plot_1} that after this plateau
region, $\tp$ sharply decreases with $x$ as a result of both
$2\ra 2$ scatterings as well as the Universe's expansion.
Finally, for large $x$ when all the scatterings are frozen-out,
$\tp$ redshifts as non-relativistic matter 
due to the effect of expansion of the Universe.

We would like to spend a few sentences on the chemical
potential of species in the dark sector. From the earlier
discussion in Section\,\,\ref{sec:dynamics}, we know that
during the number changing interactions by $3\ra 2$
processes, the chemical potential of each species
is zero. After freeze-out of these inelastic scatterings
at a temperature $\tp_c$, each species develops a
nonzero chemical potential which helps them to get
rid of the exponential suppression in number density
even if they become non-relativistic. We have derived
the expression of chemical potential ($\mu(\tp)$) for
$\tp_c>\tp>\tp_{\Gamma}$ (Eq\,\,\ref{mutp}) using the conservation
of entropy and number density in a co-moving volume. It
shows that $\mu(\tp)$ of a species first increases sharply
as $\tp$ departs from $\tp_c$ and thereafter
saturates to the mass $m$ of that particular
species when $\tp \ll \tp_c$. The exactly similar nature
for $\mu$ has been obtained from the numerical
simulations by solving the Boltzmann equations and it
has been shown in Fig.\,\,\ref{Fig:chem_pot}. Here, we have
shown the variation of $\mu$ with $x^\prime$ (inverse of $\tp$)
for three different values of $m=0.1$ GeV, 1 GeV
and 30 GeV respectively. We have computed $\mu$ from
the well known relationship between $\mu$ and $\tp$
for a species $i$ as 
$\mu = \tp\ln\left(\dfrac{n_i}{n^{\rm eq}_i}\right)$ where,
the number density $n_i$ is obtained by solving
the Boltzmann equations. We have
noticed that the analytical expression given in Eq.\,\,\ref{mutp}
remains valid between $\tp_c$ and $\tp_{\Gamma}$. However,
the results we have found from the numerical simulations
show that $\mu$ continues to maintain its saturation
value well beyond the temperature $\tp_{\Gamma}$ at
which decays of $\zp$ and $\hd$ into the SM particles
begin.

\begin{figure}[h!]
\centering
\includegraphics[height=6cm,width=8cm]
{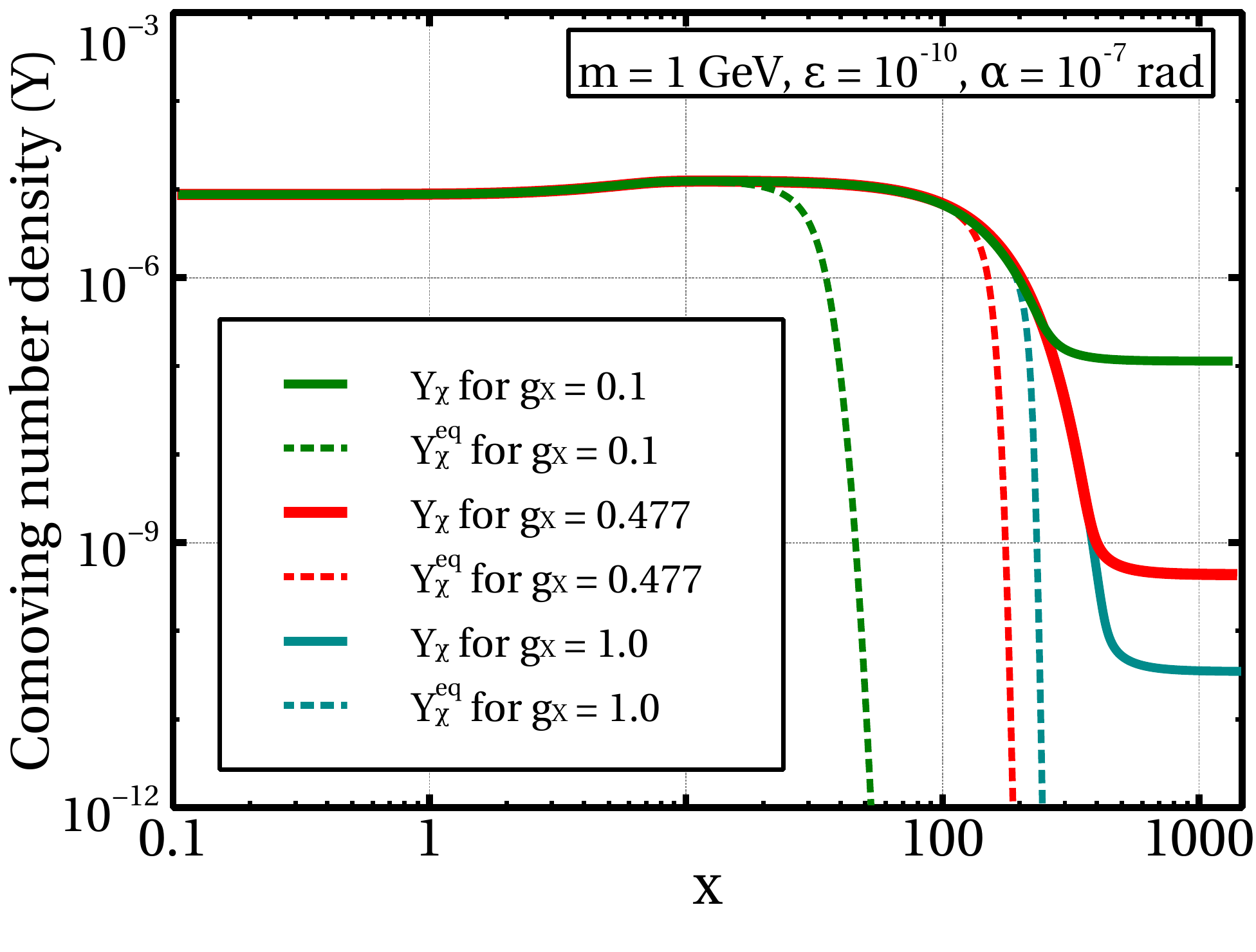}
\includegraphics[height=6cm,width=8cm]
{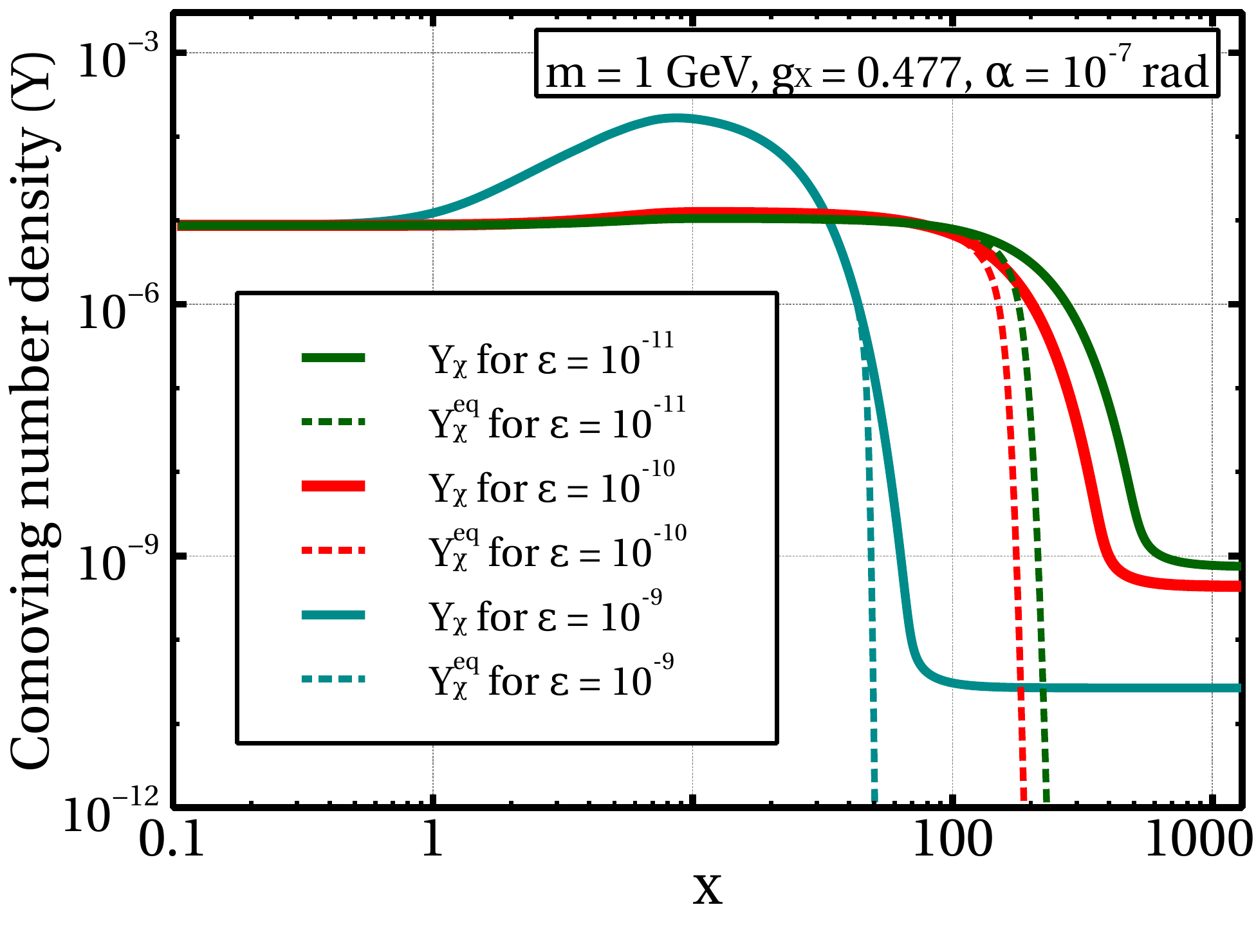}\\
\vskip 0.2in
\includegraphics[height=6cm,width=8cm]
{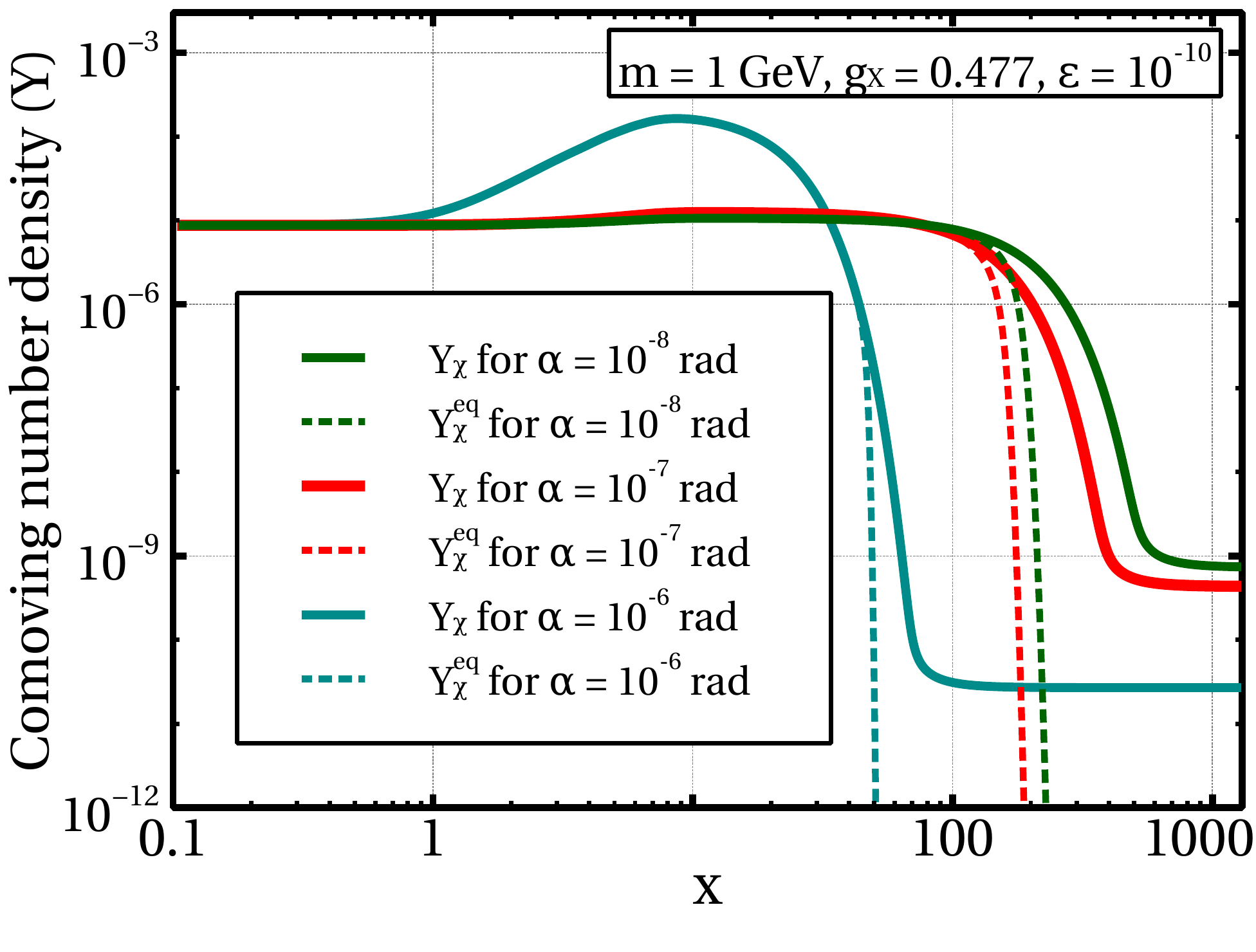}
\includegraphics[height=6cm,width=8cm]
{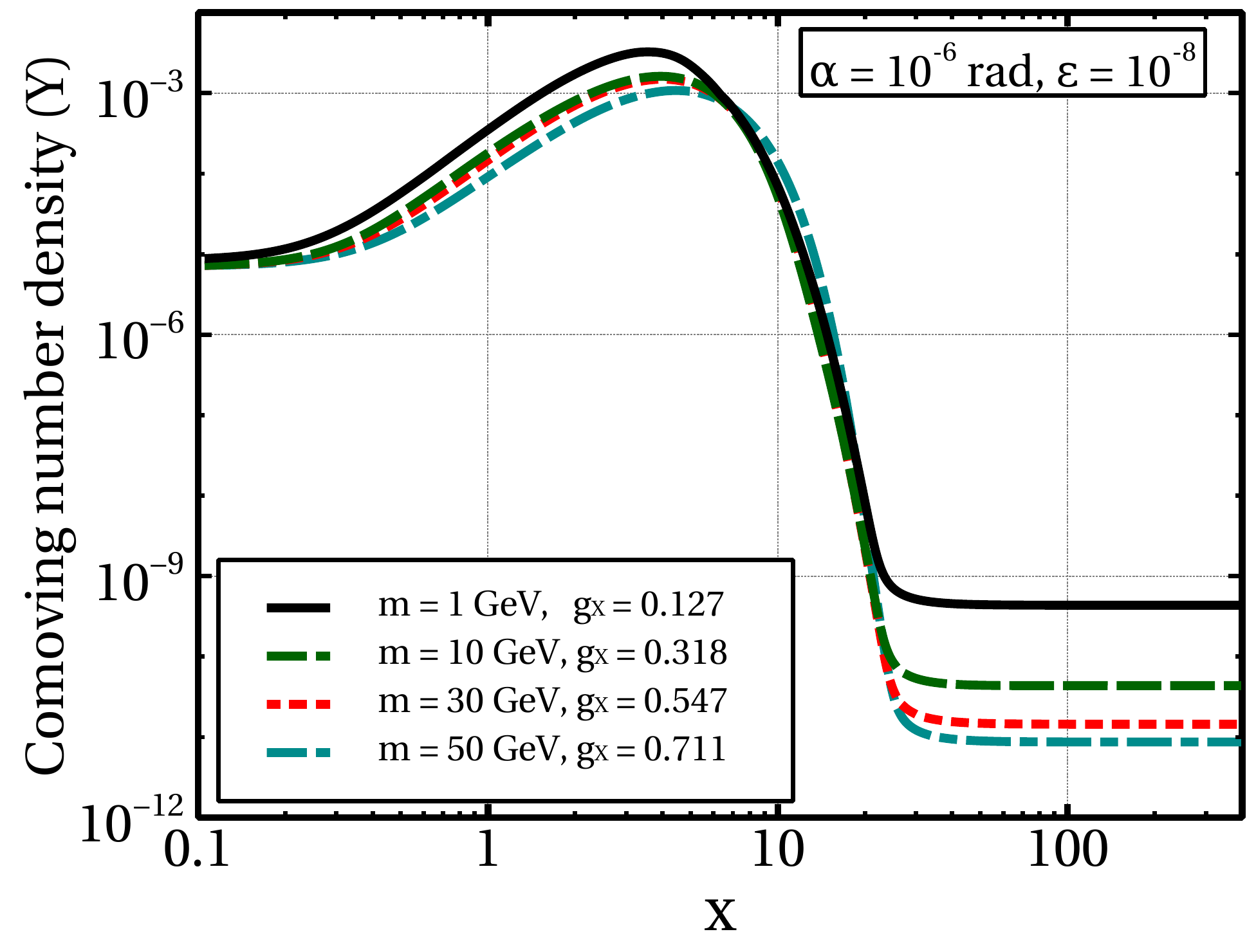}
\caption{Dependence of various model parameters on the
comoving number density.}
\label{Fig:line_plot_2}
\end{figure}
\begin{figure}[h!]
\centering
\includegraphics[height=8cm,width=12cm]{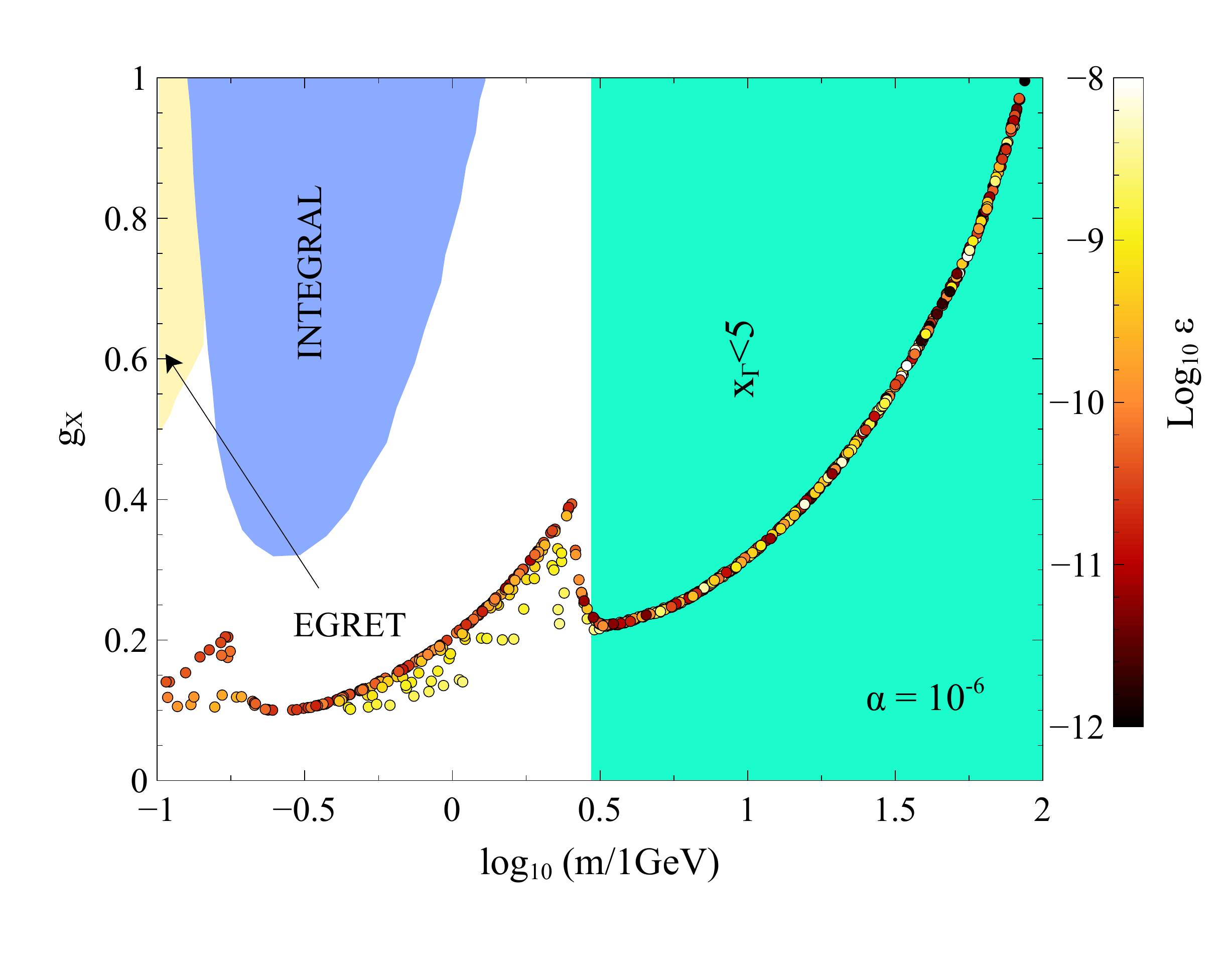}
\caption{Allowed values of $g_X$ while varying $m$ between
0.1 GeV to 100 GeV for a fixed value of scalar mixing angle
$\alpha$. The colour bar indicates the corresponding
	allowed range of $\epsilon$. The light yellow and the light blue
	regions are disallowed from indirect detection constraints coming from
	diffuse $\gamma$ ray background. A detailed
	discussion on indirect detection	is
	given in Section\,\,\ref{sec:indirect}.}
\label{Fig:scan_gx_m}
\end{figure}

In Fig.\,\,\ref{Fig:line_plot_2}, we have demonstrated the
dependence of important model parameters on the comoving
number density. In the first plot at the
top-left position, we have shown the effect of gauge
coupling $g_X$ on $Y_{\x}$ for three different
values of $g_X = 0.1$ (green solid line), 0.477 (red solid line)
and 1.0 (cyan solid line) respectively. The corresponding equilibrium
values of $Y_{\x}$ have been indicated by the dashed lines
of same colors. Note that $g_X = 0.477$ corresponds to correct relic abundance.
This plot clearly shows that increasing $g_X$ leads
to much suppressed final abundance and late freeze-out
of dark matter. For example, changing $g_X$ by one order
of magnitude ($g_X = 0.1$ to $g_X= 1.0$) reduces $Y_{\x}$
by more that three orders i.e from $Y_{\x} = 10^{-7}$ to
$Y_{\x} < 10^{-10}$. In the second plot at the
top-right position shows the effect of kinetic
mixing parameter $\epsilon$ on $Y_{\x}$. Here
also we have chosen three different values
of $\epsilon$ namely $\epsilon=10^{-11}$ (green),
$10^{-10}$ (red) and $10^{-9}$ (cyan) respectively.
From this plot one can notice that the impact of $\epsilon$
on $Y_{\x}$ depends on the value of $\epsilon$. For example,
altering $\epsilon$ from $10^{-11}$ to $10^{-10}$ does not
reflect a significant change in $Y_{\x}$. However, increasing
$\epsilon$ further by one order of magnitude (i.e. from $10^{-10}$
to $10^{-9}$) results in a noticeable difference in the
final abundance. Moreover, in the latter case much earlier
freeze-out of dark matter has occurred. This is mainly
due to the fact that increasing $\epsilon$ enhances the
decay width $\Gamma_{\zp}$ which reduces the lifetime of
$\zp$. Hence, $\zp$ decays much earlier into the SM particles
and creates enough chemical imbalance in the dark sector which
thereafter leads to earlier freeze-out of the dark matter. The
dependence of another portal coupling $\alpha$ on $Y_{\x}$ is
exactly similar to the earlier case of $\epsilon$.
It is shown in the bottom-left plot. Analogous
to the previous cases, we have adopted three different values
of scalar mixing angle $\alpha = 10^{-8}$ rad (green),
$10^{-7}$ rad (red) and $10^{-6}$ rad (cyan) respectively.
The equilibrium values of $Y_{\x}$ in each case has been
shown by the dashed line. Here also we see much earlier
dark matter freeze-out and more suppressed final abundance
when we increase $\alpha$ from $10^{-7}$ rad to $10^{-6}$ rad.
Finally, in the bottom-right plot we show dependence of
mass on $Y_{\x}$. In this plot, we have considered four
different values of $m$ as indicated in the plot legend
and portal coupling parameters are kept fixed to a
particular value as $\epsilon = 10^{-8}$ and $\alpha= 10^{-6}$ rad.
Moreover, we have also varied the gauge coupling $g_X$
so that the final value of $Y_{\x}$ for each value of $m$
reproduces the correct relic abundance. We observe
that for a fixed portal couplings as we increase $m$,
we require higher values of $g_X$ to achieve
$\Omega_{\x} h^2$ in the right ballpark value. This
parametric dependence of $Y_{\x}$ on $m$ and
$g_{X}$ will be more clear when we analyse Fig.\,\,\ref{Fig:scan_gx_m}.     
\begin{figure}[h!]
\centering
\includegraphics[height=6cm,width=8.cm]
{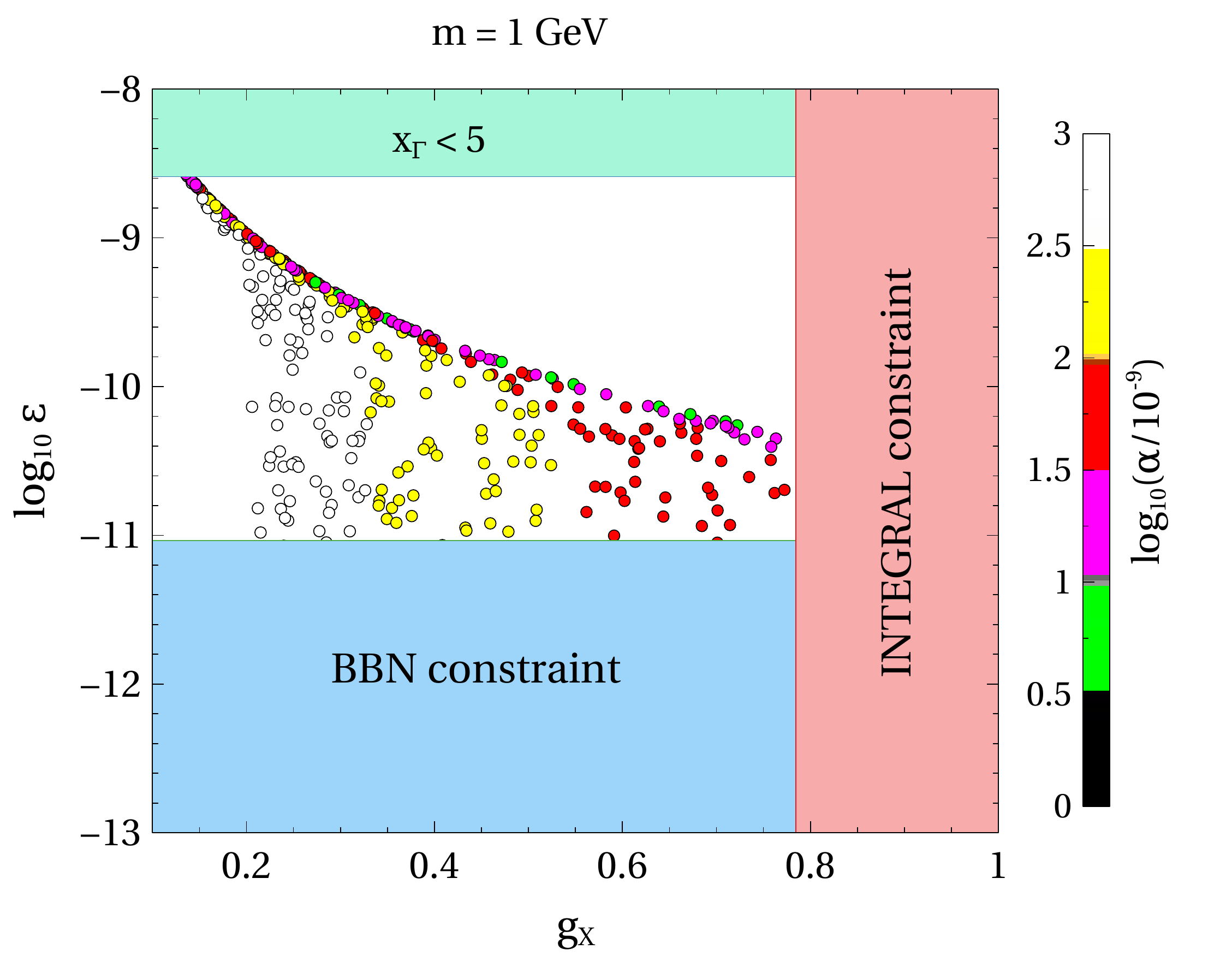}
\includegraphics[height=6cm,width=8.cm]
{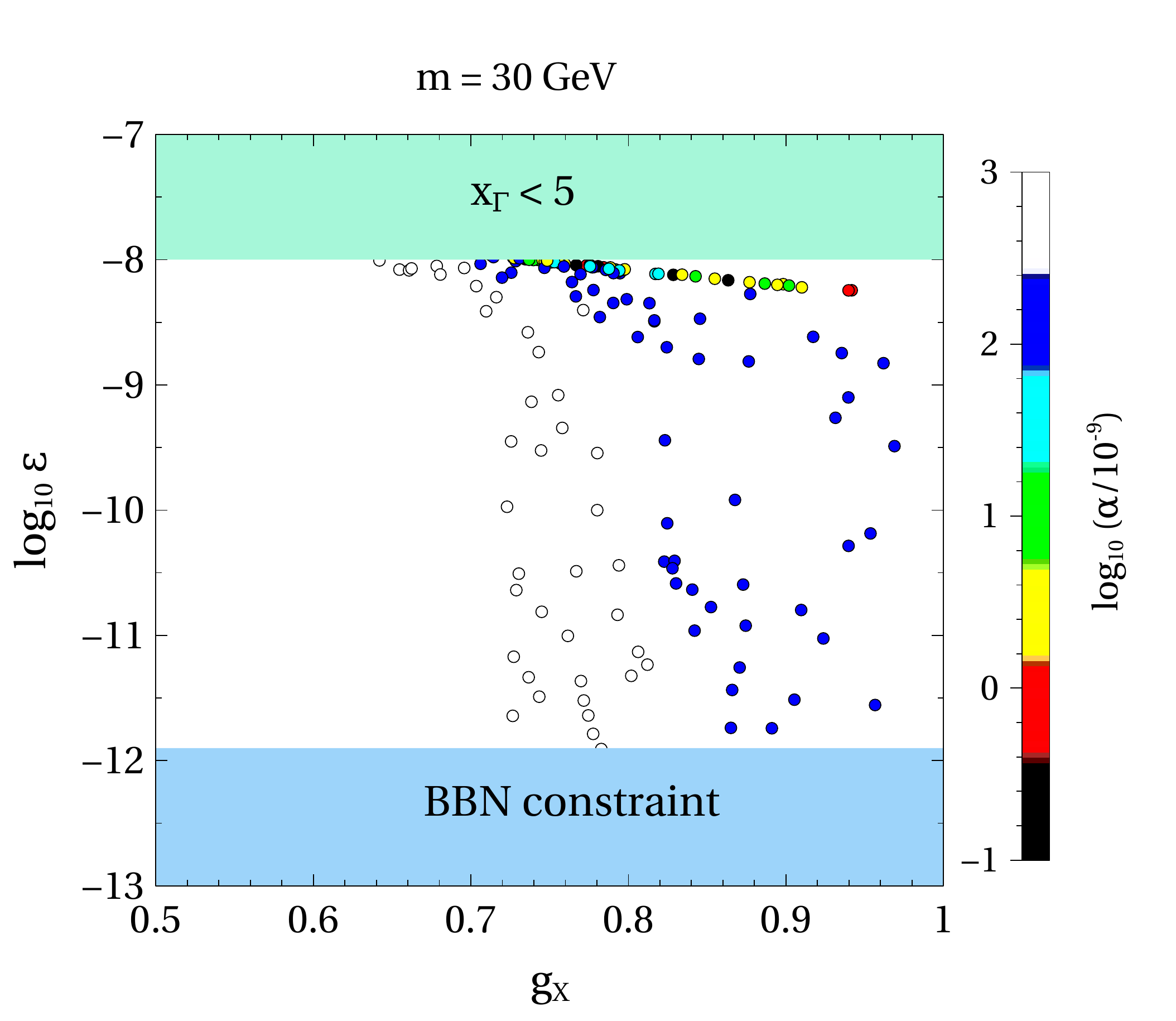}\\
\vskip 0.3in
\includegraphics[height=6cm,width=8.cm]
{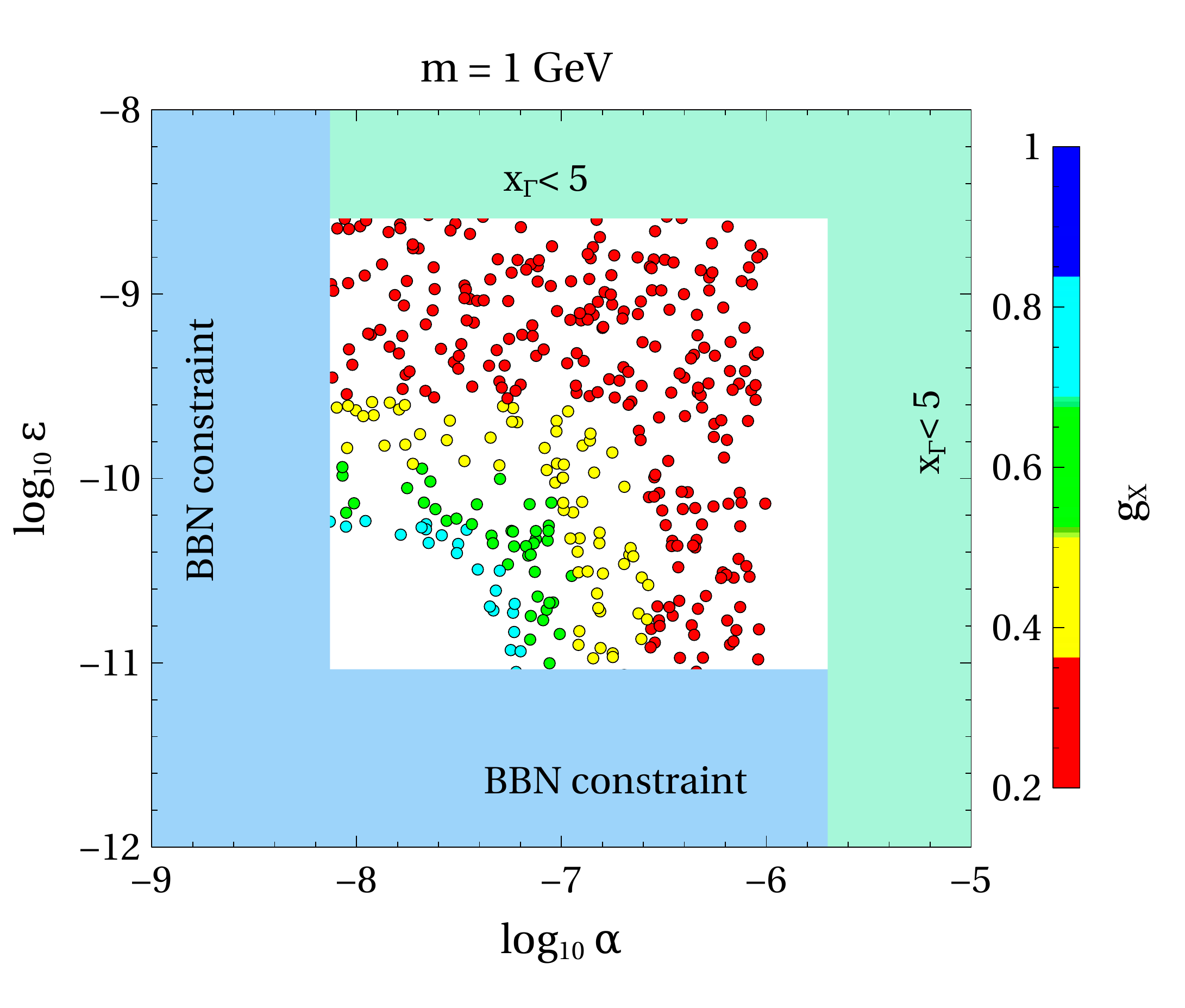}
\includegraphics[height=6cm,width=8.cm]
{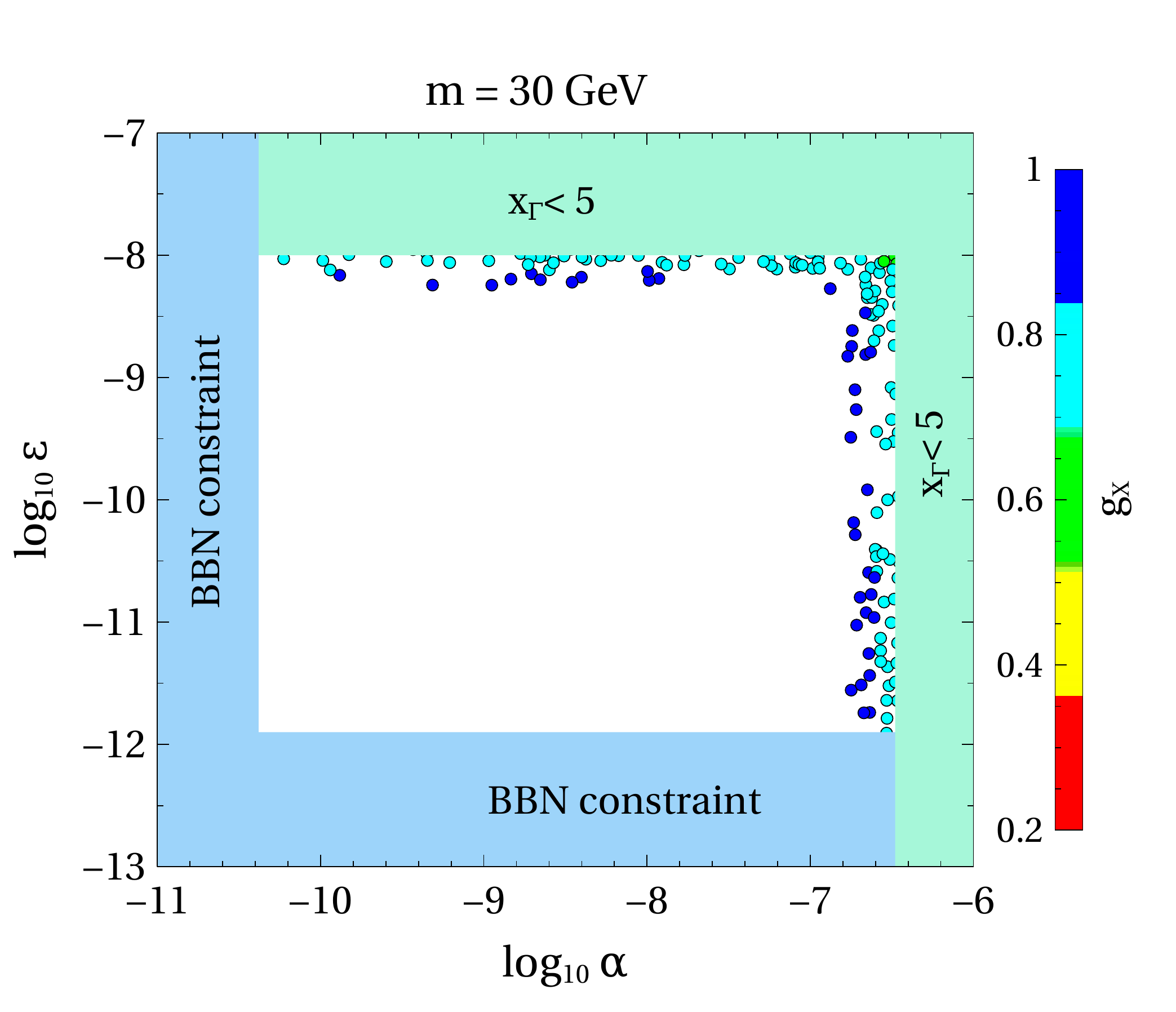}\\
\caption{\textbf{\textit{Upper panel:}} Parameter space in
$\epsilon-g_X$ plane reproducing the observed relic
density in $2\sigma$ range for $m=1$ GeV and 30 GeV.
Green and blue shaded regions are disalllowed from thermalisation
condition and BBN constraint respectively.
\textbf{\textit{Lower panel:}} Allowed $\epsilon-\alpha$ parameter
	space for $m=1$ GeV and 30 GeV. Color codes are same as upper panel.}
\label{Fig:scan_gx_eps_al}
\end{figure}

In both the panels of Fig.\,\,\ref{Fig:scan_gx_eps_al} we present
the allowed parameter space that we have obtained by solving four
coupled Boltzmann equations and comparing our results with the
observed value of relic density in $2\sigma$ range i.e. $\Omega_{\rm DM} h^2
= 0.120 \pm 0.002$ \cite{Aghanim:2018eyx}. In the upper panel,
we have shown the $\epsilon-g_{X}$ parameter space for two distinct
dark matter masses 1 GeV and 30 GeV respectively. The corresponding
variation of the third parameter $\alpha$ has been indicated by
the respective colour bar in each plot. The constraints coming
from BBN, prevention of thermalisation of the dark sector with
the SM bath and measurement of diffuse $\gamma$-ray fluxes by
the INTEGRAL space telescope have also been indicated.
From the plot in the top-left
position, we can notice that allowed range of $\epsilon$
decreases sharply as $g_{X}$ increases from $0.1$ to 1.0.
Additionally, the production of correct relic abundance
at the present era demands smaller $\alpha$ for higher $g_{X}$.
For example, we require $\alpha \simeq 10^{-9}$ rad when
$g_{X}\simeq 0.8$ while higher mixing angles $\alpha\sim 10^{-6}$ rad
are necessary for smaller couplings ($g_X\simeq 0.2$). More or less
similar feature has also been observed for the other dark matter mass ($m=30$ GeV) 
in the top-right plot. In this case, reduction of parameter space
of $\epsilon$ for increasing $g_{X}$ is not so sharp and the
variation of $\alpha$ in the scattered region mostly concentrated
around $10^{-7}\,{\rm rad}\lesssim\alpha\lesssim 10^{-6}$ rad.
Moreover, unlike the plot for $m=1$ GeV, we have not
depicted any constraint in the $\epsilon-g_X$ plane
for $m=30$ GeV from INTEGRAL as in this case the
corresponding bound is on the higher values of
gauge coupling ($g_{X}>>1$).
The allowed regions in the $\epsilon-\alpha$
parameter space have been shown in the lower panel of
Fig.\,\,\ref{Fig:scan_gx_eps_al} for $m=1$ GeV
and 30 GeV respectively. We observe a definite
pattern in the allowed values of $\epsilon$ and $\alpha$ 
in order to satisfy the correct relic abundance of dark matter. 
In the bottom-left plot, almost entire range
of $\alpha$ allowed from BBN and thermalisation condition
is also allowed from the relic density criterion
for $\epsilon\gtrsim 10^{-10}$. It implies
that for these values of $\epsilon$ the chemical
equilibrium in the dark sector is disturbed mostly
by the decay of $\zp$. In contrast,
an opposite situation is also observed
when the scalar mixing angle $\alpha\gtrsim 10^{-7}$ rad.
In this case, the entire range of $\epsilon$ satisfies
relic density in $2\sigma$ range expressing
$\Gamma_{\hd}$ dominance in the freeze-out dynamics
of dark matter. In the intermediate region, decays of
both $\zp$ and $\hd$ are responsible for the
departure of chemical equilibrium in the dark sector.
The similar feature is seen for $m=30$ GeV (bottom-right plot),
however, in this case only very small portion of $\epsilon-\alpha$
plane is allowed. In both these plots, possible
variation of gauge coupling $g_X$ is indicated in
the respective colour bar. Similar to the plots
in the upper panel, here we have also shown
the other constraints in the $\epsilon-\alpha$ plane
from BBN and thermalisation using two
different colours.

\begin{figure}[h!]
\centering
\includegraphics[height=7cm,width=9cm,angle=0]{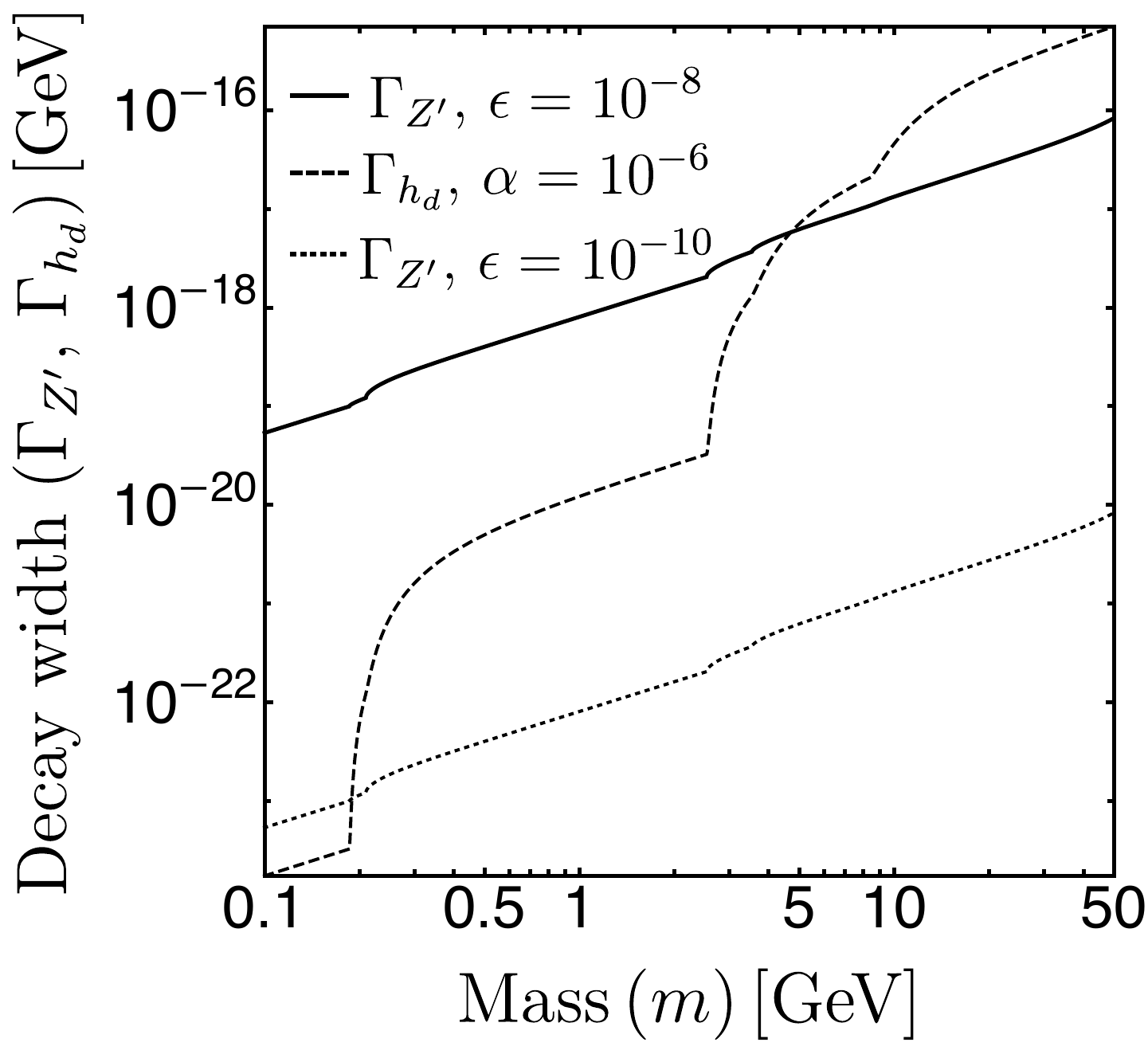}
\caption {Decay width of $\hd$ for 
$\alpha = 10^{-6} $ (dashed line) and 
$\zp$ for $\epsilon = 10^{-8}$ (solid line),
$10^{-10}$ (dotted line) as a function of mass ($m$).}
\label{Fig:decay-widths}
\end{figure}

Moreover, we have tried to find the allowed values of
$g_X$ when mass of the dark sector species varies
between 100 MeV to 100 GeV. The result has been shown in
Fig.\,\,\ref{Fig:scan_gx_m} where $\epsilon$ varies
between $10^{-12}$ to $10^{-8}$ and is indicated
by the colour bar. The green shaded region is the
disallowed as $x_\Gamma$ is less than five here
while the light yellow and the light blue shaded
regions are ruled-out from the measurements of
diffuse $\gamma$-ray background by EGRET and
INTEGRAL respectively. A detailed discussion on
the indirect detection prospects of the co-decaying
dark matter scenario has been given in Section\,\,\ref{sec:indirect}.
We have obtained this $m-g_{X}$
parameter space for a fixed value of scalar
mixing angle $\alpha= 10^{-6}$ rad. From this
plot we can see that generally as $m$ increases we need
higher values of $g_X$ (i.e. higher annihilation cross section) 
to satisfy the observed
relic abundance (within $2\sigma$ range) except for a couple 
of values of
$m$ near 200 MeV and 2 GeV respectively
where decay 
width\footnote{For $350 \, {\rm MeV}\lesssim m
\lesssim 5 \, {\rm GeV}$, the hadronic
decay modes of $\zp$ and $\hd$ are also present. 
Following \cite{Monin:2018lee, Cirelli:2016rnw}, 
we have checked the effect of these hadronic decay modes
on the relic density and we find that the impact of inclusion
of these hadronic decay modes is not very significant in $m-g_X$
plane since relic density depends on $g^4_X$.}
of $\zp$ and $\hd$ suddenly enhance
due to the crossing of kinematic thresholds for new
decay modes. In this particular case, the decay width $\Gamma_{\hd}$
increases substantially due to opening of $s {\bar s}$ and $c\bar{c}$
channels and it dominates the freeze-out process of
dark matter over $\Gamma_{\zp}$ for $\epsilon < 10^{-10}$
and $\alpha = 10^{-6}$ rad (see Fig.\,\,\ref{Fig:decay-widths}
and related discussions). The sudden enhancements
in $\Gamma_{\hd}$ are compensated by the curtailments
in dark matter annihilation cross sections
through $g_X$ as $\Omega_{\x} \sim \dfrac{m}{\langle \sigma {\rm v}\rangle\,
\sqrt{\Gamma_{\zp (\hd)}}}$ (see Appendix\,\,\ref{App:d} for
an approximate analytical expression of relic density).

Let us note that in Fig.\,\,\ref{Fig:scan_gx_m},
the threshold effects are prominent only for
$s\bar{s}$ and $c\bar{c}$ channels. This can be
understood clearly from Fig.\,\,\ref{Fig:decay-widths}.
In this figure we have plotted the total decay width of
$\hd$ ($\Gamma_{\hd}$) for $\alpha = 10^{-6}$ rad
and the total decay width of $\zp$ ($\Gamma_{\zp}$) 
for $\epsilon  = 10^{-8},\,10^{-10}$ as a function of mass $m$.
It is evident from Fig.\,\,\ref{Fig:decay-widths} that for
$m\gtrsim$ 5\,GeV, the decay width $\Gamma_{\hd}$ is dominant 
over $\Gamma_{\zp}$ for all values of $\epsilon$
we have considered and in this region the relic density
is controlled by $\Gamma_{\hd}$ only. Since the threshold
effect for the $b\bar{b}$ channel is mild in $\Gamma_{\hd}$,
the corresponding sharp feature is absent in
Fig.\,\,\ref{Fig:scan_gx_m}.
The situation becomes different for the case of
lighter dark matter where $m<5 \,{\rm GeV}$.
In this case, the freeze-out process of dark matter
is not entirely dominated by $\Gamma_{\hd}$. Instead,
the chemical imbalance prior to the freeze-out
can be created either by the decay of $\hd$ or
by $\zp$ or both depending on the portal couplings $\alpha$
and $\epsilon$. As a result, we have two distinct regimes
for $m\lesssim3 \,{\rm GeV}$ in the $m-g_X$ plane where
the allowed points forming a line corresponds to the small
values of $\epsilon < 10^{-10}$ and for these set of
allowed $g_X$ and $m$, the decay width of $\hd$ controls
the relic density. Consequently, the sharp features are
clearly visible for $\epsilon < 10^{-10}$ due to the
threshold effects of $c\bar{c}$ and $s\bar{s}$ channels.
Moreover, as the threshold effects are mild in
$\Gamma_{\hd}$ for the other decay modes like
$\mu^+\mu^-$ and $\tau^+\tau^-$, we have not observed
any peak for $m=2\,m_{\mu}$ and $2\,m_{\tau}$ respectively 
in Fig. \ref{Fig:scan_gx_m}. On the other hand,
the scattered points in the low mass regions are
for $\epsilon\gtrsim 10^{-10}$ where from Fig.\,\,\ref{Fig:decay-widths}
it is seen that $\Gamma_{\zp}\gtrsim\Gamma_{\hd}$ and
hence the relic density has been controlled by the
dynamics of $\zp$ instead of $\hd$. Additionally,
we would also like to mention that the threshold
effect in $\Gamma_{\hd}$ is more prominent in comparison
with that in $\Gamma_{\zp}$ due to
the presence of the fermion mass in the Yukawa
coupling of $\hd$ with the SM fermions.
\begin{figure}[h!]
\centering
\includegraphics[height=7cm,width=11cm,angle=0]
{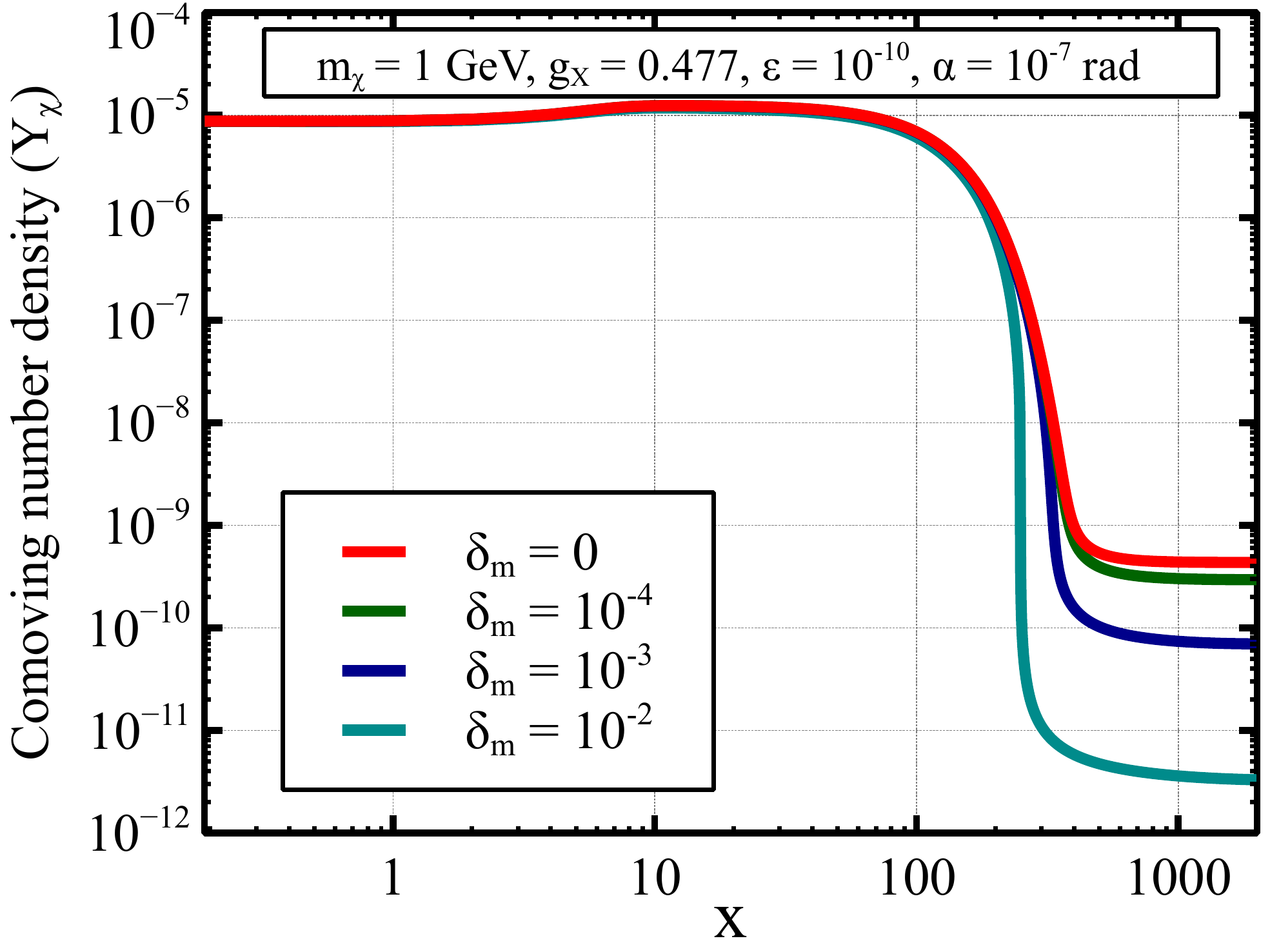}
\caption{Variation of $Y_{\x}$ with $x$ for different values
of $\delta_m$.}
\label{Fig:Y-vs-x_delM_nz}
\end{figure}

Finally, for completeness, we have solved the Boltzmann equations
(Eqs.\,\ref{Ydm-BE}-\ref{xi-BE}) for the case when there is a
finite mass splitting between the ${\x}$ (species ${\bf A}$)
and $\zp$, $\hd$ (species ${\bf B}$). Here, we have kept $m_{\zp}$
and $m_{\hd}$ fixed at 1 GeV and consider the $m_{\x}=m_{\zp}/(1-\delta_m)$.
In Fig.\,\ref{Fig:Y-vs-x_delM_nz} we have shown the variation of
$Y_{\x}$ with $x$ for four different values of $\delta_m = 0,\,10^{-4},\,10^{-3}$
and $10^{-2}$ respectively. We find that the features of the co-decaying scenario
(i.e. freeze-out at large $x$, $x_f>>20$) remains still valid for mass splitting
$\delta_m \leq 10^{-2}$. Beyond which we do not find any stable numerical
solution of $Y_{\x}$. The plots shown in Fig.\,\ref{Fig:Y-vs-x_delM_nz} are
drawn for $g_X=0.477$, $\epsilon= 10^{-10}$ and $\alpha= 10^{-7}$ rad
so that we get the observed relic abundance for $\delta_m=0$ (red line). As
we gradually increase $\delta_m$ from zero, freeze-out of $\x$
occurs earlier with a reduced final abundance. From the above figure
it is clearly seen that for increasing $\delta_m$ from 0 to $10^{-2}$,
the comoving number density of $\x$ reduces nearly two orders of magnitude
while the corresponding $x_f$ (inverse of the freeze-out temperature)
changes from $\sim 400$ to $\sim 250$.
\section{Indirect signature: $\gamma$-rays and Neutrinos from
dark matter annihilation}
\label{sec:indirect}
In this work, we also investigate the possibility
of detecting $\gamma$-ray and neutrino flux
from cascade processes\footnote{We are not considering
the direct production of $\bar{\nu}\nu$ and $\gamma$ $\gamma$
from dark matter annihilations because those processes are
suppressed by the portal couplings such as 
$\epsilon, \lambda^\prime$.} having a general structure
like $\chii \chii \ra A (A\ra X\, Y$). 

For the $\gamma$ rays, $A$, $X$, and $Y$ are $\hd$, $\gamma$ and $\gamma$
respectively whereas for the neutrinos in the final state, the intermediate
particle, instead of $\hd$, is the $U(1)_X$ gauge boson $\zp$ .
This kind of cascade process generates a polynomial box-shaped
spectrum of outgoing particles where shape of the polynomial depends
on the polarisation of the intermediate particle $A$. In the
rest frame of $A$, the energy of $X$ (or $Y$) is $m_A/2$. In the Laboratory frame,
assuming non-relativistic nature of dark matter, the energy of outgoing
particle is given by
\bea
E_{X(Y)}^{Lab} = 
\dfrac{m_{A}^2/2}{m_{\x} 
- \sqrt{m_{\x}^2 - m_{A}^2}\,\cos{\theta_{Lab}}},
\label{X-energy}
\eea
where $\theta_{Lab}$ is the angle between the direction of motion of 
$A$ and the direction of emission of $X$($Y$). It is clear
from Eq.\,\,\ref{X-energy} that the resulting spectrum has a sharp cut-off
at $\theta_{Lab} = 0$ and $\theta_{Lab}  = \pi$ and width
of the spectrum depends on mass splitting between the incoming
dark matter particle and intermediate state $A$,
i.e.\,\,$\Delta E_X = E_X^{\rm max} - E_X^{\rm min}
= \sqrt{m_\x^2 - m_A^2}$. Thus in the exact degenerate
limit of $\chii$ and $A$, this box-spectrum will be transformed into a
line. In the present case, as we have mentioned in the previous sections, 
all the dark sector species are degenerate. Therefore,
line like signal of neutrinos and $\gamma$ rays
can be originated from the annihilation of dark matter
which is frozen-out through co-decaying scenario.
In the present scenario, the final state radiation and 
the inverse Compton scattering can also result in $\gamma$-ray
flux from DM annihilation. In Section\,\,\ref{del_0},
indirect detection prospect of our DM candidate from
annihilation is discussed for mass splitting parameter
$\delta_m = (m_{\x} - m_{\zp (\hd)})/m_\x =0$
whereas the same is discussed in Section\,\,\ref{delta_ne_0}
in the light of non zero $\delta_m$.

The differential energy spectrum of $X$, produced from
the decay of an arbitrarily polarised intermediate particle $A$,
has the following general form
\bea
\label{dnde-gen}
\dfrac{d N_{X,m}}{d E_X} &= n & \dfrac{f_{m}(E_X/m_\x)}{m_\x}\,\,.
\eea
Here, $m$ is the polarisation of $A$ and $n$ is the
number of $X$ particle produced in the final state. 
Particular forms of the function
$f_{m}(E_X/m_\x)$ depends on the 
spin of the intermediate state.
The differential flux of $X$ coming from annihilation
of $\chii$ in the Milky Way halo and measured at
the earth is given by
\bea
\dfrac{d \Phi_{X}}{d E_{X}\Delta \Omega}
&=&
\dfrac{R_\odot \bar{J}_{\rm ann}}
{8 \pi m_{\x}^2} \dfrac{\rho_\odot^2}{4}
\langle \sigma {\rm v}_{\chii \chii \ra A A}\rangle
\sum_m  Br_m\,Br (A_m \ra X \, Y)\,\,
\dfrac{d N_{X,m}}{d E_{X}}\,\,, \,\, (i = 1,2)\,.
\label{diff-flux}
\eea
In the above $\langle \sigma {\rm v}_{\chii \chii \ra A_m A_m}\rangle$ is the
thermal averaged annihilation cross-section of $\chii$ into a pair of
$A$ with polarisation $m$, the quantity $ Br (A_m \ra X \, Y)$ is the branching
ratio of $A_m \ra X \, Y$ while the definition of $Br_m$ is
$Br_m = \dfrac{\langle \sigma {\rm v}_{\chii \chii \ra A_m A_m}\rangle}
{\langle \sigma {\rm v}_{\chii \chii \ra A A}\rangle}$.
Since we have two dark matter candidates
of identical properties, they will contribute equal amount
to the local dark matter density $\rho_{\odot}$ at the solar location.
Hence, an extra 1/4 factor has appeared
in the above expression of differential flux. The
total differential flux arising from annihilations of both the dark matter
candidates is noting but the above expression multiplied
by an extra factor of 2. The J-factor averaged over solid angle
$\Delta\Omega$ for dark matter annihilation is defined
in terms of the galactic co-ordinate system $(b,l)$
as
\bea
\bar{J}_{\rm ann} &=&
\dfrac{1}{\Delta \Omega}
\int_{\Delta \Omega}
\int_0^{{\rm s}_{max}}
\dfrac{1}{R_\odot}
\left(
\dfrac{\rho \left(
\sqrt{{\rm s}^2 + R_\odot^2 - 2\,{\rm s}\,R_\odot \cos b \cos l}
\right)}{\rho_\odot}
\right)^2 d{\rm s} \, d \Omega \,\,,
\label{J-fact-gal}
\eea
where, $R_{\odot}=8.5$ kpc is the distance
of the solar system from the centre of our
Milky way galaxy and ${\rm s}$ is the line of sight ($l.o.s$) distance
which is integrated over $0$ to s$_{max}$.
The expression of s$_{max}$ is given by
\bea
{\rm s}_{max} = \sqrt{R_{\rm MW}^2 - R_\odot^2 + R_{\odot}^2 \cos^2 b\cos^2 l} + 
R_\odot \cos b\cos l\,\,.
\eea

Moreover, Eq.\,\,\ref{J-fact-gal} can be framed
in terms of angle $\theta_{GC}$ where $\theta_{GC}$ being the angle between
the galactic centre and the line of sight distance. 
In this co-ordinate system $\bar{J}_{ann}$ takes
the following form,
\bea
\bar{J}_{\rm ann}
&=&
\dfrac{1}{\Delta \Omega}
\int_{\Delta \Omega}
\int_0^{{\rm s}_{max}}
\dfrac{1}{R_\odot}
\left(
\dfrac{\rho
\left(\sqrt{R_{\odot}^2 + {\rm s}^2 -
2\,{\rm s}\,R_\odot \cos \theta_{GC}}\right)}{\rho_\odot}
\right)^2
d{\rm s}\,d\Omega\,\,.
\label{J-fact-GC}
\eea
The upper limit of the $l.o.s$ distance in this co-ordinate
system is s$_{max} = \sqrt{R_{\rm MW}^2 - \sin^2 \theta_{GC} R_{\odot}^2} 
+  R_{\odot} \cos \theta_{GC}$. In Eqs.\,\,\ref{J-fact-gal} and \ref{J-fact-GC}, 
$\rho_\odot = 0.3 \,{\rm GeV/cm^3}$ is the dark matter density at the solar 
neighbourhood, $R_{\rm MW} = 40 \,{\rm kpc}$ is the size of the Milky Way (MW) galaxy halo  
and the solid angle $\Delta \Omega$ represents the field of
view of the detector. Moreover, throughout our analysis, we have assumed
the DM density follows the standard Navarro-Frenk-White (NFW)
\cite{Navarro:1995iw} density profile.

\subsection{Indirect detection for $\delta_m=0$}
\label{del_0}
\subsubsection{$\gamma$-ray signal from dark matter annihilation}
\label{sec:gamma}
In our model, $\gamma$-ray flux is composed of three
different components such as $\gamma$-ray line signal, final state
radiation (FSR) from charged particle final states and inverse Compton
scattering (ICS) of the charged particles with the CMB photons.
We have discussed all the three components in the following sections.
\begin{itemize}
\item \textbf{$\gamma$-ray line signal:}
The monochromatic photons coming from dark matter annihilation/decay
are considered as one of the important smoking gun signatures of dark matter. 
Motivated by this, we have discussed the prospect of $\gamma$-ray line
in the present scenario due to dark matter annihilation
at the galactic centre. In our model, the dark scalar $\hd$
mixes with the SM Higgs. Thus, a line shaped $\gamma$-ray
spectrum can be generated form $\chii\chii$ annihilation through
one step cascade process\footnote{The lifetime of $h_d$ is much
smaller than $t_{\rm BBN}$ in the parameter space in which we are
interested in. For example,
for $m_{\hd} = 10\,{\rm GeV}$, $\alpha = 10^{-6}$, 
$\tau_{\hd} \simeq 10^{-5} s << t_{\rm BBN}$.} like
$\chii \chii \ra \hd (\hd \ra \gamma \gamma).$
Since the intermediate state in this case is a spin 0 boson
$h_d$, the energy spectrum of the emitted photons
is a Dirac delta function for $m_{\x} = m_{\hd}$.
The differential photon spectrum is given by
\bea
\dfrac{d N^{\rm Line}_\gamma}{d E_\gamma}
&=& 2 \,\delta \left(E_\gamma - \dfrac{m_{\hd}}{2}\right)\,\,.
\label{spectra_gamma_del_0}
\eea
Here, 
the prefactor 2 in $\dfrac{d N^{\rm Line}_\gamma}{d E_\gamma}$ 
arises because for each decay of $\hd$, two photons are emitted.
Now, using Eq.\,\,\ref{spectra_gamma_del_0}, we can calculate
the $\gamma$-ray flux from Eq.\,\,\ref{diff-flux}.
The differential photon flux is given by
\bea
\dfrac{d \Phi^{\rm Line}_\gamma}{d E_{\gamma}\Delta \Omega}
&=&
\dfrac{R_\odot \bar{J}_{\rm ann}}
{8 \pi m_{\x}^2}
\dfrac{\rho_\odot^2}{4} 
\langle \sigma {\rm v}_{\chii \chii \ra \hd \hd}\rangle 
\,Br (\hd \ra \gamma \gamma)\,\,
\dfrac{d N^{\rm Line}_{\gamma}}{d E_{\gamma}}\,\,,
\label{gamma-flux}
\eea
where $Br (\hd \ra \gamma \gamma)$ is the
branching ratio of $\hd$ into a pair of $\gamma$ and 
$\langle \sigma {\rm v}_{\x \x \ra \hd \hd}\rangle$
is the dark matter annihilation cross section into a pair of $\hd$ 
which is velocity suppressed in the non-relativistic 
limit (see right panel of Fig\,\,\ref{Fig:sigmaVp}).
The analytical form of $\langle \sigma {\rm v}_{\x \x \ra \hd \hd}\rangle$
in the non-relativistic limit for exactly degenerate dark sector
is given by
\bea
\langle \sigma {\rm v}_{\x_i \x_i \ra \hd \hd}\rangle
\simeq 1.34 \times 10^{-17} {\rm cm^{3}s^{-1}} 
\left(\dfrac{g_X}{1}\right)^4 
\left(\dfrac{1 \, {\rm GeV}}{m_\x}\right)^2 v^3\,\,.
\label{NR-DMDMhdhd}
\eea
Here ${\rm v}$ is the magnitude of relative velocity
of the initial state particles and $v$ is the average thermal
velocity of DM. This is to be noted that the leading
order term is proportional to $ v^3$ instead of $v$ at threshold ($\delta_m = 0$).
This is because of the absence of $\hat{a}_0$ term
in the cross section (see Eq.\,\,\ref{thavg}
of Appendix \ref{App:F} for details).

\item \textbf{Final State Radiation:}
As discussed in \cite{Cirelli:2020bpc, Essig:2013goa},
the FSR contribution to $\gamma$-ray flux can 
constrain our parameter space in $m_\x -g_X$ plane.
To study the $\gamma$-ray flux from FSR, we have considered 
the cascade process $\x\x \ra \zp(\hd) \zp(\hd)$ and the subsequent
decay of $\zp(\hd) \ra e^+ e^-$ + FSR. The energy spectrum
of photons in the center of mass (CoM) frame of the DM
is given by\footnote{Here, we have not considered
the angular dependence in the differential FSR spectrum. However
the effect of the angular dependence is not
very significant as discussed in  \cite{Elor:2015tva}.}
\cite{Essig:2009jx}
\bea
\dfrac{dN^{\rm FSR}_{\bar{f}f}}{d x} &=& 
\dfrac{2\alpha_{EM}}{x \pi}
\left[
x^2 + 2 x \left({\rm Li}_2\left[\dfrac{m_\zp -2 m_f}
{m_\zp - m_f}\right]-{\rm Li}_2[x]\right) + (2-x^2)\log(1-x)
\right.\nonumber \\
&& \left.
+\left(\log\left[\dfrac{m_{\zp}^2}{m_f^2}\right]-1\right)
\left\{2-x^2 + 2x \log\left[\dfrac{(m_\zp -m_f)x}{m_\zp -2 m_f}\right]
-\dfrac{(m_{\zp}^2 -2 m_f^2)x}{(m_\zp -m_f)(m_\zp -2 m_f)}		
\right\}\right. \nonumber \\
&&\left. -\dfrac{x}{2m_f^2-3 m_\zp m_f+ m_{\zp}^2}
\left\{2 m_f^2 \left(2-\log\left[\dfrac{x^2 m_f^2}
{(m_\zp-2 m_f)^2(1-x)}\right]\right) \right.\right.\nonumber \\
&&\left.\left.
-3 m_f m_\zp \left( \dfrac{4}{3} - 
\log\left[\dfrac{m_f (m_\zp - m_f)x^2}{(m_\zp-2 m_f)^2 
(1-x)}\right]\right) 
\right.\right.\nn \\ && \left.\left.
+ m_{\zp}^2 \left(1-\log\left[\dfrac{(m_\zp-m_f)^2 x^2}
{(m_\zp -2 m_f)^2 (1-x)}\right]\right)
\right\}
\right]\,. 
\label{FSR spectrum}
\eea
Here $f$ denotes the final state fermion
having mass $m_f$, $\alpha_{EM}$ is the fine structure
constant and $x = E_\gamma / m_\x$ where $E_\gamma$ is the
energy of the emitted photon in the CoM frame of DM annihilation.
Therefore, using Eq.\,\,\ref{diff-flux} and Eq.\,\,\ref{FSR spectrum}
we can write down the differential photon flux from FSR as
\bea
\dfrac{d \Phi^{\rm FSR}_{\bar{f}f}}{d E_{\gamma}\Delta \Omega}
&=&
\dfrac{R_\odot \bar{J}_{\rm ann}}
{8 \pi m_{\x}^2}
\dfrac{\rho_\odot^2}{4}
\sum_{A =\hd, \zp}\langle \sigma {\rm v}_{\chii \chii \ra AA}\rangle
\,Br (A \ra \bar{f}f)\,\,
\dfrac{d N^{\rm FSR}_{\bar{f}f}}{d E_{\gamma}}\,\,.
\label{FSR flux}
\eea
In the non-relativistic limit
$\langle \sigma {\rm v}_{\chii \chii \ra \hd \hd}\rangle$ 
is given in Eq.\,\,\ref{NR-DMDMhdhd} and whereas
$\langle \sigma {\rm v}_{\chii \chii \ra \zp \zp}\rangle$
is given by the following relation.

\bea
\langle \sigma {\rm v}_{\x_i \x_i \ra \zp \zp}\rangle
\simeq 3.24 \times 10^{-18} {\rm cm^{3}s^{-1}}
\left(\dfrac{g_X}{1}\right)^4
\left(\dfrac{1 \, {\rm GeV}}{m_\x}\right)^2 v^3\,\,.
\label{NR-DMDMzpzp}
\eea
Here also the leading order term of
$\langle \sigma {\rm v}_{\x_i \x_i \ra \zp \zp}\rangle$
is proportional to $ v^3$. It is due to the fact that for $\delta_m = 0$,
the coefficient $\hat{a}_0$ in Eq.\,\,\ref{thavg} vanishes 
as it is proportional to $\delta_m^{3/2}$
(see Appendix \ref{App:F} for details).
\item \textbf{Inverse Compton Scattering:} In
the present scenario, high energy electron positron pair
can be produced from DM annihilation via one step
cascade processes through mixing parameters $\epsilon$ and $\alpha$.
The Inverse Compton scattering between these charged particles
and ubiquitous CMB photon can produce high energy gamma ray fluxes
which may be within the reach of current $\gamma$-ray detectors. 
To study this prospect, we have calculated the $\gamma$-ray
flux following \cite{Cirelli:2009vg,Cirelli:2020bpc}.
\end{itemize}

Finally, the total photon flux will have
all three contributions from different sources as
\bea
\dfrac{d\Phi_\gamma}{dE_\gamma \Delta \Omega} = 
\dfrac{d\Phi^{\rm Line}_\gamma}{dE_\gamma\Delta \Omega} 
+\dfrac{d\Phi^{\rm FSR}_{\bar{f}f}}{dE_\gamma \Delta \Omega} 
+\dfrac{d\Phi^{\rm ICS}_\gamma}{dE_\gamma\Delta \Omega}\,\,.
\label{total flux}
\eea

\begin{figure}[h!]
\centering
\includegraphics[height=5.5cm,width=8.0cm, angle =0]{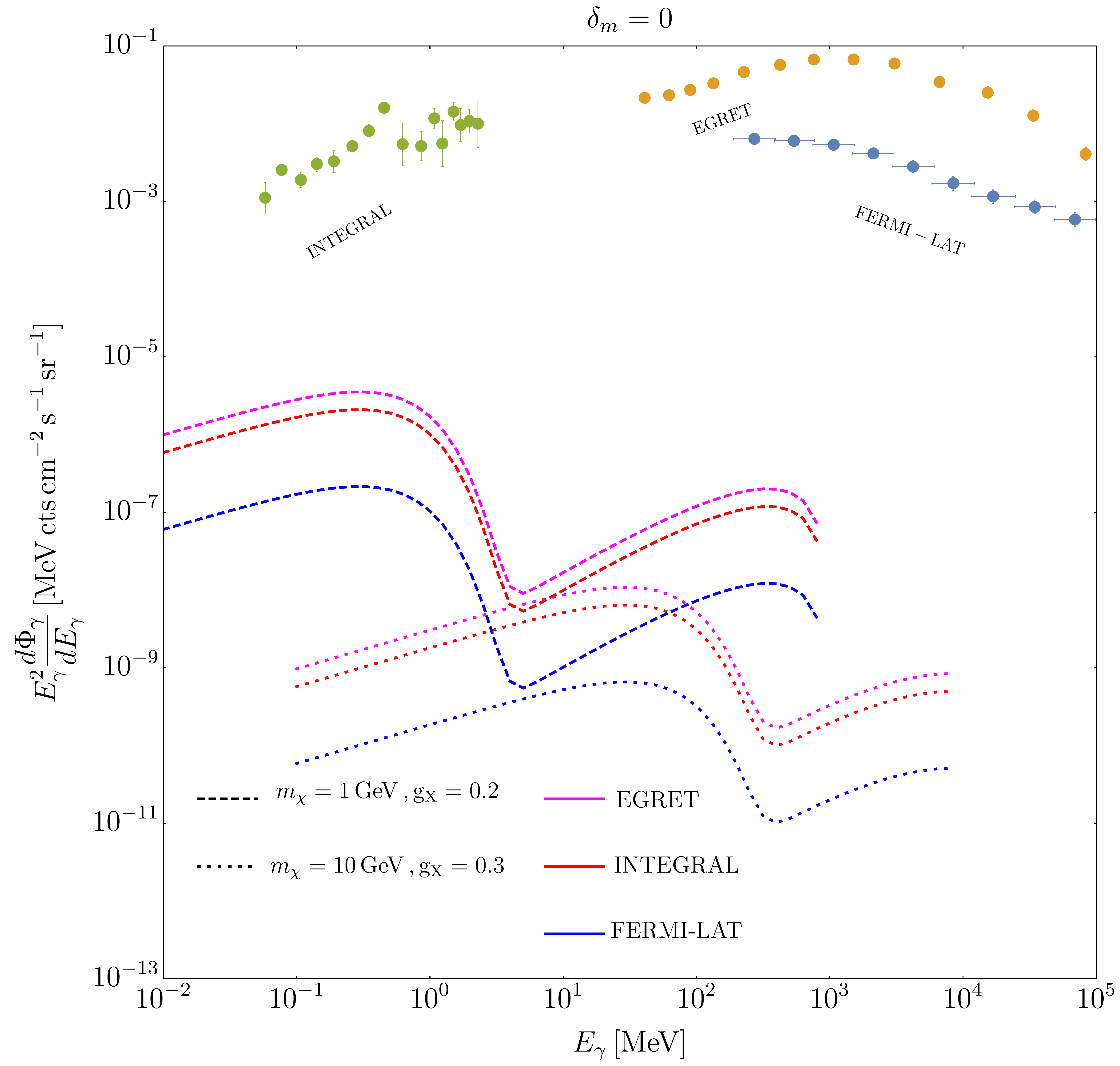}
\includegraphics[height=5.5cm,width=8.0cm,angle =0]{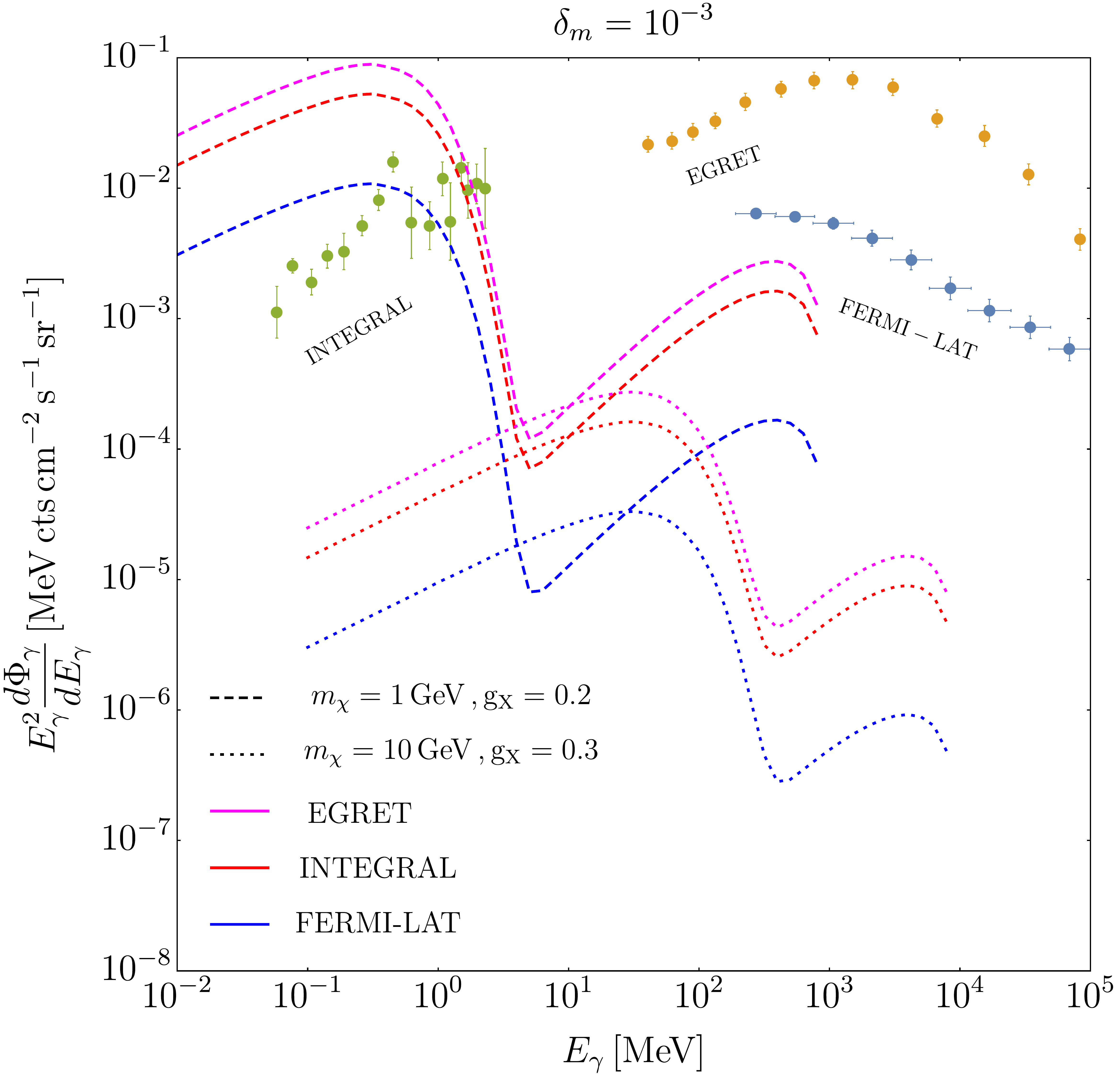}
\caption{Variation of differential $\gamma$-ray flux
from dark matter annihilation as a function of
energy of the emitted photons for two different benchmark points, allowed
from relic density constraint.
\textbf{\textit{Left panel:}} 
$\gamma$-ray flux from dark matter
annihilation is compared with the flux of diffuse $\gamma$ ray background
by EGRET, Fermi-LAT and INTEGRAL collaboration. Here magenta, red, and blue lines
indicate the flux is calculated for the ROI of EGRET, INTEGRAL and Fermi-LAT data respectively.
The flux for $m_\chi = 1 \, {\rm GeV}, g_X = 0.2$ is denoted by the dashed 
lines and the same for $m_\chi = 10 \,{\rm GeV}, g_X = 0.3$ is denoted
by the dotted lines. Here we have considered $\delta_m =(m_\x -m_\zp)/m_\x = 0$. 
\textbf{\textit{Right panel:}} The color code is same as the left panel and we have considered $\delta_m = 10^{-3}$.
}
\label{Fig:gamma-flux-fig}
\end{figure}

In Fig.\,\,\ref{Fig:gamma-flux-fig},
we have shown the $E^2_\gamma$ weighted total differential
photon flux as a function of the energy of emitted photons
for three benchmark points which satisfy the relic density constraint.
Here the fluxes are calculated for the Region of Interest (ROI)
having $(330^\circ < l < 30^\circ$, $0^\circ < |b| < 5^\circ)$, 
$(330^\circ < l < 30^\circ$, $0^\circ < |b| < 15^\circ)$, and
$(0^\circ < l < 360^\circ$, $10^\circ < |b| < 20^\circ)$.
The computed flux has been compared with the
available data for diffuse gamma-ray from
EGRET (Fig.\,\,4 of \cite{Strong:2004de}),
Fermi-LAT (Fig.\,\,6 of \cite{Abdo:2010nz}) and 
INTEGRAL(Fig.\,\,7 of \cite{Bouchet:2011fn}) collaborations.
The ROIs considered in above are adopted from
the respective experimental collaboration to be
as close as possible with the experimental data.

It is clearly seen from Fig.\,\,\ref{Fig:gamma-flux-fig} that
the $\gamma$-ray flux due to dark matter annihilation calculated
from the ROI for EGRET is larger than that for Fermi-LAT and INTEGRAL.
We know that for a cuspy density profile like the NFW profile the
value of $\bar{J}_{ann}$ increases sharply as we move towards low latitude.
Therefore, the value of $\bar{J}_{ann}$ for the ROI of
the EGRET collaboration is much larger compared to
$\bar{J}_{ann}$ for the Fermi-LAT and INTEGRAL.
In the left panel, we have considered $\delta_m = 0$ whereas
the results in the right panel are for $\delta_m = 10^{-3}$.
The enhancement in the photon flux as seen in the right panel
compared to that in the left panel has been discussed in the
light of nonzero mass splitting in Section\,\,\ref{delta_ne_0}.
\subsubsection{Neutrino line from DM annihilation}
\label{sec:nunubar_del_0}
In our model, the kinetic mixing portal can produce
an interesting neutrino signal from dark matter annihilation.
For the exact degenerate case (i.e. $\delta_m=0$), the neutrino
spectrum is a line spectrum. Here we have considered the following cascade 
process $\chii \chii \ra \zp (\zp \ra \bar{\nu}_\mu \nu_\mu)$
and the emitted neutrinos are monochromatic. The neutrino energy
spectrum is given by 
\bea
\label{neutrino_spec_del_0}
\dfrac{d N_{\nu_{\mu},m}}{d E_{\nu_{\mu}}} = 
2\, \delta \left(E_{\nu_\mu} - \dfrac{m_{\zp}}{2}\right)\,\,,
\eea
where $m$ is the polarisation of the decaying particle i.e. $\zp$.
Note that the polarisation of $\zp$ has no effect on the spectrum
of monochromatic neutrinos. However, the polarization of $\zp$
has non-trivial effect for the quasi degenerate case as discussed
later. The 2 factor in Eq.\,\,\ref{neutrino_spec_del_0} is for
$\nu_\mu$ and $\bar{\nu}_\mu$.

Using Eq.\,\,\ref{diff-flux}, we can easily write the
differential neutrino flux for $\nu_\mu$ where the
energy spectrum and $\bar{J}_{\rm ann}$ are obtained
from Eq.\,\,\ref{J-fact-GC} and Eq.\,\,\ref{neutrino_spec_del_0}
respectively.
\bea
\label{nu-flux}
\dfrac{d \Phi_{\nu_\mu}}{d E_{\nu_{\mu}}\Delta \Omega}
&=&
\dfrac{R_\odot \bar{J}_{\rm ann}}
{8 \pi m_{\x}^2} \dfrac{\rho_\odot^2}{4}
\langle \sigma {\rm v}_{\chii \chii \ra \zp \zp}\rangle
\sum_m  Br_m\,Br (\zp_m \ra \bar{\nu}_\mu \, \nu_\mu)\,\,
\dfrac{d N_{\nu_{\mu},m}}{d E_{\nu_{\mu}}}\,\,.
\eea
As the dark matter candidates are non-relativistic, 
we have found $Br_0 = 0$. Therefore, in Eq. \ref{nu-flux} we have
used \cite{Drees:1992am, Barger:2007xf} $Br_{-1} = Br_{+1} = 1/2$.
Since the annihilation cross section 
$\langle \sigma {\rm v}_{\chii \chii \ra \zp \zp}\rangle$
is velocity suppressed for $\delta_m=0$ and the branching ratio
of $\zp$ into $\bar{\nu}_\mu\nu_\mu$ is extremely small,
the resulting neutrino flux is well below the observed flux
by Super-Kamiokande detector.

\subsection{Effect of nonzero mass splitting ($\delta_m\neq 0$)
on the Indirect detection prospect}
\label{delta_ne_0}
If we consider the mass splitting parameter $\delta_m \neq$ 0,
the $\gamma$-ray and $\nu_\mu$ fluxes change significantly. 
In the following, we have discussed the $\gamma$-ray and $\nu_{\mu}$
fluxes for $\delta_m \neq 0$.
\begin{itemize}
\item \textbf{$\gamma$-ray signal for $\delta_m \neq 0$:} \\
As discussed earlier, the parameter $\delta_m \neq 0$ breaks
the degeneracy between the DM $\x_i$ and the mediators $\zp$ and $\hd$.
As a result, the $\gamma$-ray energy spectrum becomes a box like
spectrum instead of a line like one for the $\delta_m = 0$ case.
Thus, the energy spectrum of the photons from the annihilation
of DM into two $\hd$s and the subsequent decay of $\hd$ into
two photons is given by
\bea
\dfrac{d N^{\delta_m \neq 0}_\gamma}{d E_\gamma}
&=&
\dfrac{2}{m_\x}
\dfrac{1}{\sqrt{1 - r_{\hd}^2}}
\Theta\left(z - \dfrac{E_{\gamma_{\rm Lab}}^{\rm min}}{m_\x}\right)
\Theta\left(\dfrac{E_{\gamma_{\rm Lab}}^{\rm max}}{m_\x} - z\right)\,\,.
\label{spectra_gamma_del_ne_0}
\eea
Here, $r_{\hd} = m_{\hd}/ m_\x$ and $z = E_\gamma / m_\x$.
The cut-off energies for the box-spectrum are $E_{\gamma_{\rm Lab}^{\rm min}}$
and $E_{\gamma_{\rm Lab}^{\rm max}}$ and these can be
calculated from Eq.\,\,\ref{X-energy}.\,\,Origin of the factor 
2 is mentioned in Eq.\,\,\ref{spectra_gamma_del_0}.
Similar to the case of $\delta_m=0$, the total
photon flux for $\delta_m \neq 0$ also contains
three components such as $\gamma$-ray box spectrum and
$\gamma$-rays from FSR and ICS.
Therefore, using Eq.\,\,\ref{diff-flux} and Eq.\,\,\ref{total flux}, 
we calculate total photon flux for nonzero mass splitting. Note that
the line contribution in Eq. \ref{total flux} should be replaced
by the box contribution for non zero $\delta_m$.
The main difference in this case compared
to the previous one ($\delta_m=0$) is that
now the annihilation cross section
$\langle \sigma v_{\rm rel}\rangle _{\x_i \x_i \ra \zp \zp}$
is not velocity suppressed and it scales as\footnote{Note
that the annihilation cross section
$\langle \sigma v_{\rm rel}\rangle _{\x_i \x_i \ra \hd \hd}$
still remains velocity suppressed for $\delta_m\neq 0$
and subsequently has lesser contribution in the $\gamma$-ray flux.}
$\delta_m^{3/2}$.
Therefore, the final $\gamma$-ray flux also
scales as $\delta_m^{3/2}$. In the right panel
of Fig.\,\,\ref{Fig:gamma-flux-fig}, the
total photon flux from DM annihilation is plotted for
$\delta_m = 10^{-3}$. As we can see from the plot that
the computed flux is several order of magnitude larger than
the case for $\delta_m = 0$ and at some energy range
with certain choices of model parameters
it overshoots the observed flux. This is primarily because
of the presence of the s-wave term in
$\langle \sigma v_{\rm rel}\rangle _{\x_i \x_i \ra \zp \zp}$.
\item \textbf{Neutrino line from DM
annihilation for $\mathbf{\delta_m \neq 0}$}:\\
For $\delta_m\neq 0$, the neutrino spectrum
due to one step cascade process is a polynomial spectra 
with sharp cut-off at $E^{\rm min}_{\nu_{\rm Lab}} \text{ and } E^{\rm max}_{\nu_{\rm Lab}}$ and the 
shape of the polynomial depends on polarisation of $\zp$. Therefore, 
following \cite{Garcia-Cely:2016pse},
we can calculate the differential energy spectrum of
neutrinos $f_m(E_\nu/m_\x)$ for an arbitrarily polarised 
$\zp$ having polarisation $m$. The functional forms of the
spectrum for different polarisation states of
$\zp$ are given by
\bea
\label{f0}
f_{0}\left(z\right) &=&
 \dfrac{3}{2}\,\,
 \dfrac{
4 z - 4 z^2 - r_{\zp}^2 }
{\left(1 -  r_{\zp}^2\right)^{3/2}}\,\,
\Theta\left(z - \dfrac{E_{\nu_{\rm Lab}}^{\rm min}}{m_\x}\right)
\Theta\left(\dfrac{E_{\nu_{\rm Lab}}^{\rm max}}{m_\x} - z\right),\\
\label{f1}
f_{1}\left(z\right) &=&
 \dfrac{3}{4}\,\,
 \dfrac{
2 - 4 z + 4 z^2 - r_{\zp}^2 +\left(2 - 4 z\right)\sqrt{1 - r_{\zp}^2}}
{\left(1 -  r_{\zp}^2\right)^{3/2}}\,\,
\Theta\left(z - \dfrac{E_{\nu_{\rm Lab}}^{\rm min}}{m_\x}\right)
\Theta\left(\dfrac{E_{\nu_{\rm Lab}}^{\rm max}}{m_\x} - z\right),\\
\label{f-1}
f_{-1}\left(z\right) &=&
 \dfrac{3}{4}\,\,
 \dfrac{
2 - 4 z + 4 z^2 - r_{\zp}^2 -\left(2 - 4 z\right)\sqrt{1 - r_{\zp}^2}}
{\left(1 -  r_{\zp}^2\right)^{3/2}}\,\,
\Theta\left(z - \dfrac{E_{\nu_{\rm Lab}}^{\rm min}}{m_\x}\right)
\Theta\left(\dfrac{E_{\nu_{\rm Lab}}^{\rm max}}{m_\x} - z\right).
\eea
Here, $z = E_{\nu_{\mu}}/m_\x$ and $r_{\zp} = m_{\zp}/m_{\x}$.
\begin{figure}
\centering
\includegraphics[height=7cm,width=9cm,angle = 0]{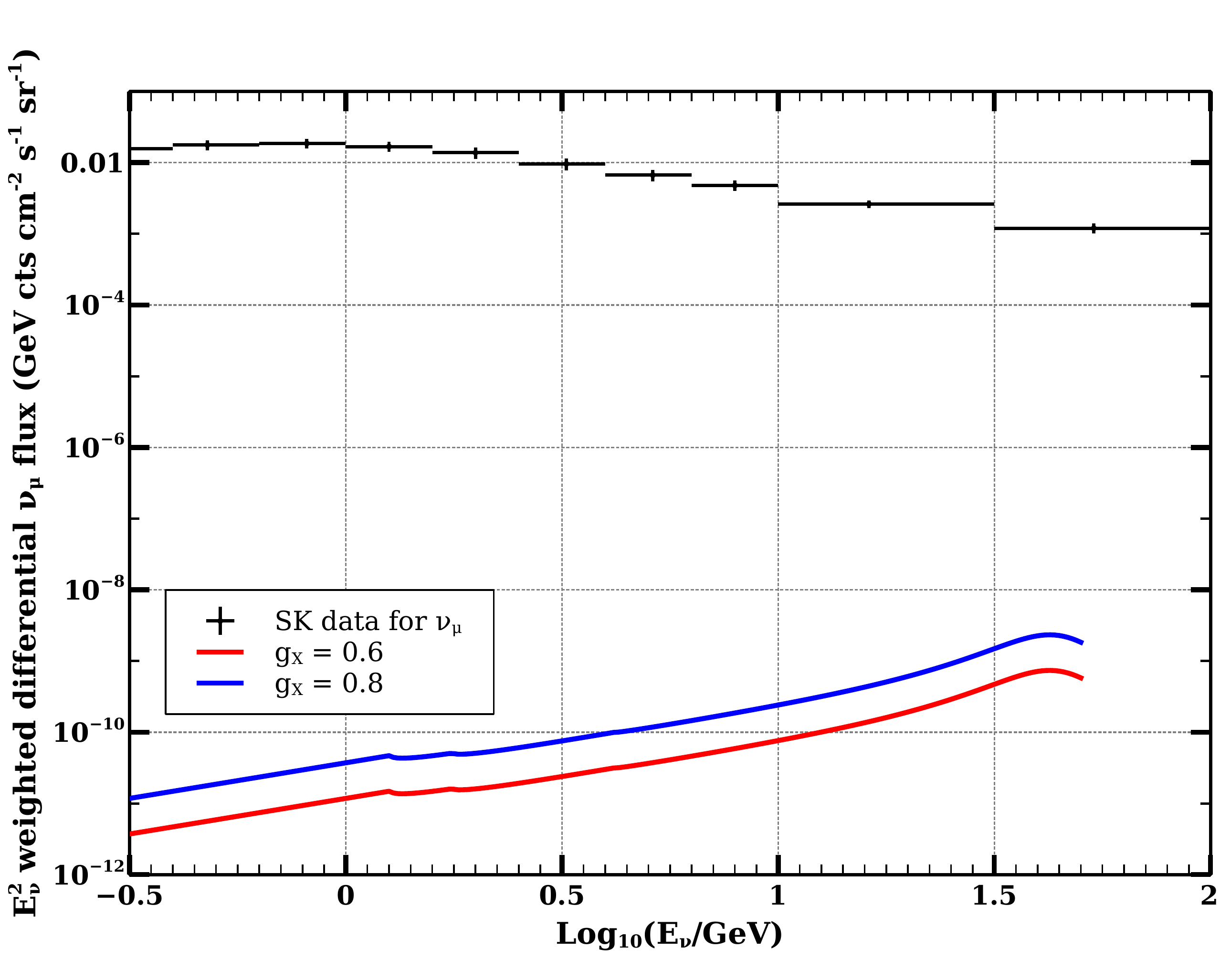}
\caption{Variation of $E_\nu^2 \dfrac{d \Phi_{\nu_\mu }}{dE_\nu}$ 
for $g_X = 0.6$ (red line), $g_X = 0.8$ (blue line). 
The data points with 1$\sigma$ errorbars are the atmospheric $\nu_\mu$ flux
measured by the SK detector	\cite{Richard:2015aua}.}
\label{Fig:flux}
\end{figure}
Using Eqs.\,\,\ref{f0}-\,\,\ref{f-1},
Eq.\,\,\ref{dnde-gen}, and Eq.\,\,\ref{diff-flux} we have
calculated the $\nu_\mu$ flux for $\delta_m \neq 0$.
As mentioned earlier, like the photon flux,
the $\nu_\mu$ flux also scales as $\delta_m^{3/2}$.

In Fig.\,\,\ref{Fig:flux}, we have shown the variation of
$E^2_\nu$ weighted differential neutrino flux for $\nu_\mu$
as a function of $E_\nu$ for $\delta_m = 10^{-3}$
and compared it with the observed $\nu_\mu$ flux
by the Super-Kamiokande (SK) detector.
We have found that the neutrino signal from dark matter
annihilation is several order of magnitude lesser
than that of the observed flux by the SK detector at low energy.
This is primarily due to the fact that branching ratio
of $\zp$ to $\bar{\nu}_\mu\nu_{\mu}$ becomes heavily suppressed at
low value of $m_\zp$ ($m_{\zp}\lesssim 10$ GeV).
Therefore, although the annihilation cross-section 
$\langle \sigma {\rm v}_{\chii \chii \ra\zp \zp}\rangle$ is not
velocity suppressed in the non-relativistic limit for $\delta_m \neq 0$
 (see left panel of Fig.\,\,\ref{Fig:sigmaVp}),
the resulting neutrino flux lies well below the observed
flux due to additional suppression in the 
branching ratio of $\zp\ra\bar{\nu} \nu$ channel.
We have divided the full range of $\cos \theta_{\rm GC}$ into ten
bins as shown by the SK collaboration \cite{Abe:2020sbr}
and combined fluxes from all such bins to show the variation
with $E_\nu$. As we have assumed the dark matter
density profile to be the cuspy NFW profile,
therefore, we would expect the neutrino flux from dark matter annihilation
has an angular variation. For the cuspy profile, 
the dark matter density and hence the J-factor 
increases as we go towards the galactic centre. Therefore,
the number of annihilations are also very large
near the galactic centre because of the huge
amount of dark matter. Thus, a large amount of neutrino
flux due to dark matter annihilation from the nearby regions
of the galactic center is expected. In the left
panel of Fig.\,\,\ref{Fig:hist-cross} we show
the variation of differential neutrino flux as a function 
$\cos \theta_{\rm GC}$ for $\delta_m = 10^{-3}$,
$m_\x = 5 \, {\rm GeV}$ and $m_\x = 50 \, {\rm GeV}$.
This plot reveals that the differential flux
$\dfrac{d\Phi_{\nu}}{dE_{\nu_{\mu}}}$
is higher for the neutrinos arising from annihilation
of 5 GeV dark matter. The energy $E_{\nu_{\mu}}$ is always larger for
neutrinos coming from $m_\x = 50 \, {\rm GeV}$ dark matter
than those are from $m_\x = 5 \, {\rm GeV}$ dark matter.
However, the differential flux proportional to
$\langle \sigma {\rm v}_{\chii \chii \ra \zp \zp}\rangle/{m^2_{\x}}$
becomes heavily suppressed as $m_{\x}$ increases.

Finally, in the right panel of Fig.\,\,\ref{Fig:hist-cross},
we have presented the variation of dark matter annihilation
cross section into neutrinos as a function of dark matter mass.
Since we have considered an one step cascaded process
to produce a pair of $\bar{\nu}_\mu \nu_\mu$, therefore we have plotted
$\langle \sigma {\rm v}_{\chii \chii \ra\zp \zp}\rangle \times 
Br(\zp \ra \bar{\nu}_\mu \nu_\mu)$ as a function of dark matter mass
($m_\x$) for two different values of the dark sector gauge
coupling ($g_X$) and compared it with the current
upper limit on DM DM $\ra \bar{\nu}\nu$ channel as
given in \cite{Frankiewicz:2015zma}. 
Here also we have considered the mass
splitting parameter $\delta_m = 10^{-3}$.
From this plot we can notice that
the effective annihilation cross section of dark matter into a pair of neutrinos
in the present scenario is several orders of magnitude smaller than
the current upper limit. Thus, our model easily avoids the indirect
detection constraint on $\bar{\nu}\nu$ channel
by the SK collaboration.
\begin{figure}[h!]
\centering
\includegraphics[height=6cm,width=8cm]{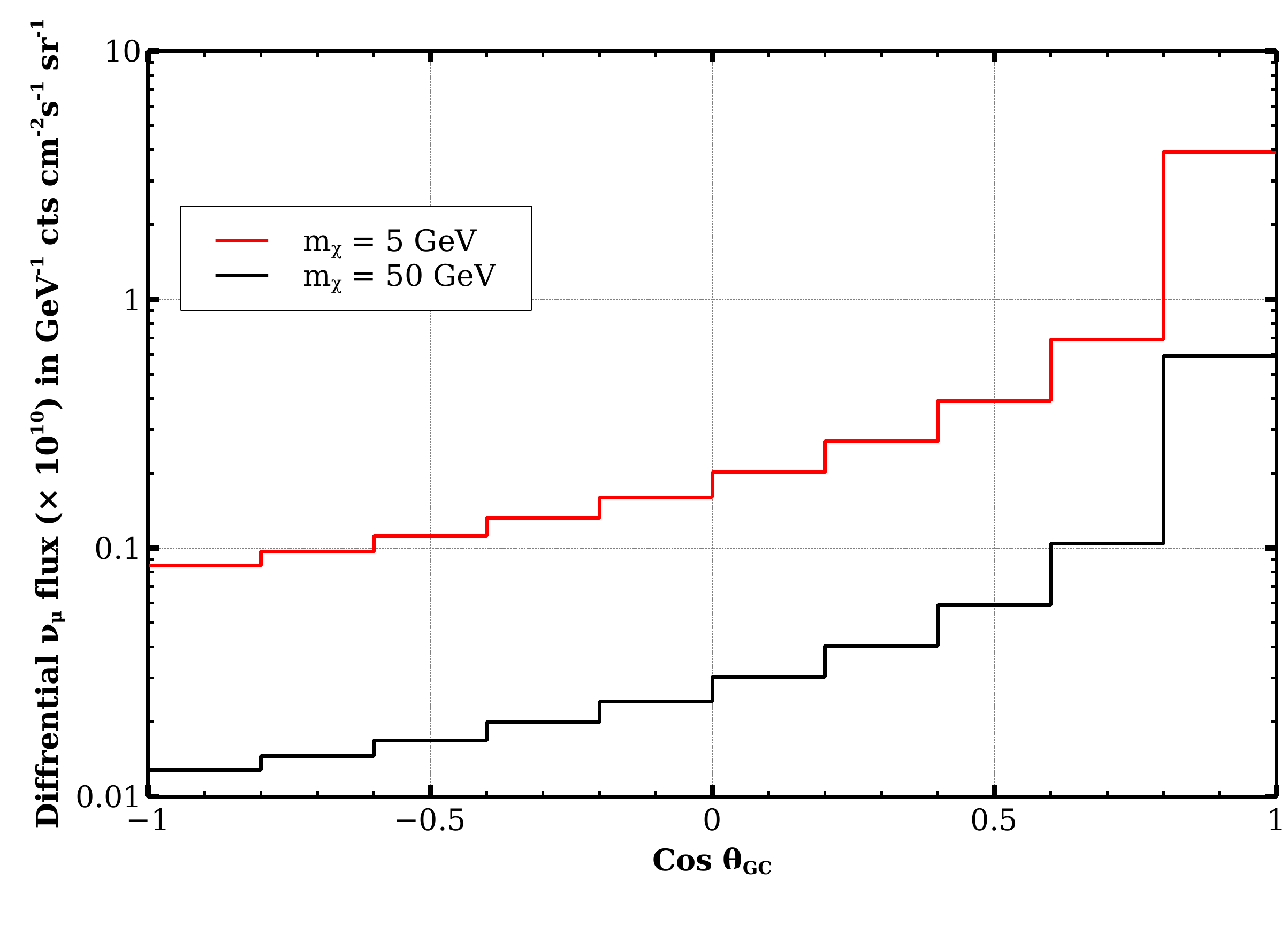}
\includegraphics[height=6cm,width=8cm]{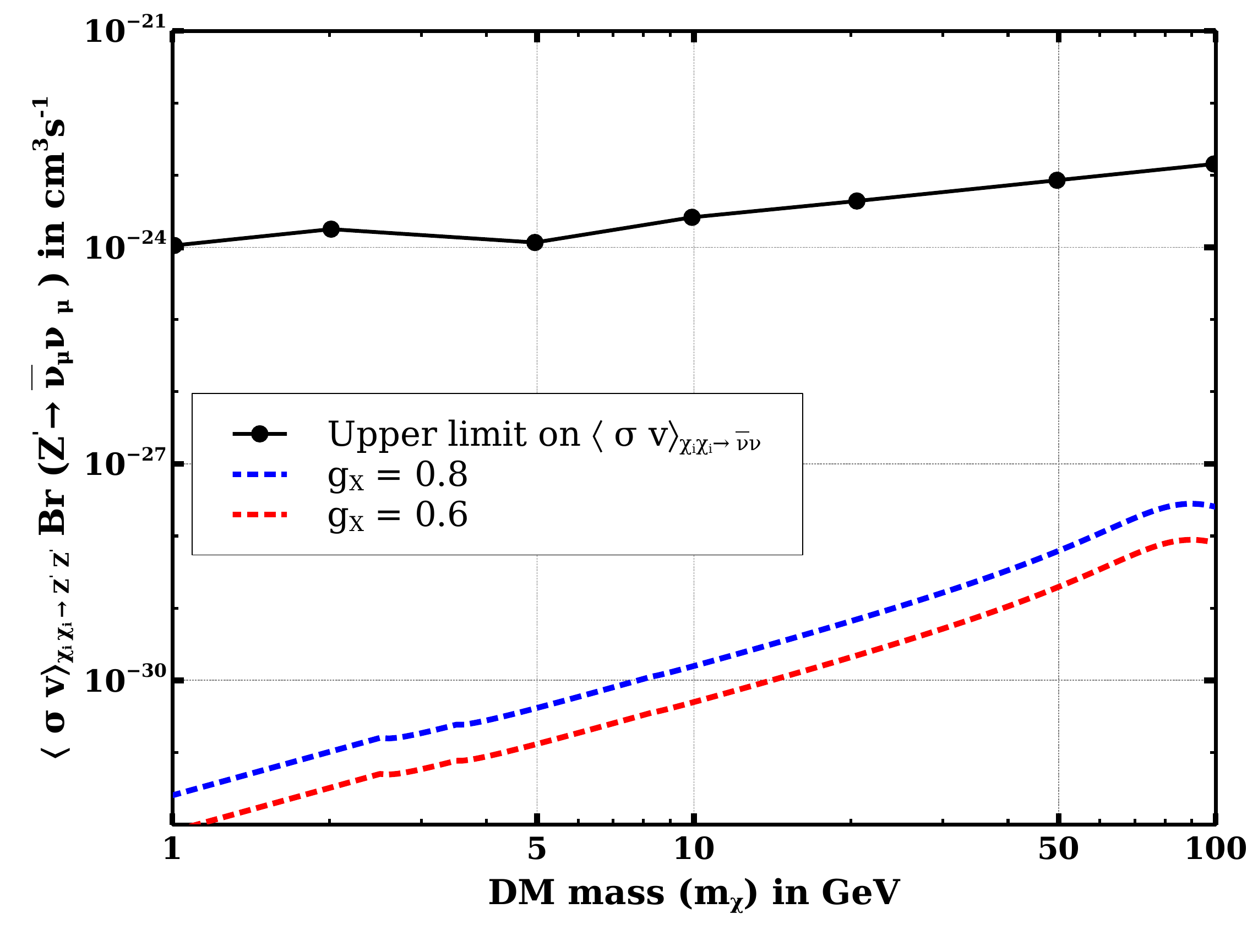}
\caption{\textbf{\textit{Left panel:}} Angular variation of differential neutrino
flux as a function of $\cos \theta_{GC}$ for $m_\x = 5 \, {\rm GeV}$ (red line) and
$m_\x = 50 \, {\rm GeV}$(black line).
\textbf{\textit{Right panel:}} Upper limit on 
$\langle \sigma v\rangle_{\x \x \ra\zp \zp}\times 
Br(\zp \ra \bar{\nu}_\mu \nu_\mu)$ as a function of $m_\x$ for 
$g_X = 0.6$ (red dashed line) and $g_X= 0.8$ (blue dashed line).
Upper limit on the annihilation cross-section from SK is shown by
the black data points. In both the plots, we have considered $\delta_m=10^{-3}$.}
\label{Fig:hist-cross}
\end{figure}
\end{itemize}
\begin{figure}[h!]
        \centering
        \includegraphics[height=7cm,width=9cm]{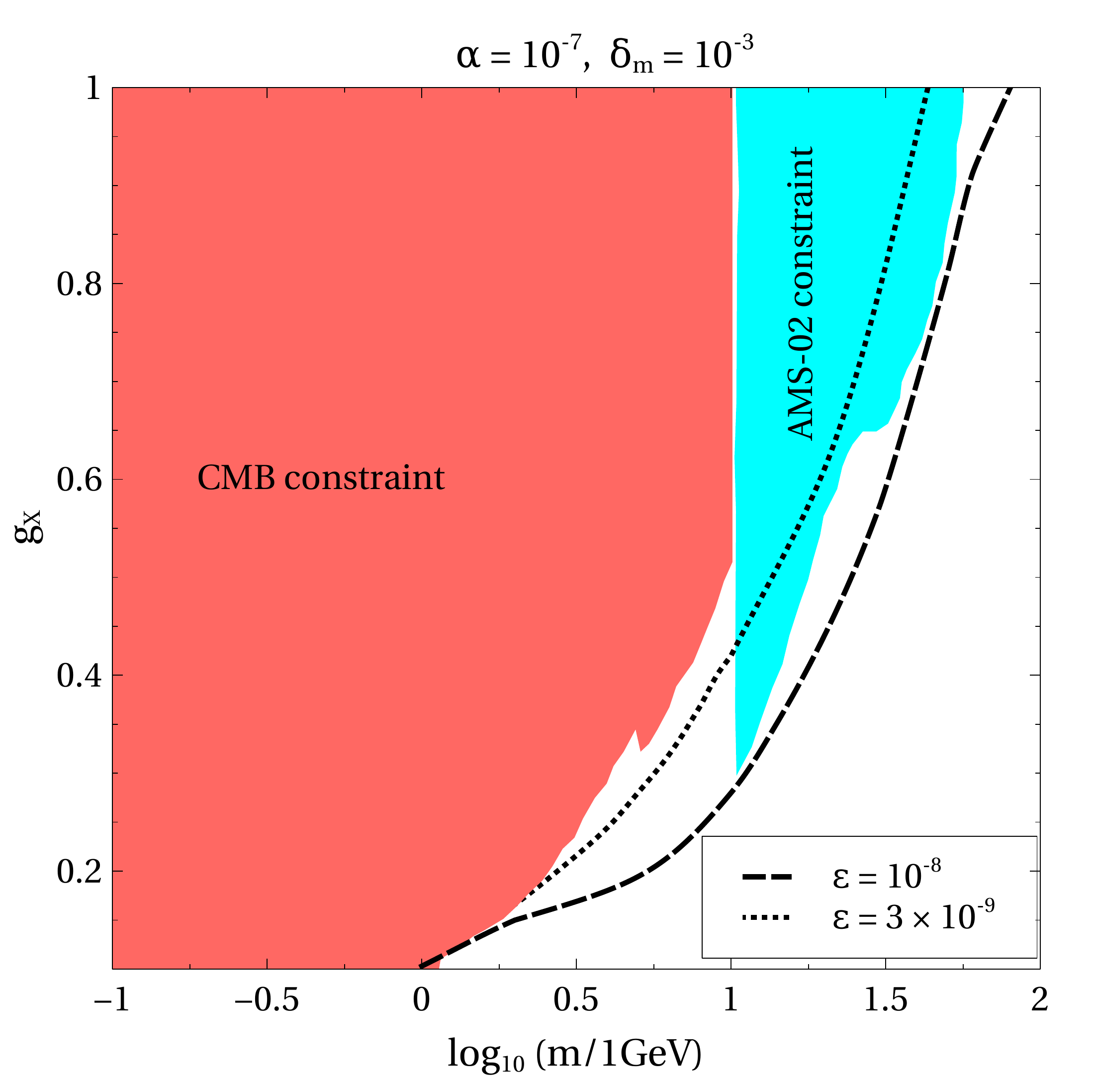}
        \caption{CMB (red) and AMS-02 (cyan) constraints
        in the $m-g_X$ plane. The relic density satisfied lines
        are shown for $\epsilon = 10^{-8}$ (dashed line)
        and $\epsilon  = 3\times 10^{-9}$ (dotted line). Here
        we have considered $\alpha = 10^{-7}$ and mass splitting
        parameter $\delta_m = 10^{-3}$. Note that
        for $\epsilon = 10^{-8}$ the entire dashed line
        satisfying correct relic density is disallowed from
        the condition
        $x_{\Gamma} > 5$.}
        \label{cross-sec-cmb-ams}
\end{figure}

It is to be noted that for $\delta_m \neq 0$, rate of production
of SM charged particles from DM annihilation is sufficient enough
to modify the ionisation history of Hydrogen and Helium
at the time of CMB. The observation from
the Planck \cite{Ade:2015xua} experiment puts
a strong constraint on DM mass as discussed in
\cite{Slatyer:2015jla}. Moreover, another constraint
is coming from the measurement of positron flux by
the AMS-02 detector \cite{Aguilar:2013qda}
as discussed in \cite{Elor:2015bho}.
In Fig.\,\,\ref{cross-sec-cmb-ams},
we show the constraints in the $m-g_X$ plane
from CMB measurements and AMS-02 experiment
for $\delta_m = 10^{-3}$. We have
also depicted two relic density satisfied 
lines in the $m-g_X$ plane for two fixed values
of the kinetic mixing parameter
$\epsilon= 10^{-8}$ (dashed line)
and $\epsilon = 3 \times 10^{-9}$ (dotted line) respectively.
From this figure one can clearly notice that
the relic density satisfied parameter space
for $\epsilon=10^{-8}$ and $m \gtrsim 1$ GeV is
allowed by the AMS-02 
measurements of positron flux in cosmic 
rays as well as CMB measurements
by Planck.
However, for $\epsilon=10^{-8}$,
the relic density satisfied parameter space
do not satisfy 
the criterion $x_{\Gamma}>5$ required for
preventing the thermalisation of the dark sector
with the SM bath
(see left panel of Fig.\,\,\ref{Fig:cosmo}).
The thermalisation criterion is more
relaxed for the portal coupling $\epsilon < 10^{-8}$. As a result,
we have found a parameter space for
$\epsilon= 3\times 10^{-9}$ (dotted line) and 
$1 \,{\rm GeV} \lesssim m \lesssim 10 \, {\rm GeV}$
which produces correct dark matter relic density
and at the same time is allowed from CMB measurements by Planck
and positron flux measured by AMS-02. 
Additionally, we would like to note that
here we have chosen $\alpha = 10^{-7}$ rad
to avoid the constraint arising from the
thermalisation criterion
(see right panel of Fig.\,\,\ref{Fig:cosmo}).
\section{Summary and Conclusion}
\label{sec:conclu}
In this work, we have considered an anomaly free 
minimal U(1)$_{X}$ extension of the SM by considering
two SM gauge singlet left chiral Weyl fermion of 
opposite U(1)$_{X}$ charge. The U(1)$_{X}$ symmetry
of the dark sector is broken by the vacuum expectation
value of a complex scalar $\eta$. Thus, the dark sector
contains a spin zero scalar $\hd$, two spin half 
Majorana fermions $\x_i\,(i=1\,,2)$ acting as
the dark matter candidates and a spin one massive
gauge boson $\zp$. Apart from that, the dark sector
can talk to the visible sector through the
kinetic mixing portal $\epsilon$ and also through the
$h-\hd$ mixing angle $\alpha$. Since, the couplings within
the dark sector particles are unknown, we choose all
the couplings in the dark sector are of the same order.
This ``democratic" choice would lead to degenerate/quasi-degenerate
dark sector which has very rich phenomenology. Under the assumption of the
feeble portal couplings, the dark sector is kinetically decoupled
and it evolves with a different temperature compared to the
visible sector. Due to the degeneracy within the dark sector,
the number density of dark species are exponentially suppressed only
when the mediator particles ($\zp$, $\hd$) start to
decay into the SM fields. Since, the decays are out-of-equilibrium,
the exponential suppressions in number densities arise
much later compared to the standard freeze-out mechanism
leading to a {\it delayed} freeze-out of dark matter and 
this is known as co-decaying dark matter.
In this scenario, we solved three coupled Boltzmann
equations (for $Y_{\chio + \chit}, Y_{\zp}, Y_{\hd}$) along
with the temperature evolution equation to study the
dynamics of kinetically decoupled dark sector taking into account
all the $2\ra2$ and $3\ra2$ processes within the dark sector.
We show that the presence of $3\ra2$ processes introduces the
\textit{cannibal} phase and it changes the dynamics
of the dark sector temperature evolution significantly.
The dark sector dynamics essentially depends on four parameters
i.e. $g_X$, $\epsilon$, $\alpha$, $m_\x$ and we have
presented the parameter spaces allowed from the relic
density constraint. Apart from that, we show parameter
space for the portal couplings allowed from the BBN observations, 
criterion for kinetically decoupled dark sector,
beam-dump experiments, SN1987A cooling, direct detection, and
Electroweak Precision Observables.

Furthermore, we have investigated the prospect of detecting
our dark matter candidates due to annihilation around the galactic
centre of the Milky-Way galaxy. Due to the presence
of the portal couplings, both $\chio$ and $\chit$ can produce
neutrino and $\gamma$-ray line spectrum from one step
cascade processes. We have calculated the neutrino flux
from a cascade process like $\chii \chii \ra \zp 
(\zp \ra \bar{\nu}_{\mu}\nu_{\mu})$
and compared our result with the observed atmospheric
neutrino flux by Super-Kamiokande neutrino detector.
We have also calculated the $\gamma$-ray flux from DM annihilation.
In our scenario, $\gamma$-ray flux composed of three
components such as $\gamma$-ray line signal from
$\chii \chii \ra \hd (\hd \ra \gamma \gamma)$, 
final state radiation from $\chii \chii \ra X (X \ra e^+ + e^-+ {\rm FSR},\, X = \hd,\,\zp)$, 
and ICS of final state charged particles with CMB photons.
The calculated total flux is then compared with the available
diffuse background $\gamma$-ray data from Fermi-LAT, EGRET, and INTEGRAL. 
We have found that the allowed parameter space obtained
after satisfying the relic density bound easily
satisfies all the constraints coming from
the indirect detection and CMB anisotropy measurement by Planck 
for the degenerate dark sector.
However, a certain region of parameter space
for the quasi degenerate dark sector has already been 
ruled-out from the measurements of
diffuse $\gamma$-ray flux by INTEGRAL, the CMB anisotropy, 
and positron flux from AMS-02.
\section{Acknowledgements}
Authors would like to acknowledge Alejandro Ibarra for a very
helpful and informative email communication regarding
indirect detection. One of the authors SG would like to acknowledge
University Grants Commission (UGC) for financial support
as a senior research fellowship.
\appendix
\section{A brief discussion on Boltzmann equation}
\label{App:A}
In this section we have derived the necessary Boltzmann equations
for the dark sector. The evolution of phase space distribution
function $f_A(p,\tp)$ of a species $A$ having $p$ and
$\tp$ as magnitude of three momentum ($p \equiv |\vec{p}|$)
and temperature respectively is governed by the Liouville
equation which has the following form
\bea
\dfrac{\partial f_A(p,\tp) }{\partial t} - \hub (t)\,
p \dfrac{\partial f_A(p,\tp) }{\partial p} 
&=& \mathcal{C}[f_A(p,\tp)] \,\,.
\label{BE_fp}
\eea
Here $\hub(t)$ is the Hubble parameter and $\mathcal{C}[f_A(p,\tp)]$
is the usual collision term for the evolution of $f_A(p,\tp)$. Due
to homogeneity and isotropy of the Universe, the distribution
function depends on $p$ only and not on the individual components.
Now, for a generic process like $A_1 + A_2 + .....+ A_n \ra B_1 +
B_2 + B_3 +.....B_m$, the moment for any quantity
(say $\mathcal{O}(p_\kappa)$) for $A_\kappa$ is given by
\bea
\hspace{-0.5cm}
\int \dfrac{g_\kappa\,d^3 \vec{p}_\kappa}{(2 \pi)^3}
\left[\dfrac{\partial f_{A_\kappa}(p_\kappa,\tp) }{\partial t} - 
\hub(t)\,p_\kappa \dfrac{\partial f_{A_\kappa}(p_\kappa,\tp) }{\partial p_\kappa}
\right]\mathcal{O}(p_\kappa)
&=&
\int\mathcal{C}[f_{A_\kappa}(p_\kappa,\tp)]\,\mathcal{O}(p_\kappa)\,
\dfrac{g_\kappa\,d^3 \vec{p}_\kappa}{(2 \pi)^3},
\label{BE_gen_op}
\eea
where $\vec{p}_\kappa$ and $g_\kappa$ are three momentum
and internal degrees of freedom of $A_\kappa$ respectively. Assuming
the phase space distribution function vanishes at the boundary,
Eq.\,\,\ref{BE_gen_op} can be written as
\bea
&&\dfrac{\partial \langle \mathcal{O}_{A_\kappa}\rangle}{\partial t} 
+ 3 \hub(t) \left(\langle \mathcal{O}_{A_\kappa}\rangle +
\langle \dfrac{p_\kappa}{3}\,\dfrac{\partial \mathcal{O}_{A_\kappa}}
{\partial p_\kappa}\rangle\right)
=
\int \prod_{\alpha=1}^{n}\prod_{\beta=1}^{m} d\Pi_{\alpha}\,d\Pi_{\beta}
\left(2 \pi \right)^4 \delta^4\left(\sum_{\alpha=1}^n P_\alpha -
\sum_{\beta=1}^m K_\beta \right)\times\nn\\
&& ~~~~~~~~~~~~~~~~~~~~~~
\overline{\left|\mathcal{M}\right|}^2
\left[f_{B_1}({k}_1,\tp) ... f_{B_m}({k}_m,\tp) - 
f_{A_1}({p}_1,\tp) ... f_{A_n}({p}_n,\tp)
\right]\mathcal{O}(p_\kappa)\,\,.
\label{BE_general_2}
\eea
In the above, we have ignored the Pauli-blocking and the stimulated
emission terms for fermions and bosons as these terms will not be
significant in our case where we have a non-relativistic dark sector
obeying the Maxwell-Boltzmann distribution. Here, $\vec{p}_\alpha$s and
$\vec{k}_\beta$s are the three momenta of the initial and the final
state particles and the corresponding four momenta are denoted by
$P_\alpha$, $K_\beta$ respectively. The Lorentz invariant phase
space measure $d\Pi_\alpha = \dfrac{g_\alpha d^3\vec{p}_\alpha}{2E_{p_\alpha}}$
where $\alpha$ stands for the initial state particles. For the
final state particles the index $\alpha$ and $p_\alpha$ in $d\Pi_\alpha$
are replaced by $\beta$ and $k_\beta$ respectively. Moreover, the Lorentz
invariant matrix amplitude square $\overline{\left
|\mathcal{M}\right|}^2$ for the process 
$A_1 + A_2 + .....+ A_n \ra B_1 + B_2 + B_3 +.....B_m$ is averaged
over spins of both initial and final state species.   
Furthermore, the quantity $\langle\mathcal{O}_{A_i}\rangle$
is a function of $\tp$ only and it is defined as
\be
\langle\mathcal{O}_{A_\kappa}\rangle = \int d\Pi_\kappa\,2E_{p_{\kappa}}\,
f_{A_\kappa}({p_\kappa},\tp)\,\mathcal{O}(p_\kappa)\,.
\label{opeator}
\ee

Let us consider a particular case when there are $x$ and
$y$ number of identical $A$ species in the initial and
the final state respectively i.e.\,the process we are 
considering is as follows\\ 
$\underbrace{ A + A + A + ....+A}_x
+\underbrace{C_{x+1} + ... + C_{n}}_{n-x}  \ra
\underbrace{ A + A + A + ....+A}_y
+ \underbrace{B_{y+1} + ... + B_{m}}_{m-y}$.
In this situation we need to keep in mind that under
exchange of $\vec{p}_1, \vec{p}_2, ...\vec{p}_x$ in the
initial state and $\vec{k}_1, \vec{k}_2, ...\vec{k}_y$
in the final state the process remains unaltered. Hence,
in order to take into account that effect the collision
term must be divided by $x!$ and $y!$ to avoid
overcounting. Now, if we want to find the time evolution
of $\langle \mathcal{O}_{A}(\tp)\rangle$ for
the species $A$, then we need to write the Boltzmann equations
(Eq.\,\,\ref{BE_general_2}) for all the $A$s (both for
initial as well as final state) and then add them. After
a few algebraic simplifications, we have arrived at the
following form
\bea
&&
{\hspace{-0.4cm}}
\sum_{\kappa =1}^{x+y}
\dfrac{\partial \langle \mathcal{O}_{A_\kappa}\rangle}{\partial t}
+ 3 \hub(t) \left(\langle \mathcal{O}_{A_\kappa}\rangle 
+ \langle \dfrac{p_\kappa}{3}\,\dfrac{\partial \mathcal{O}_{A_\kappa}}
{\partial p_\kappa}\rangle\right) =
\dfrac{1}{x!\,y!}\int \prod_{\alpha=1}^{n}\prod_{\beta=1}^{m} d\Pi_{\alpha}\,d\Pi_{\beta}
\left(2 \pi \right)^4
\delta^4\left(\sum_{\alpha=1}^n P_{\alpha} 
- \sum_{\beta=1}^m K_{\beta}
\right)
\times\nn\\
&& ~~~~~~~~~~~~~~~~~~~~
\overline{|\mathcal{M}|}^2 
\left[
f_{A}({k}_1,\tp) ... f_{A}({k}_y,\tp) 
f_{B_{y+1}}({k}_{y+1},\tp)..f_{B_m}({k}_{m},\tp)\,-
\right.\nn\\ && \left. ~~~~~~~~~~~~~~~~~~~
f_{A}({p}_1,\tp) ... f_{A}({p}_x,\tp) f_{C_{x+1}}({p}_{x+1},\tp)
..f_{C_n}({p}_{n},\tp)\right]\,
\left(x \times \mathcal{O}(\vec{p}_{r}) 
- y\times \mathcal{O}(\vec{k}_{t})\right)\,\,.
\label{BE_general_3}
\eea
In the right hand side of the above equation, we have used the
freedom of permutation symmetry of $\vec{p}_\alpha$s (for $\alpha=$ 1 to $x$),
$\vec{k}_\beta$s (for $\beta=1$ to $y$) and $1\leq r\leq x$, $1\leq t\leq y$.
Let us note in passing, if particle composition in the initial and the final
states are exactly identical (e.g. elastic scattering) then the collision
term must be divided by an extra $2!$ since under the exchange of
$\{\text{initial state}\}\leftrightarrow \{\text{final state}\}$,
the process remains unchanged. 

Now, we will derive the Boltzmann equations for our dark sector species
$\chio$, $\chit$, $\zp$ and $\hd$. Let us first consider the species $\chio$
and a particular annihilation channel $\chio(P_1)\chio(P_2)
\ra\zp(K_1)\zp(K_2)$. For this process,
the Boltzmann equation for the number density of $\chio$ follows
from Eq.\,\,\ref{BE_general_3} using $\mathcal{O} = 1$ and
$\langle\mathcal{O}\rangle = n_{\chio}(\tp)$ as
\bea
&&\dfrac{d n_{\chio}}{d t} 
+ 3 \hub(t) n_{\chio}
=
\dfrac{1}{2!\,2!}\int \prod_{\alpha=1}^{2}\prod_{\beta=1}^{2} d\Pi_{\alpha}\,d\Pi_{\beta}
\left(2 \pi \right)^4 \delta^4\left(P_1+P_2-K_1-K_2\right)
\times\nn \\ && ~~~~~~~~~~~~~~~~~~~~~~
\overline{\left|\mathcal{M}\right|}^2_{\chio\chio\ra\zp\zp}
\left[f_{\zp}({k}_1,\tp)\,f_{\zp}({k}_2,\tp) - 
f_{\chio}({p}_1,\tp)\,f_{\chio}({p}_2,\tp)
\right]\times 2\,.
\label{BE_chi1_1}
\eea
Here we consider $\chio$ is a Majorana fermion. Using
$f_{\zp}(k_\beta,\tp) = \dfrac{n_{\zp}}{n^{\rm eq}_{\zp}}\,{\rm Exp}({-E_{k_\beta}/\tp})$
and $f_{\chio}(p_\alpha,\tp) = \dfrac{n_{\chio}}{n^{\rm eq}_{\chio}}\,{\rm Exp}({-E_{p_\alpha}/\tp})$
and also the conservation of energy
in Eq.\,\ref{BE_chi1_1} we get,
\bea
&&\dfrac{d n_{\chio}}{d t} 
+ 3 \hub(t) n_{\chio}
=
-\int \prod_{\alpha=1}^{2}\prod_{\beta=1}^{2} d\Pi_{\alpha}\,d\Pi_{\beta}
\left(2 \pi \right)^4 \delta^4\left(P_1+P_2-K_1-K_2\right)
\times\nn \\ && ~~~~~~~~~~~~~~~~~~~~~~
\overline{\left|\mathcal{M}\right|}^2_{\chio\chio\ra\zp\zp}
\left[\left(\dfrac{n_{\chio}}{n^{\rm eq}_{\chio}}\right)^2
- \left(\dfrac{n_{\zp}}{n^{\rm eq}_{\zp}}\right)^2 
\right]\,{\rm Exp}\left(-\dfrac{E_{p_1}+E_{p_2}}{\tp}\right)\,\,.
\eea   
Now, using the definition of $\langle \sigma {\rm v} \rangle$
for an annihilation process as given in \cite{Gondolo:1990dk},
we can write the collision term of the above equation in
a more compact form as
\bea
&&\dfrac{d n_{\chio}}{d t} 
+ 3 \hub(t) n_{\chio}
=
- \langle {\sigma {\rm v}}_{\chio\chio\ra\zp\zp}\rangle^{\tp}
\left[\left(\dfrac{n_{\chio}}{n^{\rm eq}_{\chio}}\right)^2
- \left(\dfrac{n_{\zp}}{n^{\rm eq}_{\zp}}\right)^2 
\right]\,(n^{\rm eq}_{\chio})^2\,,
\eea
where, 
\bea
\langle {\sigma {\rm v}}_{\chio\chio\ra\zp\zp}\rangle^{\tp} =
\dfrac{1}{(n^{\rm eq}_{\chio}(\tp))^2}
\int_{4\,m^2_1}^{\infty}d\mathfrak{s}\,\sigma_{{}_{\chio\chio\ra\zp\zp}}
\sqrt{\mathfrak{s}}\,\left(\mathfrak{s}-4\,m^2_1\right)
\,{\rm K}_1\left(\dfrac{\sqrt{\mathfrak{s}}}{\tp}\right) \,,
\eea
with ${\rm K}_1\left(\frac{\sqrt{\mathfrak{s}}}{\tp}\right)$ being
the modified Bessel function of second kind and order one while
$\mathfrak{s} = (P_1 + P_2)^2$ is one of the Mandelstam variables. The
cross section $\sigma_{{}_{\chio\chio\ra\zp\zp}}$ appearing
above has the usual definition,
\bea
\sigma_{{}_{\chio\chio\ra\zp\zp}} = 
\dfrac{1}{2\,{\rm v}}\prod_{\alpha=1}^{2}\dfrac{1}
{2\,E_{p_\alpha}}\int \prod_{\beta=1}^{2} d\Pi_{\beta}
\left(2 \pi \right)^4 \delta^4\left(P_1+P_2-K_1-K_2\right)\,
\overline{\left|\mathcal{M}\right|}^2_{\chio\chio\ra\zp\zp}\,,
\eea
and $n^{\rm eq}_{\chio}(\tp)$ can be obtained from Eq.\,\ref{opeator}
by using $f_{A_\kappa}\,=\,f^{\rm eq}_{\chio}(E_{p_1},\tp)\,=\,
{\rm Exp}\left(-E_{p_1}/\tp\right)$. In our model,
$\chio$ has two annihilation channels and hence
the complete Boltzmann equation for $\chio$ is 
\bea
&&\dfrac{d n_{\chio}}{d t} 
+ 3 \hub(t) n_{\chio}
=
- \sum_{j = \zp,\hd}
\langle {\sigma {\rm v}}_{\chio\chio\ra jj}\rangle^{\tp}
\left[\left(\dfrac{n_{\chio}}{n^{\rm eq}_{\chio}}\right)^2
- \left(\dfrac{n_{j}}{n^{\rm eq}_{j}}\right)^2 
\right]\,(n^{\rm eq}_{\chio})^2\,.
\label{BE_chi1_full}
\eea
Since both the dark matter candidates $\chio$ and
$\chit$ have identical interactions, thus the Boltzmann
equation for $\chit$ is similar to that of $\chio$. Now,
if they are degenerate in mass then their number densities
will also be identical. In that case, it is needless to consider
number densities of $\chio$ and $\chit$ separately and instead
we will consider the total dark matter number density
$n_{\x}=n_{\chio} + n_{\chit}$ with $n_{\chio}=n_{\chit}$. Therefore,
in terms of total number density $n_{\x}$, the Boltzmann equation
for the dark matter can easily be obtained from Eq.\,\ref{BE_chi1_full}
as
\bea
\dfrac{d n_{\x}}{d t} 
+ 3 \hub(t) n_{\x}
=
-\dfrac{1}{4} \sum_{j = \zp,\hd}
\langle {\sigma {\rm v}}_{\x\x\ra j j}\rangle^{\tp}
\left[\left(\dfrac{n_{\chi}}{n^{\rm eq}_{\chi}}\right)^2
- \left(\dfrac{n_{j}}{n^{\rm eq}_{j}}\right)^2 
\right]\,(n^{\rm eq}_{\chi})^2\,,
\label{BE_chi}
\eea 
where, $\langle {\sigma {\rm v}}_{\x\x\ra jj}\rangle^{\tp}
=\sum_{i=1}^{2} 
\langle {\sigma {\rm v}}_{\chii\chii\ra j j}\rangle^{\tp}$.

The other two dark sector species $\zp$ and $\hd$ have
both decay and annihilation modes. Therefore, the collision
terms of both $\zp$ and $\hd$ have contributions from
annihilations and from decays as well. Therefore, following
the similar procedure as we have discussed above, one can
write the Boltzmann equations for $\zp$ and $\hd$ which
are given below,
\bea
&&\dfrac{d n_{\zp}}{d t} 
+ 3 \hub(t)\,n_{\zp}
=
\dfrac{1}{4} 
\langle {\sigma {\rm v}}_{\x\x\ra \zp \zp}\rangle^{\tp}
\left[\left(\dfrac{n_{\chi}}{n^{\rm eq}_{\chi}}\right)^2
- \left(\dfrac{n_{\zp}}{n^{\rm eq}_{\zp}}\right)^2 
\right]\,(n^{\rm eq}_{\chi})^2 - \nn \\ && ~~~~~~~~~~~
\langle {\sigma {\rm v}}_{\zp\zp\ra \hd \hd}\rangle^{\tp}
\left[\left(\dfrac{n_{\zp}}{n^{\rm eq}_{\zp}}\right)^2
- \left(\dfrac{n_{\hd}}{n^{\rm eq}_{\hd}}\right)^2 
\right]\,(n^{\rm eq}_{\zp})^2 
-\langle\Gamma_{\zp}\rangle^{\tp} n_{\zp} +
\langle\Gamma_{\zp}\rangle^{T} n^{\rm eq}_{\zp}(T)\,, \nn \\
\label{BE_zp} \\
&&\dfrac{d n_{\hd}}{d t} 
+ 3 \hub(t)\,n_{\hd}
=
\dfrac{1}{4} 
\langle {\sigma {\rm v}}_{\x\x\ra \hd \hd}\rangle^{\tp}
\left[\left(\dfrac{n_{\chi}}{n^{\rm eq}_{\chi}}\right)^2
- \left(\dfrac{n_{\hd}}{n^{\rm eq}_{\hd}}\right)^2 
\right]\,(n^{\rm eq}_{\chi})^2 + \nn \\ && ~~~~~~~~~~~
\langle {\sigma {\rm v}}_{\zp\zp\ra \hd \hd}\rangle^{\tp}
\left[\left(\dfrac{n_{\zp}}{n^{\rm eq}_{\zp}}\right)^2
- \left(\dfrac{n_{\hd}}{n^{\rm eq}_{\hd}}\right)^2 
\right]\,(n^{\rm eq}_{\zp})^2 
-\langle\Gamma_{\hd}\rangle^{\tp} n_{\hd} +
\langle\Gamma_{\hd}\rangle^{T} n^{\rm eq}_{\hd}(T)\,, \nn \\
\label{BE_hd}
\eea
where, $\langle \Gamma_{j} \rangle^{\tp}$ is the thermal averaged
total decay width of the species $j$ ($j = \zp,\hd$) and it depends
on dark sector temperature $\tp$. The expression of $\langle \Gamma_{j} \rangle^{\tp}$
in terms of total decay width $\Gamma_{j}$ is given by
\bea
\langle \Gamma_{\beta} \rangle^{\tp} = \Gamma_{j}\dfrac{{\rm K}_{1}\left(m_{j}/\tp\right)}
{{\rm K}_{2}\left(m_{j}/\tp\right)}\,.
\label{Gamaavr}
\eea
The last terms in Eqs.\,\,\ref{BE_zp} and \ref{BE_hd} represent the contribution from the
inverse decay. In our model, both $\zp$ and $\hd$ have decay
modes only into the SM particles which have temperature $T$. Hence,
these inverse decay terms depend on the SM temperature instead of
the dark sector temperature $\tp$ like the others terms. The equilibrium
number density $n^{\rm eq}_{j}(T)$ of the species $j$ at $T$
can be found from Eq.\,\ref{opeator} using
the Maxwell-Boltzmann distribution at temperature $T$. 
\subsection{Evolution of dark sector temperature ($\tp$)}
\label{App:A1}
Our next task is to derive the fourth Boltzmann equation describing
evolution of the dark sector temperature $\tp$ with respect to that
of the SM. All the dark sector species are in kinetic equilibrium
and they share a common temperature $\tp$ before
the freeze-out of dark matter candidates at $\tp_{f}$. 
Therefore, in order to compute the temperature of the dark sector,
we need to find the temperature of a particular species. Here we
consider the temperature of our dark matter candidates. As
we have seen in the previous section that for two identical dark
matter candidates, we do not need to consider them separately in
the Boltzmann equations. Similarly, in the temperature Boltzmann
equation also, we will use the total dark matter density $n_{\x}$
rather than the individual ones. The temperature $\tp$ of any
species is defined as
\cite{Bringmann:2006mu} 
\be
T^\prime = \dfrac{g_{\x}}{n_\chi(\tp)}\int \dfrac{d^3 \vec{p}}
{(2\pi)^3}\dfrac{p^2}{3E_p}f_\chi(p,\tp)\,,
\label{defn_tp}
\ee
where, $m_{\x}$ and $g_{\x}$ are the mass and internal
degrees of freedom of dark matter respectively. The energy $E_p = \sqrt{p^2 + m_\chi^2}$
and $n_\chi$ is the total number density of dark matter. 
Now considering $\mathcal{O}(p) = \dfrac{p^2}{3 E_p}$, we can easily 
derive the temperature evolution equation from Eq.\,\,\ref{BE_general_3} 
which is written below
\bea
\hspace{-1.0cm}
\dfrac{d \tp}{dt} + \left(2 - \delta(\tp)\right) \tp \hub(t) + 
\dfrac{\tp}{n_\chi}\left(\dfrac{dn_\chi}{dt} + 3\,\hub\,n_\chi \right)
&=& \dfrac{1}{n_\x}\left[\mathcal{F}(\tp)_{2\rightarrow2}
+ \mathcal{F}(\tp)_{3\rightarrow2}\right],
\label{BE_tp}
\eea
and 
\begin{equation}
\delta \left(\tp \right) = 1 -\,\dfrac{g_\x}{n_\chi(\tp) \tp}
\int \dfrac{d^3 \vec{p}}{(2\pi)^3} 
\dfrac{p^2 m_\chi^2}{3 E_p^3} f_\chi(p,\tp)\,\,.
\end{equation}
The functions $\mathcal{F}(\tp)_{2\ra2}$\,and\, 
$\mathcal{F}(\tp)_{3\ra2}$\,\,in the right hand side
of Eq.\,\,\ref{BE_tp} contain contributions from all
$2\ra2$ and $3\ra2$ processes which can affect the temperature
of the dark sector. The general form of these functions
is shown in Eq.\,\,\ref{BE_general_3}. Their actual expression
depends on the particular interaction process. We have presented
a detail discussion about these functions in Appendix \ref{App:B}.
\section{Calculation of $\mathcal{F}(\tp)_{2\ra2}$ and
$\mathcal{F}(\tp)_{3\ra2}$}
\label{App:B}
In this section, we have derived the specific form of the functions
$\mathcal{F}(\tp)_{2\ra2}$ and $\mathcal{F}(\tp)_{3\ra2}$
encoding all the information about the relevant $2\ra2$
and $3\ra2$ processes which can change the temperature of
$\chii$. Since all the species of the dark sector share a
common temperature, we have considered only those interaction
processes which have impact on the temperature of $\chii$.  
\subsection{Calculation of $\mathcal{F}(\tp)_{2\ra2}$}
\label{App:B1}
In our model, the $2\ra2$ processes within the dark sector that can
affect the evolution of $\tp$ are $\chii\chii \ra \zp \zp$
and $\chii \chii \ra \hd \hd$ $(i=1,2)$. Here we  
derive the function $\mathcal{F}_{2\ra2}(\tp)$ for a
generic process like $\chii\chii\ra j j$ where $j$
can be either $\zp$ or $\hd$. The master equation for writing the
expression of $\mathcal{F}_{2\ra2}(\tp)$ can be found from
the right hand side of Eq.\,\,\ref{BE_general_3} as
\bea
\mathcal{F}_{2\ra2}(\tp)|_{\chii\chii\ra jj} &=&\dfrac{1}{2!\,2!} 
\int \prod_{\alpha=1}^{2}\prod_{\beta=1}^{2} d\Pi_\alpha\,d\Pi_\beta
\left(2 \pi \right)^4  \delta^4 \left(\sum_{\alpha=1}^{2}
P_\alpha  - \sum_{\beta=1}^{2}K_\beta
\right) \overline{|{\mathcal{M}|^2}}_{\chii\chii \ra jj} \nn \\
&\,& \times\left[f_j (k_1,\tp) f_j(k_2,\tp) 
- f_{\chii}(p_1,\tp) f_{\chii}(p_2,\tp)
\right]  
\left(\dfrac{ p_1^2}{3 E_{p_1}} + \dfrac{ p_2^2}{3 E_{p_2}}\right),
\label{2to2Cterm_for_tp_gen}
\eea
where, as defined earlier $P_\alpha$s are the four momentum for
the initial state while that of the final state are denoted by
$K_\beta$s. The other quantities are already defined in the previous section.
Eq.\,\,\ref{2to2Cterm_for_tp_gen} is symmetric under the exchange of
$P_1 \leftrightarrow P_2$ and $K_1 \leftrightarrow K_2$ and 
using this exchange symmetry between $P_1 \leftrightarrow P_2$
one can further simplify as
\bea
\mathcal{F}_{2\ra2}(\tp)|_{\chii\chii\ra jj} &=&\dfrac{1}{2!\,2!} 
\int \prod_{\alpha=1}^{2}\prod_{\beta=1}^{2} d\Pi_\alpha\,d\Pi_\beta
\left(2 \pi \right)^4  \delta^4 \left(\sum_{\alpha=1}^{2}
P_\alpha  - \sum_{\beta=1}^{2}K_\beta
\right) \overline{|{\mathcal{M}|^2}}_{\chii\chii \ra j j} \nn \\
&\,& \times\left[f_j (k_1,\tp) f_j(k_2,\tp) 
- f_{\chii}(p_1,\tp) f_{\chii}(p_2,\tp)
\right]  
\dfrac{2\,p_1^2}{3 E_{p_1}}\,\,.
\eea
Now, following the similar procedure as we have done below
Eq.\,\,\ref{BE_chi1_1}, a more compact form of the function
$\mathcal{F}_{2\ra2}(\tp)$ is given below,
\bea
\hspace{-0.5cm}
\mathcal{F}_{2\ra2}(\tp)|_{\chii\chii\ra jj} &=& -\dfrac{2}{2!}\,
\langle \dfrac{p_1^2}{3 E_{p_1}}\,\sigma {\rm v} 
_{\chii \chii \ra j j}\rangle^{\tp}\,
\left[\left(\dfrac{n_{\chii}(\tp)}{\, n^{\rm eq}_{\chii}(\tp)}\right)^2
-\left(\dfrac{n_j (\tp)}{n^{\rm eq}_j(\tp)} \right)^2
\right]\,\, \left(n^{\rm eq}_{\chii}(\tp)\right)^2,
\label{2to2Cterm_for_chi_a}
\eea
where
\bea
\langle \dfrac{p_1^2}{3 E_{p_1}}\,\sigma {\rm v}
_{\chii\chii \ra j j}\rangle^{\tp}
&=&
\dfrac{1}
{\left(n^{\rm eq}_{\chii}(\tp)\right)^2}
\int\prod_{\alpha=1}^{2} 2E_{p_\alpha}d\Pi_\alpha\,\dfrac{p_1^2}{3 E_{p_1}}\,
\sigma {\rm v}_{\chii \chii \ra j j}\,
{\rm Exp}\left(-\dfrac{E_{p_1} + E_{p_2}}{\tp}\right)\,.
\label{sigmaVp2by3E_int}
\eea
Now, following the prescription given in \cite{Gondolo:1990dk},
the expression of $\langle \dfrac{p_1^2}{3 E_{p_1}}\,\sigma {\rm v}
_{\chii \chii \ra j j}\rangle^{\tp}$ can be
simplified further by using the new variables instead of
$E_{p_1}$, $E_{p_2}$ and $\theta$. The transformation relations are
given below 
\bea
E_+ &=& E_{p_1} + E_{p_2}\,,\nn \\
E_- &=& E_{p_1} - E_{p_2}\,,\nn\\
\mathfrak{s} &=& 2m_{i}^2 + 2E_{p_1} E_{p_2} -2|\vec{p}_1| |\vec{p}_2| \,\rm cos\theta \,,\nn
\eea
where $\theta$ is the angle between $\vec{p}_1$ and $\vec{p}_2$.
The region of integration for these new variables are
$\left|E_{-}\right| \leq \sqrt{1-\dfrac{4m^2_{i}}
{\mathfrak{s}}}\sqrt{E^2_{+}-\mathfrak{s}}$, $E_{+}\geq \sqrt{\mathfrak{s}}$
and $\mathfrak{s}\geq4m^2_{i}$. In terms of the new
variable $\mathfrak{s}$ the compact form of
$\langle \dfrac{p_1^2}{3 E_{p_1}}\,\sigma {\rm v}
_{\chii\chii \ra jj}\rangle^{\tp}$, after a few
mathematical simplifications, can be obtained as
\bea
 \langle \dfrac{p_1^2}{3 E_{p_1}}\,\sigma {\rm v}_
{\chii\chii \ra jj}\rangle^{\tp}
&=&\dfrac{1}{48\,m_{i} ^4\,{\rm K}_2\left(\dfrac{m_{i}}{\tp}\right)^2}
\int _{4m^2_{i}} ^\infty d \mathfrak{s} \left(
\sigma {\rm v}_{\chii\chii \ra jj}\right)
\sqrt{\mathfrak{s}-4m_{i}^2} 
\left[(\mathfrak{s}+2m_{i}^2)\,
{\rm K}_1\left(\dfrac{\sqrt{\mathfrak{s}}}{\tp}\right)
\right. \nn\\
&& ~~~~~~~~~~~~~~~~~~~~~~ \left.
+
\left(
\dfrac{\mathfrak{s}-4m_{i}^2}{2} \dfrac{\sqrt{\mathfrak{s}}}{\tp} +
\dfrac{4{\tp} (\mathfrak{s} + 2m_{i}^2)}{\sqrt{\mathfrak{s}}}\right)
{\rm K}_2\left(\dfrac{\sqrt{\mathfrak{s}}}{\tp}\right) 
\right]\,.
\label{sigmaVp2by3E}
\eea
This integral form of $\langle \dfrac{p_1^2}{3 E_{p_1}}\,\sigma {\rm v}
_{\chii\chii \ra jj}\rangle^{\tp}$ agrees well
with the expression given in \cite{Yang:2019bvg}. Finally, while
incorporating this term in the Boltzmann equation (i.e.\,\,in
$\mathcal{F}_{2\ra2}(\tp)$) we have to take contributions
from both the dark matter candidates (sum over the index $i$)
and replace individual densities $n_{\chii}$
by the total density $n_{\x}$. Therefore, the final expression of
$\mathcal{F}_{2\ra2}(\tp)|_{\x\x\ra jj}$ for a annihilation process $\x\x\ra jj$
is given by
\bea
\mathcal{F}_{2\ra2}(\tp)|_{\x\x\ra jj} &=& -\dfrac{1}{4}\,
\langle \dfrac{p_1^2}{3 E_{p_1}}\,\sigma {\rm v} 
_{\x \x \ra jj}\rangle^{\tp}\,
\left[\left(\dfrac{n_{\x}(\tp)}{\, n^{\rm eq}_{\x}(\tp)}\right)^2
-\left(\dfrac{n_j (\tp)}{n^{\rm eq}_j(\tp)} \right)^2
\right]\,\, \left(n^{\rm eq}_{\x}(\tp)\right)^2,
\label{F2to2term}
\eea 
with $\langle \dfrac{p_1^2}{3 E_{p_1}}\,\sigma {\rm v} 
_{\x \x \ra jj}\rangle^{\tp} = \sum_{i=1}^{2}
\langle \dfrac{p_1^2}{3 E_{p_1}}\,\sigma {\rm v} 
_{\chii \chii \ra jj}\rangle^{\tp}$.
\subsection{Calculation of $\mathcal{F}(\tp)_{3\ra2}$}
\label{App:B2}
In our case, there are several $3\ra2$ processes which can change the
dark sector temperature ($\tp$) and we have considered all of them.
\subsubsection{\underline{$\chii \chii \hd \ra \chii \chii$}}
\begin{figure}[h!]
\centering
\includegraphics[height=6cm,width=17cm,angle=0]{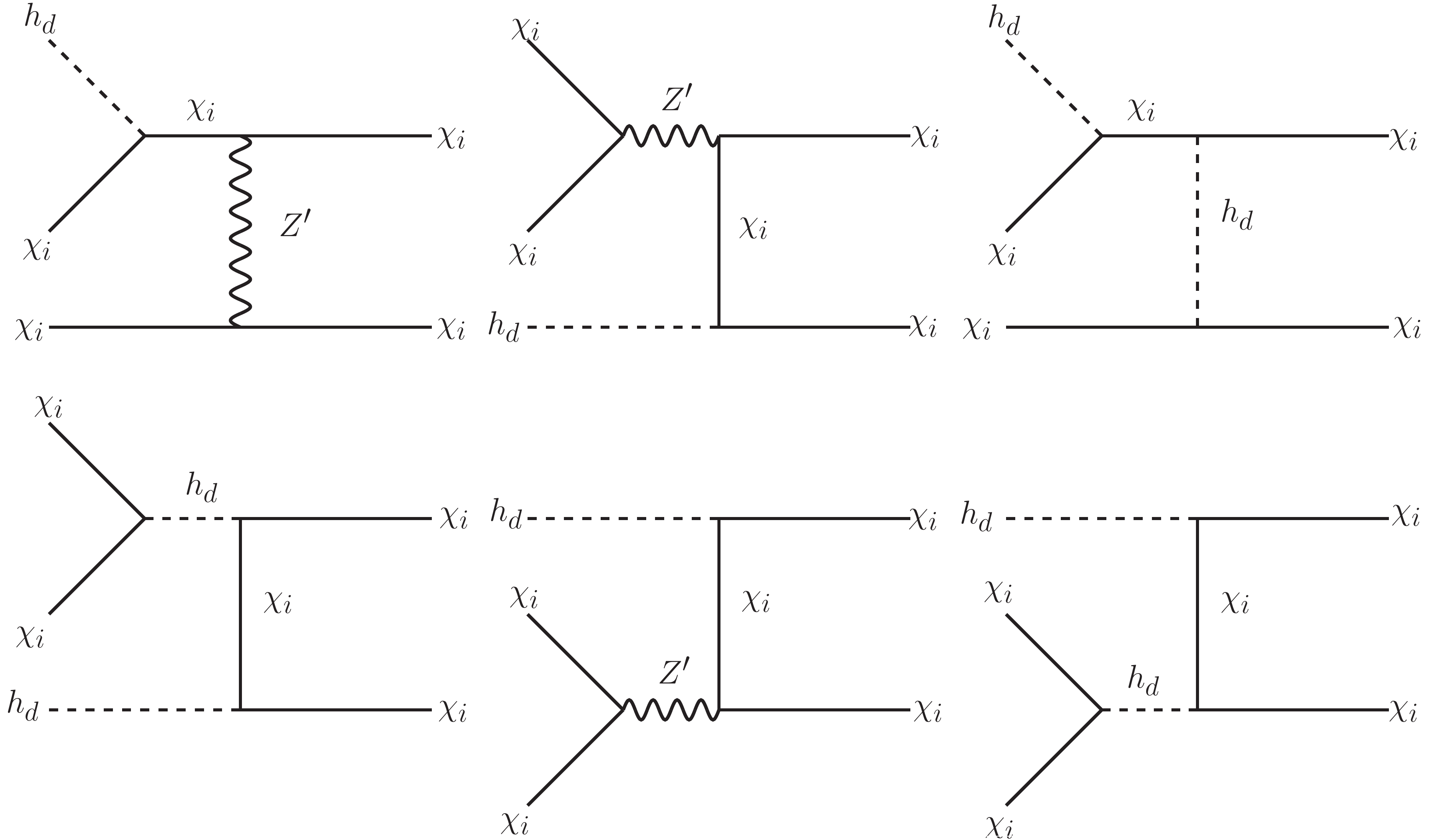}
\includegraphics[height=6cm,width=17cm,angle=0]{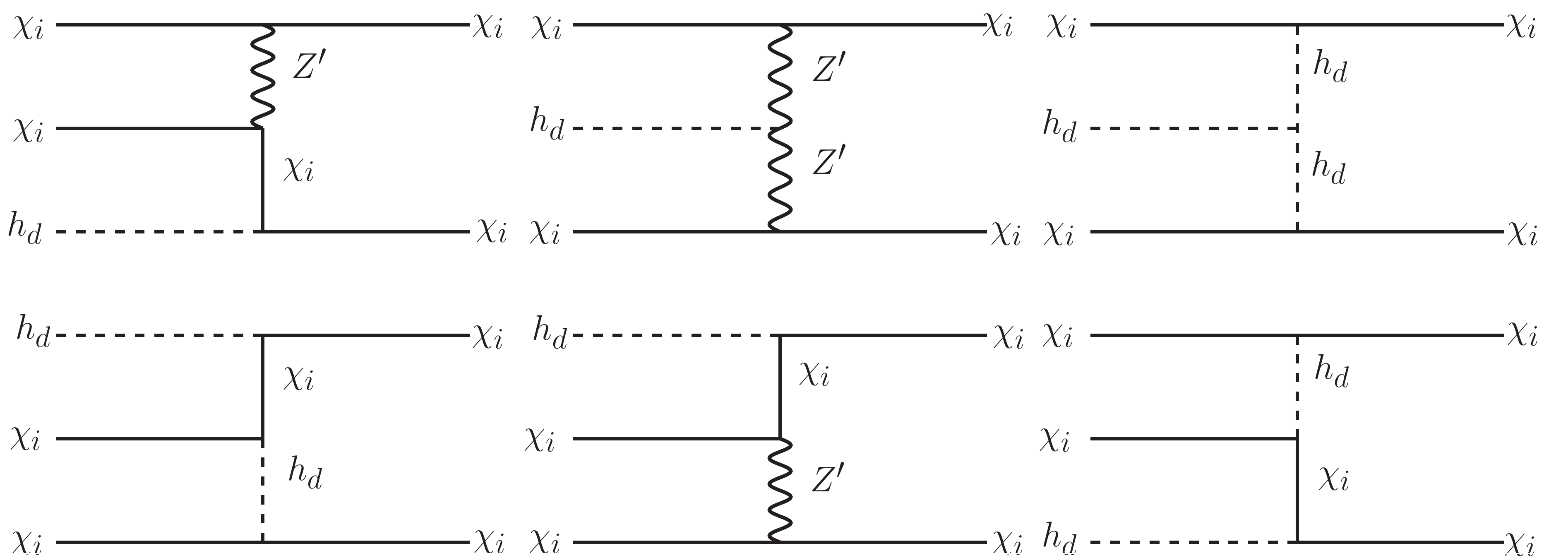}
\includegraphics[height=6cm,width=17cm,angle=0]{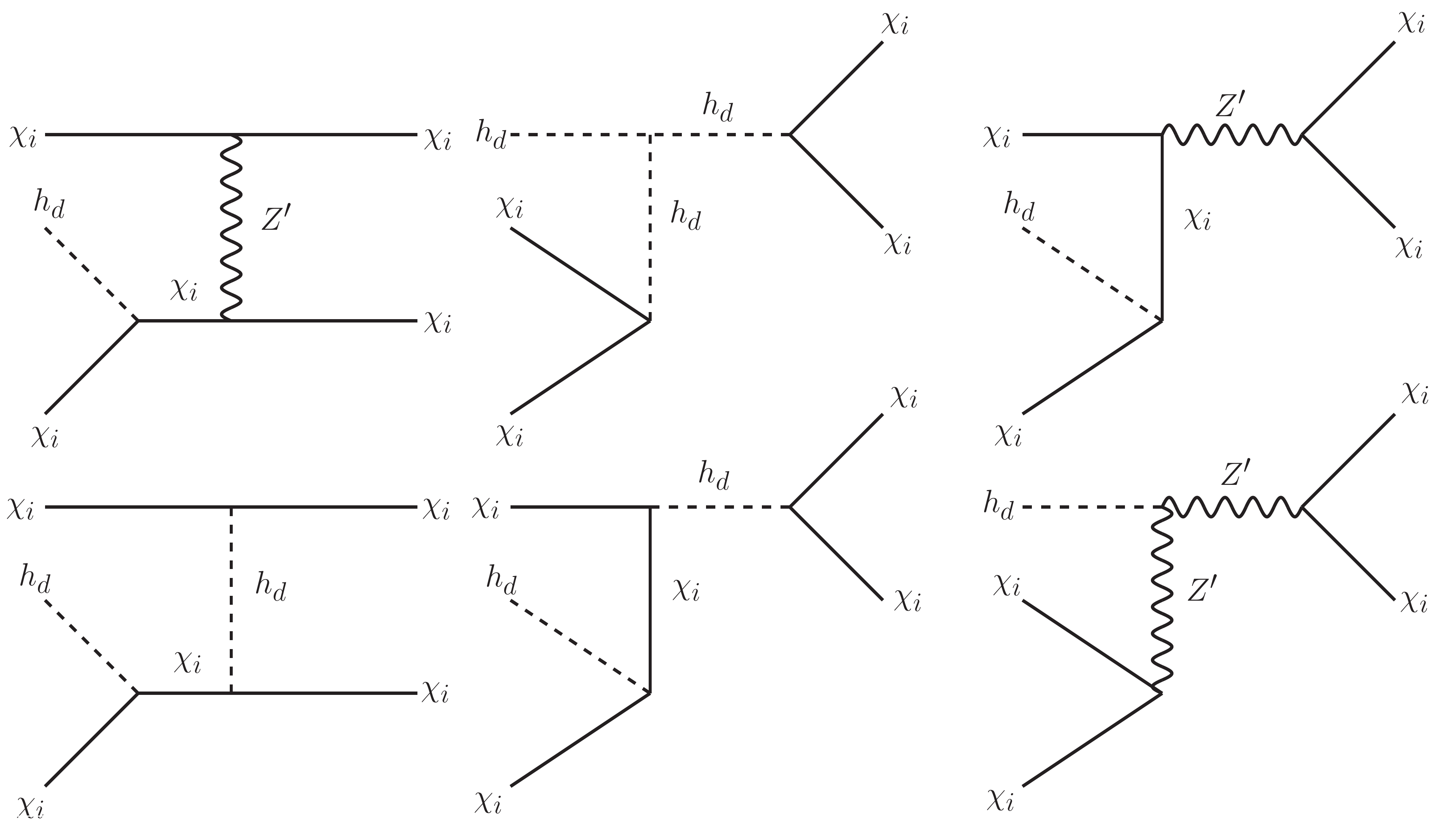}
\caption{Feynman diagrams for $\chii \chii \hd \ra \chii \chii$.}
\label{Fig:3to2_2}
\end{figure}

We first derive the contribution to $\mathcal{F}(\tp)_{3\ra2}$
from one such process namely $\chii \chii \hd \ra \chii \chii$.
Thereafter, we will present the final expressions of other $3\ra2$ processes.
The Feynman diagrams for the scattering $\chii \chii \hd \ra \chii \chii$
are shown in Fig.\,\,\ref{Fig:3to2_2}.
The contribution to $\mathcal{F}(\tp)_{3\ra2}$ from
$\chii \chii \hd \ra \chii \chii$ is given by
\bea
\mathcal{F}(\tp)_{3\ra 2}|_{\chii \chii \hd \ra \chii \chii}
&=&
\dfrac{1}{2!\,2!}\int \prod_{\alpha =1}^3
\prod_{\beta=1}^2 d\Pi_{\alpha} d\Pi_{\beta}
\left(2\pi\right)^4 \delta^4\left(\sum_{\alpha=1}^3 P_\alpha
- \sum_{\beta=1}^2 K_{\beta}\right)
\overline{|\mathcal{M}|^2}_{\chii \chii \hd \ra \chii \chii} \times
\nn\\
&&
\left[f_{\chii}(k_1,\tp) f_{\chii}(k_2,\tp)-
f_{\chii}(p_1,\tp) f_{\chii}(p_2,\tp) f_{\hd}(p_3,\tp) \right] \times \nn\\ &&
\left(
\dfrac{p_1^2}{3E_{p_1}} + \dfrac{p_2^2}{3E_{p_2}} -
\dfrac{k_1^2}{3E_{k_1}} - \dfrac{k_2^2}{3E_{k_2}} 
\right)\,\,.
\label{F3to2_gen}
\eea
As mentioned in \cite{Berlin:2016gtr}, the quantity
$\left(\sigma {\rm v}^2\right)_{a+b+c \ra d+e}$ for
a $3\ra2$ scattering process $a+b+c \ra d+e$
is defined as
\bea
\left(\sigma {\rm v}^2\right)_{a+b+c \ra d+e}
&=& \dfrac{1}{8 E_{p_1}E_{p_2}E_{p_3}\,m!}\int \prod_{\beta=1}^{2}d\Pi_{\beta}
\left(2\pi\right)^4 \delta^4(P_1 + P_2 +P_3 -K_1-K_2)\,
\overline{|\mathcal{M}|^2}_{a+b+c \ra d+e}\,\,,\nn \\
\label{sigmaV2}
\eea
where $\overline{|\mathcal{M}|^2}_{a+b+c \ra d+e}$ is the
Lorentz invariant matrix amplitude square of $a +b +c \ra d+e$
averaged over both initial and final states and the prefactor
$1/m!$ is due to $m$ number of identical particles in the final state.
In the non-relativistic limit (i.e. $E_{p_1} \simeq m_a,
E_{p_2}\simeq m_b, E_{p_3}\simeq m_c$ and $p_1=p_2=p_3\simeq 0$),
Eq.\,\,\ref{sigmaV2} can be further simplified as
\bea
\left(\sigma {\rm v}^2\right)_{a+b+c \ra d+e}
&=& \dfrac{\sqrt{\left(m_a + m_b +m_c\right)^4 - 2 \left(m_a +m_b +m_c\right)^2
\left(m_d^2 + m_e^2\right) + 
\left(m_d^2 - m_e^2\right)^2}}{m!\times 64 \pi m_a m_b m_c}
\overline{|\mathcal{M}|^2}_{a+b+c \ra d+e}\,.\nn\\
\label{sigmaV2_NR}
\eea
The expression of $\left(\sigma {\rm v}^2\right)_{\chii\chii\hd\ra\chii\chii}$
has the following form in the non-relativistic and quasi degenerate limit
\bea
\left(\sigma {\rm v}^2\right)_{\chii\chii\hd\ra\chii\chii}
&\simeq&\frac{25 \sqrt{5} g_X^6}{24 \pi  M^5}
+\frac{50 \sqrt{5} g_X^6}{9\pi M^5}\,\delta_m
+\frac{17755 \sqrt{5} g_X^6}{864 \pi M^5} \,\delta_m^2
+\frac{169387 \sqrt{5} g_X^6}{2592 \pi  M^5}\,\delta_m^3
\nn \\ &&
+\frac{29073893 g_X^6}{31104 \sqrt{5} \pi  M^5}\,\delta_m^4
+ \mathcal{O}(\delta_{m}^5)\,.
\label{ffhd2ff}
\eea
Here by quasi degenerate limit we want to mention
that the above cross section has been calculated
for $m_{1}=m_{2}=M$ and $m_{\hd}=m_{\zp} =
M\,\left(1-\delta_m \right)$ with $\delta_m \ll 1$. This
kind of quasi degenerate dark sector is necessary
for the co-decaying mechanism to work successfully \cite{Dror:2016rxc}.
  
Therefore, in the non-relativistic limit, the quantity
$\mathcal{F}(\tp)_{3\ra 2}|_{\chii\chii\hd\ra\chii\chii}$
can be written as
\bea
\mathcal{F}(\tp)_{3\ra 2}|_{\chii\chii\hd\ra\chii\chii}
&=&-\dfrac{1}{2!}
\left(\dfrac{2\,k^2_1}{3E_{k_1}}\right)_{\rm NR}
\left(\sigma {\rm v}^2\right)_{\chii \chii \hd \ra \chii \chii}
(n^{\rm eq}_{\chii}(\tp))^2n^{\rm eq}_{\hd}(\tp) \times
\nn \\ && 
\left[\left(\dfrac{n_{\chii}(\tp)}{n^{\rm eq}_{\chii}(\tp)}\right)^2
- \left(\dfrac{n_{\chii}(\tp)}{n^{\rm eq}_{\chii}(\tp)}\right)^2
\dfrac{n_{\hd}(\tp)}{n^{\rm eq}_{\hd}(\tp)}
\right],
\eea
where, $\left(\dfrac{2\,k^2_1}{3E_{k_1}}\right)_{\rm NR}$
indicates the value of the quantity within the brackets in
non-relativistic limit and it can be expressed as a function of
masses of the initial state particles as
\bea
\left(\dfrac{k^2_1}{E_{k_1}}\right)_{\rm NR}
= \dfrac{m_{\hd}\left(m_{\hd}+4m_{i}\right)}{2\,\left(m_{\hd}+2\,m_{i}\right)}\,.
\label{p1bye1_NR}
\eea
Finally, the net contribution to $\mathcal{F}(\tp)_{3\ra 2}$
from both the dark matter candidates $\chio$ and $\chit$ for
the scattering process ${\chii\chii\hd\ra\chii\chii}$ is given
by  
\bea
\mathcal{F}(\tp)_{3\ra 2}|_{\x\x\hd\ra\x\x}
= 
\dfrac{1}{2!}\left(
\dfrac{2}{3}
\dfrac{m_{\hd}\left(m_{\hd}+4m_{i}\right)}{2\,\left(m_{\hd}+2\,m_{i}\right)}
\right)
\dfrac{\left(\sigma {\rm v}^2\right)_{\x \x \hd \ra \x \x}}{4}\,
n^2_{\x}(\tp)\left(n_{\hd}(\tp)-n^{\rm eq}_{\hd}(\tp)\right). \nn\\
\label{F3to2_2}
\eea
\subsubsection{\underline{$\chii \zp \zp \ra \chii \zp$}}
\begin{figure}[h!]
\centering
\includegraphics[height=2.5cm,width=17cm,angle=0]{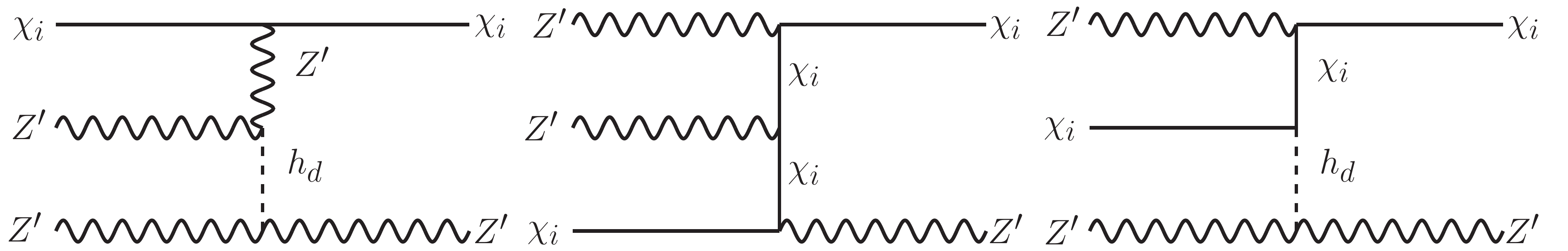}
\includegraphics[height=8cm,width=17cm,angle=0]{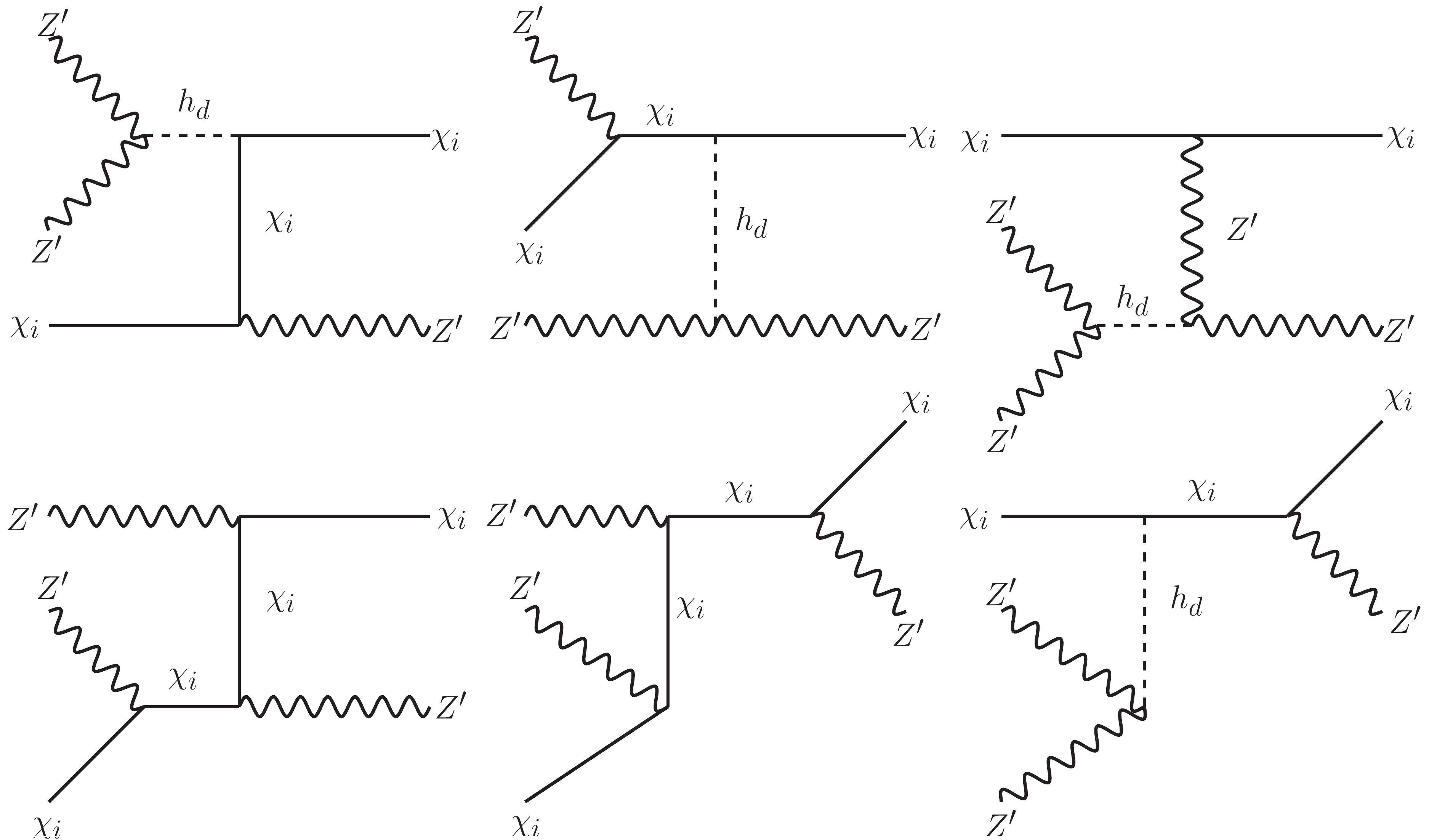}
\caption{Feynman diagrams for $\chii \zp \zp \ra \chii \zp$.}
\label{Fig:3to2_3}
\end{figure}

The Feynman diagrams for the scattering $\chii \zp \zp \ra \chii \zp$
are shown in Fig.\,\ref{Fig:3to2_3} and the expression of
$\left(\sigma {\rm v}^2\right)_{\chii\zp\zp\ra\chii\zp}$ is given by
\bea
\left(\sigma {\rm v}^2\right)_{\chii\zp\zp\ra\chii\zp} &\simeq&
   \frac{1985 \sqrt{5} g_X^6}{192 \pi  M^5}
   +\frac{94259 \sqrt{5} g_X^6}{1728 \pi  M^5}\,\delta_{m}
   +\frac{17772169 g_X^6}{20736 \sqrt{5} \pi  M^5}\,\delta_{m}^2
   +\frac{10135669 g_X^6}{4860 \sqrt{5} \pi  M^5}\,\delta_{m}^3
   \nn \\ &&
   +\frac{26951134297 g_X^6}{6220800 \sqrt{5} \pi  M^5}\,\delta_{m}^4 
   + \mathcal{O}(\delta_{m}^5).
\label{fzpzp2fzp}      
\eea
The contribution to $\mathcal{F}(\tp)_{3\ra2}$ is given by
\bea
\mathcal{F}(\tp)_{3\ra 2}|_{\x\zp\zp\ra\x\zp}
&=& 
\dfrac{1}{2!}\left(\dfrac{1}{3}
\dfrac{\lambda(M^2_{\rm in},m^2_i,m^2_{\zp})}
{2M_{\rm in}\left(M^2_{\rm in}-m^2_{\zp}+m^2_i\right)}\right)
\dfrac{\left(\sigma {\rm v}^2\right)_{\x \zp \zp \ra \x \zp}}{2}\times
\nn \\&&
n_{\x}(\tp)n_{\zp}(\tp)\left(n_{\zp}(\tp)-n^{\rm eq}_{\zp}(\tp)\right),
\label{F3to2_3}
\eea
where, $\lambda$ is the Kallen function which has the following
definition
\bea
\lambda(a^2, b^2, c^2) = a^4 +b^4 + c^4 - 2 a^2b^2 - 2b^2c^2 - 2 c^2 a^2\,,
\label{kfunc}
\eea
and $M_{\rm in} = 2\,m_{\zp} + m_i$ is the total mass of all
the initial state particles. 
\vskip 0.3in
\subsubsection{\underline{$\chii \zp \zp \ra \chii \hd$}}
\begin{figure}[h!]
\centering
\includegraphics[height=7cm,width=17cm,angle=0]{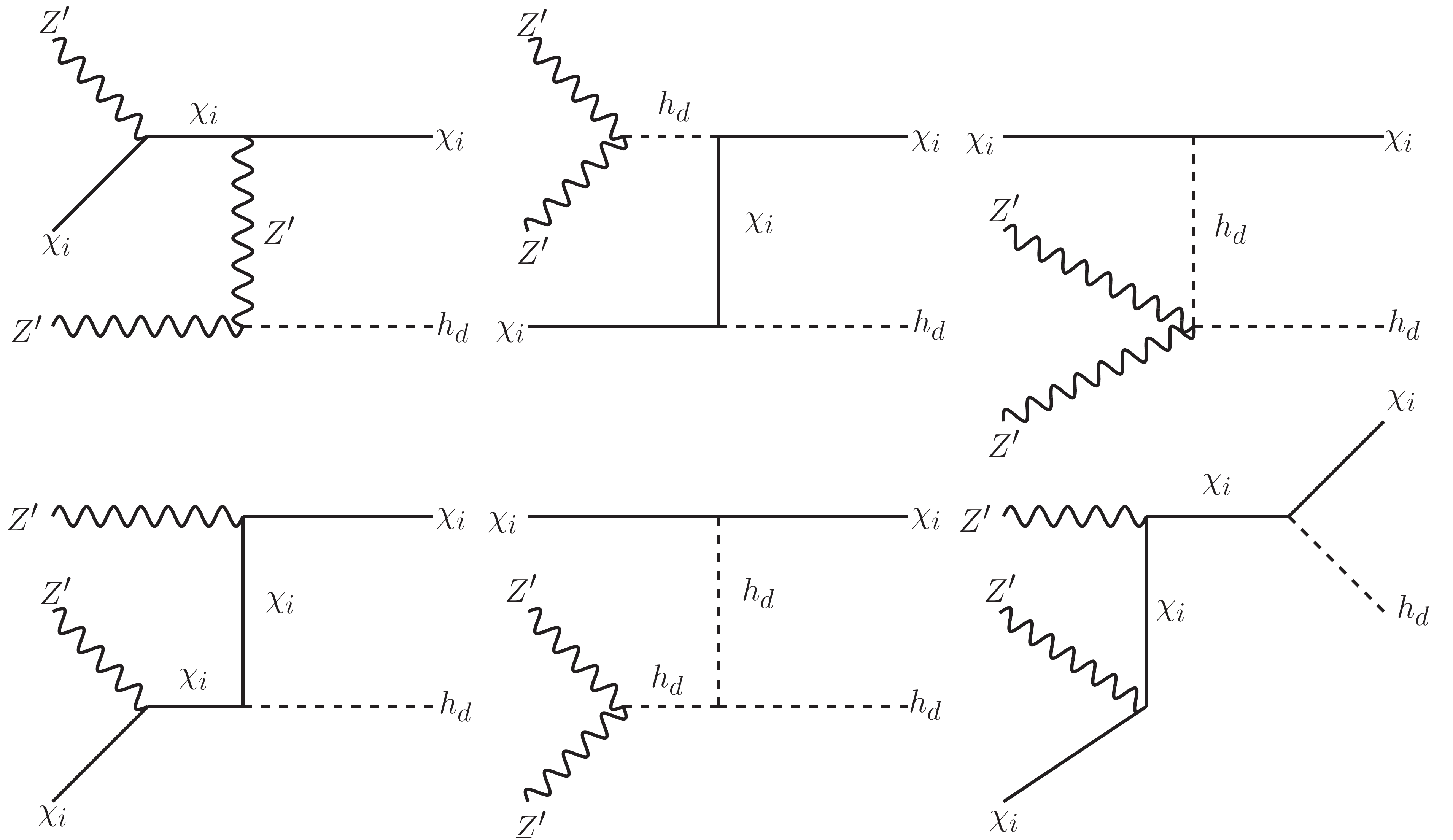}
\includegraphics[height=7cm,width=17cm,angle=0]{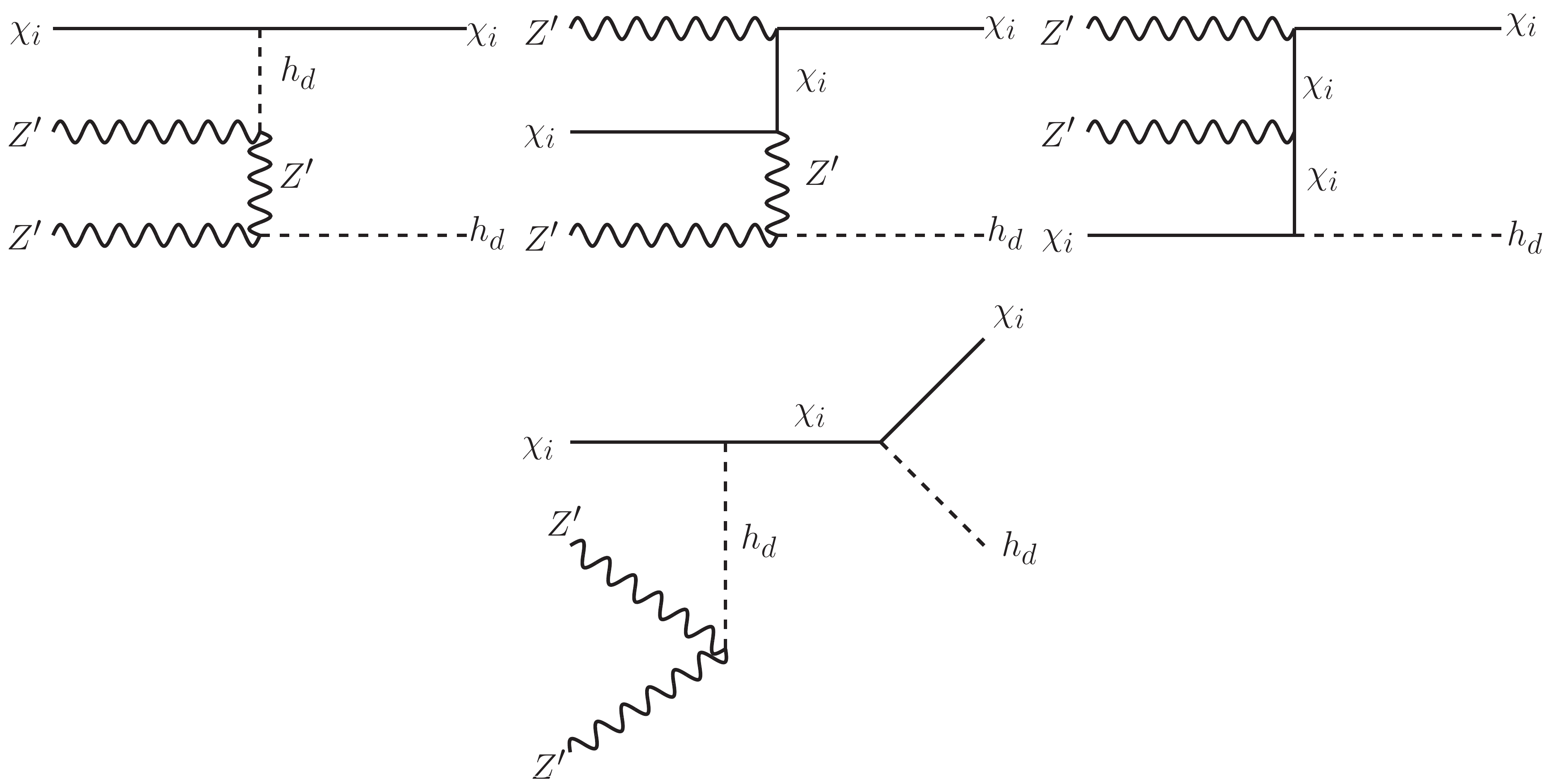}
\caption{Feynman diagrams for $\chii \zp \zp \ra \chii \hd$.}
\label{Fig:3to2_4}
\end{figure}

The Feynman diagrams for the $3\ra 2$ scattering
$\chii \zp \zp \ra \chii \hd$ are shown in Fig\,\,\ref{Fig:3to2_4}.
The corresponding expression of $\left(\sigma {\rm v^2}
\right)_{\chii \zp \zp \ra \chii \hd}$ in non-relativistic and
quasi degenerate limit is given by
\bea
\left(\sigma {\rm v^2}\right)_{\chii \zp \zp \ra \chii \hd}
&\simeq& 
   \frac{1775 \sqrt{5} g_X^6}{16 \pi  M^5}
   +\frac{148535 \sqrt{5} g_X^6}{216 \pi  M^5}\,\delta_{m}
   +\frac{1425181 \sqrt{5} g_X^6}{576 \pi  M^5}\,\delta_{m}^2
   +\frac{525450607 g_X^6}{15552 \sqrt{5} \pi  M^5}\,\delta_{m}^3
   \nn \\ && 
   +\frac{2900366015 \sqrt{5} g_X^6}{186624 \pi  M^5}\,\delta_{m}^4
   + \mathcal{O}(\delta_{m}^5).
   \label{fzpzp2fhd}
\eea 
This $3\ra2$ scattering process has the following  
contribution to $\mathcal{F}(\tp)_{3\ra2}$, 
\bea
\mathcal{F}(\tp)_{3\ra 2}|_{\x\zp\zp\ra\x\hd}
&=& 
\dfrac{1}{2!}\left(\dfrac{1}{3}
\dfrac{\lambda(M^2_{\rm in},m^2_i,m^2_{\hd})}
{2M_{\rm in}\left(M^2_{\rm in}-m^2_{\hd}+m^2_i\right)}\right)
\dfrac{\left(\sigma {\rm v}^2\right)_{\x \zp \zp \ra \x \hd}}{2} \times
\nn \\ &&
n_{\x}(\tp)
\left(n^2_{\zp}(\tp)-\dfrac{n_{\hd}(\tp)}{n^{\rm eq}_{\hd}(\tp)}
(n^{\rm eq}_{\zp}(\tp))^2\right),
\label{F3to2_4}
\eea
with $M_{\rm in} = 2\,m_{\zp} + m_i$, the total mass of all
the initial state particles. 
\subsubsection{\underline{$\chii \hd \hd \ra \chii \hd$}}
\begin{figure}[h!]
\centering
\includegraphics[height=7cm,width=17cm,angle=0]{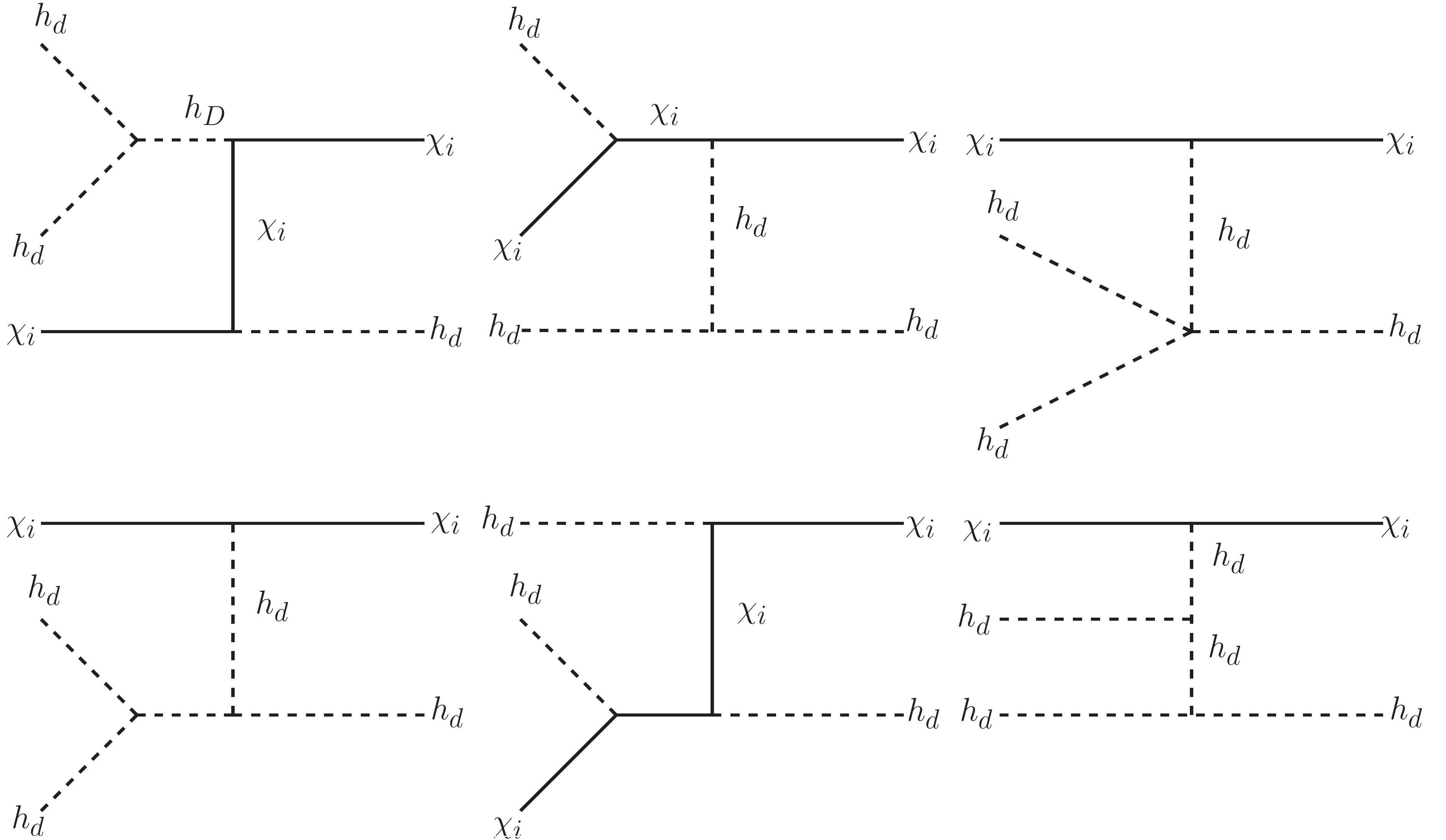}
\includegraphics[height=7cm,width=17cm,angle=0]{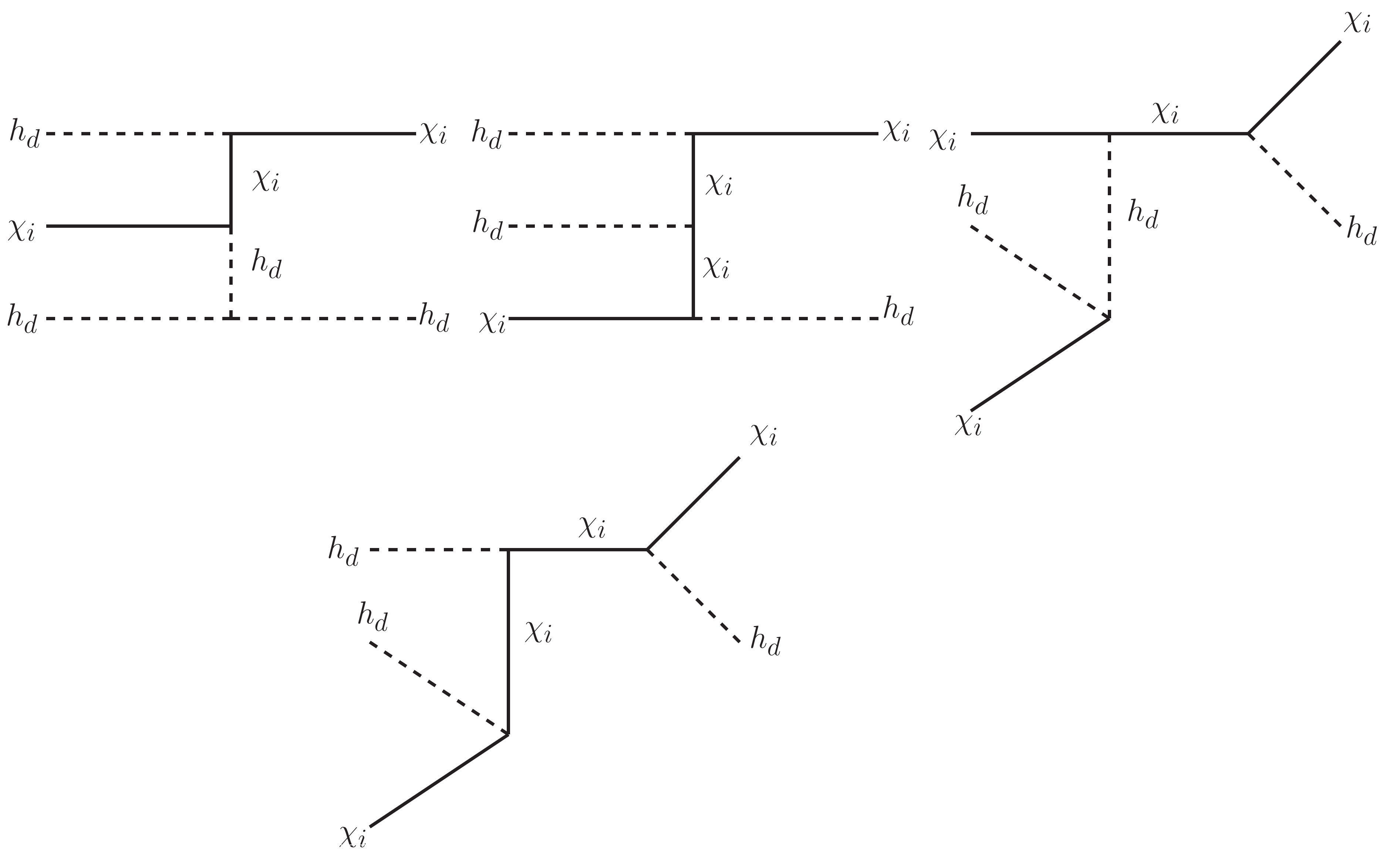}
\caption{Feynman diagrams for $\chii \hd \hd \ra \chii \hd$.}
\label{Fig:3to2_5}
\end{figure}

The Feynman diagrams for the scattering $\chii \hd \hd \ra \chii \hd$
are shown in Fig.\,\,\ref{Fig:3to2_5}. The expression of
$\left(\sigma {\rm v}^2\right)_{\chii \hd \hd \ra \chii \hd}$ is given by
\bea
\left(\sigma {\rm v}^2\right)_{\chii \hd \hd \ra \chii \hd} &\simeq&
   \frac{30 \sqrt{5} g_X^6}{\pi  M^5}
   +\frac{188 \sqrt{5} g_X^6}{\pi  M^5}\,\delta_{m}
   +\frac{20729 g_X^6}{6 \sqrt{5} \pi  M^5}\,\delta_{m}^2
   +\frac{436853 g_X^6}{45 \sqrt{5} \pi  M^5}\,\delta_{m}^3 \nn \\ &&
   + \frac{166118783 g_X^6}{7200 \sqrt{5} \pi  M^5}\,\delta_{m}^4 
   + \mathcal{O}(\delta_{m}^5)\,.
   \label{fhdhd2fhd}
\eea

This $3\ra 2$ scattering has the following contribution to
$\mathcal{F}(\tp)_{3\ra 2}$, 
\bea
\mathcal{F}(\tp)_{3\ra 2}|_{\x\hd\hd\ra\x\hd}
&=& 
\dfrac{1}{2!}\left(\dfrac{1}{3}
\dfrac{\lambda(M^2_{\rm in},m^2_i,m^2_{\hd})}
{2M_{\rm in}\left(M^2_{\rm in}-m^2_{\hd}+m^2_i\right)}\right)
\dfrac{\left(\sigma {\rm v}^2\right)_{\x \hd \hd \ra \x \hd}}{2} \times
\nn \\ &&
n_{\x}(\tp)n_{\hd}(\tp)\left(n_{\hd}(\tp)-n^{\rm eq}_{\hd}(\tp)\right)\,,
\label{F3to2_5}
\eea
with $M_{\rm in} = 2\,m_{\hd} + m_i$ is the total mass of all
the initial state particles. 
\vskip 0.5in
\subsubsection{\underline{$\chii \zp \hd \ra \chii \zp$}}
\begin{figure}[h!]
\centering
\includegraphics[height=6.5cm,width=17cm,angle=0]{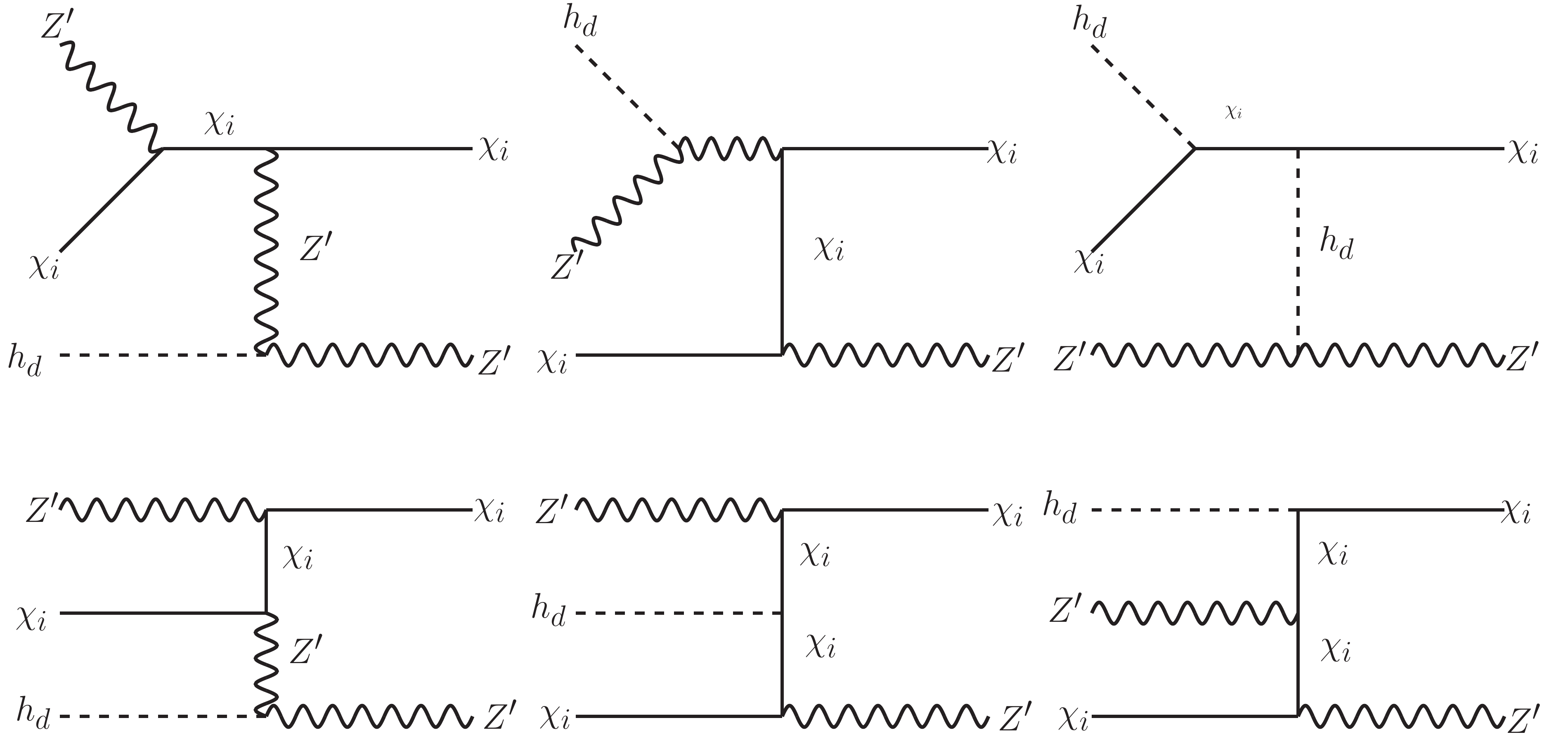}
\vskip 0.3in
\includegraphics[height=6.5cm,width=17cm,angle=0]{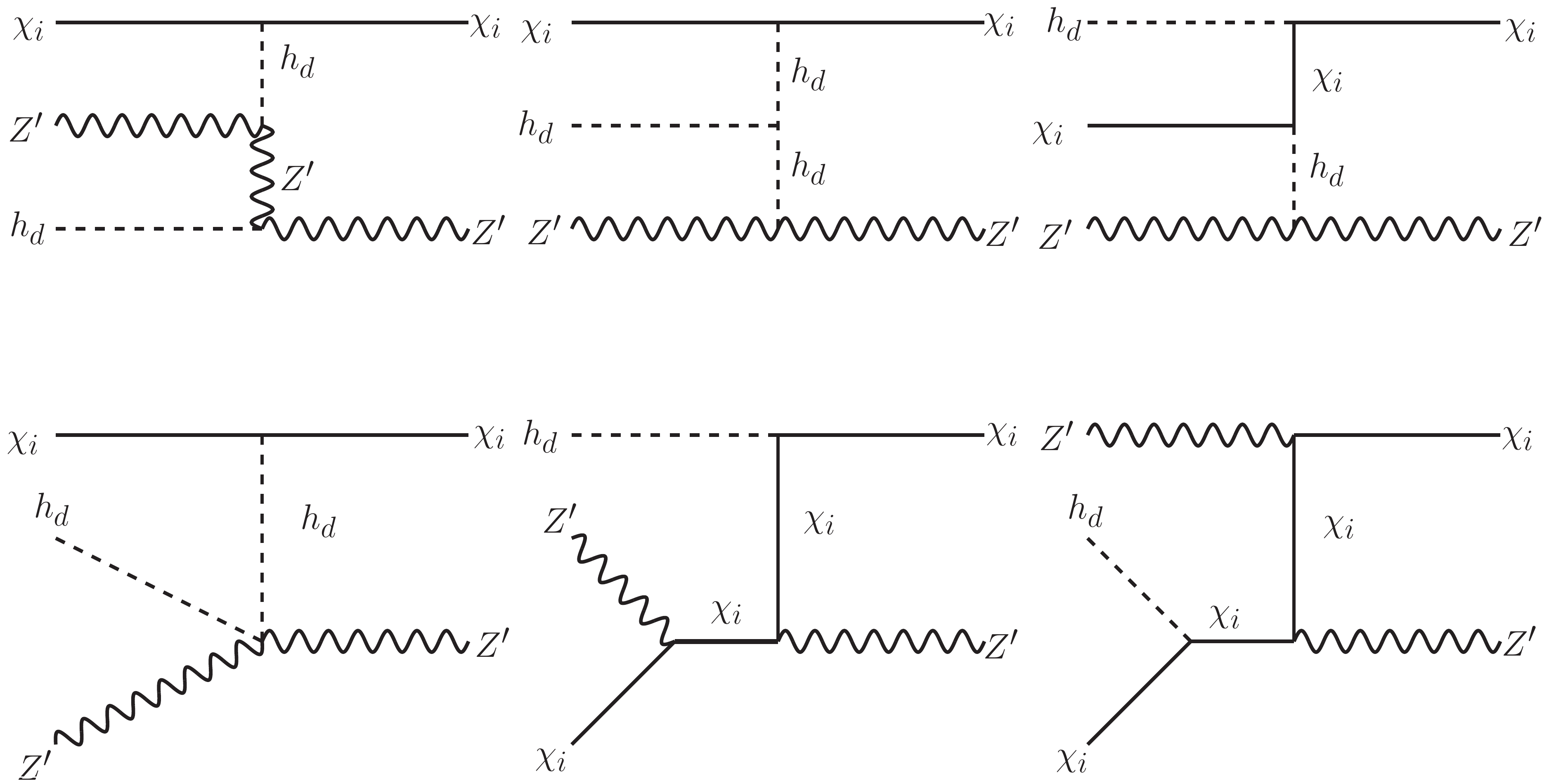}
\includegraphics[height=6.5cm,width=17cm,angle=0]{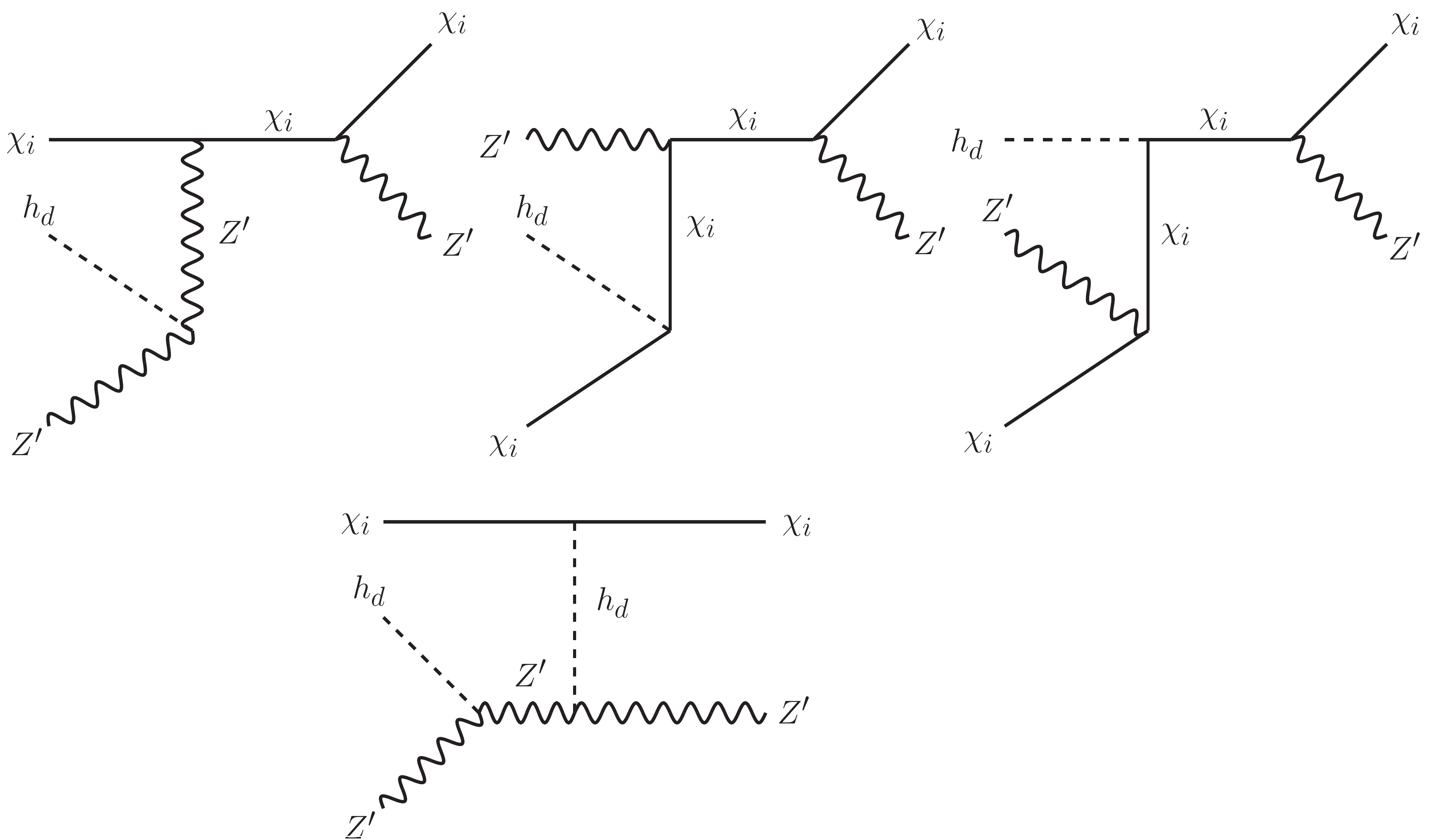}
\caption{Feynman diagrams for $\chii \zp \hd \ra \chii \zp$.}
\label{Fig:3to2_6}
\end{figure}

The Feynman diagrams for the scattering $\chii \zp \hd \ra \chii \zp$
are shown in Fig.\,\,\ref{Fig:3to2_6} and the expression of
$\left(\sigma {\rm v}^2\right)_{\chii\zp\hd\ra\chii\zp}$ in non-relativistic
and quasi degenerate limit is given by
\bea
\left(\sigma {v}^2\right)_{\chii\zp\hd\ra\chii\zp}
&\simeq&
   \frac{24235 \sqrt{5} g_X^6}{216 \pi  M^5}
   +\frac{500273 \sqrt{5} g_X^6}{648 \pi  M^5}\,\delta_{m}
   +\frac{118519543 g_X^6}{7776 \sqrt{5} \pi  M^5}\,\delta_{m}^2
   +\frac{660382819 g_X^6}{14580 \sqrt{5} \pi  M^5}\,\delta_{m}^3 
   \nn \\ &&
   +\frac{787240407727 g_X^6}{6998400 \sqrt{5} \pi  M^5}\,\delta_{m}^4
   + \mathcal{O}(\delta_{m}^5)\,.
\label{fzphd2fzp}
\eea
The contribution to the function $\mathcal{F}(\tp)_{3\ra 2}$
due to this inelastic scattering is given by
\bea
\mathcal{F}(\tp)_{3\ra 2}|_{\x \zp \hd \ra \x \zp} 
&=& 
\left(\dfrac{1}{3}
\dfrac{\lambda(M^2_{\rm in},m^2_i,m^2_{\zp})}
{2M_{\rm in}\left(M^2_{\rm in}-m^2_{\zp}+m^2_i\right)}\right)
\dfrac{\left(\sigma {\rm v}^2\right)_{\x \zp \hd \ra \x \zp}}{2} \times
\nn \\ &&
n_{\x}(\tp)n_{\zp}(\tp)\left(n_{\hd}(\tp)-n^{\rm eq}_{\hd}(\tp)\right)\,,
\label{F3to2_6}
\eea
where, $M_{\rm in} = m_{i}+m_{\zp}+m_{\hd}$ and $\lambda$ is the Kallen function
defined in Eq.\,\,\ref{kfunc}.
\vskip 0.5in
\subsubsection{\underline{$\chii \zp \hd \ra \chii \hd$}}
\begin{figure}[h!]
\centering
\includegraphics[height=6.5cm,width=17cm,angle=0]{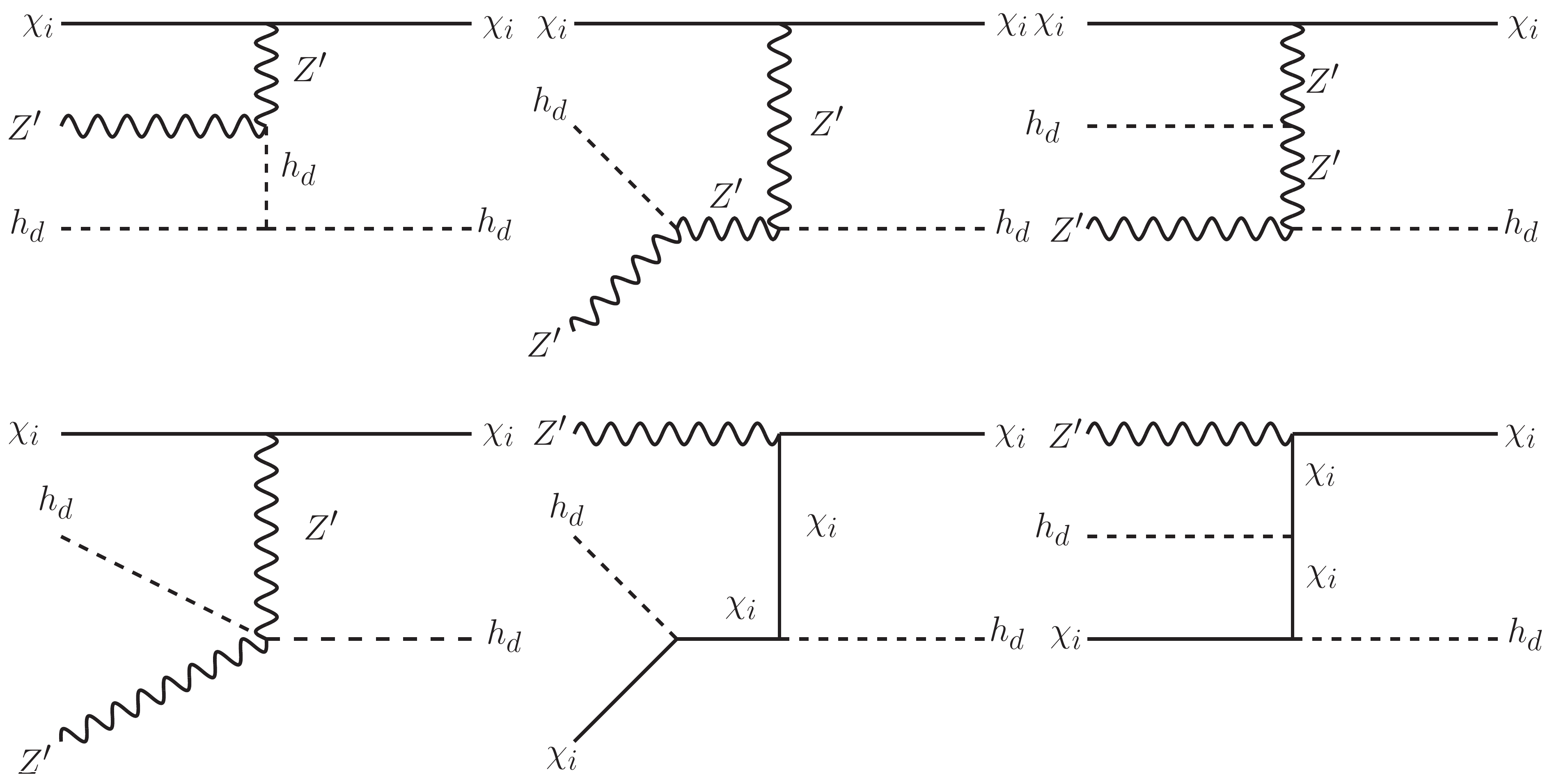}
\vskip 0.3in
\includegraphics[height=6.5cm,width=17cm,angle=0]{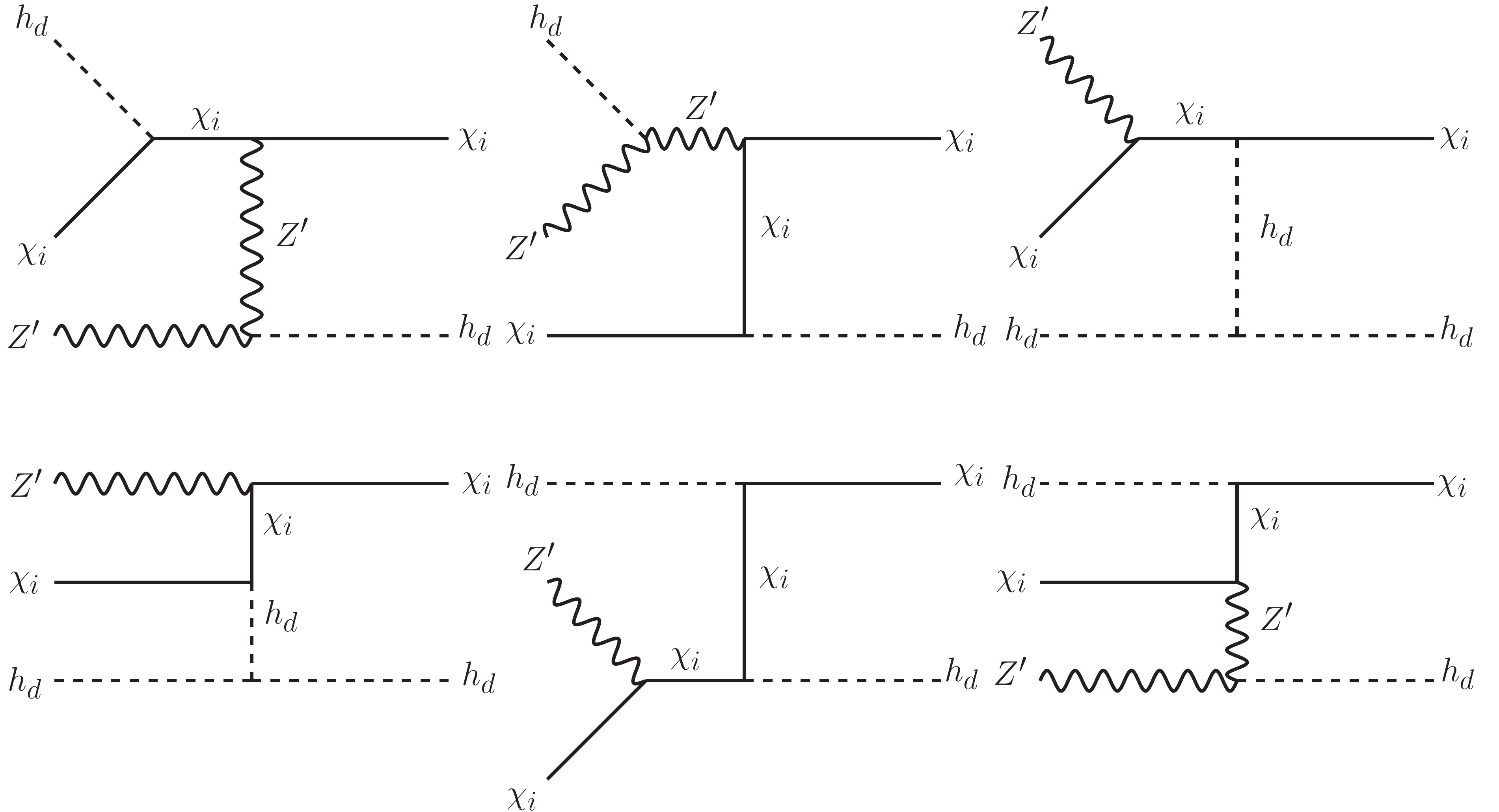}
\includegraphics[height=6.5cm,width=17cm,angle=0]{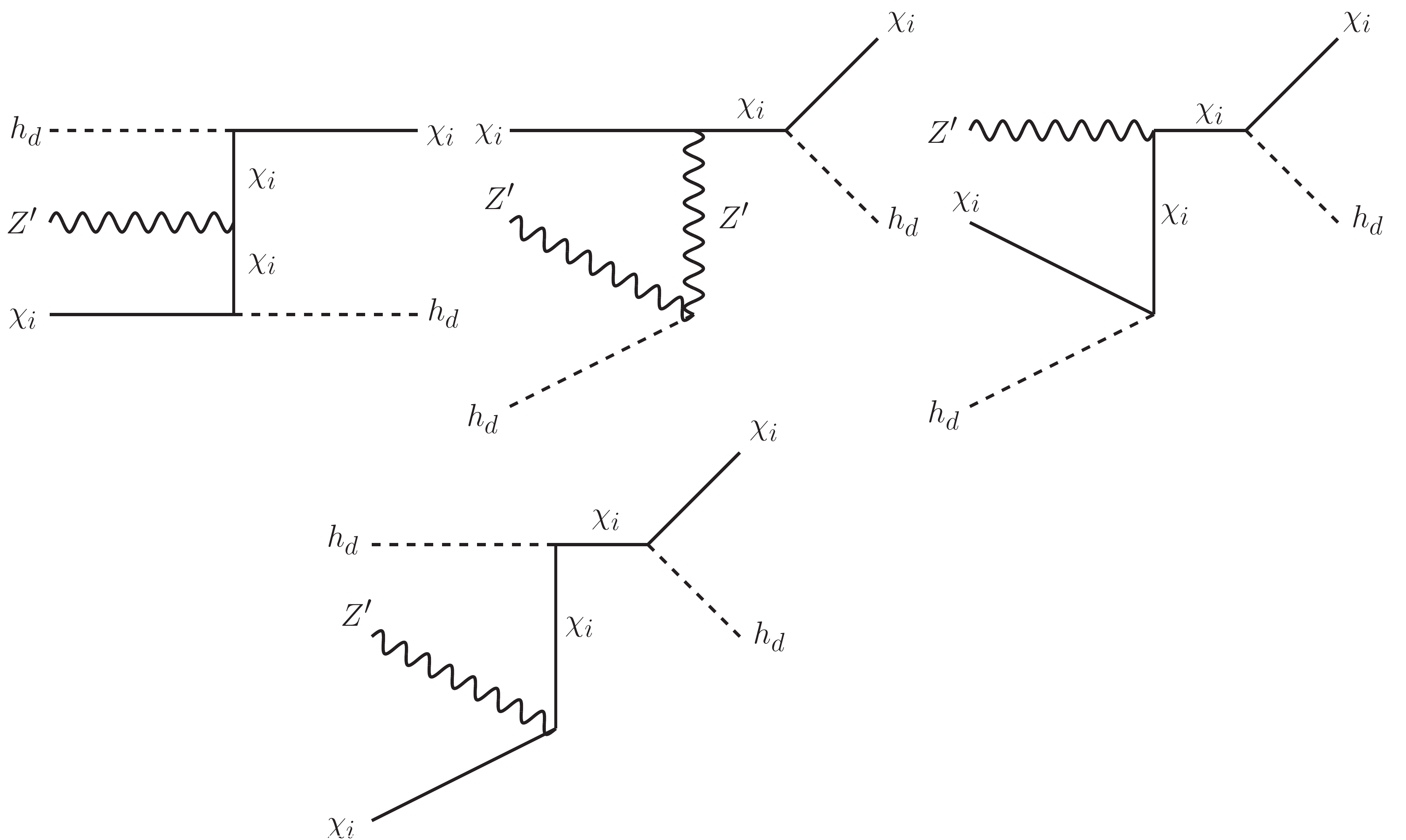}
\caption{Feynman diagrams for $\chii \zp \hd \ra \chii \hd$.}
\label{Fig:3to2_7}
\end{figure}

All the Feynman diagrams for this inelastic scattering is shown
in Fig.\,\,\ref{Fig:3to2_7} and the corresponding $\sigma {\rm v}^2$
is given by
\bea
\left(\sigma {\rm v}^2\right)_{\chii \zp \hd \ra \chii \hd} 
&\simeq& 
   \frac{7535 \sqrt{5} g_X^6}{144 \pi  M^5}
   + \frac{185357 \sqrt{5} g_X^6}{648 \pi  M^5}\,\delta_{m}
   +\frac{4029643 g_X^6}{864 \sqrt{5} \pi  M^5}\,\delta_{m}^2
   +\frac{1369755943 g_X^6}{116640 \sqrt{5} \pi  M^5}\,\delta_{m}^3 \nn \\ &&
   +\frac{352190753411 g_X^6}{13996800 \sqrt{5} \pi  M^5}\,\delta_{m}^4 
   + \mathcal{O}(\delta_{m}^5)
   \label{fzphd2fhd}
\eea
Contribution to the function $\mathcal{F}(\tp)_{3\ra 2}$
due to the $3\ra2$ scattering $\chii \zp \hd \ra \chii \hd$
is given below
\bea
\mathcal{F}(\tp)_{3\ra 2}|_{\x \zp \hd \ra \x \hd} 
&=& 
\left(\dfrac{1}{3}
\dfrac{\lambda(M^2_{\rm in},m^2_i,m^2_{\hd})}
{2M_{\rm in}\left(M^2_{\rm in}-m^2_{\hd}+m^2_i\right)}\right)
\dfrac{\left(\sigma {\rm v}^2\right)_{\x \zp \hd \ra \x \hd}}{2} \times
\nn \\ &&
n_{\x}(\tp)n_{\hd}(\tp)\left(n_{\zp}(\tp)-n^{\rm eq}_{\zp}(\tp)\right)\,,
\label{F3to2_7}
\eea
where, $M_{\rm in} = m_{i}+m_{\zp}+m_{\hd}$.
\subsubsection{\underline{$\hd \hd \hd \ra \chii \chii$}}
\begin{figure}[h!]
\centering
\includegraphics[height=7cm,width=17cm,angle=0]{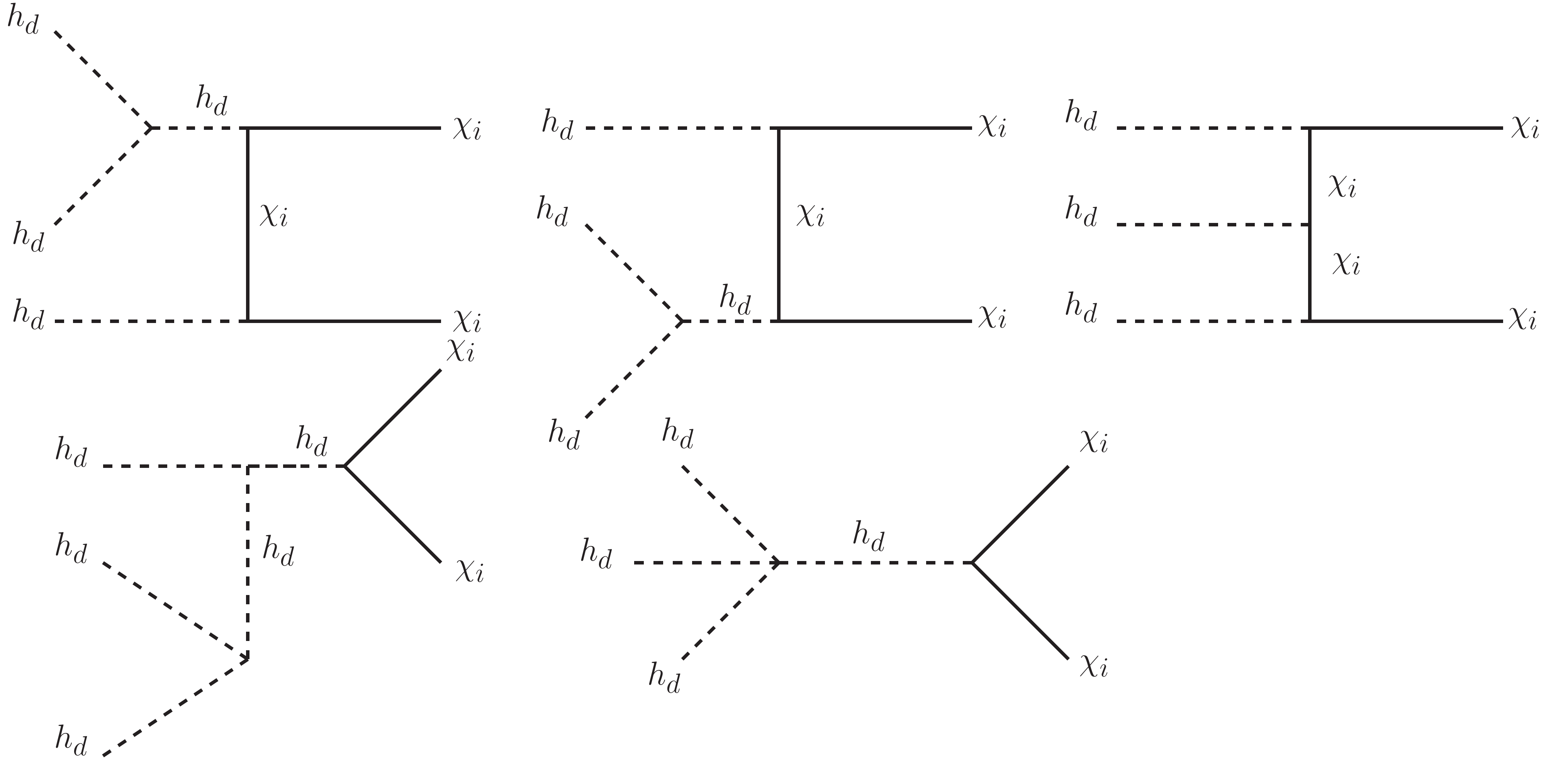}
\caption{Feynman diagrams for $\hd \hd \hd \ra \chii \chii$.}
\label{Fig:3to2_8}
\end{figure}

Feynman diagrams for $\hd \hd \hd \ra \chii \chii$
are shown in Fig.\,\,\ref{Fig:3to2_8}. In the non-relativistic
and quasi degenerate limit $\sigma {\rm v}^2$ for this
scattering has the following expression

\bea
\left(\sigma {\rm v}^2\right)_{\hd \hd \hd \ra \chii \chii} 
&\simeq& 
   \frac{15\sqrt{5} g_X^6}{16 \pi  M^5}
   -\frac{291 \sqrt{5} g_X^6}{16 \pi  M^5}\,\delta_{m}
   -\frac{2169 g_X^6}{8 \sqrt{5} \pi M^5}\,\delta_{m}^2
   +\frac{178803 g_X^6}{40 \sqrt{5} \pi M^5}\,\delta_{m}^3
   \nn \\ &&
   +\frac{9109719 g_X^6}{200 \sqrt{5} \pi M^5}\,\delta_{m}^4
   +\mathcal{O}(\delta_{m}^5)\,.
\label{hdhdhd2ff}
\eea

This $3\ra 2$ scattering has following contribution to
$\mathcal{F}(\tp)_{3\ra2}$,
\bea
\mathcal{F}(\tp)_{3\ra 2}|_{\hd \hd \hd \ra \x \x} 
&=& 
\dfrac{1}{3!}\left(\dfrac{2}{3}
\dfrac{\lambda(M^2_{\rm in},m^2_i,m^2_{i})}
{2M_{\rm in} \times M^2_{\rm in}}\right)
\left(\sigma {\rm v}^2\right)_{\hd \hd \hd \ra \x \x} \times
\nn \\ &&
\left[n^3_{\hd}(\tp)-\left(\dfrac{n_{\x}(\tp)}{n^{\rm eq}_{\x}(\tp)}\right)^2
\left(n^{\rm eq}_{\hd}(\tp)\right)^3\right]\,.
\label{F3to2_8}
\eea  
Here $\lambda$ is the Kallen function (Eq.\,\,\ref{kfunc})
and $M_{\rm in} = 3m_{\hd}$.
\subsubsection{\underline{$\hd \hd \zp \ra \chii \chii$}}
\begin{figure}[h!]
\centering
\includegraphics[height=7cm,width=17cm,angle=0]{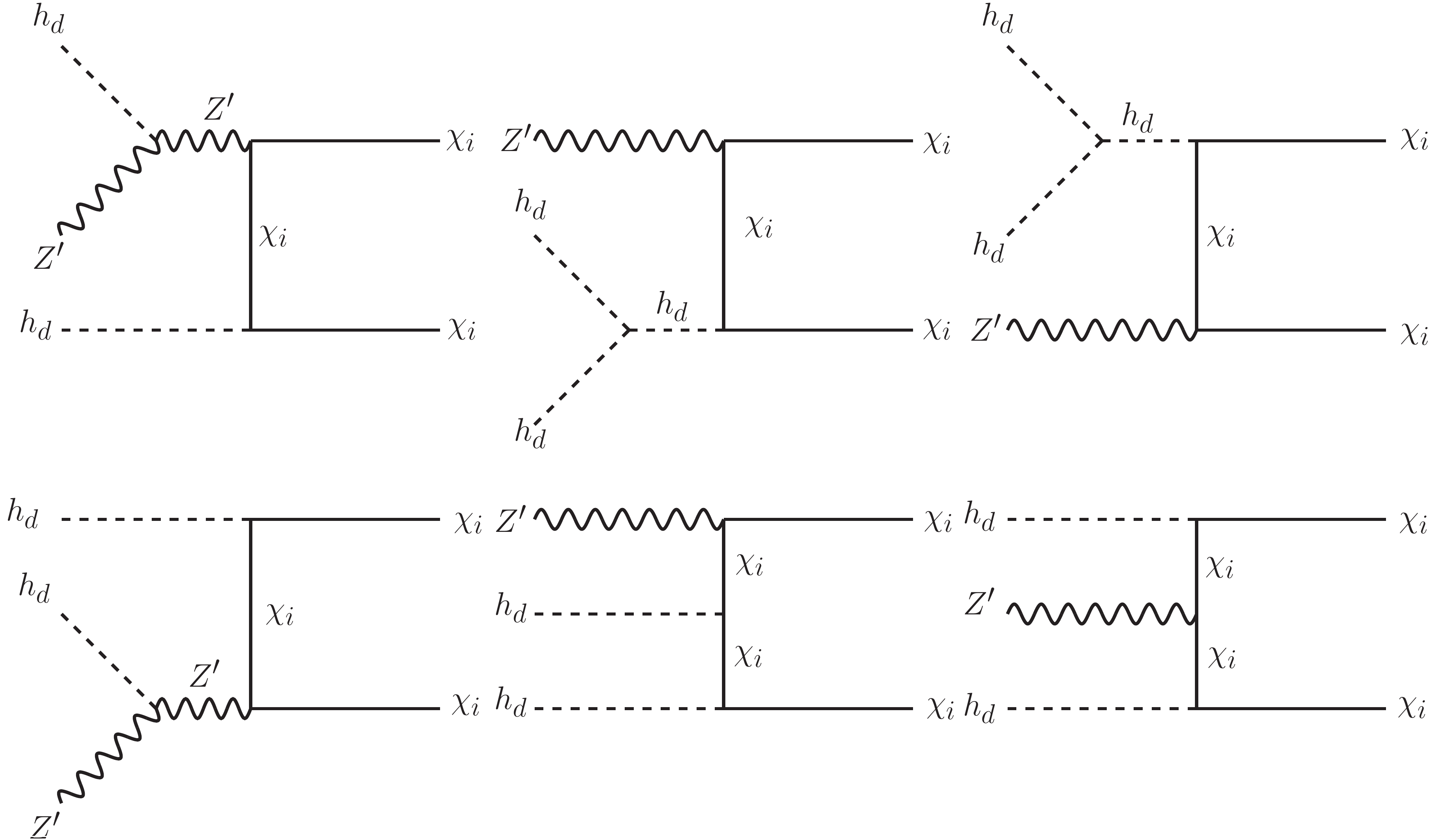}
\includegraphics[height=7cm,width=17cm,angle=0]{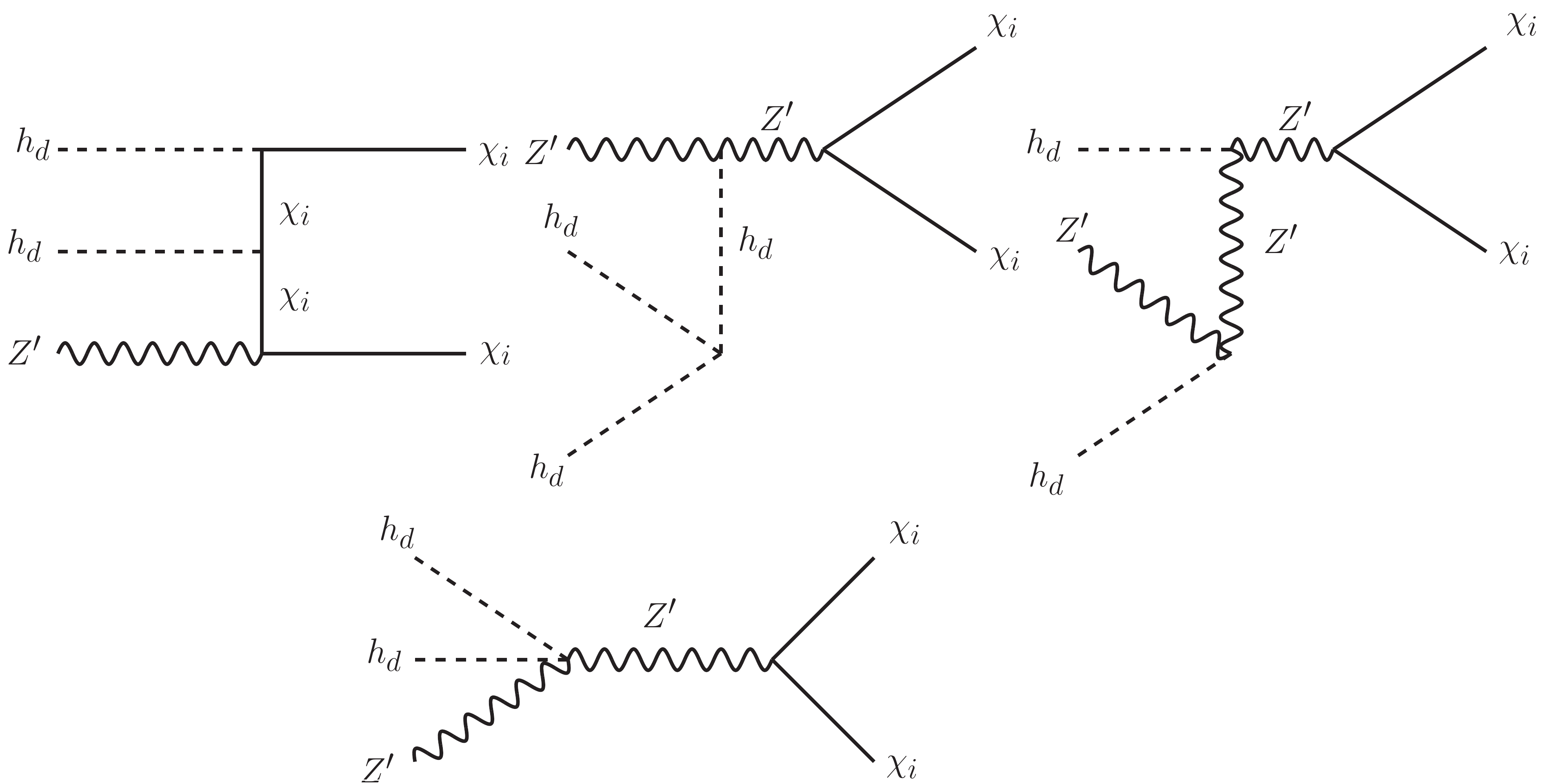}
\caption{Feynman diagrams for $\hd \hd \zp \ra \chii \chii$.}
\label{Fig:3to2_9}
\end{figure}

All the relevant Feynman diagrams for the $3\ra2$ scattering
$\hd \hd \zp \ra \chii \chii$ are shown in Fig.\,\,\ref{Fig:3to2_9}.
In the non-relativistic and quasi degenerate limit, $\sigma{\rm v}^2$
for $\hd \hd \zp \ra \chii \chii$ can be expressed as
\bea
\left(\sigma{\rm v}^2\right)_{\hd \hd \zp \ra \chii \chii}
&\simeq&
   \frac{5\sqrt{5} g_X^6}{216 \pi  M^5}
   +\frac{713 \sqrt{5} g_X^6}{216 \pi  M^5}\,\delta_{m}
   +\frac{144059 g_X^6}{216 \sqrt{5} \pi M^5}\,\delta_{m}^2
   +\frac{231821 g_X^6}{40 \sqrt{5} \pi M^5}\,\delta_{m}^3
   \nn \\&&
   +\frac{148439441 g_X^6}{5400 \sqrt{5} \pi M^5}\,\delta_{m}^4
   + \mathcal{O}(\delta_{m}^5)\,.  
   \label{hdhdzp2ff}   
\eea
In the function $\mathcal{F}_{3\ra2}(\tp)$, this particular
$3\ra2$ scattering has the following contribution
\bea
\mathcal{F}(\tp)_{3\ra 2}|_{\hd\hd\zp \ra \x \x} 
&=& 
\dfrac{1}{2!}\left(\dfrac{2}{3}
\dfrac{\lambda(M^2_{\rm in},m^2_i,m^2_{i})}
{2M_{\rm in}\times M^2_{\rm in}}\right)
\left(\sigma {\rm v}^2\right)_{\hd \hd \zp \ra \x \x}\times
\nn \\ &&
\left[\left(n_{\hd}(\tp)\right)^2 n_{\zp}(\tp) -
\left(\dfrac{n_{\x}(\tp)}{n^{\rm eq}_{\x}(\tp)}\right)^2
\left(n^{\rm eq}_{\hd}(\tp)\right)^2n^{\rm eq}_{\zp}(\tp)
\right]\,,
\label{F3to2_9}
\eea
with $M_{\rm in} = 2m_{\hd} + m_{\zp}$.
\subsubsection{\underline{$\zp \zp \hd \ra \chii \chii$}}
\begin{figure}[h!]
\centering
\includegraphics[height=7cm,width=17cm,angle=0]{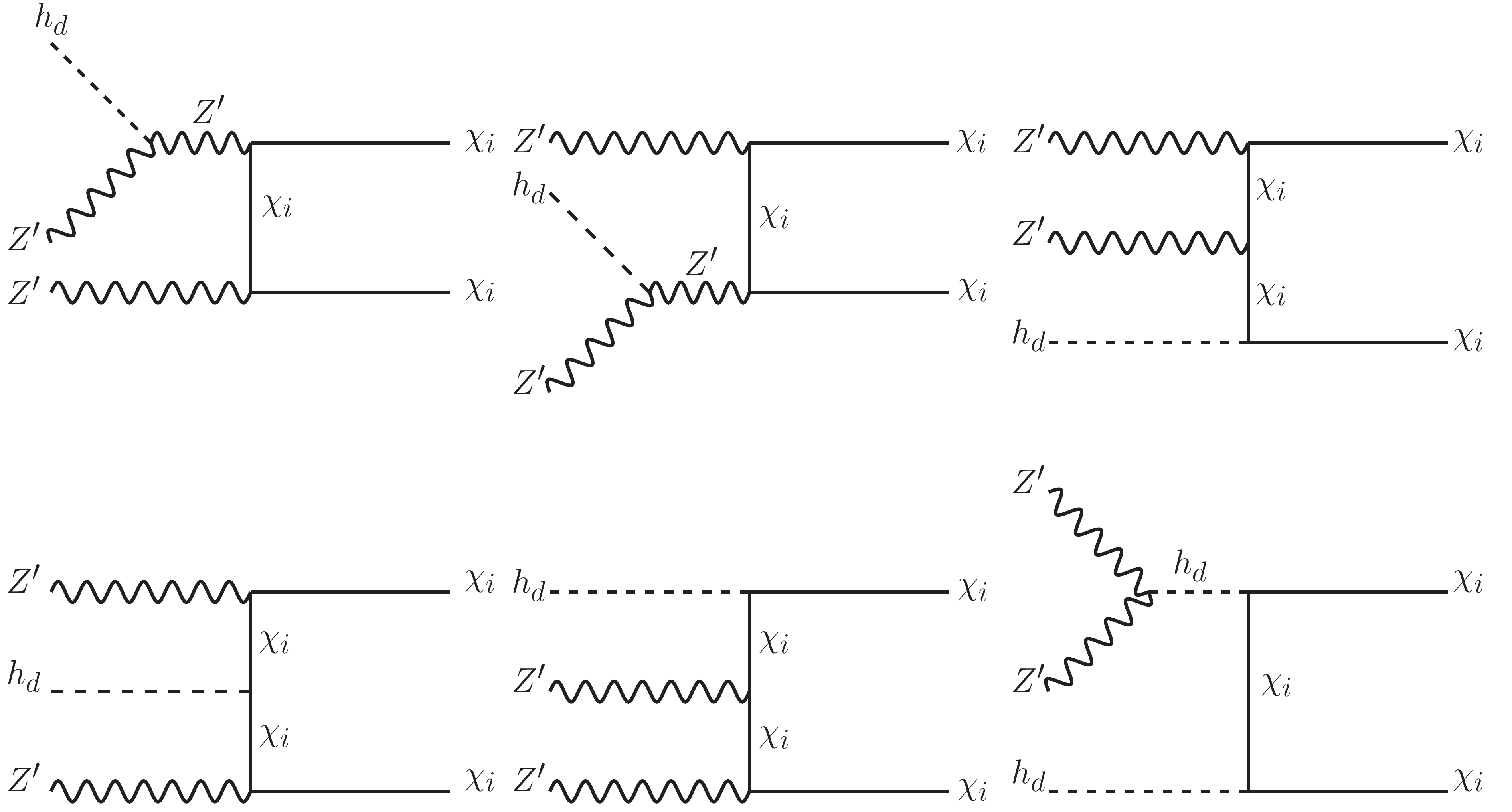}
\includegraphics[height=7cm,width=17cm,angle=0]{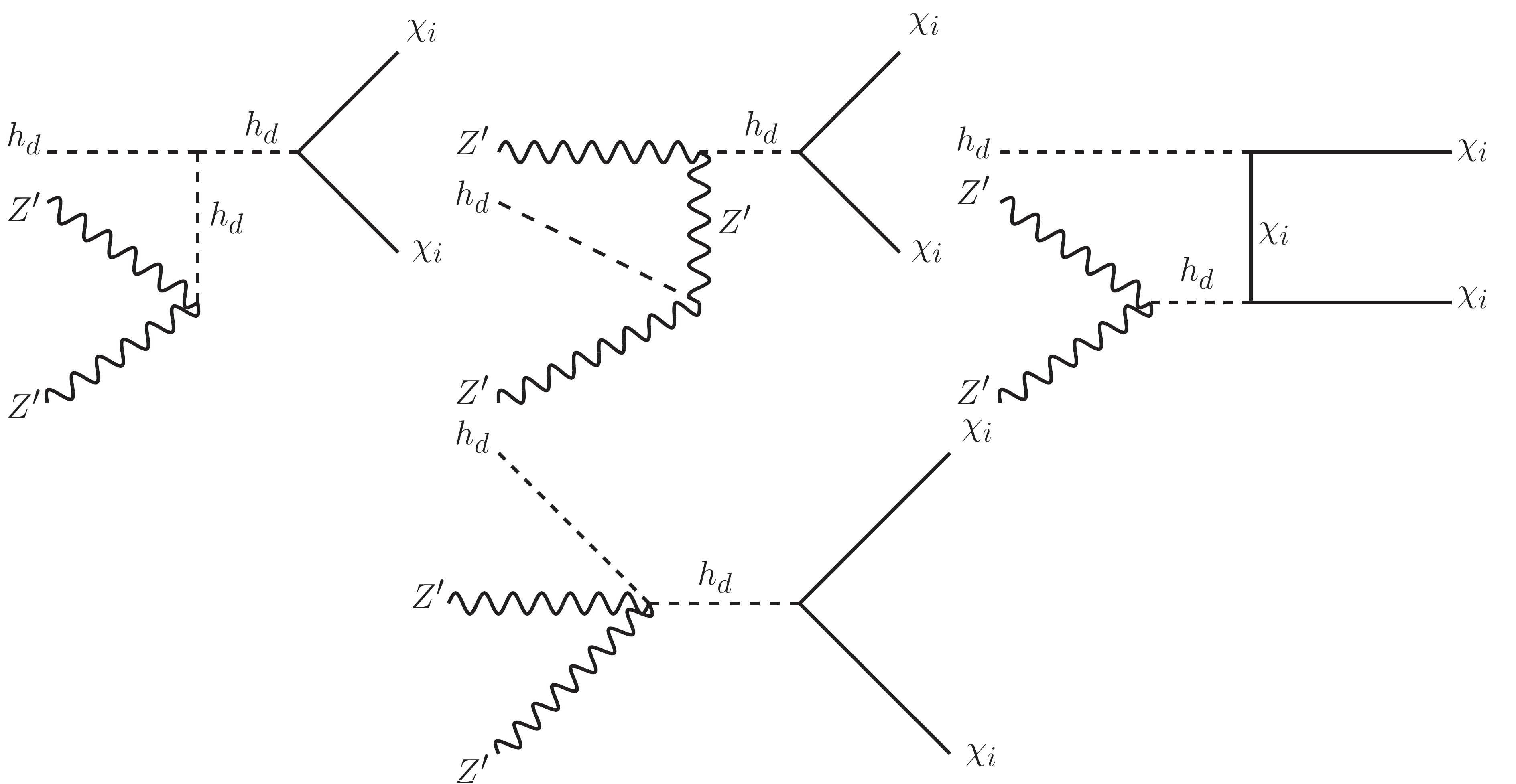}
\caption{Feynman diagrams for $\zp \zp \hd \ra \chii \chii$.}
\label{Fig:3to2_10}
\end{figure}

Feynman diagrams contributing to the scattering $\zp \zp \hd \ra \chii \chii$
are depicted in Fig.\,\,\ref{Fig:3to2_10} and corresponding
$\sigma {\rm v}^2$ in non-relativistic and quasi degenerate limit
is given below
\bea
\left(\sigma{\rm v}^2\right)_{\zp\zp\hd\ra\chii\chii}
&\simeq&
   \frac{3725 \sqrt{5} g_X^6}{864 \pi  M^5}
   +\frac{10805 \sqrt{5} g_X^6}{288 \pi M^5}\,\delta_{m}
   +\frac{23569 \sqrt{5} g_X^6}{144 \pi M^5}\,\delta_{m}^2
   +\frac{917111 g_X^6}{432 \sqrt{5} \pi M^5}\,\delta_{m}^3
   \nn \\ &&
   +\frac{592237 g_X^6}{240 \sqrt{5} \pi M^5}\,\delta_{m}^4   
   + \mathcal{O}(\delta_{m}^5)\,.
   \label{zpzphd2ff}
\eea

The expression of $\mathcal{F}_{3\ra2}(\tp)|_{\zp\zp\hd\ra \x\x}$
is given by
\bea
\mathcal{F}_{3\ra2}(\tp)|_{\zp\zp\hd\ra \x\x} =
&=& 
\dfrac{1}{2!}\left(\dfrac{2}{3}
\dfrac{\lambda(M^2_{\rm in},m^2_i,m^2_{i})}
{2M_{\rm in}\times M^2_{\rm in}}\right)
\left(\sigma {\rm v}^2\right)_{\zp \zp \hd \ra \x \x}\times
\nn \\ &&
\left[\left(n_{\zp}(\tp)\right)^2 n_{\hd}(\tp) -
\left(\dfrac{n_{\x}(\tp)}{n^{\rm eq}_{\x}(\tp)}\right)^2
\left(n^{\rm eq}_{\zp}(\tp)\right)^2n^{\rm eq}_{\hd}(\tp)
\right]\,,
\label{F3to2_10}
\eea
where, $m_{\rm in} = 2m_{\zp}+m_{\hd}$.
\subsubsection{\underline{$\zp \zp \zp \ra \chii \chii$}}
\begin{figure}[h!]
\centering
\includegraphics[height=7cm,width=17cm,angle=0]{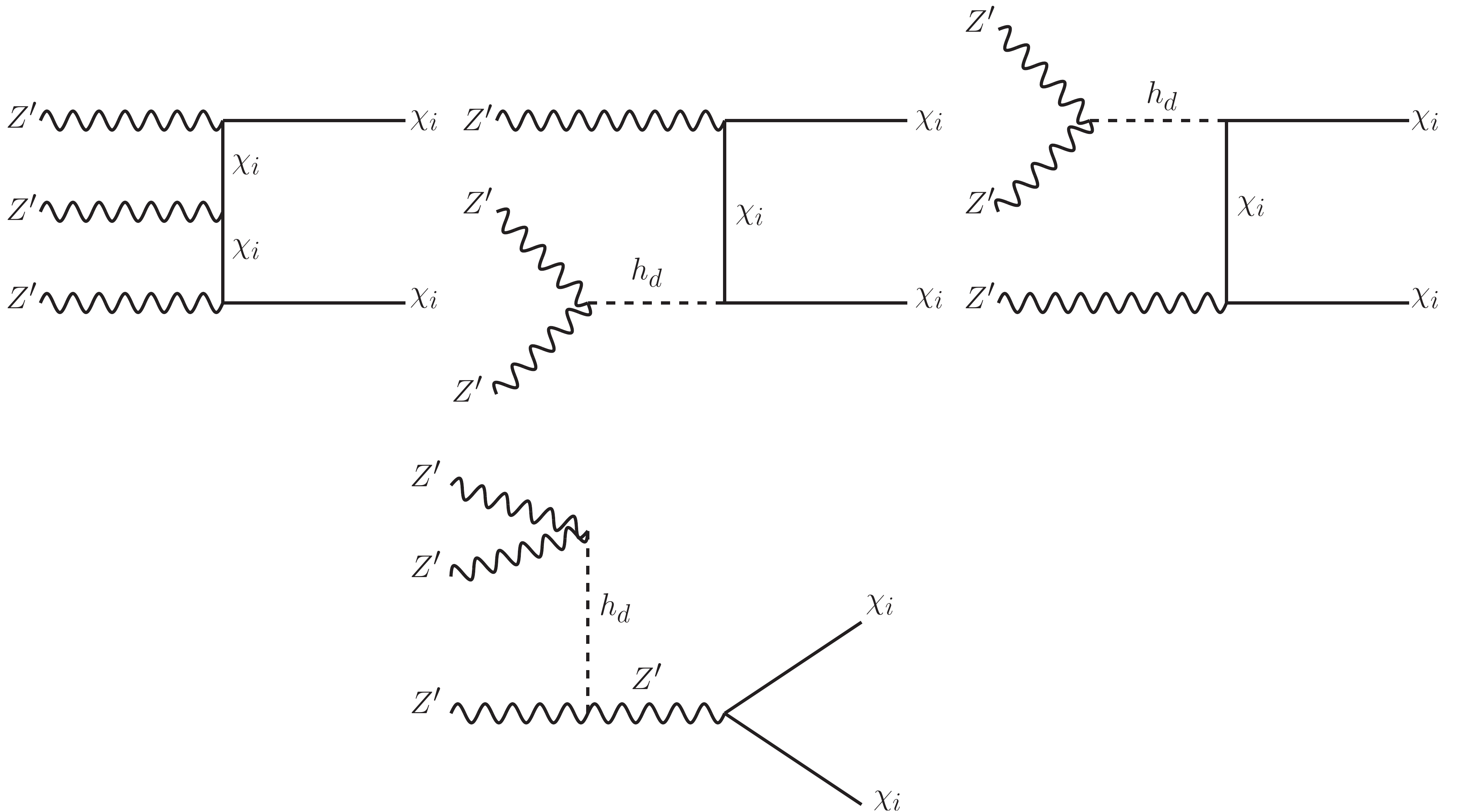}
\caption{Feynman diagrams for $\zp \zp \zp \ra \chii \chii$.}
\label{Fig:3to2_11}
\end{figure}

Feynman diagrams for this inelastic scattering are depicted
in Fig.\,\,\ref{Fig:3to2_11}. The expression of $\sigma{\rm v}^2$
for $\zp\zp\zp\ra\chii\chii$ is given by
\bea
\left(\sigma {\rm v}^2\right)_{\zp\zp\zp\ra\chii\chii}
&\simeq&
   \frac{1825 \sqrt{5} g_X^6}{288 \pi  M^5}
   +\frac{12985 \sqrt{5} g_X^6}{288 \pi M^5}\,\delta_{m}
   +\frac{47561 \sqrt{5} g_X^6}{288 \pi M^5}\,\delta_{m}^2
   +\frac{160171 g_X^6}{96 \sqrt{5} \pi M^5}\,\delta_{m}^3
   \nn \\ &&
   +\frac{97601 g_X^6}{360 \sqrt{5} \pi M^5}\,\delta_{m}^4
   +\mathcal{O}(\delta_{m}^5)\,.
\label{zpzpzp2ff}
\eea
It is needless to mention here that $\sigma{\rm v}^2$ for this
$3\ra 2$ scattering is also computed in non-relativistic
and quasi degenerate limit.
$\zp\zp\zp\ra\chii\chii$ inelastic scattering has following
contribution to $\mathcal{F}(\tp)_{3\ra2}$,
\bea
\mathcal{F}(\tp)_{3\ra 2}|_{\zp \zp \zp \ra \x \x} 
&=& 
\dfrac{1}{3!}\left(\dfrac{2}{3}
\dfrac{\lambda(M^2_{\rm in},m^2_i,m^2_{i})}
{2M_{\rm in} \times M^2_{\rm in}}\right)
\left(\sigma {\rm v}^2\right)_{\zp \zp \zp \ra \x \x} \times
\nn \\ &&
\left[n^3_{\zp}(\zp)-\left(\dfrac{n_{\x}(\tp)}{n^{\rm eq}_{\x}(\tp)}\right)^2
\left(n^{\rm eq}_{\zp}(\tp)\right)^3\right]\,,
\label{F3to2_8}
\eea  
where $\lambda$ is the Kallen function (Eq.\,\,\ref{kfunc})
and $M_{\rm in} = 3m_{\zp}$.

Moreover, there is another $3\ra 2$ scattering $\chii\chii\zp\ra\chii\chii$
involving dark matter candidates $\chii$s however, these scatterings have null
effect to the function $\mathcal{F}_{3\ra 2}(\tp)$ in the non-relativistic limit.
\section{Vertex factors}
\label{App:c}
The relevant vertex factors are listed in table \ref{Tab:vertex}.
\begin{center}
\begin{table}[h!]
\begin{tabular}{|c|c|}
\hline
Vertex & Vertex factors \\
\hline
\multirow{3}{*}{
\includegraphics[height = 1.8cm,width = 3cm]{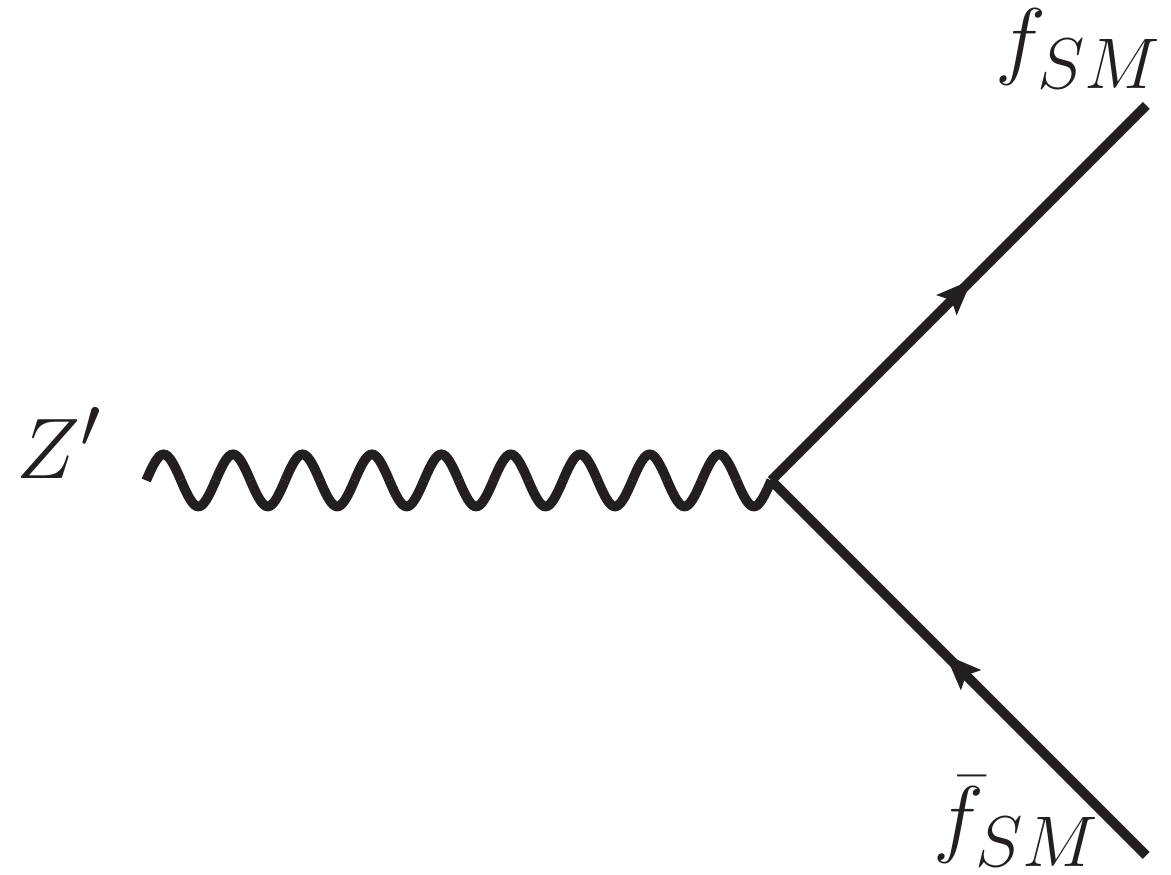}}
& 
$i \gamma_\mu  (C_V + C_A \gamma_5) \text{ where,} $ \\
&$C_V  = -\dfrac{g_2}{2\cos \theta_W}\left[\left(t_{3f} - 
2 Q_f \sin^2 \theta_W\right)\sin \theta_1-
\epsilon \left(t_{3f} - 2 Q_f \right)\sin \theta_W\cos \theta_1\right] \,\,,$\\
&$C_A =  \dfrac{g_2}{2\cos \theta_W}\left[\sin \theta_1 - 
\epsilon \sin \theta_W \cos \theta_1\right]t_{3f}\,\,.$\\
\hline	
\multirow{3}{*}{
\includegraphics[height = 1.5cm,width = 3cm]{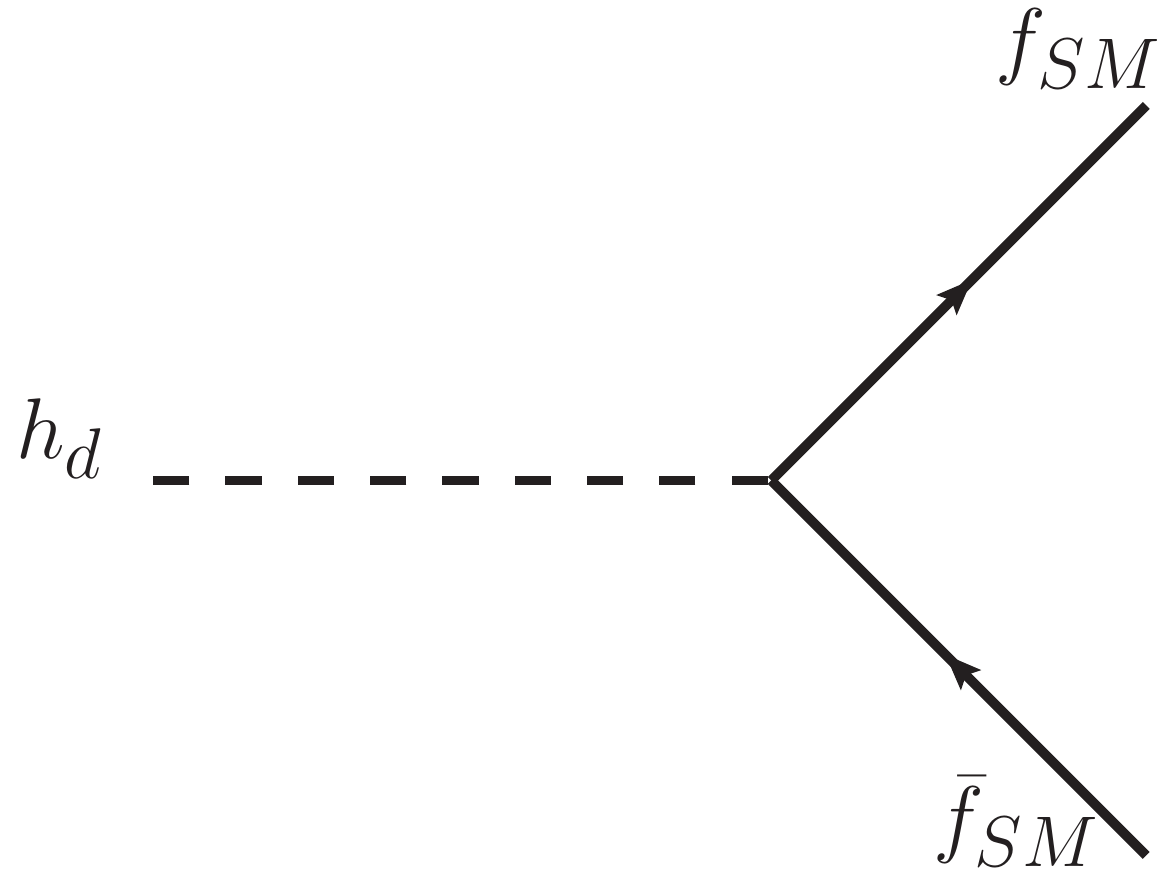}}&\\
&
$i\sin \alpha \dfrac{m_f}{v_{SM}}\,\,.$\\
&\\
\hline
\multirow{3}{*}{
\includegraphics[height = 1.5cm,width = 3cm]{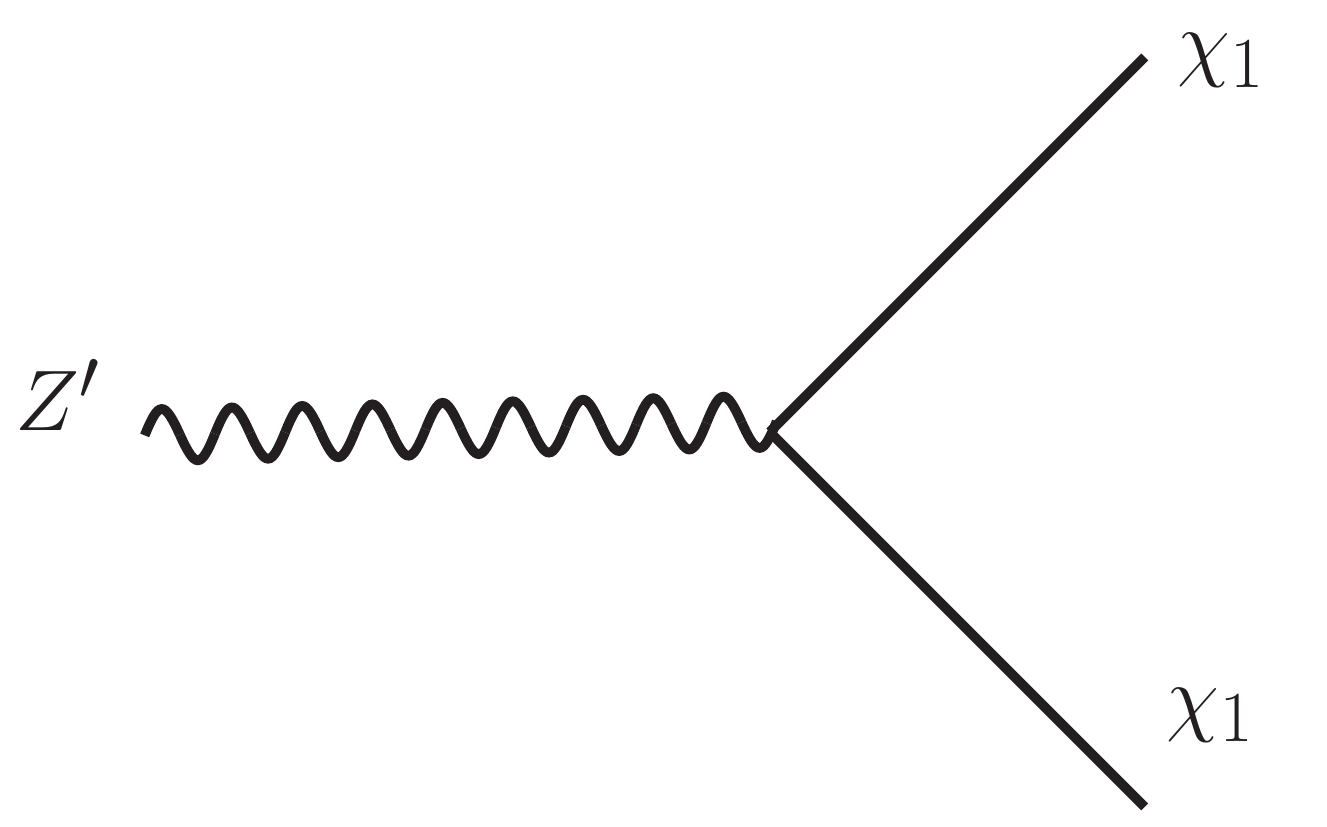}}&\\
&
$i g_X \gamma_\mu \gamma_5$\,\,.\\
&\\
\hline
\multirow{3}{*}{
\includegraphics[height = 1.5cm,width = 3cm]{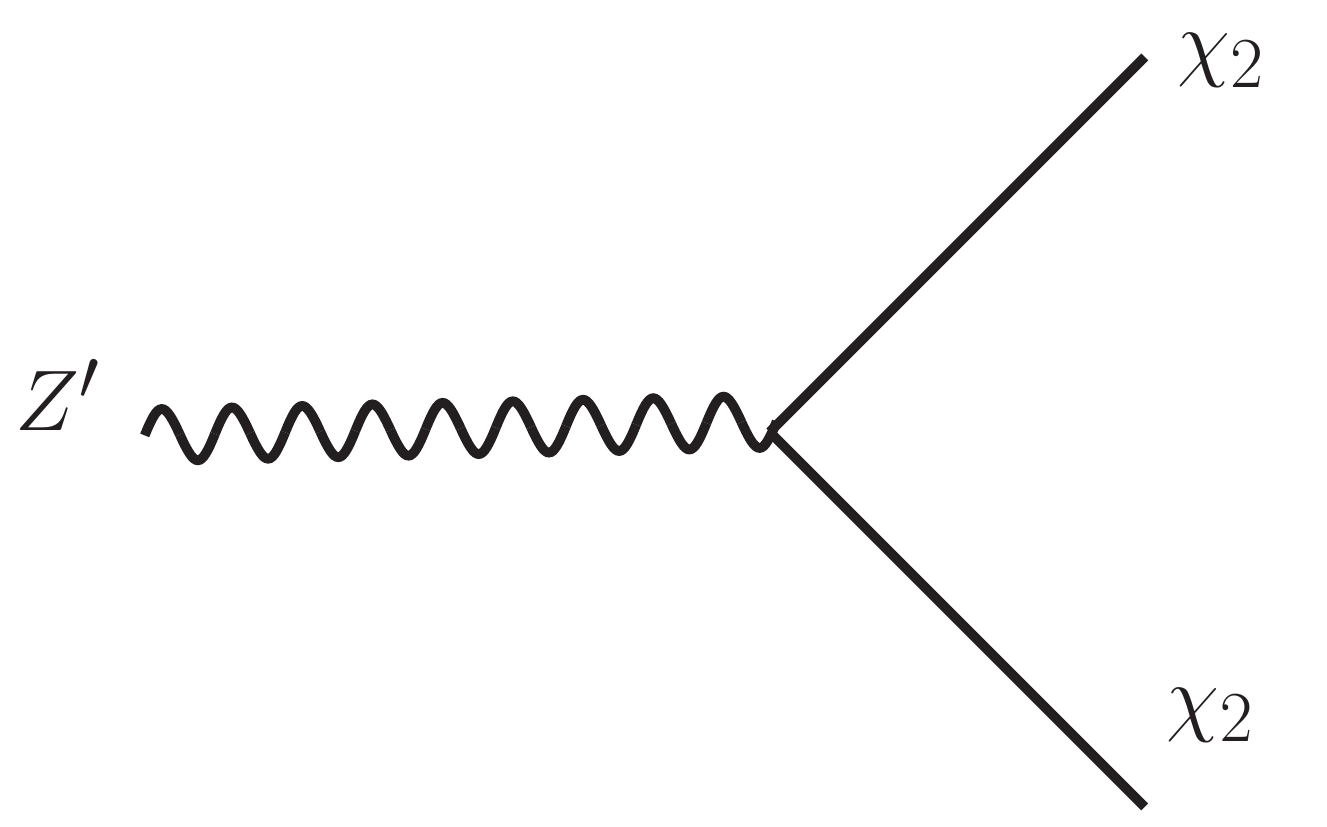}}&\\
&
$ -i g_X \gamma_\mu \gamma_5\,\,.$\\
&\\
\hline
\multirow{3}{*}{
\includegraphics[height = 1.8cm,width = 3cm]{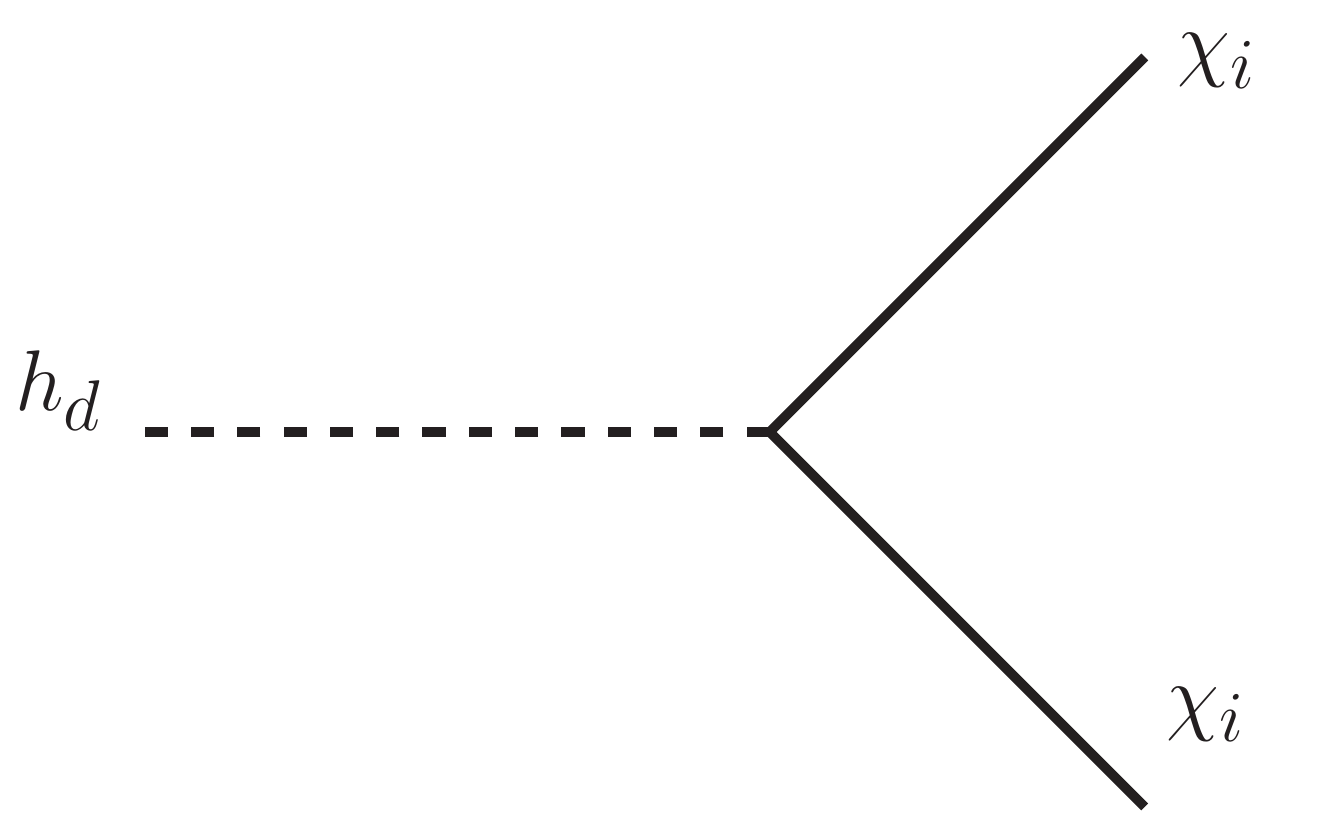}}&\\
&
$-i\cos \alpha \dfrac{m_{i}}{v_X}\,\,.$\\
&\\
\hline
\multirow{3}{*}{\includegraphics[height = 1.5cm,width = 3cm]{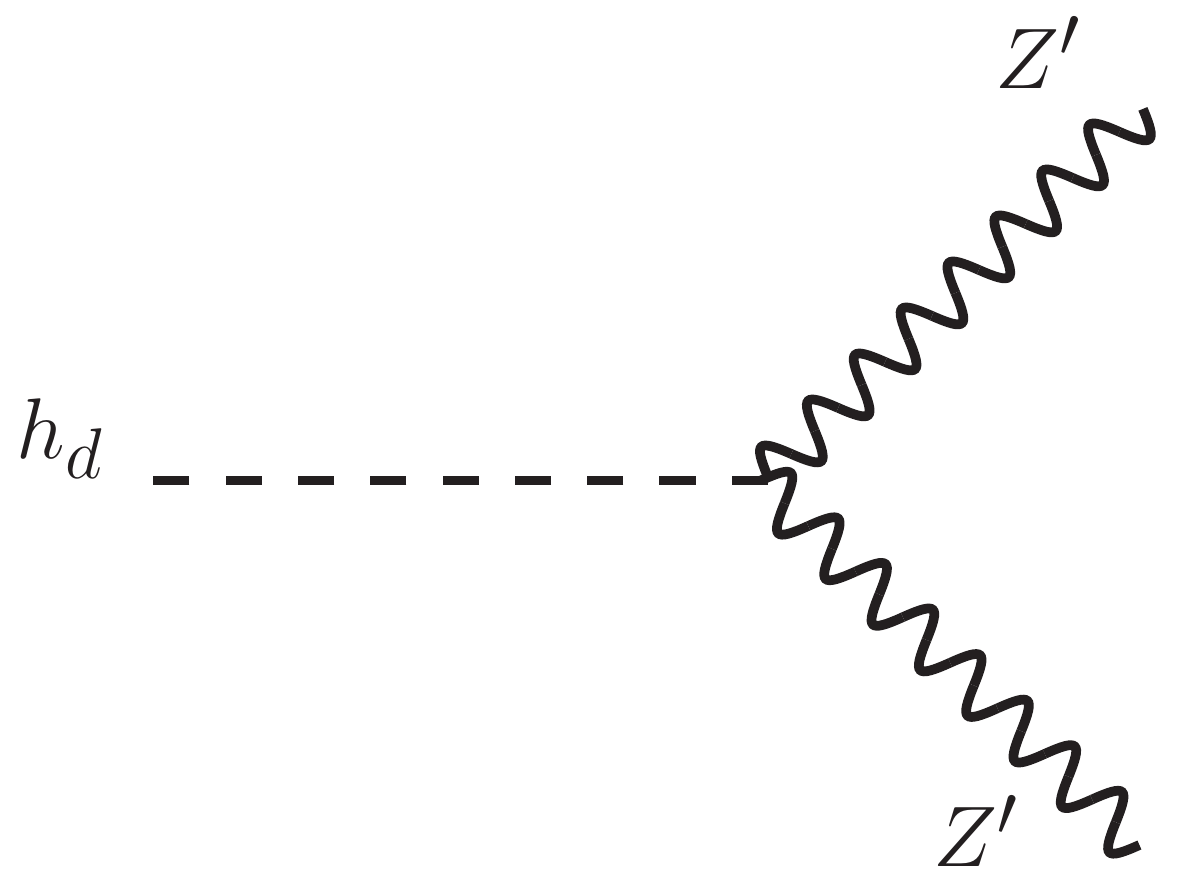}}&\\
&
$ 8 i \,\eta^{\mu \nu}v_X g_X^2\,\,.$\\
&\\
\hline
\end{tabular}
\caption{All important vertices required for computations
of $2\ra 2$, $3\ra 2$ scatterings and two-body decay widths.}
\label{Tab:vertex}
\end{table}
\end{center}
\section{Approximate analytical form of $\Omega_{\bf A}$}
\label{App:d}
In this section, we have derived an approximate analytical
expression for relic density of dark matter in co-decaying
scenario. We have considered a general scenario where the
dark sector has two degenerate species ${\bf A}$ and ${\bf B}$ as discussed
in Section\,\,\ref{sec:dynamics}. The species ${\bf A}$ is
our dark matter candidate while ${\bf B}$ can decay into the
SM particles in out of equilibrium. As mentioned in Section\,\,\ref{sec:dynamics},
$T_d$, $T_{\Gamma}$ and $T_f$ are the visible sector temperatures
corresponding to decoupling of the dark sector from the SM, beginning
of decay of the species ${\bf B}$ and the freeze-out of ${\bf A}$
respectively. The corresponding temperatures in
the dark sector are denoted by $\tp_d$, $\tp_{\Gamma}$
and $\tp_f$ with $T_d= \tp_d$. The relic density of ${\bf A}$
is defined as
\bea
\Omega_{\bf A} &=& \dfrac{m\,n^\prime_{\bf A}(T_0)}{\rho_c}, \nn \\
&=& \dfrac{m}{\rho_c} \left( \dfrac{a(T_f)}{a({T_0})} \right)^3
n^\prime_{\bf A}(\tp_f)\,, \nn \\
&=& \dfrac{m}{\rho_c} \dfrac{s(T_0)}{s(T_f)} n^\prime_{\bf A}(\tp_f)\,,
\label{OmegaA}
\eea
where, $\rho_c = 3\hub(T_0)^2/(8\pi G_N) = 3.714\times 10^{-47}$ GeV$^4$
is the critical density of the Universe and $G_N$ is the Newton's gravitational
constant. In the second step, we have used the conservation of
total number of ${\bf A}$ species per comoving volume after freeze-out
to the present era ($T_0$) while the conservation of entropy per
comoving volume between $T_f$ to $T_0$ has been utilised in the
last step. We would like to note that all the dark sector
thermodynamic variables are denoted with a prime while those
for the visible sector have no prime. Using the freeze-out
condition i.e.\,\,$n^\prime_{\bf A}(\tp_f)
\langle{\sigma {\rm v}_{\bf A A \rightarrow BB}}\rangle(\tp_f) \simeq
\hub(T_f)$ in Eq.\,\,\ref{OmegaA} we get,
\bea
\Omega_{\bf A} &\simeq& 
\dfrac{m}{\rho_c} \dfrac{s(T_0)}{s(T_f)}
\dfrac{\hub(T_f)}
{\langle{\sigma {\rm v}_{\bf A A \rightarrow BB}}\rangle(\tp_f)}, \nn \\
&\simeq&
\dfrac{s(T_0)}{\rho_c} \dfrac{\hub(m_0)}{s(m_0)}
\dfrac{m}
{\langle{\sigma {\rm v}_{\bf A A \rightarrow BB}}\rangle(\tp_f)} x_f\,,
\label{OmegaA_xf}
\eea 
where $x^{(\prime)}_f = m_0/T^{(\prime)}_f$. In order to understand
the parametric dependence of ${\Omega_{\bf A}}$ analytically, we
have neglected the variation of degrees of freedoms ($g_{\rho}$,
$g_s$) with temperature. We have taken into account these
effects while performing the numerical analyses shown in
Section\,\,\ref{sec:Nres}. Now, we need to know $x_f$
to calculate the relic density of ${\bf A}$ using Eq.\,\,\ref{OmegaA_xf}.
For that, we require the expression of $n^\prime_{\bf A}(\tp_f)$ which
is related to $n^\prime_{\bf A}(\tp_{\Gamma})$ as 
\bea
n^\prime_{\bf A}(\tp_f) a^3(T_f) &=&
n^\prime_{\bf A}(\tp_{\Gamma}) a^3(T_{\Gamma})\,
{\rm exp}\left(-\dfrac{\Gamma_{\rm B}(t_f-t_{\Gamma})}{2}\right)\,,\nn \\
&=&
n^\prime_{\bf A}(\tp_{\Gamma}) a^3(T_{\Gamma})\,
{\rm exp}\left(-\dfrac{\Gamma_{\rm B}}{4\hub(T_f)}\right)\,.
\eea
This is the unique feature of the co-decaying scenario where
only after the departure from chemical equilibrium an exponential
suppression in number density, different from the Boltzmann suppression
for a non-relativistic species, arises.    
In the last step we have used $\hub(t)\sim \dfrac{1}{2\,t}$ in the
radiation dominated era and $t_{f}>>t_{\Gamma}$. Therefore,
the number density $n^\prime_{\bf A}(\tp_f)$ is given by
\bea
n^\prime_{\bf A}(\tp_f) = 
n^\prime_{\bf A}(\tp_{\Gamma})\dfrac{s(T_f)}{s(T_{\Gamma})}\,
{\rm exp}\left(-\dfrac{\Gamma_{\rm B}}{4\hub(T_f)}\right).
\label{nAf}
\eea
The expression $n^\prime_{\bf A}(\tp_{\Gamma})$ can be found
using the second law of thermodynamics (Eqs.\,\,\ref{2ndlaw} and
\ref{sbyn_rat}) in the dark sector as
\bea
n^\prime_{\bf A}(\tp_{\Gamma}) &=&
\dfrac{\tp_{\Gamma}\,s^\prime(\tp_{\Gamma})}
{2\left(m + \frac{5}{2}\tp_{\Gamma} -\mu^\prime\right)}, \nn \\
&=& \dfrac{\tp_{\Gamma}\,\xi_{d}\,s(T_{\Gamma})}
{2\left(m + \frac{5}{2}\tp_{\Gamma} -\mu^\prime\right)}.
\label{nAgamma}
\eea
Here we have considered that the number densities and
chemical potentials of species ${\bf A}$ and ${\bf B}$
are equal at $\tp=\tp_{\Gamma}$. After $\tp_{\Gamma}$,
the species ${\bf B}$ has started to decay and thus
the number densities of ${\bf A}$ and ${\bf B}$ deviates
from each other. In the last step, we have used the
fact that entropy per comoving volume is separately
conserved in both the sectors
between $T_d$ and $T_{\Gamma}$. Moreover, $\xi_d$ is
the ratio of degrees of freedom for the dark sector
to the visible sector at the temperature $T_d$. Now,
substituting the expression of $n^\prime_{\bf A}(\tp_{\Gamma})$
in Eq.\,\,\ref{nAf} and considering the freeze-out 
for the species ${\bf A}$ once again, we get an equation for $T_f$ as
\bea
\dfrac{\tp_{\Gamma}\,\xi_{d}\,s(T_{f})}
{2\left(m + \frac{5}{2}\tp_{\Gamma} -\mu^\prime\right)}
{\rm exp}\left(-\dfrac{\Gamma_{\rm B}}{4\hub(T_f)}\right)
\simeq \dfrac{\hub(T_f)}
{\langle{\sigma {\rm v}_{\bf A A \rightarrow BB}}\rangle(\tp_f)}\,.
\eea
In terms of $x_f$, the above equation can be written as
\bea
\xi_d\dfrac{s(m_0)}{\hub(m_0)}
{\langle{\sigma {\rm v}_{\bf A A \rightarrow BB}}\rangle(x_f)}
\,{\rm exp}\left(-\dfrac{\Gamma_{\rm B}\,x^2_f}{4\hub(m_0)}\right)
\simeq 5 x_f.
\eea
It is a transcendental equation of $x_f$ which can
be solved numerically using iterative method and substituting
the exact expression of ${\langle{\sigma {\rm v}_{\bf A A \rightarrow BB}}
}\rangle$. However, an analytic solution for $x_f$ can be approximately
given if we assume that ${\langle{\sigma {\rm v}_{\bf A A \rightarrow BB}}
}\rangle$ is independent (or does not depend significantly)
of temperature. For example, the s-wave scattering at small $T$
(see the plot in left panel of Fig.\,\,\ref{Fig:sigmaVp}).
Therefore, the approximate expression of $x_f$ is
given by
\bea
x_f \simeq \sqrt{\dfrac{2\,\hub(m_0)}{\Gamma_{\rm B}}}\,
\sqrt{{W_0}\left(\dfrac{s(m_0)^2\,\Gamma_{\bf B}\,
\xi^2_d\,{\langle{\sigma {\rm v}_{\bf A A \rightarrow BB}}
}\rangle^2}{50\,\hub(m_0)^3}\right)}\,.
\label{xf}
\eea
Here $W_0$ is the principal value of Lambert $W$ function.
Finally, substituting the expression of $x_f$
in Eq.\,\,\ref{OmegaA_xf} we get an analytical
expression of relic density ($\Omega_A$)
for the co-decaying scenario as
\bea
\Omega_{\bf A} &\simeq &
\dfrac{m}
{\langle{\sigma {\rm v}_{\bf A A \rightarrow BB}}\rangle\,
\sqrt{\Gamma_{\rm B}}}
\dfrac{s(T_0)}{\rho_c}
\dfrac{\sqrt{2\,\hub(m_0)^3}}{s(m_0)}
\sqrt{W_0\left(\dfrac{s(m_0)^2\,\Gamma_{\bf B}\,
\xi^2_d\,{\langle{\sigma {\rm v}_{\bf A A \rightarrow BB}}
}\rangle^2}{50\,\hub(m_0)^3}\right)}\,.
\eea
\section{$2\ra2$ scattering cross sections}
\label{App:E}
We have listed the expressions of relevant $2\ra2$ scattering cross sections
of the dark sector species below. The corresponding Feynman diagrams are shown
in Fig.\,\ref{Fig:2to2_Feyn_dia}.

\bea
\sigma_{\x\x \ra \zp \zp}  &=&  \dfrac{-g_X^4}{8 \pi\,m_{\zp}^4
\,\ecm (\ecm-m_{\hd}^2)^2 (\ecm-4 m_\x^2)}
\left[\dfrac{\sqrt{(\ecm-4m_\x^2)(\ecm-4m_{\zp}^2)}}
{m_{\zp}^4 +m_\x^2 (\ecm-4 m_{\zp}^2)}
\Bigg[
2 m_{\zp}^8 (\ecm-m_{\hd}^2)^2
\right. \nn\\ 
&& \left.  + 8 m_{\x}^6(-48 m_{\zp}^6 + 28 m_{\zp}^4 \ecm + \ecm^3) - 
m_{\x}^2 m_{\zp}^4\bigg\{m_{\hd}^4(4m_{\zp}^2 + \ecm) + \ecm 
(8m_{\zp}^4 + 4m_{\zp}^2 \ecm-\ecm^2)
\right.\nn\\
&&\left. + 2m_{\hd}^2(8m_{\zp}^4-8m_{\zp}^2 \ecm + \ecm^2)\bigg\}
-2 m_{\x}^4\bigg\{m_{\hd}^4 (8m_{\zp}^4 -8 m_{\zp}^2 \ecm +\ecm^2)
+ 4 m_{\hd}^2 m_{\zp}^2 (-8m_{\zp}^4 + 2 m_{\zp}^2 \ecm + \ecm^2)
\right.\nn\\
&&\left.
-4 m_{\zp}^2(12 m_{\zp}^6 - 2 m_{\zp}^2 \ecm^2 + \ecm^3)
\bigg\}
\Bigg] + \dfrac{m_{\hd}^2-\ecm}{\ecm-2m_{\zp}^2}
\bigg\{
m_{\zp}^4 (m_{\hd}^2-\ecm)(\ecm^2 + 4_{\zp}^4) 
\right.\nn \\
&&\left.
- 4 m_{\x}^2 m_{\zp}^2\bigg(8 m_{\zp}^6 -8 \ecm m_{\zp}^4 -3 m_{\zp}^2 \ecm^2 + 
\ecm^3 + m_{\hd}^2(4 m_{\zp}^2-\ecm)(\ecm + m_{\zp}^2)\bigg) \right.\nn\\
&&\left.
+4 m_{\x}^4 \bigg(32 m_{\zp}^6 - 32 m_{\zp}^4 \ecm -\ecm^2(\ecm + m_{\hd}^2)
+4 m_{\zp}^2 \ecm (m_{\hd}^2 +2\ecm)\bigg) 
\bigg\}\times
\right. \nn \\ && \left.
\log\left(\dfrac{\ecm-2m_{\zp}^2 -\sqrt{(\ecm-4m_{\x}^2)(\ecm-4 m_{\zp}^2)}}
{\ecm-2m_{\zp}^2 +\sqrt{(\ecm-4m_{\x}^2)(\ecm-4 m_{\zp}^2)}}\right)
\right]\,\,.\nn\\
\label{DMDM2ZPZP}
\eea
\bea
\sigma_{\x \x \ra \hd \hd}  &=& 
\dfrac{g_X^4 m_{\x}^2}{4 \pi\,\ecm\,m_{\zp}^4 (\ecm-4m_{\x}^2)}
\left[
\Bigg(\dfrac{\sqrt{(\ecm-4m_{\x}^2)(\ecm - 4m_{\hd}^2)}}
{(\ecm-m_{\hd}^2)^2\left\{m_{\x}^2 (\ecm-4 m_{\hd}^2)+m_{\hd}^4\right\}}
\Bigg)
\bigg\{
9 s\,m_{\hd}^8 - 32 m_{\x}^6(\ecm-m_{\hd}^2)^2
\right.\nn\\
&&\left. + 3 m_{\x}^2 m_{\hd}^4(\ecm^2 -16 \ecm m_{\hd}^2 - 6 m_{\hd}^4)
+ 4 m_{\x}^4 (20 m_{\hd}^6 + 4 \ecm m_{\hd}^4 + 4 \ecm^2 m_{\hd}^2 -\ecm^3) 
\bigg\}
\right.\nn\\
&&\left.
-\left(
\dfrac{
2 m_{\x}^2
\bigg\{
18 m_{\hd}^6 + 32 m_{\x}^4 (m_{\hd}^2-\ecm) + 10 m_{\hd}^4 \ecm - 
11 m_{\hd}^2 \ecm^2 + 16 m_{\x}^2 (-5 m_{\hd}^4 + m_{\hd}^2 \ecm + \ecm^2)
\bigg\}
}{2m_{\hd}^4 - 3 m_{\hd}^2 \ecm + \ecm^2}\times
\right.\right.\nn\\
&&\left.\left.
\log\left(\dfrac{\ecm-2 m_{\hd}^2 -\sqrt{(\ecm-4m_{\hd}^2)
(\ecm - 4m_{\x}^2)}}{\ecm-2 m_{\hd}^2 +\sqrt{(\ecm-4m_{\hd}^2)
(\ecm - 4m_{\x}^2)}}\right)
\right)
\right]\,\,.
\label{DMDM2hdhd}
\eea
%

\bea
&&\sigma _{\zp \zp \ra \hd \hd} = 
\dfrac{g_X^4}
{18 \pi\,\ecm\,m_{\zp}^4 (\ecm - 4m_{\zp}^2)}
\left[
\dfrac{\sqrt{(\ecm-4m_{\hd}^2)(\ecm - 4m_{\zp}^2)}}
{(\ecm-m_{\hd}^2)^2 (m_{\hd}^4 -4 m_{\hd}^2 m_{\zp}^2 + m_{\zp}^2 \ecm)}
\bigg\{
6 m_{\hd}^{12} - 16 m_{\hd}^{10} m_{\zp}^2
\right.\nn\\
&&\left.
+ 4 \ecm^2 m_{\zp}^4 (24 m_{\zp}^4 + 3 \ecm m_{\zp}^2 + 2 \ecm^2)
+m_{\hd}^8 (32 m_{\zp}^4 - 28 \ecm m_{\zp}^2 + 3 \ecm^2) - 
4 \ecm m_{\hd}^2 m_{\zp}^4 (11 \ecm^2 + 16 \ecm m_{\zp}^2 + 48 m_{\zp}^4)
\right.\nn\\
&& \left.
-4m_{\hd}^6(64 m_{\zp}^6 + 40 \ecm m_{\zp}^4 + 7 \ecm^2 m_{\zp}^2)
+ m_{\hd}^4 (96 m_{\zp}^8 -16 \ecm m_{\zp}^6 + 60 \ecm^2 m_{\zp}^4 + 9 m_{\zp}^2 \ecm^3)
\bigg\}
\right. \nn\\
&&\left.
-4\left(
\dfrac{3 m_{\hd}^{10} + \ecm m_{\hd}^8 + 24 \ecm m_{\zp}^6 (2 m_{\zp}^2-\ecm)
m_{\hd}^6 (40 m_{\zp}^4 + 8 \ecm m_{\zp}^2 + \ecm^2) + 
2 m_{\hd}^4(64 m_{\zp}^6 - 4 \ecm m_{\zp}^4 + 5 \ecm^2 m_{\zp}^2)}
{2 m_{\hd}^4 - 3 \ecm m_{\hd}^2 + \ecm^2}
\right.\right.\nn\\
&&\left.\left.
-\dfrac{2 m_{\hd}^2 m_{\zp}^2 (24 m_{\zp}^6 + 16 \ecm m_{\zp}^4 
-6 \ecm^2 m_{\zp}^2 + \ecm^3)}
{2 m_{\hd}^4 - 3 \ecm m_{\hd}^2 + \ecm^2}
\right)
\log\left(\dfrac{\ecm-2 m_{\hd}^2 -\sqrt{(\ecm-4m_{\hd}^2)(\ecm-4 m_{\zp}^2)}}
{\ecm-2 m_{\hd}^2 +\sqrt{(\ecm-4m_{\hd}^2)(\ecm-4 m_{\zp}^2)}}\right)
\right]\,\,.
\label{zpzp2hdhd}
\eea
\section{Thermal average for degenerate initial and final state}
\label{App:F}
In the non-relativistic limit, the matrix amplitude square for
the process $\x_i (P_1) \x_i (P_2)\ra X (P_3) X (P_4)$ ($X = \zp, h_d$)
can be written as,
\bea
\overline{|\mathcal{M}|^2}
&\simeq&
a_0 + a_1 \kappa + a_2 \kappa^2 + ......\,\,.
\eea
Where, the coefficients $a_i = a_i (\delta_m, \theta, m_{\x_i})$ and
$\delta_m = \dfrac{m_{\x_i}-m_{X}}{m_{\x_i}}$, 
$\cos \theta = \dfrac{\vec{p}_1 . \vec{p}_3}{|\vec{p}_1||\vec{p}_3|}$.
In the above we have used $\kappa = \dfrac{s - 4 m_{\x_i}^2}{4 m_{\x_i}^2}<<1$
(definition of $\kappa$ is same as $\epsilon$ in \cite{Gondolo:1990dk})
for the non-relativistic regime.

Therefore the cross section of $\x_i \x_i \ra XX$ is given by
\bea
\sigma &=& \dfrac{1}{32 \pi s} \sqrt{\dfrac{s-4m_{X}^2}{s-4m_{\x_i}^2}}\int_0^\pi 
\left( a_0 + a_1 \kappa + ... \right) \sin \theta d\theta\,\,.
\eea
Now, using the definition of $\kappa$
and the magnitude of relative velocity
of the initial state particles in CM frame, $\rm v =
\dfrac{2\sqrt{\kappa}}{\sqrt{1+\kappa}}$, we can write
$\sigma \rm v$ as
\bea
{\sigma \rm v}_{\x_i \x_i \ra X X}
&\simeq&
\hat{a}_0 + \hat{a}_1 \kappa + ...... \,\,\,\, \text{for } \delta_m \neq 0\,\,\nn\\
{\sigma \rm v}_{\x_i \x_i \ra X X}
&\simeq&
\hat{a}_0|_{\delta_m = 0}\,\sqrt {\kappa}+ \hat{a}_1|_{\delta_m= 0}\,{\kappa}^{3/2} 
+ ...... \,\,\,\, \text{for } \delta_m = 0\,\,\,.
\eea
In the above, the dependence of $\kappa$
in ${\sigma \rm v}_{\x_i \x_i \ra X X}$
is different for two scenarios. This is because
the prefactor in $\sigma$ i.e. $\sqrt{\dfrac{s-4m_{X}^2}{s-4m_{\x_i}^2}}
\propto \kappa^{-1/2}$ for $\delta_m \neq 0$
whereas it becomes independent of $\kappa$ for $\delta_m = 0$.
Let us note that we have absorbed the phase space factors into $a_i$s and
define a new quantity $\hat{a}_i$. Thus following the prescription given
in \cite{Gondolo:1990dk}, we have arrived at the following
expression of the thermal average for $\x_i \x_i \ra X X$ in the
non-relativistic limit.
\bea
\langle{\sigma \rm v}_{\x_i \x_i \ra X X} \rangle
&\simeq&
\hat{a}_0 + \dfrac{3 \hat{a}_1}{2 x}  + ....\,\,\, \text{ for } \delta_m \neq 0 \,\,, \\
\langle {\sigma \rm v}_{\x_i \x_i \ra X X} \rangle 
&\simeq&
\dfrac{2 \hat{a}_0|_{\delta_m = 0}}{\sqrt{\pi} \sqrt{x}} + 
\dfrac{4 \hat{a}_1|_{\delta_m = 0}}{\sqrt{\pi} x^{3/2}}  + ....\,\,\, \text{ for } \delta_m = 0 \,\,.
\label{thavg}
\eea

Now using $x =m_{\x_i}/T \simeq 2/v^2$ where $v$
is the average DM thermal velocity, it is clearly seen that
the thermal averaged annihilation cross section
depends on either the even powers of $v$ for
$\delta_m \neq 0$ (corresponds to the away from threshold condition)
or the odd powers of $v$ for $\delta_m = 0$
(corresponds to the threshold condition).

In our model, the coefficient $\hat{a}_0$ for the process
$\x_i \x_i \ra \zp \zp$ vanishes when $\delta_m = 0$
(since $\hat{a}_0 \propto \delta_m^{3/2}$).
Therefore, $\langle{\sigma {\rm v}}_{\x_i \x_i \ra \zp \zp}\rangle$
is proportional to $v^3$ at threshold whereas it has a velocity independent
leading order term away from threshold ($\delta_m \neq 0$).
In contrast, the other annihilation channel of DM $\x_i \x_i \ra h_d h_d$
has the coefficient $\hat{a}_0 = 0$ for all values of $\delta_m$,
which implies that the thermal averaged annihilation cross section
$\langle{\sigma {\rm v}}_{\x_i \x_i \ra h_d h_d}\rangle$ 
is proportional to $v^2$ and $v^3$ for $\delta_m \neq 0$ and
$\delta_m =0$ respectively.
\newpage
\bibliographystyle{JHEP}
\bibliography{co-decay} 
\end{document}